\documentstyle[amsmath,amssymb,graphicx]{article}

\def\be{\begin{eqnarray}}
\def\ee{\end{eqnarray}}
\def\nn{\nonumber}

\def\Tr{{\rm Tr}\,}

\def\l[{\phantom.[}

\def\P{{\mathfrak{P}}}
\def\R{{\mathfrak{R}}}
\def\H{H}

\def\openC{{\mathbb{C}}}

\def\cong5{\ \stackrel{5}{\cong}\ }

\def\lar{&\longrightarrow &}
\def\Lar{&\Longrightarrow &}

\def\pr{B}
\def\pl{A}

\hoffset= - 1.0in         

\begin{document}


\hfill ITEP/TH-07/14

\bigskip


\centerline{\Large{Towards ${\cal R}$-matrix construction of Khovanov-Rozansky  polynomials
}}
\centerline{\Large{I. Primary $T$-deformation of HOMFLY}}

\bigskip

\centerline{A.Anokhina and A.Morozov}

\bigskip

\centerline{\it ITEP, Moscow, Russia}

\bigskip

\centerline{ABSTRACT}

\bigskip

{\footnotesize
We elaborate on the simple alternative
\cite{DM3}
to the matrix-factorization construction of Khovanov-Rozansky (KR)
polynomials for arbitrary knots and links
in the fundamental representation of arbitrary $SL(N)$.
Construction consists of two steps:
first, with every link diagram with $m$ vertices one associates an $m$-dimensional hypercube
with certain $q$-graded vector spaces, associated to its $2^m$ vertices.
A generating function for $q$-dimensions of these spaces is what we suggest
to call the primary $T$-deformation of HOMFLY polynomial --
because, as we demonstrate, it can be  explicitly reduced to calculations
of ordinary HOMFLY polynomials,
i.e. to manipulations with quantum $R$-matrices, what brings the story
completely inside the ordinary Chern-Simons theory.
The second step is a certain minimization of residues of this new polynomial
with respect to  $T+1$.
Minimization is ambiguous and is actually specified by the choice of commuting
cut-and-join morphisms, acting along the edges of the hypercube -- this promotes it
to Abelian quiver, and KR polynomial is a Poincare polynomial of
associated complex, just in the original Khovanov's construction at $N=2$.
This second step is still somewhat sophisticated -- though incomparably simpler
than its conventional matrix-factorization counterpart.
In this paper we concentrate on the first step, and provide just a
mnemonic treatment of the second step.
Still, this is enough to demonstrate that all
the currently known examples
of KR polynomials in the fundamental representation
can be easily reproduced in this new approach.
As additional bonus we get a simple description of the DGR relation between
KR polynomials and superpolynomials and demonstrate that the
difference between reduced and unreduced cases, which looks essential
at KR level, practically  {\it disappears} after transition to superpolynomials.
However, a careful {\it derivation} of all these results from cohomologies
of cut-and-join morphisms remains for further studies.
}

\bigskip

\bigskip

\tableofcontents

\section{Introduction}

$3d$ Chern-Simons theory \cite{CS} can be considered as straightforward generalization
of $2d$ conformal theory \cite{CFT}: conformal blocks describe its Hilbert space,
and evolution is intimately related to their monodromies and modular transformations.
Thus the study of this theory is a natural step after recent breakthrough
in instanton calculus \cite{ica1}-\cite{ical} and establishing its relation to
CFT \cite{AGT1}-\cite{AGTl} -- and it indeed attracts a lot of new attention.
The main observables in the theory are Wilson averages along knots and links,
which in the case of the simply-connected space-time are just polynomials \cite{knp}
(called HOMFLY polynomials)
of the variables $q=\exp(g_s)$ and $A=q^N$ ($N$ parameterizes the gauge group $SL(N)$
and polynomial also depends on representation of this group),
which can be interpreted as instanton sums in certain related models \cite{Witlast},
what again makes the story close to CFT/string dualities.
The study of knot polynomials is just at its infancy, but already a lot of
hidden structure is discovered.
In particular, HOMFLY allow additional "$T$-deformation" to Khovanov-Rozansky (KR)
polynomials  \cite{Witlast}-\cite{DM1},
depending on additional variable $T$ and possessing certain
cohomological interpretation -- anticipated in \cite{GV,OV} and further generalized
in \cite{GSV} and \cite{DGR}.

In \cite{DM3} a simple and practical construction was suggested to reproduce the
KR polynomials in the fundamental representation --
which can substitute the sophisticated matrix-model factorization procedure
of \cite{KhR} and make calculations for generic $N$ as simple as they are made now
for the ordinary HOMFLY and superpolynomials \cite{TR}-\cite{HOMFLYl}
and the Jones-Khovanov ($N=2$) \cite{Kho,BN,DM1} polynomials.

In fact \cite{DM3} outlines the program, but leaves many details of the
algorithm obscure (nothing to say about the proofs).
The purpose of this paper is to make a next step and provide a more
clear and constructive formulation of the procedure.
The goal of \cite{DM3} was to make the procedure for arbitrary $N$
as close as only possible to that of \cite{Kauff} and \cite{Kho} for $N=2$,
in the version of \cite{BN} and \cite{DM1}
(which we suppose the reader to be familiar with).
Accordingly there are two steps:

(i) With a link diagram ${\cal L}$ with $n_\bullet$ and $n_\circ$
black and white vertices (see \cite{DM3,DM1} for notational details)
one associates a hypercube ${\cal H}^{{\cal L}}$ with $q$-graded vector spaces
over each of its $2^{n_\bullet+n_\circ}$ vertices.
${\cal L}$ itself is associated with a particular "initial" vertex of the hypercube,
what converts edges into arrows, directed away from initial vertex.

(ii) With each arrow one associate a "morphism" -- a linear maps of degree $-1$ between
the spaces at the ends of the edge.
Morphisms commute, therefore the hypercube acquires a structure of Abelian quiver
-- and KR is the Poincare polynomial
of the associated complex.

The difference from conventional KR construction of \cite{KhR} is that
for arbitrary $N$ there are still vector spaces, not cyclic complexes,
what makes unnecessary the matrix-factorization routine and allows calculations
by modest means in reasonable time.
Moreover, the results emerge as explicit functions of $N$,
while in the matrix-factorization
approach one gets them for each particular $N$ and after that looks for an
interpolation formula, see \cite{CM} for detailed presentation and the most
extensive set of examples within this standard framework.
Comparing the {\it shapes} of our formulas with their counterparts in \cite{CM},
one can easily appreciate the power and relevance of our new approach.

Another advantage is explicit understanding of vector spaces as associated
with the cycles of the resolutions of the link diagram --
exactly as at $N=2$.
The only deviation from the $N=2$ case is that
one of the two resolutions is somewhat different and
vector spaces are rather {\it factor}-spaces
(what makes the story a part of generic non-linear algebra studies
\cite{NLA} and seems conceptually close
to the cyclic-complex viewpoint,
still technically is considerably simpler).
As at $N=2$, morphisms are associated with the cut-and-join operators,
which either glue the two cycles into one, or cut one into two
(what puts the
story into the context of matrix models and seems conceptually close to
matrix-factorization ideas, but again, is much simpler technically).

Both steps were only outlined in \cite{DM3} and additional effort is
needed to make these ideas into a universally applicable and rigorous approach.
In this paper we do this with the first step (i),
encoding the information about the graded vector spaces at the hypercube vertices
in the generating function, depending on the choice of initial vertex.
We call it the
"{\bf primary $T$-deformation $\P^{\cal L}(q,T,N)$ of HOMFLY polynomial
$\H^{\cal L}(q,N)$}".
What is important, the recent advances in HOMFLY calculus allow one to
calculate $\P^{\cal L}$ in quite a number of sufficiently representative examples.
In addition, already at this step it gets clear, what is so special for the fundamental
representation -- and this understanding can help in the search for the definition
of  the colored Khovanov-Rozansky polynomials.

As to the second step (ii), it is
still just illustrated with examples --
however, they are now less trivial and more systematic
than in the original presentation of \cite{DM3}.
As in the modern HOMFLY calculus, it looks convenient to
represent knots and links by braids, and study first the 2-strand
examples, then the 3-strand ones and so on.
It turns out that there is
a clear structure, depending on the number $m$ of strands in the braid,
and already at the level of $m=4$ this knowledge provides a drastic
simplification of calculations.

While already practical (as compared to matrix-factorization technique),
this second step remains time-consuming,
and one can think about alternative
procedures of "minimization" of the primary polynomial $\P^{\cal L}(q,T,N)$,
which was the outcome of the first step.
This hope provides to the step (i) an additional value of its own,
but we are still far from developing this hope into a well-defined
and universally applicable method.

The paper is focused on the calculational side of the story,
and the step (ii) still suffers from conceptual loopholes
and ambiguities,
but it already provides an extremely powerful technique to calculate
Khovanov-Rozansky polynomial for arbitrary {\it given} knot or link.
In particular, we immediately reproduce all
the known answers\footnote{
Agreement is with the third version of that paper, where misprints
are corrected.
} from \cite{CM},
as well as those for many torus and twist knots/links.
In our presentation, we assume whenever needed some familiarity with
the by-now-standard approaches to evaluation of link/knot polynomials,
surveyed in detail in \cite{TR}-\cite{HOMFLYl}, \cite{DM1} and \cite{DM3}.

\section{Construction of KR polynomials}

\subsection{Generalization of Kauffman  ${\cal R}$-matrix
and the "primary deformation" $\P$ of HOMFLY polynomial}

\setcounter{equation}{0}

Conceptual origin of Khovanov's approach is in the general categorification program.
However, operationally it begins from the much simpler step:
a splitting of ${\cal R}$-matrix into two pieces, see, for example, \cite{DM1}.
After that one of them is weighted with additional deformation parameter $T$,
and this provides a naive ("primary") $T$-deformation of the HOMFLY polynomial
(actually, in the case of $N=2$ it is the Jones polynomial).
Alternatively, one can consider this as a natural version of $\beta$-deformation,
which plays a big role in the contexts of Jack or Mac Donald polynomials
and of AGT relations \cite{betadefo}.
Non-deformed (HOMFLY\footnote{
Usually, HOMFLY (or, HOMFLY-PT) polynomial is defined as an expansion in powers
of $q$ and $A=q^N$
(strictly speaking,it is then indeed a polynomial only for knots and only
in reduced case).
In the present paper we consider it instead as a (Laurent) polynomial of $q$,
with $N$ defining the spectrum of exponents
-- and then it is a polynomial both for knots and links
and both in reduced and unreduced cases.
Since this is the only meaning, ascribed to the term HOMFLY polynomial in the text below,
this should not cause too much confusion.
Accordingly, positivity will be understood as the properties of the coefficients
of {\it this} expansion -- and it is not literally the same as positivity of
the coefficients of the double expansion in $q$ and $A$.
In particular, all quantum numbers and their products are positive, e.g.
$[N] = \frac{q^N-q^{-N}}{q-q^{-1}}=q^{N-1}+q^{N-3}+\ldots +q^{1-N}$,
while the corresponding difference $A-1/A$, as a (Laurent) polynomial in $A$, is not.
To restore positivity one usually makes an additional change of variables $A^2 = -a^2T$,
see \cite{DGR} and \cite{DMMSS}.
})
case is associated with $\beta=1$ or $T=-1$ -- the latter convention has
its origin in cohomological interpretation of the procedure.

Quantum ${\cal R}$-matrices are important in knot theory, because HOMFLY polynomials
can be calculated by putting ${\cal R}$ and its inverse ${\cal R}^{-1}$
at the black and white vertices of link diagram ${\cal L}$ respectively
and summing over indices with certain grading corrections (take {\it quantum} traces),
see \cite{TR,MMMkn1,Anopaths} for details of this by now standard procedure.
A possible approach to KR and superpolynomial calculus is to do just the same,
by $T$-deforming the ${\cal R}$-matrix, its inverse and quantum traces.
This idea works nicely for torus knots \cite{AgSh,DMMSS,Che}, but in a slightly
indirect and badly generalizable way (actually, not ${\cal R}$, but only
its power ${\cal R}^m$ is deformed there so that ${\cal R}^{-1}$ and ${\cal R}$
never appear simultaneously).
It works universally for all link diagrams at $N=2$,
where the $T$-deformation of ${\cal R}$-matrix  and its inverse is
provided by Kauffman's matrix \cite{Kauff}, which
-- for appropriate tracing convention --
satisfies all the three Reidemeister invariances.
The idea of \cite{DM3} was to sacrifice explicit invariances at generic $N$,
but preserve the calculus: surprisingly or not, in the final answers
the topological invariance is restored.
The purpose of this paper is to demonstrate this in rather representative
examples, but precise formulations and proofs are still lacking.

As already mentioned, the $T$-deformation of ${\cal R}$-matrix is dictated
by its decomposition into two items.
While for $N=2$ the relevant splitting of the ${\cal R}$-matrix {\it in the
fundamental representation} is well established \cite{Kauff},
its direct analogue for arbitrary $N$ is not so familiar.
However, it is equally simple:
\be
{\cal R} = q^{N-1}\Big(I\otimes I-q\cdot [2]P_{11}\Big),\nn \\
{\cal R}^{-1} = q^{1-N} \Big(I\otimes I- \frac{1}{q}[2]P_{11}\Big) =
-\frac{1}{q^N}\Big([2]P_{11}-q\cdot I\otimes I\Big)
\label{genR}
\ee
where
\be
P_{11} = \frac{\frac{1}{q}\cdot I\otimes I - \frac{1}{q^N}{\cal R}}{[2]}
\ee
with the property $P_{[11]}^2=P_{[11]}$
is the projector on the $[11]$ (antisymmetric)
constituent of representation product of two fundamental representations
$[1]\otimes [1] = [2] + [11]$.
This ${\cal R}$ the standard ${\cal R}$-matrix\footnote{
In the standard notation from \cite{MMMkn1} it would be rather
$$
\frac{1}{A}{\cal R} = q^{-N}\Big(q - (q-{\cal R})\Big) =
q^{1-N}\Big(I\otimes I-\frac{1}{q}[2]P_{11}\Big)
$$
where ${\cal R}$ has eigenvalues $q$ and $-q^{-1}$, corresponding to representations
$[2]$ and $[11]$ respectively, and projector $P_{11} = \frac{q-R}{q+q^{-1}}$.
It satisfies $P_{11}^2=P_{11}$ because of the skein (Hecke) relation
${\cal R} - {\cal R}^{-1} = q-q^{-1}$.
In the present text we absorb $A=q^N$ into ${\cal R}$ and
change $q\longrightarrow q^{-1}$ to match the notation of \cite{DM3}
-- and, actually, to simplify the formulas for KR polynomials.
\label{footq}}
in the fundamental representation $[1]=\Box$, which satisfies the skein relation
$q^{-N}{\cal R}- q^N{\cal R}^{-1} = -q+q^{-1}$.
Eqs.(\ref{genR}) decompose both ${\cal R}$ and its inverse ${\cal R}^{-1}$ into linear combinations of
two operators -- unity $I\otimes I$ and $[2]P_{11}$, and this gives rise to representation of
HOMFLY polynomials as a sum over vertices of the hypercube with selected initial vertex\footnote{
Note that the factor $[2]$ in front of the projector is independent of $N$
-- it is rather the $q$-deformation of $2$ in the relation $(I-2P)^2=I$ for $P^2=P$
between projectors and the roots of unity.}
-- exactly as described in \cite{BN,DM1} and \cite{DM3}.

Following \cite{DM3}, we define the $T$-deformation as follows:
\be
{\cal R} \ \longrightarrow\  & q^{N-1} \Big( I\otimes I + (qT)\cdot [2]P_{11} \Big)
& =\ -T\cdot{\cal R}  \ +\ q^{N-1}(1+T)\cdot I\otimes I
\nn \\
{\cal R}^{-1} \ \longrightarrow \ &
\frac{1}{q^{N}T} \Big( [2]P_{11} + (qT) \cdot I\otimes I \Big)
& =\ (-T\cdot{\cal R})^{-1} \ + \ q^{1-N}(1+T^{-1})\cdot I\otimes I
\label{Rdef}
\ee
When $T=-1$,  we return to (\ref{genR}), but for $T\neq -1$ all the three
Reidemeister properties are lost: these quantities do not satisfy Yang-Baxter
equations and are {\it not} even inverse of each other.
Still, for the link diagram ${\cal L}$
{\bf calculate the polynomial $\P_{_\Box}^{\cal L}(q,T,N)$ with the so deformed expressions
by the above-described usual method of \cite{TR,MMMkn1,Anopaths}}.
Then decompose this {\it primary $T$-deformation of HOMFLY} as follows:
\be
\P(q,T,N) = {\cal P}(q,T,N)  \ +\  (T+1)\cdot {\cal Q}(q,T,N)
\label{Pdeco}
\ee
so that {\it both} the ${\cal P}$ and ${\cal Q}$ are {\it positive polynomials}
i.e. all the coefficients of expansion in powers of $q$ and $T$ --
after quantum numbers are expanded in powers of $q$ --
are {\it positive} integers.

\subsection{Primary polynomial as a combination of HOMFLY's for
sub-diagrams of ${\cal L}$}

Of course, modulo $T+1$, the primary deformation $\P$ is just a power of $T$ times the ordinary
HOMFLY polynomial, a function of $q$ and $N$ only:
\be
\P^{\cal L}_\Box(q,T,N) =
(-T)^{n_\bullet-n_\circ}\cdot H^{\cal L}_\Box(q,N)
+ \sum_{i=0}^{n_\bullet}\sum_{j=0}^{n_\circ}(T+1)^{i+j}T^{n_\bullet-n_\circ-i} \cdot
Q_{ij}^{{\cal L}}(q,N)
\label{expinH}
\ee
where $Q_{ij}$ are just the fundamental HOMFLY polynomials,
for the link diagrams  with some $i$ black  and $j$ white vertices
of original ${\cal L}$ resolved in a trivial way.
Original HOMFLY of ${\cal L}$ is actually the $i=j=0$ term of this sum.

But this does {\it not} mean that ${\cal P}$ coincides with $H$.
 If
\be
\P = \sum_{i\ge0} a_i(q,N) T^i
\ee
then the simplest solution to (\ref{Pdeco}) could seem to be just
the alternated sum of the coefficients $a_i$, i.e. the ordinary
$T$-independent HOMFLY polynomial
\be
H(q,N) = \sum_{i\ge0} (-)^ia_i
\label{altsum}
\ee
Indeed, the difference is divisible by $(T+1)$:
\be
\P-H = (1+T)\left(\sum_{j\ge0} T^j \cdot \sum_{i\ge1} (-)^i a_{i+j}(q,N)\right)
\label{PTdiff}\ee
This indeed provides a solution to (\ref{Pdeco}),
so that KR polynomial does not depend on $T$ and just
coincides with HOMFLY, ${\cal P} = H$, only if the whole set of
positivity conditions is satisfied:
\be
{\cal P} = H \ \ \ \ \ \Leftrightarrow \ \ \ \ \ \ \
\forall j\geq 0\ \ \ \ \ \
\sum_{i\ge1} (-)^i a_{i+j}(q,N) \ > 0 \ \ \ ({\rm as\ polynomials\ in}\ q)
\label{altsums}
\ee
i.e. all the coefficients in (\ref{PTdiff}) are positive integers.
Such examples exist (e.g., unknot), but usually this is not so.
We emphasize that HOMFLY itself {\it can} be positive
(Hopf link is the simplest example), but the would be $Q$ is usually not --
and this makes the problem far less trivial.

\subsection{A concept of KR polynomial}

\setcounter{equation}{0}

Informally, KR polynomial is the "minimal possible" ${\cal P}$.
The need for minimization is that
deformed matrices in (\ref{Rdef}) are no longer inverse of each other and do not
satisfy Yang-Baxter relations,
therefore $\P(q,T,N)$ is not Reidemeister invariant and
thus is not really a link/knot polynomial.
However, minimization -- if uniquely defined -- could cure this problem
and provide a topologically invariant ${\cal P}$.

The problem is that it is not so simple to define an unambiguous minimization --
the simplest example below of such ambiguity will be
the unreduced polynomial for the Hopf link, see sec.\ref{altHopf}.
At present as a substitute (or, perhaps, a healthy realization) of the minimization procedure
one needs to apply a rather involved cohomological construction
to fix the ambiguities -- in a way, which is, perhaps, artificial.
As reviewed in \cite{DM1,DM3} in the well understood case of $N=2$,
this involves reinterpretation of primary polynomial
as a generating function of $q$-dimensions of graded vector spaces at the
resolution-hypercube vertices and then evaluation of cohomologies of the differentials,
made from the cut-and-join morphisms along the hypercube edges.
Graded spaces and morphisms can be constructed explicitly and the problem
is essentially that of the elementary linear algebra.
This is what we referred to as the "second step" (ii) in the introduction.
Since in the approach of \cite{DM3} the construction of morphisms is still not fully specified,
below we continue to provide examples -- which hopefully will help to understand
a rigorous definition, to be further transformed into a computer algorithm,
like it was already done at $N=2$.

\subsection{The task of this paper}

The main result of this paper is the construction of the primary polynomial
$\P^{\cal L}_\Box$ --  it is now fully clear (it was not quite in \cite{DM3}).

It is also clear, what is so special about the fundamental representation
(when necessary we denote it by $ _\Box$ subscript, but omit this symbol in most other formulas):
only in this case the product $\Box\otimes\Box = [2]+[11]$ contains just {\it two}
irreducible representations -- and one can naturally split ${\cal R}$-matrix
in two constituents, which provide two types of resolutions at every vertex
of the knot/link diagram ${\cal L}$.
This is what gives rise to the hypercube of resolutions.
The same logic, applied to higher representations would require
a generalization, e.g. in the spirit of gerbe theory.

Besides listing primary polynomials in many examples, we demonstrate, how
the step (ii) works.
Actually, in most cases we explicitly describe not the {\it morphisms}, but only
the {\it differentials} whose cohomologies provide the KR polynomials.
We demonstrate that these are quite easy to build, once the primary polynomial
is known, and ambiguity is in fact rather low.
Moreover, ambiguity can actually be eliminated if we consider a whole set of link
diagrams, for example, made from the braids with a given number of strands $m$ --
this is in fact very close conceptually to the {\it evolution method} of
\cite{DMMSS} and \cite{evo}.
The basic reason for this is of course that differentials are made from
the underlying morphisms, which are in turn prescribed by the action of
cut-and-join operators.

However, for generic $N$ morphisms are in fact a little less simple than
in the $N=2$ case:
they are associated with the action of cut-and-join operators on the factor
spaces and are therefore sensitive to entire path to the given edge from
the initial vertex of the hypercube (while they were fully ``localized'' on
the given edge in the $N=2$ case).
Construction and understanding of morphisms is of course crucial for
putting the approach of \cite{DM3} on solid ground and for complete proof
of its topological (Reidemeister) invariance.
We continue to fill these conceptual gaps in the subsequent publications.

However, this paper already provides a rather reliable {\it practical method}
to calculate KR polynomials and superpolynomials for arbitrary knots and links
with a given number $m$ of strands -- we demonstrate this explicitly
for many {\it representative} examples at $m=2,3,4$, and generalization
to each particular higher $m$ seems straightforward.
Of course, as a byproduct we provide many {\it examples} of topological invariance,
since we encounter the same knot and link in several different braid representations.
Thus, even in the absence of exhaustive theory, one has a practical method
to calculate superpolynomials, which was not available so far.
This also opens a new way to think about colored superpolynomials --
though no immediate idea to define and/or construct them follows from
our consideration
(see \cite{Dan} for a related comment on the base of  \cite{Wu}).

\section{Plan of the paper}

\setcounter{equation}{0}

The rest of the paper consist of examples:
for different kinds of link diagrams ${\cal L}$ we provide
the primary polynomials $\P^{\cal L}$ and discuss both their (often ambiguous)
minimization and the chain \
\be
{\rm hypercube}\ \longrightarrow \ {\rm morphisms}\ {\it a\ la}\
\cite{DM3}\ \longrightarrow \ {\rm Abelian\ quiver} \ \longrightarrow \
{\rm KR\ complex}\ \longrightarrow \ {\rm its\ Poincare/KR\ polynomial}
\nn
\ee

Examples cover all the known KR polynomials: they were found for knots and links
with up to six intersections in \cite{CM}, what generalizes some previous
calculations in \cite{RJ} and \cite{GIKV}.
We use also important examples from \cite{DGR}.
Our formulas are valid for arbitrary $N$, while some of the results in \cite{CM}
were found only for $N\leq 4$ or even $N\leq 3$.

Formulas for particular knots/links from this list can be located in the present paper
with the help of the following table:

\newpage

\centerline{{\bf The list of simplest prime (non-composite)
knots and links}}

\bigskip

\bigskip

\centerline{
$
\begin{array}{|c|l|c|c||c|}
\hline
\multicolumn{2}{|c|}{}&&&\\
\multicolumn{2}{|c|}{\mbox{ Name}}&\mbox{Braid word}&\mbox{Orientation}&\mbox{Our\ answer}\\
\multicolumn{2}{|c|}{}&&&\\[-3mm]
\cline{1-2}
&&&&\\[-3mm]
\cite{katlas} &\cite{CM}&&&\\
&&&&\\[-3mm]
\hline
L2a1& 2_1^2\!\!&\{1, 1\} \ {\rm or}\ \{-1, -1\}&\{0\}\&\{1\}
& (\ref{KRr22})\&(\ref{KRur22}) \\
& {\rm Hopf}&&&\\
\hline
&3_1
&\{1, 1, 1\}\ {\rm or}\ \{1, 2, 1, 2\}& & (\ref{KR23})\&(\ref{KR[23]ur})\& \\
& {\rm Trefoil}
&& & (\ref{KR32r})\&(\ref{KR32})\phantom{\&} \\
\hline
&4_1&\{1, -2, 1, -2\}& & (\ref{KR41r}) \& (\ref{KR41ur}) \\
&{\rm Figure-eight}&& & \\
\hline
L4a1&4_1^2\ v1&
\{1,2,-1,2,1\}  \ {\rm or}&\{0\} & (\ref{KR421v1})  \\
&&
\{1, -2, 3, -2, -1, -2, -3, -2\}&&\\
\cline{2-5}
&\phantom{4_1^2}\ v2&\{1, 1, 1, 1\}&\{1\}
& (\ref{KRo24})\&(\ref{KR421})\\
&{\rm Torus}\ [2,4]
&&&\&(\ref{KR421v2ur})\\
\hline
&5_1&\{1, 1, 1, 1, 1\}& & (\ref{KR51}) \\
& {\rm Fivefoil}&\{1, 1, 1, 1, 1\}& & \\
\hline
&5_2&\{1, 1, 1, 2, -1, 2\}& &(\ref{KR52}) \\
\hline
L5a1&5_1^2
&\{-1, 2, -1, 2, -1\}&\{0\} & (\ref{KR521r})\& (\ref{KR521ur})\\
&{\rm Whitehead}
&&&\\
\hline
&6_1&\{1, 1, 2, -1, -3, 2, -3\}& & (\ref{KR61r})\&(\ref{KR61ur})\\
\hline
&6_2&\{1, 1, 1, -2, 1, -2\}& & (\ref{KR62r})\&(\ref{KR62ur}) \\
\hline
&6_3&\{1, 1, -2, 1, -2, -2\}& & (\ref{KR63r})\&(\ref{KR63ur})\\
\hline
L6a3& 6_1^2\ v1&\{1, 1, 1, 1, 1, 1\}&\{0\} & (\ref{KR[26]}) \\
&{\rm Torus}\ [2,6]&&&\\
\cline{2-5}
&\phantom{6_1^2}\ v2&\{1, -2, 3, -4, 5, 4, 3, 2, -1, 3, -4, 3, 2, 3, -5, 4, 3, -2\}&\{1\}
& (\ref{KR621v2r})\&(\ref{KR621v2ur})\\
\hline
L6a5&6_1^3\ v1&\{1, -2, -3, -2, 4, 3, -2, -1, -2, -3, -2, -4, 3, -2\}&\{0,0\}
& (\ref{KR631v1r})\&(\ref{KR631v1ur})\\
\cline{2-5}
&\phantom{6_1^3}\ v2&\{1,1, -2, 1, 1, -2\}&\{0,1\},\{1,0\} & (\ref{KR631r})\&(\ref{KR631ur})\\
 \hline
L6a2&6_2^2& \{1,1,1,1,-2,-2,1,1,2\}& \{0\},\{1\}&(\ref{KR622r})\&(\ref{KR622ur}) \\
 \hline
L6a4&6_2^3
&\{1, -2, 1, -2, 1, -2\}&\{0,0\},\{1,1\} &
(\ref{KRborr_r})\&(\ref{KRborr_ur}) \\
&
{\rm Borromean\ rings} &&&\\
 \hline
L6a1&6_3^2\ v1&\{1, -2, 3, -2, 1, -2, -3, -2\}&\{0\}
& (\ref{KR623v1r1})\&(\ref{KR623v1ur1}) \\
\cline{2-5}
&\phantom{6_3^2}\ v2& \{1, 2, 3, 2, 2, -1, 2, 2, -3, 2\}&\{1\}\ & (\ref{KR623v2r})\&(\ref{KR623v2ur})\\
\hline
L6n1&6_3^3\,v2&\{1, -2, 1, 2, -1, 2\}&\{0,0\} & (\ref{KR633v2})\\
 \cline{2-5}
&\phantom{6_3^3\,}v1&\{1, 2, 1, 2, 1, 2\}&\{1,0\}
& (\ref{KR33true})\&(\ref{KR33true1})\\
&{\rm Torus}\ [3,3] &&
&\\
 \hline
\end{array}
$
}

\bigskip

\bigskip

Fundamental {\it superpolynomials} \cite{GSV,DGR,diff} are also known for a number of more
complicated examples, including all torus   \cite{AgSh,DMMSS,Che,Gor}
and twisted \cite{IMMMfe,evo,AENV} knots/links.
Relation between superpolynomials and KR polynomials can seem somewhat
subtle \cite{DGR}: the former depend on additional parameter $A$ (and the claim
is that they can be 3-graded \cite{GSV} and, perhaps, even 4-graded \cite{GGS,diff}
generalizations of the 2-graded KR polynomials),
while KR polynomials depend on the group size $N$ (for $SL(N)$) through
the quantum numbers $[N-p]$, and powers $q^N$.
Still, at least in simple examples (going far beyond the "thin" knots),
relation {\it for generic $N$} is straightforward -- through a change of variables
\be
\boxed{
q^N=a, \ \ \ \ \ \ \l[N-p\,] = \frac{q^{2p}+a^2T}{q^{p-1}(1-q^2)a}
}
\label{KRtosup0}
\ee
(note that our $q$ and $a$ here are ${\bf q}$ and ${\bf a}$ from \cite{DMMSS}).
In this sense evaluation of KR polynomials in these examples is equivalent
to evaluation of superpolynomials.
Of course,
additional simplifications in true KR polynomials arise when $N-p=0$.

Since our approach is based on the use of the primary $T$-deformation
(presumably, an avatar of generic $\beta$-deformation \cite{betadefo})
of HOMFLY, the complexity of our calculations naturally depends on the number $m$
of strands -- and in this paper we do not go beyond $m=2,3,4$
(this is enough to handle most of the examples from \cite{CM},
and the two exceptions can be studied by alternative means).

Arbitrary two-strand knots and links (they are all torus)
are analyzed in sec.\ref{2stra}, eqs.(\ref{P2odd}) and (\ref{P2even}).
For three-strand torus knots and links, see secs.\ref{3stratorus} and \ref{3stratorus2}.

Already at the level of $m=3$ strands, we encounter a lot of composite knots and links --
and we demonstrate that their reduced KR polynomials are nicely factorized
(as HOMFLY did), while unreduced KR polynomials do not seem to
decompose in any simple way.
For some examples of composites, see sec.\ref{compos0}.

We also consider in sec.\ref{sec942} the celebrated example of the knot $9_{42}$
\be
9_{42}: \ \ \ \
(1,1,1,-2,-1,-1,3,-2,3)
\nn
\ee
and demonstrate how the non-minimal KR polynomials  (\ref{KR942r}) \& (\ref{KR942ur})
emerge in this case.

Note that in \cite{katlas} the HOMFLY polynomials for links are currently
provided only for one of a few possible orientations --
for other orientations see \cite{indiana} and \cite{CM}.

\bigskip

Our calculations get really simple and concise in sec.\ref{cofo},
where we begin directly at the level of KR complex, omitting construction
of the hypercube.
However, we provide before that  a detailed and slow hypercube-based description
in secs.\ref{2stra}-\ref{4stra}, because it establishes clear links to original
Khovanov formalism and is also needed to construct and study the morphisms
along lines of \cite{DM3}.

\section{The 2-strand examples
\label{2stra}}

\setcounter{equation}{0}

All the 2-strand knots and links are torus, and therefore they are labeled
as $[2,n]$. For odd $n=2k+1$ we get knots, for even $n=2k$ -- two-component links.
Exhaustive description of the corresponding superpolynomials
(which in the 2-strand case coincide with KR polynomials) can be found, for example,
in \cite{AgSh,DMMSS,Che}.

\subsection{Unknot $[2,1]$}

\subsubsection{Primary deformation $\P^{[2,1]}$ of HOMFLY polynomial $H^{[2,1]}$}

Making use of the definitions of quantum dimensions and traces,
\be
\Tr_{[1]\otimes[1]}\ I\otimes I = [N]^2\nn \\
\l[2]\Tr_{[1]\otimes [1]}\ P_{11} = [N][N-1]
\ee
with $[N] = \frac{q^N-q^{-N}}{q-q^{-1}}$,
we get from the prescription (\ref{Rdef}) the following {\it primary}
$T$-deformation of
HOMFLY polynomial $H^{[2,1]} = \Tr_{[1]\otimes [1]}\ {\cal R}$:
\be
\P^{[2,1]} = \Tr_{[1]\otimes [1]}\   q^{N-1} \Big( I\otimes I + (qT)[2]P_{11} \Big)
= q^{N-1}\Big( [N]^2 + qT\, [N][N-1]\Big)
\label{P21}
\ee
This deformation, however, is not topological invariant and does not
really deserve a name of knot polynomial -- the second step (\ref{Pdeco})
is still needed to get a reasonable quantity: the KR polynomial ${\cal P}$.

Still, before proceeding to this step, let us  note that
alternatively the same $P^{[2,1]}$ in (\ref{P21})
can be obtained from expansion like (\ref{expinH}):
\be
\P^{[2,1]} = \Tr_{[1]\otimes [1]} \Big(-q^{N}T{\cal R}  + q^{N-1}(1+T)\cdot I\otimes I\Big)
= -T\, H^{[2,1]} + (1+T)\,q^{N-1} H^{[2,0]} = \nn \\ =
-T\underbrace{q^N \left(q^{-1}\frac{[N][N+1]}{[2]}-q\frac{[N][N-1]}{[2]}\right)}_{[N]} +
(1+T)\,q^{N-1}[N]^2 = \boxed{-T\,[N]} + (1+T)\, q^{N-1}[N]^2
= \nn \\
= q^{N-1}[N]^2 + T\,q^N [N][N-1] \equiv [N]\cdot\P^{[2,1]}_r
\ee
Boxed is the contribution of the HOMFLY polynomial $H^{[2,1]}$,
and $P^{[2,1]}$ differs from it by a contribution, proportional to $(1+T)$.
Note, however, that in this particular case the boxed polynomial is {\it not}
positive -- therefore it can {\it not} play the role of KR polynomial.

\subsubsection{The basic morphism and reduced KR polynomial}

The primary $T$-deformation
\be
\P^{[2,1]}_r  = q^{N-1}[N] + T\,q^N [N-1]
\label{P21or}
\ee
is a combination of two quantum dimensions, $[N]$ and $[N-1]$  -- of the two graded
vector spaces $\openC^N$ and $\openC^{N-1}$, standing at the two points of the hypercube,
which in this case is just a segment with a single edge.
As a simplification and a minor abuse of notation,
in what follows we denote the vector spaces by their
dimensions $\openC^{N-k} \equiv [N-k]$ and deviate from this rule only when this can cause a
confusion.

\begin{picture}(100,90)(-170,-35)
\put(0,0){\line(1,0){107}}
\put(0,0){\circle*{6}}
\put(110,0){\circle{6}}
\put(-22,26){\mbox{$\openC^N = [N]$}}
\put(92,26){\mbox{$\openC^{N-1} = [N-1]$}}
\put(52,35){\mbox{$\alpha$}}
\put(33,30){\vector(1,0){44}}
\put(77,25){\vector(-1,0){44}}
\put(52,13){\mbox{$\beta$}}
\put(28,-20){\mbox{{\rm HCube}$([2,1])_r$}}
\end{picture}

In order to define the KR polynomial we need a morphism between these two spaces,
associated with the edge of the hypercube.
This morphism should decrease grading by one, and it is quite obvious, what it is:
\be
\openC^N \ \stackrel{\alpha}{\longrightarrow}\ \openC^{N-1}: \ \ \ \ \ \
\left\{
\begin{array}{ccc}
\alpha\left(q^{N+1-2k}\right) = q^{N-2k} && {\rm for}\ 1\leq k \leq N-1\\    \\
\alpha\left(q^{1-N}\right) = 0
\end{array}
\right.
\label{alphadef}
\ee
Since $[N]=q[N-1] + q^{1-N}$,
this morphism has non-trivial kernel, but a trivial coimage:
\be
\begin{array}{c}
{\rm Ker} \Big(\alpha_{[N]\longrightarrow [N-1]}\Big) = q^{1-N} ,\\ \\
{\rm CoIm} \Big(\alpha_{[N]\longrightarrow [N-1]}\Big) = \emptyset
\end{array}
\ee
Since for a one-dimensional hypercube there is just one differential and
it coincides with the morphism, $d_0=\alpha$,
these formulas imply that the KR polynomial is
\be
{\cal P}^{[2,1]}_r = q^{N-1} \left\{ {\rm dim}_q
{\rm Ker} \Big(\alpha_{[N]\longrightarrow [N-1]}\Big)
+ (qT)\cdot  {\rm dim}_q
{\rm CoIm} \Big(\alpha_{[N]\longrightarrow [N-1]}\Big)
\right\} = 1 = {\cal P}^{{\rm unknot}}_r
\ee
as it should be for the unknot.

In fact, this is the reduced polynomial, defined for $\bullet$ as initial vertex.
If we consider instead $\circ$ as initial vertex, then we need another morphism, $\beta$
acting in the opposite direction, but still decreasing grading by one.
In this case it is equally obvious, what it is:
\be
\openC^{N-1} \ \stackrel{\beta}{\longrightarrow}\ \openC^{N}: \ \ \ \ \ \
\beta\left(q^{N-2k}\right) = q^{N-2k-1} \ \ \ \  {\rm for}\ 1\leq k \leq N-1
\label{betadef}
\ee
This time it has vanishing kernel, but non-vanishing coimage $q^{N-1}\!\in\! [N]$,
so that the corresponding KR polynomial is
\be
{\cal P}^{[2,1]}_r = \frac{1}{q^NT} \left\{ {\rm dim}_q
{\rm Ker} \Big(\beta_{[N-1]\longrightarrow [N]}\Big)
+ (qT)\cdot  {\rm dim}_q
{\rm CoIm} \Big(\beta_{[N-1]\longrightarrow [N]}\Big)
\right\} = \nn \\
= \frac{1}{q^NT}\Big(0+(qT)\cdot q^{N-1}\Big)= 1 = {\cal P}^{{\rm unknot}}_r
\label{KR21}
\ee
and again reproduces the right answer for the unknot.

\subsubsection{Unreduced KR polynomial}

In this case we need morphisms of degree $-1$ between the bigger vector spaces
$\openC^N \otimes \openC^N = [N]^2$ and $\openC^N \otimes \openC^{N-1} = [N][N-1]$.
In fact, for
$\openC^N \otimes \openC^N = [N]^2 \ \longrightarrow \ \openC^N \otimes \openC^{N-1} = [N][N-1]$
we now have two obvious choices $\alpha_1=\alpha\otimes I$ and $\alpha_2=I\otimes \alpha$.
However, both choices produce the same answer: for $\bullet$ as initial vertex
\be
{\cal P}^{[2,1]} = q^{N-1}
\left\{ {\rm dim}_q
{\rm Ker} \Big(\alpha_{[N]\longrightarrow [N-1]}\Big)\cdot[N]
+ (qT)\cdot  {\rm dim}_q
{\rm CoIm} \Big(\alpha_{[N]\longrightarrow [N-1]}\Big)\cdot[N]
\right\} =\nn \\
= q^{N-1}\left\{ [N]\cdot{\rm dim}_q
{\rm Ker} \Big(\alpha_{[N]\longrightarrow [N-1]}\Big)
+ (qT)\cdot  [N]\cdot {\rm dim}_q
{\rm CoIm} \Big(\alpha_{[N]\longrightarrow [N-1]}\Big)
\right\} = [N] = {\cal P}^{{\rm unknot}}
\ee

However, for the morphism in the opposite direction
$\openC^N \otimes \openC^{N-1} = [N][N-1] \ \longrightarrow \ \openC^N \otimes \openC^N = [N]^2$
we have just a single options $I\otimes \beta$, providing for KR polynomial with initial vertex
$\circ$
\be
{\cal P}^{[2,1]} = \frac{1}{q^NT} \left\{ [N]\cdot {\rm dim}_q
{\rm Ker} \Big(\beta_{[N-1]\longrightarrow [N]}\Big)
+ (qT)\cdot[N]\cdot  {\rm dim}_q
{\rm CoIm} \Big(\beta_{[N-1]\longrightarrow [N]}\Big)
\right\} =  [N] = {\cal P}^{{\rm unknot}}
\ee

Note that
\be
{\cal Q}^{[2,1]} = q^N[N][N-1]
\ee
is also a positive polynomial
and -- in this particular case -- neither ${\cal P}$ nor ${\cal Q}$ depends on $T$.

Moreover, in this case
\be
{\cal P}^{[2,1]}= [N]\cdot {\cal P}_{r}^{[2,1]}
\ee
though in general  $\P$ and $\P_{r} \equiv [N]^{-1}\cdot \P\ $
can provide KR polynomials, which are not related in such a simple way
(though reduced and unreduced HOMFLY always are).

\subsection{The main rules of $T$-reduction}

Conversion of the primary polynomial (\ref{P21or})
into KR polynomial (\ref{KR21}) can be considered
as certain {\it reduction} of its dependence on $T$.

The elementary $T$-reduction rule
\be
\boxed{
\l[N] + (qT)\cdot [N-1] =q^{1-N} + (1+T)\cdot q[N-1] \sim q^{1-N}
}
\label{mainredrule1}
\ee
together with another one,
\be
\boxed{
\l[N-1] +(qT)\cdot q[N-1] = q^{2-N}+q^NT + (1+T)\cdot q[N-2] \sim q^{2-N}+q^{N}T
}
\label{mainredrule2}
\ee
will play the central role in all our calculations.

For unreduced polynomials there will be an additional
$T$-reduction rule -- to {\bf reduce
factor $[N]$ rather than $[N-1]$} whenever the product $[N][N-1]$ appears,
we will see this for the first time in eq.(\ref{congNzero}) below.

Note that we need to use the term "reduction" in two different meanings:
for $T$-reduction and for elimination of an extra $[N]$ factor in the
primary polynomial -- and these two reductions do not commute:
$T$-reduction of reduced polynomial in general leads to
"reduced KR polynomials" -- something
rather different from "unreduced KR polynomial",
the $T$-reduction of unreduced polynomial, which usually is just not
divisible by  $[N]$ (it is rather a linear combination of
$[N]$ and $(1+q^{2N}T)$, which also gets proportional to $[N]$ when $T=-1$).
In what follows we often omit $T$- when referring to $T$-reduction
and hope that this abuse of terminology (like "reduction of reduced polynomial")
will not cause additional misunderstanding.

\subsection{Hopf link}

\subsubsection{Primarily deformed HOMFLY polynomial}

In this case
\be
\P^{[2,2]} = \Tr_{[1]\otimes [1]}   q^{2N-2} \Big( I\otimes I + (qT)[2]P_{11} \Big)^2 =
\Tr_{[1]\otimes [1]}   q^{2N-2} \Big( I\otimes I + (qT)(2+qT[2])[2] P_{11}   \Big) = \nn \\
= q^{2N-2}\Big( [N]^2 + qT\,(2+qT[2]) [N][N-1]\Big) =
[N]\cdot\underbrace{q^{2N-2}\Big( [N] + 2[N-1]\cdot(qT)+[2][N-1]\cdot(qT)^2\Big)}_{\P^{[2,1]}_{r}}
\label{KRT22}
\ee
where we used the projector property $P_{11}^2 = P_{11}$.


\subsubsection{Reduced KR polynomial}

The hypercube is now two-dimensional,
and morphisms in the case of initial vertex $\bullet\bullet$ are:
\be
\begin{array}{ccccc}
\l[N] & \stackrel{d_0}{\longrightarrow} & 2\times [N-1]
& \stackrel{d_1}{\longrightarrow} & [2][N-1] \\ \\ \\
&& [N-1] && \\
&\stackrel{\alpha}{\nearrow} && \stackrel{\beta}{\searrow} & \\
\l[N] &\stackrel{d_0}{\longrightarrow}\!\!\!\!\!\!\!\!\!\!\! & \oplus &
\!\!\!\!\!\!\!\!\!\!\!\stackrel{d_1}{\longrightarrow} &
\begin{array}{c}q[N-1]  \\ \frac{1}{q}[N-1] \\ \end{array}
 \\
 &\stackrel{\ \alpha}{\searrow} && \stackrel{\!\!\!\!-\beta}{\nearrow} & \\
 && [N-1]    \\
 \end{array}
 \ee
In other words, the two differentials are
\be
d_0 = \left(\begin{array}{c} \alpha \\ \alpha \end{array}\right)
= \left(\begin{array}{cc||cc}
&&& \\ \alpha &&& \l[N-1]  \\ \alpha &&& \l[N-1] \\  \\
\hline\hline   \\ \l[N] &&   &\stackrel{\nearrow}{}\ \ \ \ \ \ \
\end{array}  \right) \ \ \ \ \ \ \ \ \ \ \ \ \ \ \nn \\ \nn \\
d_1 = \left(\begin{array}{cc}
0 & 0 \\
\beta & -\beta \end{array}\right) =
\left(\begin{array}{ccc||cc}
&&&&\\
0 & 0 &&& q[N-1] \\
\beta & -\beta &&& q^{-1}[N-1]   \\ \\
\hline\hline      \\
\l[N-1] & \l[N-1] &&&\stackrel{\nearrow}{}\ \ \ \ \ \ \end{array} \right)
\ee
Clearly, $d_1d_0=0$, and
\be
{\rm Ker}(d_0) = {\rm Ker}\Big(\alpha_{[N]\longrightarrow [N-1]}\Big) = q^{1-N}, \nn \\  \nn \\
{\rm Im}(d_0) = d_0\Big(a[N]\Big) = \left(\begin{array}{c} a[N-1] \\ a[N-1] \end{array}\right)
= {\rm Ker}(d_1), \nn \\  \nn\\
{\rm Im}(d_1) = d_1\left(\begin{array}{c} a[N-1] \\ b[N-1] \end{array}\right) =
\left(\begin{array}{c} 0 \\ (a-b)\frac{1}{q}[N-1]\end{array}\right)
\ \ \Longrightarrow \ \
{\rm CoIm}(d_1) =  \left(\begin{array}{c}  {q}[N-1] \\ 0 \end{array}\right)
\ee
It is important here, that $\beta_{[N-1]\longrightarrow q^{-1}[N-1]}$ is
cohomologically trivial. Parameters $a$ and $b$ are arbitrary,
but they are the {\it same} in the two constituents of $\ {\rm Im}(d_0)$.

It follows, that
\be
{\cal P}^{[2,2]}_r = q^{2N-2} \left\{ {\rm dim}_q {\rm Ker}(d_0) +
(qT) \Big({\rm dim}_q{\rm Ker}(d_1) - {\rm dim}_q{\rm Im}(d_0)\Big)
+(qT)^2 {\rm dim}_q{\rm CoIm}(d_1)\right\} = \nn \\
= q^{2N-2} \Big\{q^{1-N} + (qT)^2\cdot q[N-1]\Big\} =
\boxed{q^{N-1} + q^{2N+1}T^2[N-1]}
\label{KRr22}
\ee

This can be compared with the reduced superpolynomial, eq.(136) of \cite{DMMSS}:
\be
\sim\ 1+q^2T^2\,\frac{q^2+a^2T}{1-q^2}
\ee
and matching implies that while $q^N=a$, the quantum number should be substituted
by a different rule:
\be
\l[N-1] \longrightarrow \frac{q^2+a^2T}{a(1-q^2)}
\ee
instead of the naive, but {\it not} positive
$\frac{\frac{a}{q}-\frac{q}{a}}{q-\frac{1}{q}}= \frac{q^2-a^2}{a(1-q^2)}$.

\subsubsection{Reduced KR polynomial for initial vertex $\bullet\circ$}

The hypercube is the same, but two of the four morphisms are different:
\be
\begin{array}{ccccccccc}
&& [N-1] &&   &&&&  \l[N-1]\\
&\stackrel{\beta}{\swarrow} &\ \ \  \downarrow{d_0}
& \stackrel{\beta}{\searrow} & &&&& \downarrow{d_0} \\
\l[N] &    & \oplus &  &
\begin{array}{c}q[N-1]  \\ \frac{1}{q}[N-1] \\ \end{array}
&&&& [N] \oplus [2][N-1]  \\
 &\stackrel{\ \beta}{\searrow} &\ \ \  \downarrow{d_1}
 & \stackrel{\!\!\!\!-\beta}{\swarrow} &  &&&& \downarrow{d_1}\\ \\
 && [N-1] && &&&& [N-1]   \\
 \end{array}
 \ee
Accordingly
\be
d_0 = \left(\begin{array}{c} \beta \\ 0 \\ \beta \end{array}\right)
= \left(\begin{array}{cc||cc}
&&& \\ \beta &&& \l[N]  \\ 0 &&& q[N-1] \\
\beta &&&  \frac{1}{q}[N-1]\\ \\
\hline\hline \\ \l[N-1] &&& \stackrel{\nearrow}{}\ \ \ \ \ \ \
\end{array}  \right) \ \ \ \ \ \ \ \ \ \ \ \ \ \ \nn \\ \nn \\
d_1 = \left(\begin{array}{ccc}
\alpha & 0 &-\beta \end{array} \right) =
\left(\begin{array}{cccc||cc}
&&&&\\
\alpha & 0 & -\beta  &&& \l[N-1] \\ \\
\hline\hline      \\
\l[N] & q[N-1] & \frac{1}{q}[N-1] &&&\stackrel{\nearrow}{}\ \ \ \ \ \ \ \\ \end{array} \right)
\ee
With this choice $d_1d_0 = 0$, and $q[N-1]$ belongs to coimage of $d_0$ and to the kernel of $d_1$.
For the other spaces we have:
$$
\begin{array}{ccccc}
&& \begin{array}{c} \\ q^{N-1} \\ \left\{\begin{array}{c} \l[N-2] \\ q^{1-N} \end{array}\right.
\end{array}
\!\!\!\!\!\!\!\!\!\!\left.\begin{array}{c} \phantom. \\ \phantom. \end{array}\right\}&& \\
&\nearrow &\ \ \ \ \ \ \ \ \stackrel{\searrow}{} & \!\!\!\!\!\!\!\!\!\!\!\searrow & \\
\l[N-1] &&\ \ \ \ \ \ \ \ \ \ \ \ \ \ \ \ \ \stackrel{0}{} && \l[N-1] \\
&\searrow &&  \!\!\!\!\!\!\!\!\!\!\!\stackrel{-}{\nearrow} & \\
&&\frac{1}{q}[N-1] = \left\{\!\!\! \begin{array}{c} \l[N-2] \\ q^{1-N} \end{array}\right. &&\\
&&& \!\!\!\!\!\!\!\!\!\!\! \!\!\!\!\!\!\!\!\!\!\!\stackrel{\searrow}{} &\\
&&& \!\!\stackrel{0}{}& \\
\end{array}
$$
From this picture it is clear that
\be
{\rm Im}(d_1) = d_1\Big(a[N-1]\Big) = \left(\begin{array}{ccc}
0 && \\ a[N-2] && \\ bq^{1-N} && \\
\hline
a[N-2] && \\
bq^{1-N} && \end{array}\right) \nn \\
{\rm Ker}(d_2) =
\left(\begin{array}{ccc}
0 && \\ a[N-2] && \\ bq^{1-N} && \\
\hline
a[N-2] && \\
\underline{cq^{1-N}} && \end{array}\right) \oplus \underline{q[N-1]}
\ee
i.e. that ${\rm Ker}(d_1)/{\rm Im}(d_0)$ consists of the two underlined items
and has dimension $q^{1-N} + q[N-1] = [N]$.
Also it is clear that the coimage of $d_1$ is empty.
Therefore in this case the reduced KR polynomial is
\be
\frac{q^{N-1}}{q^NT} \Big( 0 + (qT)\cdot[N] + (qT)^2\cdot 0\Big) = [N] =
{\cal P}^{{\rm unknot}} = {\cal P}^{{\rm unknot}\ \cup\ {\rm unknot}}_r
\ee
Similarly one can study the initial vertex $\circ\circ$, which reproduces
the Hopf KR polynomial ${\cal P}^{[2,2]}_r$.

\subsubsection{Unreduced ${\cal P}^{[2,2]}$, initial point $\bullet\bullet$}

Now the hypercube is
\be
\begin{array}{ccccc}
\l[N]^2 & \stackrel{d_0}{\longrightarrow} & 2\times [N][N-1]
& \stackrel{d_1}{\longrightarrow} & [2][N][N-1] \\ \\ \\
&& \l[N][N-1] && \\
&\stackrel{\alpha_2}{\nearrow} && \stackrel{\beta_2}{\searrow} & \\
\l[N]^2 &\stackrel{d_0}{\longrightarrow}\!\!\!\!\!\!\!\!\!\!\! & \oplus &
\!\!\!\!\!\!\!\!\!\!\!\stackrel{d_1}{\longrightarrow} &
\begin{array}{l}  \   ^{q[N][N-1] \oplus}  \\  \frac{1}{q}[N][N-1] \\ \phantom{5} \end{array}   \\
 &\stackrel{\ \alpha_2}{\searrow} && \stackrel{\!\!\!\!-\beta_2}{\nearrow} & \\
 && \l[N][N-1]    \\
 \end{array}
 \label{hypercubeur22}
 \ee
and the morphisms can be understood just as $\alpha_2=I\otimes\alpha$
and $\beta_2=I\otimes\beta$.
Since in this case the vector space $\openC^N = [N]$ acts just as a sterile factor,
this is enough to obtain the well-known answer from \cite{CM,DM3}
\be
\boxed{
{\cal P}^{[2,2]} = [N]\cdot {\cal P}^{[2,2]}_r
= q^{N-1}[N] + q^{2N+1}T^2[N][N-1]
}
\label{KRur22}
\ee
Decomposition (\ref{Pdeco}) becomes
\be
\P^{[2,2]} = {\cal P}^{[2,2]} + (1+T)\cdot{\cal Q}^{[2,2]},\ \ \ \ \ \
(1+T)\cdot{\cal Q}^{[2,2]} = (1+T)^2\cdot q^{2N-1} [N][N-1] \ > \ 0
\label{Qur22}
\ee

\subsubsection{Alternative unreduced KR polynomial: also minimal, but incorrect
\label{altHopf}}

Still, it is instructive to consider another option
(this makes sense already because for all other knots and links
${\cal P}\neq [N]\cdot {\cal P}_r$).
Instead of
\be
d_0 = \left(\begin{array}{c} I\otimes\alpha \\ I\otimes\alpha \end{array}\right)
= \left(\begin{array}{c} \alpha_2 \\ \alpha_2 \end{array}\right), \ \ \ \ \ \
d_1 = \Big( I\otimes \beta, \ \ \ -I \otimes\beta\Big) = \Big(\beta_2, \ \ \ -\beta_2\Big)
\ee
in (\ref{hypercubeur22}) one could try
\be
\tilde d_0 =
\left(\begin{array}{c} \alpha\otimes I \\ I\otimes\alpha   \end{array}\right)
= \left(\begin{array}{c} \alpha_1 \\ \alpha_2 \end{array}\right), \ \ \ \ \
\tilde d_1 = \Big( I\otimes \alpha, \ \ \ -\alpha \otimes I\Big) = \Big(\alpha_2, \ \ \ -\alpha_1\Big)
\ee
Note that to have the property $\tilde d_1\tilde d_0=0$ preserved, we need to use $\alpha$
rather than $\beta$ in $\tilde d_1$.

What happens at the first step can look optimistic:
\be
{\rm Ker}(\tilde d_0) = {\rm Ker}(\alpha_1)\ \bigcap\ {\rm Ker}(\alpha_2)
= q^{1-N}[N]\ \bigcap \ [N]q^{1-N} = q^{2-2N}
\label{kerd0ur22}
\ee
gets much smaller than ${\rm Ker}(d_0)= q^{1-N}[N]$ in (\ref{KRur22}).
However, decrease of particular cohomology  can not lead
to a "smaller" version of KR polynomial, because (\ref{KRur22})
already saturates the Euler characteristic: has the same number of
terms as HOMFLY polynomial, which can not be further diminished.
Alternative polynomial can be either "bigger" or of the "same size" as
${\cal P}^{[2,2]}$.

To understand (\ref{kerd0ur22}) better,
it can be convenient to introduce two different variables $q_1$ and $q_2$,
so that $[N]^2$ is substituted by $[N]_1[N]_2$, and $\alpha_1$, $\alpha_2$ act
on $[N]_1$ and $[N]_2$ respectively.
Then the intersection of kernels looks more explicit:
$$q_1^{1-N}[N]_2 \ \bigcap\ q_2^{1-N}[N]_1 = (q_1q_2)^{1-N}$$
The action of constituents of $\tilde d_1$ is also easily defined in these terms:
\be
\begin{array}{ccccc}
\l[N]^2 & \stackrel{\tilde d_0}{\longrightarrow} & 2\times [N][N-1]
& \stackrel{\tilde d_1}{\longrightarrow} & [2][N][N-1] \\ \\ \\
&& \l[N-1]_1[N]_2 && \\
&\stackrel{\alpha_1}{\nearrow} && \stackrel{\alpha_2}{\searrow} & \\
\l[N]_1[N]_2 &\stackrel{\tilde d_0}{\longrightarrow}\!\!\!\!\!\!\!\!\!\!\! & \oplus &
\!\!\!\!\!\!\!\!\!\!\!\stackrel{\tilde d_1}{\longrightarrow} &
\begin{array}{c} \ ^{q[N][N-1] \oplus}  \\ \l[N-1]_1[N-1]_2 \\ \ _{\oplus q^{-N}[N-1]}
\end{array}   \\
 &\stackrel{\ \alpha_2}{\searrow} && \stackrel{\!\!\!\!-\alpha_1}{\nearrow} & \\
 && \l[N]_1[N-1]_2    \\
 \end{array}
 \label{hypercubetildeur22}
 \ee
Clearly,
\be
{\rm Im}(\tilde d_1) = [N-1]_1[N-1]_2, \ \ \ \ \ \
{\rm CoIm}(\tilde d_1) = q[N][N-1] + q^{-N}[N-1] = [N+1][N-1]
\ee
and in the middle of the complex we have
\be
{\rm Im}(\tilde d_0) = \tilde d_0\Big(a\cdot q_1q_2[N-1]_1[N-1]_2
+ b\cdot q_1^{1-N}q_2[N-1]_2 + c\cdot q_1[N-1]_1q_2^{2-N} + d\cdot q_1^{1-N}q_2^{1-N}\Big)
= \nn \\
= \left(\begin{array}{ccc} q_2[N-1]_1\Big(a[N-1]_2 + cq_2^{1-N}\Big) \\
q_1\Big(a[N-1]_1 + bq_1^{1-N}\Big)[N-1]_2 \end{array}\right)
= {\rm Ker}(\tilde d_1) \ \ \ \ \ \ \ \ \ \ \ \ \ \ \
\ee
Thus $\ {\rm Ker}(\tilde d_1)\Big/{\rm Im}(\tilde d_0) = \emptyset\ $ and
\be
\boxed{
\tilde{\cal P}^{[2,2]} = q^{2N-2}\Big(q^{2-2N} + 0\cdot(qT) + [N+1][N-1]\cdot (qT)^2\Big)
= 1 + q^{2N}T^2[N+1][N-1]
}
\label{KRtildeur22}
\ee
This time decomposition (\ref{Pdeco}) is
$\ \P^{[2,2]} = \tilde {\cal P}^{[2,2]} + (1+T)\cdot\tilde{\cal Q}^{[2,2]}\ $ with
\be
\!\!\!\!
\tilde{\cal Q}^{[2,2]}
= q^{2N-1}[N-1]\Big([N](1+T) +q^{1-N}(1-T)\Big)
= q^{2N-1}[N-1]\Big(\big([N]+q^{1-N}\big) +q[N-1]T\Big) \ > \ 0
\ee
Clearly, this structure, though positive, is more involved than in (\ref{Qur22}).
Also $\tilde{\cal P}^{[2,2]}\neq {\cal P}^{[2,2]}$ even for $N=2$,
therefore $\tilde{\cal P}^{[2,2]}$ does {\it not} reproduce the answer \cite{katlas}
for the Jones-Khovanov polynomial.
However, internal reasons to prefer (\ref{KRur22}) over (\ref{KRtildeur22})
are unclear at the moment.

\subsection{Trefoil as a torus knot $[2,3]$}

In this case
\be
\P^{[2,3]} = \Tr_{[1]\otimes [1]} \  q^{3N-3} \Big( I\otimes I + (qT)[2]P_{11} \Big)^3 = \nn \\
= \Tr_{[1]\otimes [1]}\   q^{3N-3}
\left\{ I\otimes I + \Big(3qT+3q^2T^2[2]+q^3T^3[2]^2\Big)[2] P_{11} \right\}
= \nn \\
= q^{3N-3}\left\{ [N]^2 + (3qT+3q^2T^2[2]+q^3T^3[2]^2\Big) [N][N-1]\right\}, \nn \\ \nn \\
\P^{[2,3]}_{r} \equiv [N]^{-1}\cdot \P^{[2,3]}
=q^{3N-3}\left\{ [N] + (3qT+3q^2T^2[2]+q^3T^3[2]^2\Big) [N-1]\right\}
\ee

\subsubsection{Reduced KR polynomial}

In reduced case the primary polynomial is decomposed as follows:
\be
q^{3-3N}\P_r^{[2,3]} = [N]+3qT[N-1]+3(qT)^2[2][N-1]+(qT)^3[2]^2[N-1] =\nn \\ \nn \\
= \underbrace{\Big([N]+qT[N-1]\Big)}_{\sim q^{1-N}}
+ 2qT\underbrace{\Big([N-1]+qT\cdot\frac{1}{q}[N-1]\Big)}_{\sim 0}
+ (qT)^2\underbrace{\Big(\frac{1}{q}[N-1]+qT\cdot\frac{1}{q^2}[N-1]\Big)}_{\sim 0} + \nn \\
+ 2(qT)^2\underbrace{\Big(q[N-1]+qT[N-1]\Big)}_{\sim 0}
+ (qT)^2\underbrace{\Big(q[N-1]+qT\cdot{q^2}[N-1]\Big)}_{\sim q(q^{2-N} + q^{N} T)}
\ee
This decomposition corresponds to the following pattern
of morphisms in the hypercube:
\be
\begin{array}{cccccccccccc}
\\
\l[N] & \stackrel{d_0}{\longrightarrow} & 3[N-1] &
 \stackrel{d_1}{\longrightarrow} & 3[2][N-1] &
  \stackrel{d_2}{\longrightarrow} & [2]^2[N-1]\\
\\   \hline \\
&&&&3\times \frac{1}{q}[N-1]&& \frac{1}{q^2}[N-1]\\
\l[N]&\longrightarrow &3\times [N-1]&\longrightarrow && \longrightarrow &2\times [N-1]   \\
&&&&3\times {q}[N-1]&& {q^2}[N-1]\\
\\  \hline  \\
\l[N] & \Longrightarrow & [N-1]&& \frac{1}{q}[N-1] &\longrightarrow & \frac{1}{q^2}[N-1] \\ \\
&&2\times [N-1] &\longrightarrow & 2\times \frac{1}{q}[N-1] \\  \\
&&&& 2\times q[N-1] &\longrightarrow & 2\times[N-1] \\     \\
&&&& q[N-1] &\Longrightarrow & {q^2}[N-1] \\
\\
\end{array}
\ee
In the bottom part of the table all arrows are one-to-one, i.e. have neither kernels
nor coimages,  except for just two, denoted by double lines.
In more details, this is described by the following shape of the differentials:
\be
d_2 = \left(\begin{array}{cc|cc|cc}
0 & 0 & 0 & 0 & \beta & 0 \\ \hline
\beta & 0 & 0 & 0 & 0 & 0 \\
0 & 0 & \beta & 0 & 0 & 0 \\ \hline
0 & \beta & 0 & \beta & 0 & \beta  \end{array}\right), \ \ \ \ \
d_1 = \left(\begin{array}{ccc} 0 & 0 & 0 \\ \beta & -\beta & 0 \\ \hline
0 & 0 & 0 \\ 0& \beta & -\beta  \\ \hline
0 & 0 & 0 \\ -\beta & 0 & \beta  \end{array}\right), \ \ \ \ \
d_0 = \left(\begin{array}{c} \alpha \\ \alpha \\ \alpha \end{array}\right)
\ee
Thus we obtain for reduced KR polynomial:
\be
\boxed{
{\cal P}^{[2,3]}_{r} = q^{3N-3}\Big(q^{1-N} + (qT)^2\cdot q(q^{2-N} + q^{N} T)\Big) =
q^{2N-2}\Big(1+q^4T^2+q^{2N+2}T^3\Big)
}
\label{KR23}
\ee
what is the standard answer \cite{CM,DM3}.

\subsubsection{Unreduced KR polynomial}

In unreduced case the  only difference will be with the second double-line arrow:
$$
q[N][N-1] \Longrightarrow q^2[N][N-1]
$$
There are now two options for the morphism
(see s.6.6.1 of \cite{DM3} for a picture):
one can either use the same
(\ref{mainredrule2}), as in the reduced case:
$$
\l[N]\cdot \Big(q[N-1] \Longrightarrow q^2[N-1]\Big)
$$
or apply instead its analogue for the $N$-dimensional space:
$$
q[N-1]\cdot \Big([N] \Longrightarrow q[N]\Big)
$$
i.e.,
\be
\l[N] + qT\cdot q[N] = q^{1-N} + q^{N+1} T + (1+T)\cdot {q} [N-1]
\sim  q^{1-N} + q^{N+1} T
\label{congNzero}
\ee
Clearly, the second option {\it decreases} the resulting polynomial,
and thus we should apply it get the minimal ${\cal P}^{[2,3]}$:
\be
{\cal P}^{[2,3]}  =q^{3N-3} \Big(q^{1-N}[N] + (qT)^2\cdot q(q^{1-N} + q^{N+1} T)[N-1]\Big)
= \nn \\
\boxed{
= q^{2N-2}\Big([N]+q^3[N-1]\cdot T^2+q^{2N+3}[N-1]\cdot T^3\Big)
}
= q^{2N-2}[N] +q^{2N+1} T^2(1+q^{2N}T)[N-1]
\label{KR[23]ur}
\ee
what is  the standard answer from \cite{CM,DM3}.

Note that ${\cal P}^{[2,3]} \neq [N]\cdot{\cal P}^{[2,3]}_r$, but
this property is restored for HOMFLY at $T=-1$, $H^{[2,3]} = [N]\cdot H^{[2,3]}_r$
-- and the factor $[N]$
arises from the factor $(1+q^{2N}T)[N-1]$ at the r.h.s. of (\ref{KR[23]ur}).
This will work just the same way for all other 2-strand knots and links.

\subsection{Generic knot $[2,2k+1]$}

This time
\be
\P^{[2,2k+1]}=
 q^{(2k+1)(N-1)} \Big( I\otimes I + (qT)[2]P_{11} \Big)^{2k+1} =
[N]^2 + \left(\sum_{i=1}^{2k+1} (qT)^iC^i_{2k+1}[2]^{i-1}\right)[N][N-1]
\label{2str1}
\ee
The pattern of morphisms in this general situation
is described in s.6.7 of \cite{DM3}.
It is absolutely clear from the representative example of the $5$-foil $(k=2)$:
\be
\begin{array}{cccccccccccccc}
\\
\l[N] &\Longrightarrow & 5[N-1] &\longrightarrow &  \frac{10}{q}[N-1]
&\longrightarrow &  \frac{10}{q^2}[N-1]&\longrightarrow &  \frac{5}{q^3}[N-1]
&\longrightarrow &  \frac{1}{q^4}[N-1] \\   \\
&&&& {9}{q}[N-1] &\longrightarrow &   20[N-1] &\longrightarrow & \frac{15}{q}[N-1]
&\longrightarrow &  \frac{4}{q^2}[N-1] \\  \\
&&&& q[N-1] &\Longrightarrow & {10}{q^2}[N-1]&\longrightarrow & 15q[N-1]
&\longrightarrow &  6[N-1]\\  \\
&&&&&&&&  4q^3[N-1]  &\longrightarrow &  4q^2[N-1]\\   \\
&&&&&&&& q^3[N-1] &\Longrightarrow & q^4[N-1] \\ \\
\end{array}
\label{25morph}
\ee
In each line the alternated sum of multiplicities is zero,
therefore it is further decomposed into one-to-one maps an obvious way,
e.g. the first line is in fact
\be
\begin{array}{cccccccccccccc}
\\
\l[N] &\Longrightarrow & [N-1] &  &  \frac{6}{q}[N-1]
&\longrightarrow &  \frac{6}{q^2}[N-1]& &  \frac{1}{q^3}[N-1]
&\longrightarrow &  \frac{1}{q^4}[N-1] \\   \\
  &  & 4[N-1] &\longrightarrow &  \frac{4}{q}[N-1]
& &  \frac{4}{q^2}[N-1]&\longrightarrow &  \frac{4}{q^3}[N-1]
 \\   \\
\end{array}
\ee
The three double arrows in (\ref{25morph})  stand for the three maps with non-trivial cohomologies,
which contribute to KR polynomial
\be
{\cal P}^{[2,5]}_r = q^{5N-5}\left(\underbrace{q^{1-N}}_{\sim [N]+qT\cdot q[N-1]}
+(qT)^2\cdot \underbrace{q\Big(q^{2-N}+qT\cdot q^{N-1}\Big)}_{\sim q[N-1]+qT\cdot{q^2}[N-1]}
+(qT)^4\cdot \underbrace{q^3\Big(q^{2-N}+qT\cdot q^{N-1}\Big)}_{\sim q^3[N-1]+qT\cdot{q^4}[N-1]}
\right)
\label{KR25r}
\ee
\be
\!\!\!\!\!\!\!\!\!\!\!\!
{\cal P}^{[2,5]} = q^{5N-5}\left(\underbrace{q^{1-N}[N]}_{\sim [N]^2+qT\cdot q[N][N-1]}
+(qT)^2\cdot \underbrace{q\Big(q^{1-N}+qT\cdot q^N\Big)[N-1]}_{\sim q[N][N-1]+qT\cdot{q^2}[N][N-1]}
+(qT)^4\cdot \underbrace{q^3\Big(q^{1-N}+qT\cdot q^N\Big)[N-1]}_{\sim q^3[N][N-1]
+qT\cdot{q^4}[N][N-1]}
\right)
\label{KR25ur}
\ee
Of course, these two formulas are in accord with \cite{CM} and \cite{DM3}.

\bigskip

For general $k$ decomposition has just the same structure as in this example:
from  (\ref{2str1})
\be
\P^{[2,2k+1]}_r=
[N] + \left(\sum_{i=1}^{2k+1} (qT)^i \sum_{j=0}^{i-1}q^{1-i+2j}C^i_{2k+1}C^j_{i-1}\right)[N-1]
= \nn \\
= q^{1-N} + q[N-1] \sum_{j=0}^{2k} q^{2j} \left(\delta_{j0}+
\sum_{i=j+1}^{2k+1}C^i_{2k+1}C^j_{i-1} T^i\right)
\ee
Thus there are $2k+1$ lines, which we label by index $j=0,\ldots,2k$.
However, the alternated sum of coefficients in the line $j$, i.e.
the value of the sum in brackets, evaluated at $T=-1$, does not vanish,
but is rather equal to $(-)^{j+1}$:
it is given by the coefficient in front of $x^j$ in the sum
\be
\sum_{i=1}^{2k+1} (-)^iC^i_{2k+1}(1+x)^{i-1} = \frac{-1+\Big(1-(1+x)\Big)^{2k+1}}{1+x}
= -1+ x-x^2 + \ldots -x^{2k}
\ee
In order to convert these sums to zero we just move one item from the lines
with odd $j$ to adjacent line $j+1$, exactly as in (\ref{25morph}).
Now each line can be decomposed into a combination of one-to-one maps,
with the exception of the very first mapping in each {\it even} line. Hence
$$
{\cal P}^{[2,2k+1]}_r = q^{(2k+1)(N-1)}\left(\underbrace{q^{1-N}}_{\sim [N]+qT\cdot q[N-1]}
+\sum_{{\rm even}\ j = 2}^{2k} (qT)^j\cdot
\underbrace{q^{j-1}\Big(q^{2-N}+q^{N-1}\cdot qT\Big) }_{\sim q^{j-1}[N-1]
+(qT)\cdot q^{j}[N-1]}
\right) =
$$
\vspace{-0.3cm}
\be
\boxed{
= q^{2k(N-1)}\left(1 + \Big(1+q^{2N-2}T\Big)\sum_{j'=1}^{k} (q^2T)^{2j'}\right)
}
\label{KR2strknotr}
\ee

Unreduced case differs just by the chain of equivalences,
from
\be
[N-1]+(qT)\cdot q[N-1] \ \sim\ q^{2-N} + q^NT
\ee
to
\be
[N-1]\Big([N][N-1]+(qT)\cdot q[N][N-1]\ \sim\  \Big(q^{1-N}+q^{N+1}T\Big)[N-1]
\ee
so that
$$
{\cal P}^{[2,2k+1]} = q^{(2k+1)(N-1)}\left(\underbrace{q^{1-N}[N]}_{\sim [N]^2+qT[N][N-1]}
+\sum_{{\rm even}\ j = 2}^{2k} (qT)^j\cdot
\underbrace{q^{j-1}\Big(q^{1-N}+q^{N}\cdot qT\Big)[N-1]}_{\sim q^{j-1}[N][N-1]
+(qT)\cdot q^{j}[N][N-1]}
\right) =
$$
\vspace{-0.3cm}
\be
\boxed{
= q^{2k(N-1)}\left([N] + \frac{1}{q}\sum_{j'=1}^{k} (q^2T)^{2j'}\Big(1+q^{2N}T\Big)[N-1]\right)
}
\label{KR2strknotur}
\ee
Note that the second piece of this formula does not contain a factor $[N]$,
thus ${\cal P}^{[2,2k+1]}\neq [N]{\cal P}^{[2,2k+1]}_r$.
However, this property should be restored when we return to HOMFLY polynomial, i.e. at $T=-1$.
This is guaranteed by emergence of the new factor $(1+q^{2N}T)$ at the place of $[N]$:
at $T=-1$ we have
\be
(1-q^{2N})[N-1] = q[N](1-q^{2N-2})
\ee

\subsection{Generic link $[2,2k]$}

The only difference from the case of 2-strand links is that now the
number of lines is of different parity,
and in the last line  with $j=2k-1$ and there is an additional mismatch.
This is clear from any example. Say, for $k=3$
we have instead of (\ref{25morph})
\be
\begin{array}{cccccccccccccc}
\\
\!\!\!\!\!\!\!\!\!\!
\l[N] &\Longrightarrow & 6[N-1] &\longrightarrow &  \frac{15}{q}[N-1]
&\longrightarrow &  \frac{20}{q^2}[N-1]&\longrightarrow &  \frac{15}{q^3}[N-1]
&\longrightarrow &  \frac{6}{q^4}[N-1]  &\longrightarrow &  \frac{1}{q^5}[N-1]\\   \\
&&&& {14}{q}[N-1] &\longrightarrow &   40[N-1] &\longrightarrow & \frac{45}{q}[N-1]
&\longrightarrow &  \frac{24}{q^2}[N-1] &\longrightarrow &  \frac{5}{q^3}[N-1] \\  \\
&&&& q[N-1] &\Longrightarrow & {20}{q^2}[N-1]&\longrightarrow & 45q[N-1]
&\longrightarrow &  36[N-1] &\longrightarrow &  \frac{10}{q}[N-1]\\  \\
&&&&&&&&  14q^3[N-1]  &\longrightarrow &  24q^2[N-1] &\longrightarrow & 10q[N-1]\\   \\
&&&&&&&& q^3[N-1] &\Longrightarrow & 6q^4[N-1] &\longrightarrow &  5q^3[N-1] \\ \\
&&&&&&&&&&&& \boxed{q^5[N-1]} \\    \\
\end{array}
\nn
\ee
In addition to non-one-to-one mappings marked by $\Longrightarrow$
the last line is a new item in the box, not involved in any mappings at all.
It provides an additional contribution, moreover, the one proportional to $[N-1]$.
In this particular example we get:
\be
{\cal P}^{[2,6]}_r = q^{5N-5}\left(\underbrace{q^{1-N}}_{\sim [N]+qT\cdot q[N-1]}
+(qT)^2\cdot \underbrace{q\Big(q^{2-N}+qT\cdot q^{N-1}\Big)}_{\sim q[N-1]+qT\cdot{q^2}[N-1]}
+(qT)^4\cdot \underbrace{q^3\Big(q^{2-N}+qT\cdot q^{N-1}\Big)}_{\sim q^3[N-1]+qT\cdot{q^4}[N-1]}
\right.+\nn\\ \left.\boxed{(qT)^6\cdot q^5[N-1]}\phantom{\underbrace{q}_{q}}\right), \nn \\ \nn \\
{\cal P}^{[2,6]} = q^{6N-6}\left(\underbrace{q^{1-N}[N]}_{\sim [N]^2+qT\cdot q[N][N-1]}
+(qT)^2\cdot \underbrace{q\Big(q^{1-N}+qT\cdot q^N\Big)[N-1]}_{\sim q[N][N-1]+qT\cdot{q^2}[N][N-1]}
+(qT)^4\cdot \underbrace{q^3\Big(q^{1-N}+qT\cdot q^N\Big)[N-1]}_{\sim q^3[N][N-1]
+qT\cdot{q^4}[N][N-1]}  \right.+ \nn \\
\left.+\ \boxed{(qT)^6\cdot q^5[N][N-1]}\phantom{\underbrace{1}_{\Big(}} \right)
\ \ \ \ \ \ \ \ \ \ \
\ee
Note that all the terms are exactly the same as in (\ref{KR25r}) and (\ref{KR25ur})
-- except for additional contribution in the box.

This is of course true for arbitrary $k$ and
\be
{\cal P}^{[2,2k]}_r = q^{N-1}\cdot {\cal P}^{[2,2k-1]}_r + q^{2k(N-1)}(qT)^{2k}\cdot q^{2k-1}[N-1]
\ \stackrel{(\ref{KR2strknotr})}{=} \nn \\
= q^{2k(N-1)}\left( q^{1-N}
+\sum_{{\rm even}\ j = 2}^{2k-2}
 q^{2j-1}T^j\Big(q^{2-N}+q^{N-1}\cdot qT\Big)
  + (qT)^{2k}\cdot q^{2k-1}[N-1]
\right) = \nn \\
= q^{(2k-1)(N-1)}\left(1
+\Big(1+q^{2N-2}T\Big)\sum_{j' = 1}^{k-1}
 (q^{2}T)^{2j'} + (q^2T)^{2k}q^{N-2}[N-1]\right)
\label{KR2strlinkr}
\ee
and similarly in unreduced case,
\be
{\cal P}^{[2,2k]} = q^{N-1}\cdot {\cal P}^{[2,2k-1]} + q^{2k(N-1)}(qT)^{2k}\cdot q^{2k-1}[N][N-1]
\ \stackrel{(\ref{KR2strknotur})}{=} \nn \\
 \!\!\!\!\!\!\!\!\!\!\!\!\!\!
 = q^{2k(N-1)} \left( q^{1-N}[N]
+\sum_{{\rm even}\ j = 2}^{2k-2}
 q^{2j-1}T^j\Big(q^{1-N}+q^{N}\cdot qT\Big)[N-1]
+ (qT)^{2k}\cdot q^{2k-1}[N][N-1]\right)= \nn \\
= q^{(2k-1)(N-1)}\left([N]
+\frac{1}{q}\sum_{j' = 1}^{k-1} (q^{2}T)^{2j'} \Big(1+q^{2N}T\Big)[N-1]
+ (q^2T)^{2k}q^{N-2}[N][N-1]\right)
\label{KR2strlinkur}
\ee
For $k=1$ we reproduce (\ref{KRr22}) and (\ref{KRur22}) for the Hopf link, for $k=2$ we
obtain:
\be
{\cal P}^{[2,4]}_r = q^{3N-3}\Big(1+ q^4T^2+q^{2N+2}T^3+q^{N+6}T^4[N-1]\Big), \nn \\
{\cal P}^{[2,4]} = q^{3N-3}\Big([N]+ q^3T^2(1+q^{2N}T)[N-1]+q^{N+6}T^4[N][N-1]\Big)
\label{KRo24}
\ee

\subsection{Evolution in $k$, MacDonald dimensions  and $\gamma$-factors
\label{2strator}}

\subsubsection{Conversion to evolution representation}

Eqs.(\ref{KR2strknotr}) and (\ref{KR2strlinkr})
can be rewritten in more familiar forms:
\be
{\cal P}^{[2,2k+1]}_{r} = q^{(2k+1)(N-1)}\Big(
q^{1-N} + q^{5-N}T^2 + q^{9-N}T^4 + \ldots + q^{4k+1-N}T^{2k}
+  \ \ \ \ \ \ \ \ \ \  \nn \\
+ q^{N+3}T^3 + q^{N+7}T^5 + \ldots + q^{N+4k-1}T^{2k+1}\Big) =
\label{P2odd}
\ee
\vspace{-0.5cm}
$$
= q^{2k(N-1)}\left( \frac{1-q^{4k+4}T^{2k+2}}{1-q^4T^2}
+ q^{2N+2}T^3\,\frac{1-q^{4k}T^{2k}}{1-q^4T^2}\right)
= \boxed{
q^{2Nk}\left( q^{-2k}\,\frac{1+ q^{2N+2}T^3}{1-q^4T^2}-
(qT)^{2k}\,\frac{q^{2N+2}T^3+q^{4}T^{2}}{1-q^4T^2}\right)
}
$$
and
\be
{\cal P}^{[2,2k]}_r = q^{N-1}\cdot{\cal P}^{[2,2k-1]}_r+ q^{2k(N-1)}(qT)^{2k}\cdot q^{2k-1}[N-1]
= \nn \\
=q^{(2k-1)(N-1)}\left(\frac{1-(q^4T^2)^{k}}{1-q^4T^2} + q^{2N+2}T^3\,\frac{1-(q^4T^2)^{k-1}}{1-q^4T^2}
+ (q^4T^2)^k\cdot q^{N-2}[N-1]\right) = \nn \\
\boxed{
= q\cdot  q^{(2k-1)N}\left\{q^{-2k}  \frac{1+q^{2N+2}T^3}{1-q^4T^2} +
(qT)^{2k} \left(q^{N-2}[N-1] - \frac{1+q^{2N-2}T}{1-q^4T^2}\right)
\right\}
}
\label{P2even}
\ee
what are the right answers from \cite{DMMSS}.

The crucial point here is that the polynomials for torus knots are just fragments of
the geometrical progression in powers of $q^2T$, with the length which grows linearly in $k$
-- and this the only place (except for a common power of $q^k$) where $k$ enters
the formulas.
As to the additional term, added for links, it also depends on $k$ through the power
of the same $q^2T$.
This is what is responsible for reemergence of the Rosso-Jones-like formula \cite{RJ,DMMSS}
at the level of KR polynomial -- and justifies the simple form of the $k$-evolution,
suggested in \cite{DMMSS,evo}.

\subsubsection{Conversion to MacDonald dimensions and DGR-trick}

Note, however, that formulas in \cite{DMMSS} have very different structure:
there reduced polynomials, as functions of $N$,
were linear combinations of MacDonald dimensions
\be
M_1^* = \frac{\{A\}}{\{q\}} = \frac{q}{A}\cdot\frac{1-A^2}{1-q^2}
= \frac{q}{A}\cdot\frac{1+a^2T}{1-q^2} = \frac{q}{A}\cdot\frac{1+q^{2N}T}{1-q^2}, \nn \\
\frac{M_2^*}{M_1^*}=\frac{\{AqT\}}{\{q^2T\}} = \frac{q}{A}\cdot\frac{1-A^2q^2T^2}{1-q^4T^2}
= \frac{q}{A}\cdot\frac{1+a^2q^2T^3}{1-q^4T^2}= \frac{q}{A}\cdot\frac{1+q^{2N+2}T^3}{1-q^4T^2},
\nn \\
\frac{M_{11}^*}{M_1^*} = \frac{\{A/q\}}{\{q^2\}}=\frac{q}{A}\cdot\frac{q^2-A^2}{1-q^4}
= \frac{q}{A}\cdot\frac{q^2+a^2T}{1-q^4}= \frac{q}{A}\cdot\frac{q^2+q^{2N}T}{1-q^4}
\ee
where $\{x\}= x-x^{-1}$ and $A^2=a^2T=q^{2N}T$.
Clearly, MacDonald dimensions are not polynomials, and their denominators
depend essentially on the Young diagram.
At the same time, in  (\ref{P2odd}) and (\ref{P2even}) there are common denominators,
moreover, the same ones for knots and links.
From (\ref{P2odd}) it immediately follows that
\be
{\cal P}^{[2,2k+1]}_{r} = q^{2Nk}\cdot\frac{A}{q}\left(q^{-2k} \frac{M^*_2}{M^*_1}
- \widetilde{\gamma_{\rm knot}}\cdot(qT)^{2k} \frac{M^*_{11}}{M^*_1}\right)
= {Aq^{2Nk+2}}\left(q^{-2k-1} \frac{M^*_2}{M^*_1}
- \gamma_{\rm knot}\cdot (qT)^{2k+1} \frac{M^*_{11}}{M^*_1}\right)
\label{P2odd1}
\ee
where
\be
\gamma_{\rm knot} = T\,\frac{1-q^4}{1-q^4T^2}, \ \ \ \
\widetilde{\gamma_{\rm knot}} = \gamma_{\rm knot}\cdot q^2T = q^2T^2\,\frac{1-q^4}{1-q^4T^2},
\ee
in agreement with \cite{DMMSS}.

\bigskip

However, for links the situation is more interesting.
Like in (\ref{P2odd}), the r.h.s. of (\ref{P2even}) also has all $k$-dependence
concentrated in two powers, also of $q^{-1}$ and of $qT$ and is nicely described
by $k$-evolution {\it a la} \cite{evo}.
Moreover, the coefficient in front of $q^{-2k}$ is again MacDonald dimension $M^*_2/M^*_1$.
But for the coefficient in front of $(qT)^{2k}$ the story is different:
\be
{\rm for} \ \ \ \  [N-1]=q^{-N}\frac{q^2-q^{2N}}{1-q^2}\ \ \ \ \
{\rm there\ is\ no\ match:} \ \ \ \
q^{N-2}[N-1] - \frac{1+q^{2N-2}T}{1-q^4T^2} \ \neq \
\gamma_{{\rm link}}\cdot\frac{M_{11}^*}{M_1^*}
\ee

with $N$-independent $\gamma$-factor.
An equality would hold, if we substitute on the place of $[N-1]$ another
cohomologically-equivalent expression:
\be
\l[N-1] \ \longrightarrow \ q^{-N}\frac{q^2+q^{2N}T}{1-q^2}
= q^{-N}\frac{q^2-q^{2N}}{1-q^2} + \frac{q^N(1+T)}{1-q^2}
\label{KRtosup00}
\ee
Then
\be
q^{N-2}[N-1] - \frac{1+q^{2N-2}T}{1-q^4T^2} \ \sim\
\frac{q^2+q^{2N}T}{q^2(1-q^2)}  - \frac{1+q^{2N-2}T}{1-q^4T^2}
= \frac{(1-q^2T^2)(q^2+q^{2N}T)}{(1-q^2)(1-q^4T^2)}
\ee
and we would get from (\ref{P2even})
\be
{\cal P}^{[2,2k]}_{r} \sim q\cdot q^{(2k-1)N}\cdot\frac{A}{q}\left(q^{-2k} \frac{M^*_2}{M^*_1}
+ {\gamma_{\rm link}}\cdot(qT)^{2k} \frac{M^*_{11}}{M^*_1}\right)
\label{P2even1}
\ee
with
\be
\gamma_{\rm link} = \frac{(1+q^2)(1-q^2T^2)}{1-q^4T^2},
\ee
again in perfect agreement with \cite{DMMSS}.

In general, the rule to be applied instead of (\ref{KRtosup00}) is (\ref{KRtosup0}),
\be
\boxed{
q^N=a, \ \ \ \ \ \ \l[N-p\,] \sim \frac{q^{2p}+a^2T}{q^{p-1}(1-q^2)a}
}
\label{KRtosup1}
\ee
and it is the way the DGR relation \cite{DGR} between superpolynomials
and KR polynomials is realized in general situation.

\subsubsection{An origin of the $\gamma$-factors}

The second "mystery" which we now understand a little better --
is that of the $\gamma$-factors \cite{DMMSS}.
We see that the origin of $\gamma$-factors is exactly the desire to express the
answer through MacDonald dimensions -- in "natural" variables $\gamma$-factors
are not needed.
The question is what are these natural variables.

We see, that KR polynomials are naturally expressed not through MacDonald dimensions,
but through quantum numbers like $[N-1]$ -- which, moreover, have only negative
shifts. In this respect they are more similar to $M^*_{11}$ -- where shift of $N$
is also negative, -- but denominator in KR polynomials is $1-q^4T^2$, not $1-q^4$,
what is a feature of $M_2^*$ rather than $M^*_{11}$.

Thus MacDonald polynomials once again appeared to be {\it partly} adequate to
description of knot polynomials: they clearly are related, especially for torus knots and links
\cite{DMMSS,Che}, and they are clearly not {\it absolutely} adequate --
as we just saw and as it is well known in the story of {\it colored} polynomials
(see \cite{Slelast} and references therein).

\subsubsection{Unreduced KR polynomials}

Now we can check that unreduced polynomials can also be expressed
through Mac-Donald dimensions -- moreover, after the substitution
(\ref{KRtosup1}) the difference from reduced polynomials disappears(!):
it becomes again just a simple factor $M_1^*$ -- a natural substitute
of the factor $[N]$ for HOMFLY.

Applying (\ref{KRtosup1}) to (\ref{KR2strknotur}) and (\ref{KR2strlinkur}), we get
respectively:
\be
{\cal P}^{[2,2k+1]}
= q^{2k(N-1)}\left([N] + \frac{1}{q}\sum_{j'=1}^{k} (q^2T)^{2j'}\Big(1+q^{2N}T\Big)[N-1]\right)\nn \\
\stackrel{(\ref{KRtosup1})}{\sim}\
q^{(2k-1)(N-1)}\frac{1+q^{2N}T}{1-q^2}\left(1
+ (q^2T^2)\cdot (q^2+q^{2N}T)\frac{1-(q^2T)^{2k}}{1-q^4T^2}\right) = \nn \\
= q^{(2k-1)(N-1)}\frac{A}{q}M_1^*\left\{
\left(1 + \frac{q^2T^2(q^2+q^{2N}T)}{1-q^4T^2}\right)
-(q^2T)^{2k}\cdot\frac{q^2T^2(1-q^4)}{1-q^4T^2}\frac{M_{11}^*}{M_1^*}\frac{A}{q}
\right\} = \nn \\
= q^{(2k-1)(N-1)}\frac{A^2q^{2k}}{q^2}\Big(q^{-2k}M_2^*
- \widetilde{\gamma_{\rm knot}}\cdot(qT)^{2k}M_{11}^*\Big)
\ \stackrel{(\ref{P2odd1})}{=}\
\frac{A}{q^N}\cdot M_1^*P^{[2,2k+1]}_r
\ee
and
\be
{\cal P}^{[2,2k]} =
q^{(2k-1)(N-1)}\left([N]
+\frac{1}{q}\sum_{j' = 1}^{k-1} (q^{2}T)^{2j'} \Big(1+q^{2N}T\Big)[N-1]
+ (q^2T)^{2k}q^{N-2}[N][N-1]\right)
\nn \\
\stackrel{(\ref{KRtosup1})}{\sim}\ q^{2(k-1)(N-1)}\,\frac{1+q^{2NT}}{1-q^2}\left(
\frac{1+q^{2N+2}T^3}{1-q^4T^2} + (q^2T)^{2k}\frac{(1-q^2T^2)(q^2+q^{2N}T)}{(1-q^2)(1-q^4T^2)}
\right) = \nn \\
= q^{2(k-1)(N-1)}\frac{A^2q^{2k}}{q^2}\left(q^{-2k}M_2^*
+ (qT)^{2k}\frac{(1+q^2)(1-q^2T^2)}{1-q^4T^2}M_{11}^*\right)
\ \stackrel{(\ref{P2even1})}{\sim}\
\frac{A}{q^N}\cdot M_1^*P^{[2,2k]}
\ee
i.e. indeed
\be
{\cal P}^{[2,2k+1]}\ \stackrel{(\ref{KRtosup1})}{\sim}\ \boxed{
q^{-N}A\cdot M_1^*{\cal P}^{[2,2k+1]}_r
}, \nn \\
{\cal P}^{[2,2k]}\ \stackrel{(\ref{KRtosup1})}{\sim}\ \boxed{
q^{-N}A\cdot M_1^*{\cal P}^{[2,2k]}_r \ \ \ \ \ \
}
\ee
Note that $A^2=a^2T$ and it is $a$, not $A$, that is substituted by $q^N$ according to
(\ref{KRtosup1}).

\newpage

\section{The $3$-strand braids
\label{3stra}}

\setcounter{equation}{0}

\subsection{Combinations of projectors}

In the 3-strand case there are two different ${\cal R}$-matrices,
associated with the crossings of the first two and the last two strands,
we naturally denote them as ${\cal R}\otimes I$ and $I\otimes {\cal R}$.
These two ${\cal R}$-matrices do not commute, thus HOMFLY polynomial
for generic $3$-strand braid is
\be
H_{_\Box}^{(a_1,b_1|a_2,b_2|\ldots|a_n,b_n)}
\sim \Tr_{[1]\otimes[1]\otimes [1]}\
\Big\{({\cal R}\otimes I)^{b_n}(I\otimes {\cal R})^{a_n}\ \ldots\
({\cal R}\otimes I)^{b_2}(I\otimes {\cal R})^{a_2} \ldots
({\cal R}\otimes I)^{b_1}(I\otimes {\cal R})^{a_1}\Big\}
\ee
For HOMFLY itself the proportionality coefficient is $q^{(a_1+b_1+\ldots a_n+b_n)N}$,
but after the $T$-deformation (\ref{Rdef}) contributions from positive and negative
$a_i$ and $b_i$ will be different.

After the substitution (\ref{Rdef}) we obtain a linear combination of terms,
which include traces of products of alternating projectors:
\be
\Tr_{[1]^{\otimes 3}}\ \ldots (I\otimes P_{11})(P_{11}\otimes I)
\ee
with up to $n$ factors in the product.
To deal with them we should follow the standard procedure, outlined in detail
in \cite{MMMkn1}: decompose the product of three fundamental representations
in two ways:
\be
\begin{array}{cccccc}
\\
&& [1]\otimes[1]\otimes[1] && \\
&\swarrow && \searrow & \\  \\
([1]\otimes[1])\otimes[1] \!\!\!\!\!\!\! &&&& \!\!\!\!\!\!\![1]\otimes([1]\otimes[1])\\
||&&&&|| \\
([2]+\underline{[11]})\otimes [1] \!\!\!\!\!\!\! &&&& \!\!\!\!\!\!\!
\l[1]\otimes([2]+\underline{[11]})\\
|| &&&& || \\
\l[3]+[21]' + \underline{[21]+[111]}\!\!\!\!\!\!\! &&&& \!\!\!\!\!\!\!
\l[3]+\widetilde{[21]'}+\underline{\widetilde{[21]}+[111]}\\
\\ \\
\end{array}
\ee
Projectors $P_{11}\otimes I$ and $I\otimes P_{11}$ are on the underlined
subspaces in the left and the right columns respectively,
where $[111]$ is the same, but $\widetilde{[21]}$ is a linear combination
of $[21]$ and $[21]'$, explicitly described by Racah coefficients,
see \cite{MMMkn1} for details.
Therefore
\be
(I\otimes P_{11})(P_{11}\otimes I) = P_{111} + |\widetilde{[12]}> c <[12]|, \nn \\
<[21]|\widetilde{[21]}>\ = c = \frac{1}{[2]}
\ee
and so on:
\be
(P_{11}\otimes I)(I\otimes P_{11})(P_{11}\otimes I) = P_{111} + |{[21]}> c^2 <[21]|, \nn \\
(I\otimes P_{11})(P_{11}\otimes I)(I\otimes P_{11})(P_{11}\otimes I) =
P_{111} + |\widetilde{[21]}> c^3 <[21]|, \nn \\
\ldots
\ee
Here
\be
\l[21] = [11]\otimes [1] \in \Big([1]\otimes [1]\Big)\otimes [1],\nn \\
\widetilde{[21]} = [1]\otimes [11] \in [1]\otimes \Big([1]\otimes [1]\Big)
\ee

\bigskip

For the $3$-strand torus knots/links $[3,n]$ the braid word is
$(a_1,b_1|a_2,b_2|\ldots|a_n,b_n) = (\underbrace{1,1|1,1|\ldots|1,1}_{n\ {\rm pairs}\ 1,1})$,
and
\be
\P^{[3,n]} = q^{2n(N-1)}\cdot \Tr_{[1]^{\otimes 3}} \
\Big(I\otimes I\otimes I\ + \ (qT)[2](P_{11}\otimes I + I\otimes P_{11})
\ + \ (qT)^2[2]^2 (I\otimes P_{11})(P_{11}\otimes I)\Big)^n
\label{mP3links}
\ee

\subsection{Dimensions $v_{n,k,}(q,N)$ of spaces for three-strand braids}

In the case of three strands we can associate with
each vertex of the hypercube (i.e. with a resolution of
original link/knot diagram) a string
$A^{i_0} B^{j_1} A^{i_1} B^{j_2} \ldots B^{j_k}$,
where all powers $i_0,j_1,i_1,\ldots, j_k>0$ and the sum
$i_0+j_1+i_1+\ldots +j_k = n$, i.e.
$n$ is the number of non-trivially resolved (white) crossings,
while $k$ is the number of jumps from $A$ to $B$ resolutions.
Since in our calculations $A$ and $B$ are actually substituted by
projectors times $[2]$, expression for above string is almost
independent of the powers $i_0,\ldots,j_k$; it equals $[2]^m$ times the expression for $(AB)^k$,
which depends only on $k$. We denote the contribution of this string by $v_{n,k}$.

There are $s_{n,k}^{\cal L}$ vertices of each type,
and the contribution of each vertex is $v_{n,k}$,
so that the primary deformed reduced polynomial is
\be
\boxed{
\P^{\cal L}_r(q,T,N) = \sum_{n,k} s_{n,k}^{\cal L} v_{n,k}(q,N)\cdot (qT)^n
= \sum_n (qT)^n \cdot {\rm dim}_n^{\cal L}(q,N)
}
\label{Psv}
\ee
Quantum (graded) dimensions $v_{n,k}$ depend on parameters
$(q,T,N)$, but are independent of
the choice of the $m=3$-strand knot/link.
They are equal to:
\be
\boxed{\begin{array}{l}
v_{0,0}=[N]^2,\\[1.5mm]
v_{n,0}=[2]^nD_{11} = [2]^{n-1}[N][N-1],\ \ \ \ \ n>0,\\[1.5mm]
v_{n,k}=   [2]^n \frac{ D_{111} +[2]^{-2k}D_{21}}{[N]}
= [2]^{n-2k}[N-1]\,\frac{[2]^{2k-1}[N-2]+[N+1]}{[3]} =  \\
\ \ \ = [2]^{n-2k}[N-1]\left\{[N-1] \ +\
 \left(\sum_{i=0}^{k-2}\ [2]^{2i}\right)\cdot [2][N-2]\right\},
\ \ \ \ \  0<2k\le n
\end{array}}
\label{dim3strands}
\ee

\subsection{Torus links and knots:
KR deformation of $H^{[3,n]}=\Tr_{\Box^{\otimes 3}}\Big({\cal R}_1{\cal R}_2\Big)^n$
and alike
\label{3stratorus}
}

\subsubsection{Cycles diagram}

In the case of three-strand torus knots all
the hypercube vertices, i.e. colorings of the knot/link diagram,
can be labeled by telling the positions of white vertices on the
intersections of the first two and the last two strands:
hypercube vertices are labeled by $\Big(i_1<\ldots <i_k\Big| j_1<\ldots <j_l\Big)$

As explained in \cite{DM3},
the classical (at $q=1$) values of dimensions $v$ are defined by the numbers of connected
cycles, appearing in the corresponding resolution of the link diagram --
and in our present case these numbers depend only on $k$ and $l$, on the
numbers of white vertices at two kinds of intersections:
$$
\begin{array}{c|ccc}
k & l && \#({\rm cycles}) \\
\hline
{\rm even} &  {\rm even}&& 3 \\
{\rm even}&{\rm odd}&& 2 \\
{\rm odd}&  {\rm even}&&2  \\
{\rm odd}&{\rm odd}&&1       \\
\end{array}
$$
Dimensions $v$ are constructed from these numbers  by taking appropriate
linear combinations, reflecting the structure of the corresponding factor-spaces
\cite{DM3}.
Moreover, for torus knots and links, these numbers can be easily quantized.
In the following simple examples, we show how the numbers of cycles and then
the dimensions $v$ are ascribed to  hypercube vertices.
A universal technique providing the rigorous quantization rules
will be described in the next subsection \ref{rigquant}.

\bigskip

{\bf The unknot $\l[3,1]:$}
$$
\begin{array}{ccccc}
&&2&& \\
3&&&& 1 \\
&& 2 &&\\
\end{array}
\ \ \ \ \ \ \ \ \ \ \ \ \ \
\begin{array}{ccccc}
&&\l[N]^2[N-1]&& \\
\l[N]^3&&&& \l[N][N-1]^2 \\
&& \l[N]^2[N-1] &&\\
\end{array}
$$

\bigskip

{\bf The trefoil $\l[3,2]:$}
$$
\arraycolsep=0.5mm
\begin{array}{cccccccccc}
&&&&3  \\
&&2&&1&&2 \\
&&2&&1&&2 \\
3&&&&&&&& 1 \\
&&2&&1&&2 \\
&&2&&1&&2 \\
&&&&3  \\
\end{array}
\ \ \ \ \
\begin{array}{cccccccccccc}
&&&&\l[2][N]^2[N-1] \\
&&\l[N]^2[N-1]&&\l[N][N-1]^2 && \l[2][N][N-1]^2 \\
&&\l[N]^2[N-1]&&\l[N][N-1]^2 && \l[2][N][N-1]^2 \\
&&&&&&&&  \l[N] [N-1]^2  \\
\l[N]^3 &&&&&&&& \oplus \\
&&&&&&&&  [2][N][N-1][N-2] \\
&&\l[N]^2[N-1]&&\l[N][N-1]^2 && \l[2][N][N-1]^2 \\
&&\l[N]^2[N-1]&&\l[N][N-1]^2 && \l[2][N][N-1]^2 \\
&&&&\l[2][N]^2[N-1] \\
\end{array}
$$



{\bf The 3-component link $[3,3]$}

The cycle diagram:

\bigskip

$$
\arraycolsep=1mm
\begin{array}{ccccccccccccccccccc}
&&&&&&2\\
\\
&&&&&&2\\
&&&&&&2\\
&&&&3&&2&&1  \\
&&&&3&&2&&1  \\
&&&&3&&2&&1  \\
\\
&&&&1&&2&&1  \\
&&2&&1&&2&&1&&2 \\
&&2&&1&&2&&1&&2 \\
&&2&&1&&2&&3&&2 \\
3&&&&1&&&& 3&&&&3 \\
&&2&&1&&2&&3&&2 \\
&&2&&1&&2&&1&&2 \\
&&2&&1&&2&&1&&2 \\
&&&&1&&2&&1  \\
\\
&&&&3&&2&&1  \\
&&&&3&&2&&1  \\
&&&&3&&2&&1  \\
&&&&&&2\\
&&&&&&2\\
\\
&&&&&&2\\
\end{array}
\ \ \ \ \ \ \ \ \ \ \ \
\begin{array}{ccccccccccccccccccc}
&&&&&&\pl^3\\
\\
&&&&&&\pl^2\pr\\
&&&&&&\pl^2\pr\\
&&&&\pl^2&&\pl^2\pr&&\pl^3\pr  \\
&&&&\pl^2&&\pl^2\pr&&\pl^3\pr  \\
&&&&\pl^2&&\pl^2\pr&&\pl^3\pr  \\
\\
&&&&\pl\pr&&\pl^2\pr&&(\pl\pr)^2  \\
&&\pl&&\pl\pr&&\pl^2\pr&&(\pl\pr)^2&&\pl(\pl\pr)^2 \\
&&\pl&&\pl\pr&&\pl^2\pr&&(\pl\pr)^2&&\pl(\pl\pr)^2 \\
&&\pl&&\pl\pr&&\pl^2\pr&&\pl^2\pr^2&&\pl(\pl\pr)^2 \\
I&&&&\pl\pr&&&& \pl^2\pr^2&&&&(\pl\pr)^3 \\
&&\pr&&\pl\pr&&\pl\pr^2&&\pl^2\pr^2&&(\pl\pr)^2\pr \\
&&\pr&&\pl\pr&&\pl\pr^2&&(\pl\pr)^2&&(\pl\pr)^2\pr \\
&&\pr&&\pl\pr&&\pl\pr^2&&(\pl\pr)^2&&(\pl\pr)^2\pr \\
&&&&\pl\pr&&\pl\pr^2&&(\pl\pr)^2  \\
\\
&&&&\pr^2&&\pl\pr^2&&\pl\pr^3  \\
&&&&\pr^2&&\pl\pr^2&&\pl\pr^2  \\
&&&&\pr^2&&\pl\pr^2&&\pl\pr^3  \\
&&&&&&\pl\pr^2\\
&&&&&&\pl\pr^2\\
\\
&&&&&&\pr^3\\
\end{array}
$$

\bigskip

\noindent
In the right table we explicitly wrote the sequences of resolutions
$\pl = P_{11}\otimes I$ and $\pr = I\otimes P_{11}$,
associated with the hypercube vertices (up to cyclic order,
in order not to further overload the notation).
An exhaustively detailed notation would be
$(\pl\cdot I_r\cdot \pl\cdot \pr \cdot I_l\cdot I_r)$ where
brackets denote the cyclic class.
Then edges of hypercube connect vertices, differing by a single substitution
$I_l \leftrightarrow \pl$ or $I_r\leftrightarrow \pr$.
However, we abbreviate the notation to just $\pl^2\pr$,
so that the pattern of edges gets a little more obscure in this table.
Note that $\pl$ and $\pr$ are resolutions, not projectors,
therefore one can not substitute $A^2$ by $A$.

The next step is the table of quantum dimensions of graded spaces
at the hypercube vertices:

\bigskip

\be
&&\l[N]^3 \ \ +\ \  \l[N][N-1]\times \nn\\[2mm]&&\times\
\boxed{
\begin{array}{ccccccccccccccccccc}
\\
&&&&&&[2]^2[N]\\
\\
&&&&&&[2][N-1]\\
&&&&&&[2][N-1]\\
&&&&[2][N]&&[2][N-1]&&[2]^2[N-1]  \\
&&&&[2][N]&&[2][N-1]&&[2]^2[N-1] \\
&&&&[2][N]&&[2][N-1]&&[2]^2[N-1]  \\
\\
&&&&[N-1]&&[2][N-1]&&2[N-1]+[N-3]  \\
&&[N]&&[N-1]&&[2][N-1]&&2[N-1]+[N-3]&&[2]\big(2[N-1]+[N-3]\big) \\
&&[N]&&[N-1]&&[2][N-1]&&2[N-1]+[N-3]&&[2]\big(2[N-1]+[N-3]\big) && \\
&&[N]&&[N-1]&&[2][N-1]&&[2]^2[N-1]&&[2]\big(2[N-1]+[N-3]\big) && [N+1]\oplus\\
&&&&[N-1]&&&& [2]^2[N-1]&&&&    5[N-1]\oplus
\\
&&[N]&&[N-1]&&[2][N-1]&&[2]^2[N-1]&&[2]\big(2[N-1]+[N-3]\big)&& 4[N-3]\oplus\\
&&[N]&&[N-1]&&[2][N-1]&&2[N-1]+[N-3]&&[2]\big(2[N-1]+[N-3]\big)&& [N-5] \\
&&[N]&&[N-1]&&[2][N-1]&&2[N-1]+[N-3]&&[2]\big(2[N-1]+[N-3]\big) \\
&&&&[N-1]&&[2][N-1]&&2[N-1]+[N-3]  \\
\\
&&&&[2][N]&&[2][N-1]&&[2]^2[N-1]  \\
&&&&[2][N]&&[2][N-1]&&[2]^2[N-1]  \\
&&&&[2][N]&&[2][N-1]&&[2]^2[N-1]  \\
&&&&&&[2][N-1]\\
&&&&&&[2][N-1]\\
\\
&&&&&&[2]^2[N]\\  \\
\end{array}}
\nn\ee

\bigskip

One can see that already in these examples some dimensions $v$ are quantized
in a rather sophisticated way.
The general technique for doing this is exactly the main suggestion of
the present paper, and it and is provided by consideration of primary $T$-deformed
polynomials.

\subsubsection{Primary $T$-deformation $\P^{[3,n]}$ for generic $3$-strand torus knot/link
\label{rigquant}}

The primary polynomial is given by a trace of the following product with $x=qT[2]$:
\be
q^{-2n(N-1)}\P^{[3,n]} = \Tr_{[1]^{\otimes 3}} \ \left\{
\Big(I^{\otimes 3}+x\cdot I\otimes P_{11}\Big)
\Big(I^{\otimes 3}+x\cdot P_{11}\otimes I\Big)\right\}^n = \nn \\
= D_{3}\ +\ 2D_{21}\cdot\left(\sum_{j=0}^n  \frac{\prod_{i=0}^{j-1} (n^2-i^2)}{(2j)!}(cx)^{2j}
(1+x)^{n-j}\right)   \ +\ D_{111}\cdot(1+x)^{2n} =
\label{P3n}
\ee
\be
= [N]\left\{\frac{[N+1][N+2] + \big(1+qT[2]\big)^{2n}[N-1][N-2]}{[2][3]}
\ +\right.\nn\\\left.+\ \frac{2[N-1][N+1]}{[3]} \cdot\left(\sum_{j=0}^n  \frac{n(n+j-1)!}{(2j)!(n-j)!}(qT)^{2j}
\big(1+qT[2]\big)^{n-j}\right)\right\}=
\nn\ee
\be
= [N]\left\{ [N]^2\ +\ 2n[N][N-1](qT)\ +\
\underbrace{n\Big((n-1)[N+1]+(2n-1)[N-1]\Big)}_{n(n-1) [2][N]+n^2[N-1]}[N-1](qT)^2\ +
\phantom{5^{5^{5^{5^{5^5}}}}}\right.
\nn\ee
\be
+ \underbrace{\frac{n(n-1)}{3}\Big((n-2)[N+1]+ 2(2n-1)[N-1]\Big)}_{
\frac{n(n-1)(n-2)}{3}[2][N]+n^2(n-1)[N-1]}[2][N-1](qT)^3 +\nn\\
+ \frac{n(n-1)}{12}\Big( (n-2)(n-3)[N+3] + 9(n-1)(n-2)[N+1] +
\big(18(n-1)^2 - n(n+1)\big)[N-1] +\nn\\+ 2(2n-1)(2n-3)[N-3]\Big) [N-1](qT)^4 +
\ldots \Big\}
\nn\ee


For $T=-1$ this coincides with \cite{RJ,DMMSS}
\be
q^{2n}[N]\left\{
q^{-2n}\frac{[N+1][N+2]}{[2][3]}
- \frac{[N+1][N-1]}{[3]}
+ q^{2n}\frac{[N-1][N-2]}{[2][3]}\right\}
& {\rm for\ knots}, & n = 3k\pm 1, \nn \\
q^{2n}[N]\left\{ q^{-2n}\frac{[N+1][N+2]}{[2][3]} + 2\cdot\frac{[N+1][N-1]}{[3]}
+ q^{2n}\frac{[N-1][N-2]}{[2][3]}\right\}
& {\rm for\ links}, & n = 3k
\ee
because $[N\pm k]= \frac{\{Aq^{\pm k}\}}{\{q\}}$ and $[k] = \frac{\{q^k\}}{\{q\}}$,
where $\{x\} = x - x^{-1}$.


\subsubsection{Multiplicities $s_{m,k}^{\cal L}$ for the {\it torus}
links/knots $[3,n]$}

In variance with $v_{k,m}$, the
multiplicities $s_{m,k}^{\cal L}$ depend on ${\cal L}$.
Now we describe them for the torus links/knots $[3,n]$,
thus they depend on $n$:

\be
s_{m,0} = C^m_nC^0_n + C^0_nC^m_n = 2\,C^m_n = \frac{2\cdot n!}{m!(n-m)!},
& 0\leq m \leq n, \nn \\
s_{m,1} = n^2\,C^{m-2}_{n-1} = \frac{n\cdot n!}{(m-2)!(n-m+1)!},
& 2\leq m \leq n+1, \nn \\
s_{m,2} = \frac{n^2(n^2-1)}{12}\,C_{n-2}^{m-4}, & 4\leq m \leq n+2, \nn \\
s_{m,3} = \frac{n^2(n^2-1)(n^2-4)}{360} \, C_{n-3}^{m-6}, & 6 \leq m \leq n+3, \nn \\
\ldots \nn \\ \nn \\
\boxed{\ s_{m,k}
= 2\,\frac{\prod_{i=0}^{k-1} (n^2-i^2)}{(2k)!} \, C_{n-k}^{m-2k}
= \frac{2n\cdot(n+k-1)!}{(2k)!\,(m-2k)!\,(n+k-m)!},\ } &2k\leq m \leq n+k\
\ee

\bigskip

Thus dimensions (\ref{Psv}) of the spaces in KR complex for 3-strand torus knots are:
\be
{\rm dim}_m = \sum_{k=n-m}^{m/2} s_{m,k}v_{m,k} =
[N-1]\left\{ [N-1]\cdot \left(2q[2]^{m-1}C^m_n + \sum_{k=1}^{m/2} [2]^{m-2k}s_{m,k}\right) +
\right. \nn \\ \left.+
[N-2]\cdot\left(\sum_{k=2}^{m/2} [2]^{m+1-2k}\left(\sum_{i=0}^{k-2}\ [2]^{2i}\right)s_{m,k}\right)
+ q^{-N} \cdot 2q[2]^{m-1}\, C^m_n\right\},
\ \ \ \ \ m>0
\label{dims3n}
\ee
For $m=0$ the dimension is just ${\rm dim}_0 = [N]^2 = [N-1]^2 + 2q^{2-N}[N-1] +q^{2-2N}$.
With these values of $s_{m,k}$ eq.(\ref{Psv}) reproduces the primary polynomial (\ref{P3n}).
However, now it is decomposed into three parts, what is convenient to reveal
the general structure of cohomologies for all $n$.

\subsubsection{Unknot $[3,1]$, $n=1$}

In this case the dimensions of spaces in KR complex
$\ \ v_{00} \ \stackrel{d_0^{\bullet\bullet}}{\longrightarrow}\ 2v_{10}
\ \stackrel{d_1^{\bullet\bullet}}{\longrightarrow}\  v_{21}\ \ $ are:
\be
\dim_0=&v_{0,0}&=[N]^2\nn\\[1.5mm]
\dim_1=&2v_{1,0}&=2[N][N-1]\nn\\[1.5mm]
\dim_2=&v_{2,1}&=[N-1]^2\nn\\[1.5mm]
\ee
(with a minor abuse of notation we denote the constituent spaces of the complex and
their dimensions by the same letters $v_{mk}$, and identify $\oplus$ with $+$).

Since $[N]^2 = q^2[N-1]^2 + 2q^{2-N}[N-1] + q^{2-2N}$
and $[N][N-1] = q[N-1]^2 + q^{1-N}[N-1]$, we have the following
decomposition of $\P^{[3,1]}_r$:

\be
q^{2-2N}\P^{[3,1]}_r  = v_{00}+2v_{10}(qT) +v_{21}(qT)^2
= [N]^2 + 2[N][N-1]\,(qT) + [N-1]^2\,(qT)^2 = \nn \\
\nn \\
= \boxed{q^{2-2N}}\ +\ [N-1]\times
\left.\begin{array}{|cc|cccccccc }
\hline
&&\\
\l[N-1]&\times & q^2  \lar 2q  \lar 1 \\ &&\\
q^{-N}&\times  & 2q^{2 } \lar 2q
\\ && \\
\hline
\end{array}\right|
\ee
where the columns of the table are multiplied by powers of $(qT)$
and arrows stand for the cohomologically trivial sub-complexes of graded spaces,
while  cohomologies are boxed.
In more complicated examples below we denote by $\Longrightarrow$ the differentials,
which are not cohomologically trivial, and separate by horizontal lines
the cohomologically trivial and non-trivial chains.
Double horizontal lines will separate transformation steps of the set of complexes,
targeted at localizing its cohomologies.

Since in the case of $[3,1]$ the only cohomologically non-trivial item is
the boxed $q^{2-2N}$, we obtain:
\be
{\cal P}^{[3,1]}_r = q^{2N-2}\cdot q^{2-2N} = 1
\ee
for reduced KR polynomial.

Since cohomologically trivial complex remains cohomologically trivial
after multiplication by $[N]$,  we get
\be
{\cal P}^{[3,1]} = [N]
\ee
for the unreduced one.

\subsubsection{Another initial vertex}

In fact, one and the same hypercube describes link diagram ${\cal L}$
with an arbitrary coloring of vertices -- the graded vector spaces
at the hypercube vertices are the same.
What differs is the choice of initial vertex and morphisms along the edges,
which all point away from initial vertex.
In particular, the hypercubes for the two series
$\Big({\cal R}_1{\cal R}_2\Big)^n$ and $\Big({\cal R}_1{\cal R}_2^{-1}\Big)^n$
are the same -- and it is instructive to consider them together.

In particular, the just discussed representation for the unknot $[3,1] = (1,1)$
can be compared with another representation for the unknot, $(1,-1)$,
which gives rise to by the same hypercube, but different morphisms, KR complex
and differentials.
This time the complex  is
$\ \ v_{10} \ \stackrel{d_0^{\bullet\circ}}{\longrightarrow}\  v_{00}\oplus v_{21}
 \ \stackrel{d_1^{\bullet\circ}}{\longrightarrow}\  v_{10}\ \ $
and
\be
\frac{q^NT}{q^{N-1}}\,\P^{1,-1}_r =
q^{2-2N}(qT)\ +\ [N-1]\times
\left.\begin{array}{|cc|cccccccc }
\hline
&&\\
\l[N-1]&\times & q\lar  q^2+1  \lar q  \\ &&\\
q^{-N}&\times  & q\Lar 2q^{2 } \lar q
\\ && \\
\hline \hline
&&\\
q^{-N}&\times & q \Lar q^2
\\ && \\
\hline
\end{array}\right|
\ee
Now, $q[N-1] \Longrightarrow q^2[N-1]$ is cohomologically equivalent to
$q^{3-N} \Longrightarrow q^N$, this should be further multiplied by $q^{-N}$,
Afterwards we add the term $q^{2-2N}$ and get:
$q^{3-2N} \Longrightarrow  1 + q^{2-2N}$, which is equivalent to just $0 \Longrightarrow 1$.
Thus we obtain
\be
{\cal P}^{1,-1}_r = \frac{1}{qT}\, \Big(1\cdot (qT)\Big) = 1
\ee
while unreduced ${\cal P}^{1,-1}=[N]$.

\subsubsection{Trefoil $[3,2]$}

The KR complex is
\be
v_{00}\ \stackrel{d_0}{\longrightarrow}\ 4v_{10} \ \stackrel{d_1}{\longrightarrow}\
2v_{20} + 4v_{21} \ \stackrel{d_2}{\longrightarrow}\  4v_{31}
\ \stackrel{d_3}{\longrightarrow}\  v_{42}
\label{32comp}
\ee
and the generating function of its graded dimensions
\be
\dim_0=&v_{0,0}&=[N]^2\nn\\[1.5mm]
\dim_1=&4v_{1,0}&=4[N][N-1]\nn\\[1.5mm]
\dim_2=&2v_{2,0}+4v_{2,1}&=2[2][N][N-1]+4[N-1]^2\nn\\[1.5mm]
\dim_3=&4v_{3,1}&=4[2][N-1]^2\nn\\[1.5mm]
\dim_4=&v_{4,2}&=[N-1]^2+[2][N-1][N-2]\nn
\ee
-- the primary $T$-deformation of HOMFLY -- can be written in the form
of the following table:
\be
q^{4-4N}{\cal P}^{[3,2]}_r =
\boxed{q^{2-2N}}\ +\ [N-1]\times
\left|\begin{array}{cc|ccccccccccc}
\hline && \\
\l[N-1] & \times & && 2q^2 & \underline{4q} & 1 &&&& (-1) \\
\l[N-2] & \times & && && \underline{q} &&&& (+1) \\ && \\
\l[N-1] & \times & q^2 & 4q & 6 & \underline{\frac{4}{q}} & &&&&(-1)\\
\l[N-2] & \times &&&&& \underline{\frac{1}{q}} &&&& (+1) \\ && \\
q^{-N} & \times & 2q^2 & 4q & 2 && &&&& (0) \\ &&\\
q^{-N} & \times &&& \underline{\underline{2q^2}} && &&&& (2) \\ &&\\
\hline
\end{array}\right|
\ee
We added the last column -- it lists the alternated sums of coefficients
in every line.
Then we underlined the terms, which are selected to
contribute to the cohomologies
(multiplicity of the contribution is expressed by the number of underlines).
This selection is not always fully unique.
We adjust it to match the right answers, when they are available
(for the simple links and knots).
After that we require that selections follow some general
rule within particular series of links/knots --
as it is usually done  in the evolution method \cite{evo}.
The freedom is in fact severely restricted by the minimality of KR
polynomial: it requires the absence of gaps between non-vanishing
cohomologies.
Thus the underlined items from adjacent lines are in adjacent columns,
what allows to substitute $[N-1] \Longrightarrow [N-2]$   by
$q^{2-N}\Longrightarrow 0$.
In fact this also restricts the choice of underlined
items in non-adjacent lines
(we shall see examples below).

Now we iteratively extract the cohomologically non-trivial part.
First -- pick it up from the underlined items:
\be
\boxed{q^{2-2N}}\ +\ [N-1]\times
\left|\begin{array}{cc|ccccccccccc}
\hline &&\\
\l[N-1] &\times &&& 2q^2 &3q & 1 \\ && \\
\l[N-1] &\times q^2 &4q & 6 & \frac{3}{q} \\&& \\
q^{-N}&\times & 2q^2 & 4q & 2 \\ && \\
\hline &&\\
\l[N-1] &\times &&&& [2] &  \\
\l[N-2] & \times &&&&& [2] \\
q^{-N} & \times & && 2q^2 \\ && \\
\hline
\end{array}\right|
\ee
The upper part of the table is cohomologically trivial,
the lower is not.
Still, since $\ [N-1] \Longrightarrow [N-2]\ $ is cohomologically equivalent
to $q^{2-N} \Longrightarrow 0$, it can be further diminished: first to
\be
\boxed{q^{2-2N}}\ +\ [N-1]\times
\left|\begin{array}{cc|ccccccccccc}
\hline &&\\
q^{-N} & \times & && 2q^2 & q^2[2] & \\ && \\
\hline
\end{array}\right|
\ee
and then to
\be
\boxed{q^{2-2N}}\ +\ [N-1]\times
\left|\begin{array}{cc|ccccccccccc}
\hline && \\
q^{-N} & \times & && q^2 & q^3 &  \\ && \\
\hline
\end{array}\right|
\ee
At the last step we recall the common factor $[N-1]$
(and $[N][N-1]$ in unreduced case)
and further substitute
\be
\l[N-1]\Big(q^2 \Longrightarrow q^3\Big) \ \ \ {\rm by} \ \ \
q^{4-N} \Longrightarrow q^{N+1}
\label{2n3n}
\ee
and
\be
\l[N][N-1]\Big(q^2 \longrightarrow q^3\Big)  \ \ \ {\rm by} \ \ \
q^{3-N}[N-1] \longrightarrow q^{N+2}[N-1]
\label{2n3nur}
\ee
Thus
\be
\boxed{{\cal P}^{[3,2]}_r= q^{4N-4} \Big\{
q^{2-2N} + q^{4-2N}(qT)^2 +q\cdot(qT)^3\Big\} =
q^{2N-2}\Big\{1 + q^4T^2 + q^{2N+2}T^3\Big\}}
\label{KR32r}
\ee
and
\be
\boxed{{\cal P}^{[3,2]}=
q^{2N-2} \Big\{[N] + q^3[N-1]\,T^2 + q^{2N+3}[N-1]\,T^3\Big\} }
\label{KR32}
\ee
Note that ${\cal P}^{[3,2]}\neq [N]{\cal P}^{[3,2]}_r$ -- this is because
after multiplication by $[N]$ the irreducible cohomologically non-trivial
complex can acquire a cohomologically trivial sub-complex,
and thus cohomologies can get smaller -- and KR polynomial gets "smaller":
the difference $[N]{\cal P}_r - {\cal P}$ is a non-vanishing positive polynomial.

\subsubsection{The figure-eight knot $4_1$, $\Big({\cal R}_1{\cal R}_2^{-1}\Big)^2$}

This knot has the same three-strand knot diagram as the trefoil,
but with two vertices of different color.

The KR complex is
\be
v_{20}\ \stackrel{d_0}{\longrightarrow}\ 2v_{10}+2v_{31} \ \stackrel{d_1}{\longrightarrow}\
v_{00} + 4v_{21} +v_{42} \ \stackrel{d_2}{\longrightarrow}\  2v_{10}+2v_{31}
\ \stackrel{d_3}{\longrightarrow}\  v_{20}
\ee
with the same dimensions $v_{mk}$, and
\be
\left(\frac{q^{N-1}}{q^N T}\right)^{-2}\!\!\!\cdot
{\cal P}^{4_1}_r =
\boxed{q^{2-2N}}\cdot (qT)^2\ +
\ \ \ \ \ \ \ \ \ \ \ \ \ \ \ \ \ \ \ \ \ \ \ \ \ \ \ \ \ \ \ \ \ \ \
\ \ \ \ \ \ \ \ \ \ \ \ \ \ \ \ \ \ \ \ \ \ \ \ \ \ \ \ \ \ \ \ \ \ \
\nn \\
+ [N-1]\times
\left|\begin{array}{cc|ccccccccccc}
\hline && \\
\l[N-1] & \times & q^2&2q&1&2q&q^2       &&&& (-1) \\
\l[N-2] & \times & &&{q}&&    &&&&         (+1) \\ && \\
\l[N-1] & \times & 1 &2q+\frac{2}{q}&q^2+4&2q+\frac{2}{q}&1    &&&&(-1)\\
\l[N-2] & \times & &&\frac{1}{q}&&             &&&& (+1) \\ && \\
q^{-N} & \times &  1&2q&2q^2&2q&1           &&&& (0) \\ &&\\
q^{-N} & \times &  q^2 &&&&q^2      &&&& (2) \\ &&\\
\hline
\end{array}\right|
\ee
Since the table differs by permutation of columns,
the alternated sums are the same as for the trefoil.
Now the doubling of lines does not make much sense, and we begin from rewriting
the table as
\be
\l[N-1]\times
\left|\begin{array}{cc|ccccccccccc}
\hline && \\
\l[N-1] & \times & q^2+1 &4q+\frac{2}{q}&q^2+5&4q+\frac{2}{q}&q^2+1    &&&&(-2)\\
\l[N-2] & \times & &&q+\frac{1}{q}&&             &&&& (+2) \\ && \\
q^{-N} & \times &  q^2+1&2q&2q^2&2q&q^2+1           &&&& (2) \\ &&\\
\hline
\end{array}\right|
\ee
Extracting the trivial exact subsequences, we get
\be
\l[N-1]\times
\left|\begin{array}{cc|ccccccccccc}
\hline && \\
\l[N-1] & \times & q^2&3q&2\\
\l[N-1] & \times & && q^2 &2q & 1 \\
\l[N-1] & \times & 1 & \frac{1}{q} \\
\l[N-1] &\times & && 2 &\frac{2}{q}\\&&\\
\hline
&& \\
\l[N-1] & \times & &q+\frac{1}{q}&1&2q &q^2   \\
\l[N-2] & \times & &&q+\frac{1}{q}&&              \\ && \\
q^{-N} & \times &  q^2+1&2q&2q^2&2q&q^2+1           \\ &&\\
\hline
\end{array}\right|
\ee
Cohomologically non-trivial contributions are in the lower part of the table.
The next transformations involve the $N-1$ factor, which is inside the table:
$(q+1/q)[N-1]\Longrightarrow (q+1/q)[N-2]$ is cohomologically equivalent to
$(q+1/q)\cdot q^{2-N}\Longrightarrow 0$,
while $[N-1]\Big(1 \Longrightarrow 2q \Longrightarrow q^2\Big)$  -- to
$\ q^{2-N} \Longrightarrow q^{N-1} + q^{3-N} \Longrightarrow q^N$.
This means that the next iteration is
\be
\l[N-1]\times
\left|\begin{array}{cc|ccccccccccc}
\hline &&\\
q^N & \times &&&& q^{-1} & 1\\&&\\
q^{-N} & \times &  q^2+1&2q+q^2\left(q+\frac{1}{q}\right)&2q^2+q^2&2q+q^3&q^2+1 \\ &&\\
\hline
\end{array}\right|
\ee
and the second line is immediately reduced to
\be
\l[N-1]\times
\left|\begin{array}{cc|ccccccccccc}
\hline &&\\
q^{-N}&\times & q^2 & q \\
q^{-N}&\times &&q^3 & 2q^2 & q \\
q^{-N} &\times &&& q^3+q & q^2+1 \\ &&\\
\hline &&\\
q^N & \times &&&& q^{-1} & 1\\&&\\
q^{-N} & \times &  1&2q&q^2  \\ &&\\
\hline
\end{array}\right|
\ee
Substituting once again
\be
\l[N-1]\Big(1 \Longrightarrow 2q \Longrightarrow q^2\Big)
\ \ \ \ \ \ {\rm  by} \ \ \ \ \ \
 q^{2-N} \Longrightarrow q^{N-1} + q^{3-N} \Longrightarrow q^N
 \nn \\
\frac{1}{q}[N-1] \Longrightarrow [N-1]
\ \ \ \ \ \ {\rm by}\ \ \ \ \ \ q^{1-N}\longrightarrow q^{N-2}
\label{fesubsr}
\ee
we obtain:
\be
(qT)^2 {\cal P}^{4_1}_r \sim
q^{2-2N} (qT)^2 +
\left|\begin{array}{cc|ccccccccccc}
\hline &&\\
q^N & \times &&&& q^{1-N} & q^{N-2}\\&&\\
q^{-N} & \times &  q^{2-N}&q^{N-1}+q^{3-N}&q^N  \\ &&\\
\hline
\end{array}\right| = \nn \\ \nn \\
= \underline{q^{2-2N}} (qT)^2 +
q^{2-2N} + \left(q^{-1}+\underline{q^{3-2N}}\right)(qT)+(qT)^2 + q\cdot(qT)^3 + q^{2N-2}(qT)^4
\label{Pferprom}
\ee
It remains to note that the two underlined items form a cohomologically trivial pair.
Throwing it away, we finally obtain
\be
\boxed{{\cal P}^{4_1}_r = \frac{1}{q^{2N}T^2} + \frac{1}{q^2T} + 1 + q^2T + q^{2N}T^2}
= 1 + \left(q^{N+1}T^{3/2}+\frac{1}{q^{N+1}T^{3/2}}\right)
\left(q^{N+1}T^{1/2}+\frac{1}{q^{N+1}T^{1/2}}\right)
\label{KR41r}
\ee
The last expression is nothing but the superpolynomial $1+\{Aq\}\{A/t\}$ of \cite{IMMMfe,evo},
expressed in our current notation.

\bigskip

In the case of unreduced polynomial we need to change the substitution rules
(\ref{fesubsr}), as usual: now we substitute
\be
\l[N][N-1]\Big(1 \Longrightarrow 2q \Longrightarrow q^2\Big)
\ \ \ \ \ \ {\rm  by} \ \ \ \ \ \
 \l[N-1]\Big(q^{1-N} \Longrightarrow q^{N} + q^{2-N} \Longrightarrow q^{N+1}\Big)
 \nn \\
\frac{1}{q}[N][N-1] \Longrightarrow [N][N-1]
\ \ \ \ \ \ {\rm by}\ \ \ \ \ \ [N-1]\Big(q^{-N}\longrightarrow q^{N-1}\Big)
\label{fesubs}
\ee
The outcome acquires factors $[N-1]$ rather than $[N]$, and also the powers of $q$
are shifted by $\pm 1$.
At the same time the item $q^{2-2N}$ in front of the table gets multiplied by $[N]$.
This means that instead of (\ref{Pferprom}) for unreduced polynomial we get
\be
(qT)^2 {\cal P}^{4_1} \sim
q^{2-2N}[N] (qT)^2 + [N-1]\times
\left|\begin{array}{cc|ccccccccccc}
\hline &&\\
q^N & \times &&&& q^{-N} & q^{N-1}\\&&\\
q^{-N} & \times &  q^{1-N}&q^{N}+q^{2-N}&q^{N+1}  \\ &&\\
\hline
\end{array}\right| = \nn \\ \nn \\
= \underline{q^{2-2N}}[N] (qT)^2 +
\l[N-1]\Big(q^{1-2N} +
(1+\underline{q^{2-2N}} )(qT)+q\cdot(qT)^2 + (qT)^3 + q^{2N-1}(qT)^4 \Big)
\label{feprom}
\ee
The two underlined terms  $\ q^{2-2N}\Big( [N-1] \Longrightarrow [N]\Big)\ $
are cohomologically equivalent to $\ 0 \Longrightarrow q^{1-N}\ $,
and $q^{1-N}+q[N-1] = [N]$, so that finally
\be
\boxed{{\cal P}^{4_1} = [N]
\ +\ [N-1]\left(\frac{1}{q^{2N+1}T^2} + \frac{1}{qT} + qT + q^{2N+1}T^2\right)}
\label{KR41ur}
\ee
This is in accordance with \cite{CM}.

\subsubsection{Other initial vertices: three more representations for unknot}

There are three other possible choices for initial vertex: $v_{21}$, $v_{10}$
and $v_{31}$ -- all the three lead to unknot. If $v_{42}$ is taken for initial
vertex, we get a "dual" representation of the trefoil with initial vertex $v_{00}$.
It is instructive to briefly present these examples.

\bigskip

\noindent
\underline{Initial vertex $v_{21}$}

\bigskip

\be&&{\footnotesize
q^{2-2N}(qT)^2 + [N-1]\times}\nn\\[1mm]&&{\footnotesize\times\
\left|\begin{array}{cc|cccccccccccccccccccc}
\hline && \\
&&  1 && (qT) && (qT)^2 && (qT)^3 && (qT)^4\\ && \\
&& v_{21}  \lar 2v_{10}+2v_{31} \lar v_{00}+2v_{21} +2v_{20} +v_{42}
\lar 2v_{10}+2v_{31}\lar v_{21} \\ &&\\
\hline && \\
\l[N-1]& \times & 1 && 4q+\frac{2}{q} && q^2+2+2(q^2+1)+1 && 4q+\frac{2}{q} && 1 \\
\l[N-2]&\times & && && q+\frac{1}{q} \\
&& \lar \lar \lar \lar \\
q^{-N} & \times & && 2q && 2q^2+0+2(q^2+1)+0 && 2q \\
& & \\ \hline
\end{array}\right|
}\nn\ee

\bigskip

\noindent
The first line with $[N-1]$ is cohomologically equivalent to
$$\ \Big(\ 0 \ \ \ q+\frac{1}{q}  \ \ \ 0 \ \ \ 0 \ \ \ 0\ \Big)\ $$
it is combined with the second line to produce
$$ \ \Big(\ 0 \ \ \ q^{2-N}\left(q+\frac{1}{q}\right)  \ \ \ 0 \ \ \ 0 \ \ \ 0\ \Big)\ $$
what adds to the last line and gives:
\be
\underline{q^{2-2N}}(qT)^2\ +\ q^{-N}[N-1]\,
\Big( \ 0 \ \ \ 2q+q^2\left(q+\frac{1}{q}\right) \ \ \ 4q^2+2\ \ \ 2q \ \ \ 0 \ \Big)=\nn \\
= \underline{q^{2-2N}}(qT)^2 + q^{-N}[N-1]\Big(\ 0 \ \ \ q \ \ \ q^2 \ \ \ 0 \ \ \ 0 \ \Big)
=   \Big(\ 0 \ \ \ q^{3-2N} \ \ \ 1+\underline{q^{2-2N}} \ \ \ 0 \ \ \ 0 \ \Big) = 1
\ee
as it should be for the unknot.

\bigskip \bigskip

\noindent
\underline{Initial vertex $v_{10}$}

\bigskip

\be&&{\footnotesize
q^{2-2N}(qT) + [N-1]\times}\nn\\&&\times\ {\footnotesize
\left|\begin{array}{cc|cccccccccccccccccccc}
\hline && \\
&&  1 && (qT) && (qT)^2 && (qT)^3 && (qT)^4\\ && \\
&& v_{10}  \lar v_{00} + v_{20}+2v_{21} \lar 3v_{10}+3v_{31} \lar v_{20} +2v_{21}+v_{42}
\lar v_{31} \\ &&\\
\hline && \\
\l[N-1]& \times & q && q^2+q^2+1+2 && 6q+\frac{3}{q} && q^2+1+2+1 && q+\frac{1}{q} \\
\l[N-2]&\times & && && && q+\frac{1}{q} \\
&& \lar \lar \lar \lar \\
q^{-N} & \times & q && 2q^2+q^2+1 && 3q && q^2+1 \\
& & \\ \hline
\end{array}\right|
}\nn\ee

\bigskip

\noindent
The first line is cohomologically equivalent to $q+\frac{1}{q}=[2]$ at the middle place,
which is multiplied then by $[N-1]$ and further combined with $q+\frac{1}{q}=[2]$
in the second line, multiplied byu $[N-2]$. Cohomologically non-trivial remnant
contributes $q^{2-N}[2]$ to the middle place in the last line, converting it into
$$
\Big( \ q \ \ \ 3q^2+1 \ \ \ q^2[2] + 3q \ \ \ q^2+1 \ \ \ 0 \ \Big)
$$
which is cohomologically trivial.
Thus the KR polynomial is reduced to the single term $q^{2-2N}$ outside the table.
Since in this example original vertex is $v_{10}$, i.e. has $n_\bullet = 3$
and $n_\circ = 1$, we finally get:
$$
\frac{q^{3N-3}}{q^NT} \cdot q^{2-2N}(qT) = 1
$$
-- exactly as needed for the unknot.

\bigskip \bigskip

\noindent
\underline{Initial vertex $v_{31}$}

\bigskip

\be
&&{\footnotesize
q^{2-2N}(qT)^3 + [N-1]\times}\nn\\[1mm]&&\times\
\left|\begin{array}{cc|cccccccccccccccccccc}
\hline && \\
&&  1 && (qT) && (qT)^2 && (qT)^3 && (qT)^4\\ && \\
&& v_{31} \lar  v_{20} +2v_{21}+v_{42} \lar 3v_{10}+3v_{31} \lar v_{00} + v_{20}+2v_{21}
\lar v_{10} \\ &&\\
 \hline && \\
\l[N-1]& \times &  q+\frac{1}{q} && q^2+1+2+1 && 6q+\frac{3}{q} && q^2+q^2+1+2 && q   \\
\l[N-2]&\times & &&   q+\frac{1}{q} \\
&& \lar \lar \lar \lar \\
q^{-N} & \times &    && q^2+1 && 3q && 2q^2+q^2+1 && q \\
& & \\ \hline
\end{array}\right|
\nn\ee

{\centerline{\footnotesize
$
=q^{2-2N}(qT)^3 + [N-1]\times
\left|\begin{array}{cc|cccccccccccccccccccc}
\hline && \\
&&  1 && (qT) && (qT)^2 && (qT)^3 && (qT)^4\\ && \\
 \hline && \\
\l[N-1]& \times &  q+\frac{1}{q} &&  1 && 2q  && q^2  &&    \\
\l[N-2]&\times & &&   q+\frac{1}{q} \\
&& \lar \lar \lar \lar \\
q^{-N} & \times &    && q^2+1 && 3q && 2q^2+q^2+1 && q \\
& & \\ \hline
\end{array}\right|
$}}

\centerline{{\footnotesize
$
=q^{2-2N}(qT)^3 + [N-1]\times
\left|\begin{array}{cc|cccccccccccccccccccc}
\hline && \\
&&  1 && (qT) && (qT)^2 && (qT)^3 && (qT)^4\\ && \\
 \hline && \\
 q^N & \times & && && q^{-1}&& 1 \\
&& \lar \lar \lar \lar \\
q^{-N} & \times &  \underline{q^3+q}  &&
q^2+1 + \underline{q^2} && 3q + \underline{q^3}&& 2q^2+q^2+1 && q \\
& & \\ \hline
\end{array}\right|
$}}

\bigskip

\noindent
Underlined items in the last line came as cohomologically non-trivial remnant
of the first two.

Together with the common factor $[N-1]$ the remaining table is
cohomologically equivalent to
$$
\left|\begin{array}{cc|cccccccccccccccccccc}
\hline && \\
&&  1 && (qT) && (qT)^2 && (qT)^3 && (qT)^4\\ && \\
  \hline && \\
 q^N & \times & && && q^{1-N}&& q^{N-2} \\
&& \\
q^{-N} & \times &    && && q^{3-N} && q^{N} &&  \\
& & \\ \hline
\end{array}\right| =
$$
$$
 \left|\begin{array}{cccccccccccccccccccc}
\hline && \\
&  1 && (qT) && (qT)^2 && (qT)^3 && (qT)^4\\ && \\
  \hline && \\
 & && && q + q^{3-2N} &&q^{2N-2} + 1 \\
& & \\ \hline
\end{array}\right| =
\left|\begin{array}{cccccccccccccccccccc}
\hline && \\
&  1 && (qT) && (qT)^2 && (qT)^3 && (qT)^4\\ && \\
  \hline && \\
 & && &&  \underline{q^{3-2N}} &&q^{2N-2} \\
& & \\ \hline
\end{array}\right|
$$

\bigskip

\noindent
The underlined item combines in a cohomologically trivial pair
with the term $q^{2-2N}(qT)^3$ outside the table, and the answer
for KR polynomial is made from the remaining last item.
Since in this example $n_\bullet = 1$ and $n_\circ = 3$,
the result is
\be
\frac{q^{N-1}}{(q^NT)^3}\cdot q^{2N-2}(qT)^3 = 1
\nn
\ee
-- as it should be for the unknot.

\bigskip\bigskip

\noindent
\underline{Initial vertex $v_{42}$}

\bigskip
This is the same trefoil $[3,2]$, but   represented by "inverted" complex
(\ref{32comp}):
\be
v_{00}\ \stackrel{d_3}{\longleftarrow}\ 4v_{10} \ \stackrel{d_2}{\longleftarrow}\
2v_{20} + 4v_{21} \ \stackrel{d_1}{\longleftarrow}\  4v_{31}
\ \stackrel{d_0}{\longleftarrow}\  v_{42}
\label{32compinv}
\ee
with absolutely different morphisms and differentials -- they act in the opposite
direction, but still decrease grading by one.
\be
q^{2-2N}(qT)^4\ +\ [N-1]\times
\left|\begin{array}{cc|ccccccccccc}
\hline && \\
\l[N-1] & \times & 1  &4q+\frac{4}{q} & 2q^2+6 &  4q & q^2  \\ && \\
\l[N-2] & \times & q+\frac{1}{q} && &&     \\ && \\
q^{-N} & \times &  &   & 2q^2 + 2 &4q& 2q^2   \\ &&\\
 \hline
\end{array}\right|
\ee
This is the lengthiest example of all so far, because we need to handle
"increasing" sequences may times.

First, eliminate all the "obvious" cohomologically trivial pieces --
they are shifted to the upper part of the table:
\be
 q^{2-2N}(qT)^4\ +\ [N-1]\times
\left|\begin{array}{cc|ccccccccccc}
\hline && \\
&&    1 & \frac{1}{q} \\
\l[N-1]] &\times &&3q & 3 \\
&&&& 2q^2 & 2 \\
&&\\
\hline && \\
\l[N-1] &&& \frac{2}{q} &3 & 2q & q^2 \\
\l[N-1] & \times & &q+\frac{1}{q}   \\
\l[N-2] & \times & q+\frac{1}{q} && &&     \\ && \\
q^{-N} & \times &  &   & 2q^2 + 2 &4q& 2q^2   \\ &&\\
 \hline
\end{array}\right|
\label{32inv1}
\ee
In what follows we drop the upper part of the table.

In the lower part we have out familiar diagonal
$(q+1/q)\Big([N-2]\Longrightarrow [N-1]\Big)$, but now it is
in inverted order: the smaller space is embedded into the bigger one,
and this is cohomologically equivalent to $(q+1/q)\Big( 0 \Longrightarrow q^{N-2}\Big)$.

Increasing sequence in the first line
$$\ \frac{2}{q} \Longrightarrow 2+1 \Longrightarrow q+q \Longrightarrow q^2\ $$
after multiplication by $[N-1]$ gets equivalent to
$$\ \frac{2}{q}q^{2-N} \Longrightarrow 2q^{N-2} +q^{2-N} \Longrightarrow q\cdot q^{N-2}+
q\cdot q^{2-N} \Longrightarrow q^2\cdot q^{N-2} $$

Thus the lower part of above table is in fact equivalent to
\be
 q^{2-2N}(qT)^4\ +\ [N-1]\times
\left|\begin{array}{cc|ccccccccccc}
\hline && \\
q^N & \times & 0 & q^{-2}\left(q+\underline{q^{-1}}\right) & \underline{2}q^{-2}
& q^{-1} & 1 \\ & \\
q^{-N} & \times & 0 & \underline{2q}  & \underline{\underline{3q^2}} + {\underline 2}
&\underline{\underline{\underline{q^3}}}+\underline{\underline{4}}q
& \underline{\underline{\underline{2}}}q^2   \\ &&\\
 \hline
\end{array}\right|
\label{32inv2}
\ee
where underlined items can be immediately eliminated to give
\be
 q^{2-2N}(qT)^4\ +\ [N-1]\times
\left|\begin{array}{cc|ccccccccccc}
\hline && \\
q^N & \times & 0 & q^{-3}  &  q^{-2} & q^{-1} & 1 \\ & \\
q^{-N} & \times & 0 &  0 & 0
& q& q^2   \\ &&\\
 \hline
\end{array}\right|
\label{32inv3}
\ee
The two increasing sequences, after multiplication y $[N-1]$, get equivalent to
\be
\left|\begin{array}{cc|ccccccccccc}
\hline && \\
1 & \times &&&&& q^{2-2N} \\ & \\
q^N & \times & 0 & q^{-3}q^{2-N}  &  q^{-2}q^{N-2} & q^{-1}q^{2-N} & q^{N-2} \\ & \\
q^{-N} & \times & 0 &  0 & 0
& q\cdot q^{2-N}& q^2q^{N-2}   \\ &&\\
 \hline
\end{array}\right|
\label{32inv4}
\ee
where we also included the outside-the-table term with $q^{2-2N}$.
After obvious reordering we get
$$
\left|\begin{array}{cc|ccccccccccc}
\hline && \\
q^{2N}& \times & 0 & 0 & q^{-4} & 0 & q^{-2} \\ && \\
1 & \times & 0 & q^{-1}  & 0 & \underline{q} & \underline{1} \\ & \\
q^{-2N} & \times & 0 &  0 & 0 &\underline{q^3} & \underline{q^2}   \\ &&\\
 \hline
\end{array}\right|
$$
The underlined terms form two cohomologically trivial pairs
and the resulting KR polynomial contains just three terms:
\be
\frac{1}{(q^NT)^4}\Big(q^{-1}(qT) + q^{2N-4}(qT)^2 + q^{2N-2}(qT)^4\Big)
= q^{2-2N}\Big( 1 + q^{-4}T^{-2} + q^{-2N-2}T^{-3}\Big) =
{\cal P}^{[3,2]}_r(q^{-1},T^{-1},N)
\ee
where ${\cal P}^{[3,2]}_r$ is the KR polynomial for the trefoil,
which we already reproduced in (\ref{KR23}) and (\ref{KR32r}).
As usual \cite{DM3}, reversion of complex means inversion of $q$ and $T$ in the answer.

\bigskip

For unreduced KR polynomial the difference appears
in the transition from (\ref{32inv3}), multiplied by a common $[N]$,
to (\ref{32inv4}): we should now use $[N]$ instead of $[N-1]$ in eliminating
the cohomologically trivial pieces and get instead of (\ref{32inv4}):
\be
 q^{2-2N}(qT)^4[N]\ +\ [N-1]\times
\left|\begin{array}{cc|ccccccccccc}
\hline && \\
q^N & \times & 0 & q^{-3}q^{1-N}  &  q^{-2}q^{N-1} & q^{-1}q^{1-N} & q^{N-1} \\ & \\
q^{-N} & \times & 0 &  0 & 0
& q\cdot q^{1-N}& q^2q^{N-1}   \\ &&\\
 \hline
\end{array}\right|
\nn
\ee
or, after reordering:
\be
 \underline{\underline{{q^{2-2N}(qT)^4[N]}}}\ +\ [N-1]\times
\left|\begin{array}{cc|ccccccccccc}
\hline && \\
q^{2N} & \times & 0 & 0 &  q^{-3} & 0 & q^{-1} \\ & \\
1 & \times &0 & q^{-2} & 0 & \underline{1} & \underline{q} \\ && \\
q^{-2N} & \times & 0 &  0 & 0 & \underline{\underline{q^2}} & 0\\ &&\\
 \hline
\end{array}\right|
\nn
\ee
Underlined are the terms, which were cohomologically trivial in reduced case.
Now things are a little more difficult.
For single-underlined terms we substitute
$[N-1]\Big(1 \Longrightarrow q\Big)$ by $ \Big(\underline{ q^{2-N}} \Longrightarrow q^{N-1}\Big)$,
for doubled-underlines terms we substitute
$\Big(q^{2-2N}[N-1] \Longrightarrow q^{2-2N}[N]\Big)$ by
$\Big(0 \Longrightarrow q^{2-2N}q^{N-1}\Big) = \Big(0 \Longrightarrow \underline{q^{1-N}}\Big) $.
The two underlined terms in these new contributions form a cohomologically
trivial pair, which can be eliminated.
After that the KR polynomial becomes
\be
\frac{1}{q^{4N}T^4}\left\{([N-1]\Big((q^{-2}(qT) + q^{2N-3}(qT)^2 + q^{2N-1}(qT)^4\Big)
+ q^{N-1}(qT)^4\right\} = \nn \\
= q^{2-2N}\Big\{[N] + q^{-3}T^{-2}[N-1]+q^{-2N-3}T^{-3}[N-1]\Big\}
= {\cal P}^{[3,2]}(q^{-1},T^{-1},N)
\ee
i.e. is appropriately related to (\ref{KR32}) and (\ref{KR23}).

\subsubsection{The 3-component 3-braid torus link $[3,3]$ ($6^3_3(v1)$ of \cite{CM})}

The braid diagrams for the 3-strand links can be represented as two strings of
black or white dots, standing at positions of the ${\cal R}$-matrices ${\cal R}_1^{\pm 1}$
and ${\cal R}_2^{\pm 1}$ -- without drawing the strands themselves (what simplifies
the drawing a lot). If restored, the strands would go from the left to the right.
In the particular case of the torus link $[3,3]$ this picture looks
as follows:
$$
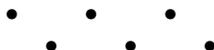
\begin{array}{ccccccccccc}
\bullet && \bullet && \bullet &\\
& \bullet && \bullet && \bullet \\
\end{array}
$$
or, with the strands restored:

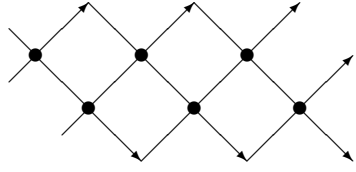
\begin{picture}(100,80)(-170,-30)
\put(0,20){\circle*{5}}
\put(20,0){\circle*{5}}
\put(-10,30){\vector(1,-1){50}}
\put(-10,10){\vector(1,1){30}}
\put(10,-10){\vector(1,1){50}}
\put(40,20){\circle*{5}}
\put(60,0){\circle*{5}}
\put(20,40){\vector(1,-1){60}}
\put(80,-20){\vector(1,1){40}}
\put(60,40){\vector(1,-1){60}}
\put(80,20){\circle*{5}}
\put(100,0){\circle*{5}}
\put(40,-20){\vector(1,1){60}}
\end{picture}

\noindent
The KR complex in this case consists of the space with dimensions
\be
\dim_0=&v_{0,0}&=[N]^2\nn\\[1.5mm]
\dim_1=&6v_{1,0}&=6[N][N-1]\nn\\[1.5mm]
\dim_2=&6v_{2,0}+9v_{2,1}&=6[2][N][N-1]+9[N-1]^2\nn\\[1.5mm]
\dim_3=&2v_{3,0}+18v_{3,1}&=2[2]^2[N][N-1]+18[N-1]^2[2]\nn\\[1.5mm]
\dim_4=&9v_{4,1}+6v_{4,2}&=9[2]^2[N-1]^2+6[N-1]^2+6[2][N-1][N-2]\nn\\[1.5mm]
\dim_5=&6v_{5,2}&=6[2][N-1]^2+6[2]^2[N-1][N-2]\nn\\[1.5mm]
\dim_6=&v_{6,3}&=[N-1]^2+([2]+[2]^3)[N-1][N-2]\nn
\label{dimstor33}
\ee

\bigskip

From this we read:

\be
q^{6-6N}{\P}^{[3,3]}_r =
\boxed{q^{2-2N}}\ +\ [N-1]\times
\left|\begin{array}{cc|ccccccccccc}
\hline && \\
\l[N-2] & \times &  &&&&&& \underline{q^3} &&&&(1)\\ & \\
\l[N-1] & \times & &&& 2q^3 & \underline{\underline{9q^2}} & 6q & 1 &&&&(-2)\\
\l[N-2] & \times & &&&&& \underline{\underline{6q^2}} & 4q &&&& (2) \\ && \\
\l[N-1] & \times & && 6q^2 & {22q} & \underline{\underline{24}} & \frac{6}{q} & &&&& (2) \\
\l[N-2] & \times & && && {6q} & \underline{\underline{12}} & \frac{4}{q} &&&& (-2) \\ && \\
\l[N-1] & \times & q^2 & 6q & 15 & \underline{\frac{20}{q}} & \frac{9}{q^2}
 &&  &&&&(-1)\\
\l[N-2] & \times &&&&&  \underline{\frac{6}{q}} & \frac{6}{q^2} &
\frac{1}{q^3} &&&& (1) \\ && \\
q^{-N} & \times & 2q^2 & 6q & 6 & \frac{2}{q} &&& &&&& (0) \\ &&\\
q^{-N} & \times &&& \underline{\underline{ 6q^2}} & 4q&&& &&&& (2) \\ &&\\
q^{-N} & \times &&&& \underline{\underline{2q^3}} &&& &&&& (2) \\ && \\
\hline\hline &&\\
\l[N-2] & \times &&&&&&&q^3 \\ && \\
\l[N-1] &\times &&&&& 2[2]q &  \\
\l[N-2] & \times &&&&&& 2[2]q&\\ && \\
\l[N-1] &\times &&&& \frac{1}{q} &  \\
\l[N-2] & \times &&&&& \frac{1}{q} &\\
q^{-N} & \times & && 2q^2 & 2q^3 \\ && \\
\hline\hline &&\\
\l[N-2] & \times &&&&&&&q^3 \\ && \\
q^{-N} & \times & && 2q^2 & q \\ && \\
q^{-N} &\times &&&& 2q^3 & 2[2]q^3 \\ && \\
\hline\hline && \\
\l[N-2] & \times &&&&&&&q^3 \\ && \\
q^{-N} & \times & && q^2 &q^3& q^2+2q^4 \\ && \\
\hline
\end{array}\right|
\nn
\ee

\bigskip

\noindent
To reproduce the answer from \cite{CM},
the term $q^3\ \Longrightarrow \ q^2$  should be preserved in the last line
-- but only once, its second copy is sent to the cohomologically trivial part.
Now, restoring the factor $[N-1]$ and  making use of (\ref{2n3n}),
we obtain for the KR polynomials:
$$
{{\cal P}^{[3,3]}_r=
q^{6N-6}\Big(q^{2-2N} + q^{4-2N}(qT)^2 + q\,(qT)^3 + (q^{2-N}+2q^{4-N})[N-1]\,(qT)^4
+ q^3[N-1][N-2]\,(qT)^6\Big) }
$$
\vspace{-0.5cm}
\be
= \boxed{q^{4N-4}\Big(1 +
q^4T^2+q^{2N+2}T^3 + (q^{N+4}+2q^{N+6})[N-1]\,T^4+q^{2N+7}[N-1][N-2]\,T^6\Big)}
\label{KR33true}
\ee
Similarly, using (\ref{2n3nur}),
\be
\boxed{\begin{array}{r}{\cal P}^{[3,3]}=
q^{4N-4}\Big([N] + q^3[N-1]\,T^2 + q^{2N+3}[N-1]\,T^3 + (q^{N+4}+2q^{N+6})[N][N-1]\,T^4+\\+ q^{2N+7}[N][N-1][N-2]\,T^6\Big)\end{array}}
\nn
\ee

This reproduces the two answers, obtained for $6^3_3(v1)$
in \cite{CM} for $N=2$ and $N=3$:

reduced case
\be
N=2: & q^4\Big(1+q^4T^2 + q^6T^3 +(q^6+2q^8)T^4\Big) \nn \\ \nn \\
N=3: & q^8\Big( 1 +q^4T^2 + q^8T^3 +(\underbrace{q^6+3q^8+2q^{10}}_{[2](q^7+2q^9)})T^4
+(q^{12}+q^{14})T^6\Big)
\ee

unreduced case
\be
N=2: & q^4\Big([2]+q^3T^2 + q^7T^3 +(q^6+2q^8)[2]T^4\Big) \nn \\ \nn \\
N=3: & q^8\Big( [3] +(q^2+q^4)T^2 + (q^8+q^{10})T^3 +(q^6+3q^8+2q^{10})[3]T^4
+(q^{12}+q^{14})[3]T^6\Big)
\ee


\bigskip

Relation  of reduced polynomial to the superpolynomial from eq.(148) of \cite{DMMSS}
\be
\frac{1}{(1-q^2)^2}
\Big((1-2q^2+q^4)  + (q^4-2q^6+q^8)T^2 +
(q^2-2q^4+q^6)T^3a^2 +(q^6+q^8-2q^{10})T^4+ \nn \\
+ (q^4+q^6-2q^8)T^5a^2 +q^{12}T^6  +q^9[2] T^7a^2
+q^6T^8a^4\Big)=\nn \\
= 1 + q^4T^2+q^2T^3a^2+ q^4(1+2q^2)T^4\,\frac{q^2 +a^2T}{1-q^2}
+q^6T^6\,\frac{(q^2+a^2T)(q^4+a^2T)}{(1-q^2)^2}
\ee
is somewhat more involved.
To obtain a superpolynomial, which is a function of $a$ instead of $N$,
one should substitute $a=q^N$ instead of powers $q^N$  in (\ref{KR33true}),
while quantum numbers are substituted by
\be
\l[N-1]  \rightarrow \frac{q^2+Ta^2}{a(1-q^2)},\nn \\
\l[N-2]  \rightarrow \frac{q^4+Ta^2}{aq(1-q^2)}
\label{KRtoSUP}
\ee
-- this is instead of the naive
$\frac{q^{N-1}-q^{1-N}}{q-q^{-1}} = \frac{\frac{a}{q} -\frac{q}{a}}{q-\frac{1}{q}}
= \frac{q^2-a^2}{a(1-q^2)}$ and
$\frac{q^{N-2}-q^{2-N}}{q-q^{-1}}= \frac{q^4-a^2}{aq(1-q^2)}$,
which would not provide a positive polynomial in $a$.
The same rule applies to unreduced polynomials
(for {\it knots} the reduced KR polynomials do not contain quantum numbers).
Note that these substitutions increase the full degree of the polynomial in $T$ .
Some comments on this issue can be found in \cite{DGR} and \cite{DMMSS}.

\subsubsection{Another coloring: the 3-component Borromean link ($L6a4$ of \cite{katlas}
or $6^3_2$ of \cite{CM})}

If instead of ${\cal R}_1{\cal R}_2{\cal R}_1{\cal R}_2{\cal R}_1{\cal R}_2$
we consider ${\cal R}_1{\cal R}_2^{-1}{\cal R}_1{\cal R}_2^{-1}{\cal R}_1{\cal R}_2^{-1}$, then
the braid is
$$
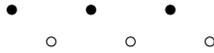
\begin{array}{ccccccccccc}
\bullet && \bullet && \bullet &\\
& \circ && \circ && \circ \\
\end{array}
$$
The positions of points (and thus the hypercube) are the same,
but the colorings (and thus morphisms and the KR complex) are different.
Actually, the KR complex
\be
v_{00} \stackrel{d_0}{\longrightarrow} 6v_{10}
\stackrel{d_1}{\longrightarrow} 6v_{20} + 9v_{21}
\stackrel{d_2}{\longrightarrow} 2v_{30} +  18v_{31}
\stackrel{d_3}{\longrightarrow} 9v_{41} + 6v_{42}
\stackrel{d_4}{\longrightarrow} 6v_{52}
\stackrel{d_5}{\longrightarrow} v_{63}
\ee
is changed for
\be
v_{30}
\stackrel{\tilde d_0}{\longrightarrow} 3v_{20}+3v_{41}
\stackrel{\tilde d_1}{\longrightarrow}  3v_{10}+9v_{31}+3v_{52}
\stackrel{\tilde d_2}{\longrightarrow}  v_{00} + 9v_{21}+3v_{41}+6v_{42} + v_{63}
\stackrel{\tilde d_3}{\longrightarrow}
\nn \\
\stackrel{\tilde d_3}{\longrightarrow} 3v_{10}+9v_{31}+3v_{52}
\stackrel{\tilde d_4}{\longrightarrow} 3v_{20}+3v_{41}
\stackrel{\tilde d_5}{\longrightarrow} v_{30}
\ee
and therefore
\be
\left(\frac{q^{3N-3}}{q^{3N}T^3}\right)^{-1}\cdot{\P}^{6^3_2}_r \ =\
q^{2-2N}\cdot(qT)^3\ +\ [N-1]\times \ \ \ \ \ \ \ \ \ \ \ \ \ \ \ \ \ \ \ \ \ \
\ \ \ \ \ \ \ \ \ \ \ \ \ \ \ \ \ \ \ \ \ \ \ \ \ \ \ \ \ \ \ \ \ \ \ \ \ \ \ \ \ \ \ \
 \nn \\ \nn \\
\left|\begin{array}{c|ccccccccccc}
\hline && \\
\l[N-1]\times & q[2]^2 & 3q[2]+3[2]^2& 3q+9[2]+3[2]&q^2+9+3[2]^2+6+1
&3q+9[2]+3[2]& 3q[2]+3[2]^2&q[2]^2\\&&\\
\l[N-2]\times & &&3[2]^2 & 6[2]+[2](1+[2]^2) & 3[2]^2 \\ &&\\
q^{-N}\ \times & q[2]^2 & 3q[2] & 3q & 2q^2 & 3q & 3q[2] & q[2]^2 \\
&&\\ \hline\end{array}\right|
\nn
\ee
In more detail the table is:
\be
q^{2-2N}\ +\ (qT)^{-3}[N-1]\times
\left|\begin{array}{c|llllllllccccccc}
\hline && \\
\l[N-1]\times & &&&&&& \underline{q^3} &&&& (1) \\ &&\\
\l[N-1]\times & &&&&&6\underline{\underline{\underline{\underline{q^2}}}}&2q &&&& (-4) \\ && \\
\l[N-1]\times & &&&4q^2&15\underline{\underline{\underline{q}}}&9&q^{-1} &&&& (3) \\ && \\
\l[N-1]\times & q^3&6q^2&15q&22\cdot\underline{\underline{\underline{1}}}
&12q^{-1}&3q^{-2} & &&&&(-3) \\ &&\\
\l[N-1]\times & 2q & 9 & 12\underline{\underline{q^{-1}}}&3q^{-2} &&& &&&&(2) \\ && \\
\l[N-1]\times & q^{-1} & 3\underline{\underline{q^{-2}}} &&&&& &&&&(-2) \\ &&\\
&\\
\l[N-2]\times & &&&q^3&3\underline{\underline{q^2}} &&&&(2) \\ &&\\
\l[N-2]\times & &&3q^2 & 10\underline{q} & 6 &&&&(-1) \\ &&\\
\l[N-2]\times & &&6 & 10\underline{q^{-1}} &3q^{-2} &&&& (-1) \\ &&\\
\l[N-2]\times & &&3\underline{\underline{q^{-2}}}& q^{-3} & &&&&(2) \\ && \\
&&\\
q^{-N}\ \times & &&&&&& q^3 &&&& (1) \\ && \\
q^{-N}\ \times & &&&& & 3q^2 & 2q &&&& (-1) \\ &&\\
q^{-N}\ \times & &&& 2q^2 & 3q & 3 & q^{-1} &&&&(-1)\\ &&\\
q^{-N}\ \times & q^3 & 3q^2 & 3q &  &  && &&&& (1) \\ &&\\
q^{-N}\ \times & 2q & 3 &&&&& &&&&(-1) \\ && \\
q^{-N}\ \times & q^{-1} &&&&&& &&&&(1) \\
&&\\ \hline\end{array}\right|
\nn
\ee
The cohomologically non-trivial part is restricted by the values of alternated sums
in lines and at the next step reduces to
\be
q^{2-2N}\ +\ (qT)^{-3}[N-1]\times
\left|\begin{array}{c|lllcllllccccccc}
\hline && \\
\l[N-1]\times & 0& 2q^{-2}\!\!\! & \ \ \ 2q^{-1}\!\!\!\! & 3 & \! 3q & \!\!\!4q^2 & q^3 \\
& &&\!\!\!\!\!\!\nwarrow&\nwarrow\ \ \ \ \ \ \ \ \ \ \ \ \nearrow&\ \ \ \ \ \ \nearrow \\
\l[N-2]\times & 0 & 0 &2q^{-2}&q^{-1}+q&2q^2 \\ && \\
q^{-N}\ \times & q^{-1} & 1 & q&q^2 & 0&q^2 & q^3 \\
&&\\ \hline \end{array}\right|
\nn \\ \nn \\
\sim q^{2-2N}\ +\ (qT)^{-3}[N-1]\times
\left|\begin{array}{c|lllcllllccccccc}
\hline && \\
\l[N-1]\times &&&& \boxed{1} \\ && \\ && \\
q^N\ \times & &&&& q^{-1}&2&0\\ &&\\
q^N\ \times & 0&0&0&\boxed{q^{-2}}&q^{-1}&1&q \\ && \\
&& \\
q^{-N}\ \times &0&0&q&\boxed{q^2}&q^3&q^4&0        \\ && \\
q^{-N} \ \times &0 &2 &q &
& \ \ \ \searrow \!\!\!\!\!\!&\  \ \ \searrow \!\!\!\!\!\!\!   \\  &&\\
q^{-N} \ \times &q^{-1} &1 &q& q^2 & 0 &\  q^2 & \ q^3 \\  &&\\
\hline \end{array}\right|
\nn \\ \nn \\
\sim \ \boxed{[N-1]^2+(q^{N-2}+q^{2-N})[N-1]}
+\ (qT)^{-3}[N-1]\times
\left|\begin{array}{c|lllcllllccccccc}
\hline && \\
q^N\ \times & &&&&2q^{-1}&3&q \\ && \\
q^{-N} \ \times &q^{-1} &3 &2q &&&& \\  &&\\ &&\\
q^{-N} \ \times & & &q &q^2 &&&       \\  &&\\
\hline \end{array}\right| \ +\  \underline{q^{2-2N}}
\nn \\ \nn \\
\sim [N-1]^2+(q^{N-2}+q^{2-N})[N-1]
+\ (qT)^{-3}\times
\left|\begin{array}{c|lllcllllccccccc}
\hline && \\
q^{2N}\ \times & &&&&&2q^{-2}&q^{-1} \\ && \\
1\ \times & &&&&2q & q^2 &  \\ && \\
1\ \times&  & q^{-2} & 2q^{-1} &  &  &  &  \\ && \\
q^{-2N}\ \times & q & 2q^2 &  &  &  &  &  \\ && \\ && \\
1 \ \times & & & &1&&& \\  &&\\ \hline &&\\
q^{-2N} \ \times & & &\underline{q^3} &\underline{q^2} &&&       \\  &&\\
\hline \end{array}\right|
\nn
\ee
The underlined term, standing after the next-to-the-last table is absorbed into
the last table, where it is eliminated together with another underlined term
in the last line.

Thus reduced KR polynomial is
\be
{\cal P}^{6^3_2}_r \ =\
\frac{1}{q^{2N+2}T^3} + \frac{q^{N-2}+2q^{2-N}}{q^{N+2}T^2} +
\frac{2}{q^2T} + \nn \\
+\Big([N-1]+q^{N-2}\Big)\Big([N-1]+q^{2-N}\Big)
+ 2q^2T + (2q^{N-2}+q^{2-N})\cdot q^{N+2}T^2  + q^{2N+2}T^3
\label{KRborr_r}
\ee

In unreduced case the deviation begins from the next-to-the-last table:
\be
{\cal P}^{6^3_2} \
\sim \ [N][N-1]^2+(q^{N-2}+q^{2-N})[N][N-1] +
\ \ \ \ \ \ \ \ \ \ \ \ \ \ \ \ \ \ \ \ \ \ \ \nn \\
+\ (qT)^{-3}[N][N-1]\times
\left|\begin{array}{c|lllcllllccccccc}
\hline && \\
q^N\ \times & &&&&2q^{-1}&3&q \\ && \\
q^{-N} \ \times &q^{-1} &3 &2q &&&& \\  &&\\ &&\\
q^{-N} \ \times & & &q &q^2 &&&       \\  &&\\
\hline \end{array}\right| \ +\  \underbrace{\underline{q^{2-2N}[N]}}_{q^{1-2N}[N-1]+q^{1-N}}
\nn \\ \nn \\
\sim [N][N-1]^2+(q^{N-2}+q^{2-N})[N][N-1] + q^{1-N}
+\nn \\ \nn \\+\ (qT)^{-3}[N-1]\times
\left|\begin{array}{c|lllcllllccccccc}
\hline && \\
q^{2N}\ \times & &&&&&2q^{-1}&1 \\ && \\
1\ \times & &&&&2 & q &  \\ && \\
1\ \times&  & q^{-1} & 2 &  &  &  &  \\ && \\
q^{-2N}\ \times & 1 & 2q &  &  &  &  &  \\ && \\ && \\
1 \ \times & & & &q&&& \\  &&\\ \hline &&\\
q^{-2N} \ \times & & &\underline{q^2} &\underline{q} &&&       \\  &&\\
\hline \end{array}\right|
\nn
\ee
and, absorbing $q^{1-N}$ back into $q^{1-N}+q[N-1] = [N]$, we obtain unreduced polynomial:
\be
{\cal P}^{6^3_2} \ =\
\left(\frac{1}{q^{2N+3}T^3} + \frac{q^{N-1}+2q^{1-N}}{q^{N+2}T^2}+
\frac{2}{qT}\right)[N-1]  + \nn \\
+\Big([N-1]+q^{N-2}\Big)\Big([N-1]+q^{2-N}\Big)[N]
+ \Big(2qT + (2q^{N-1}+q^{1-N})\cdot q^{N+2}T^2  + q^{2N+3}T^3\Big)[N-1]
=
\label{KRborr_ur}
\ee
\vspace{-0.3cm}
$$
=\left(\frac{2}{qT}+\frac{1}{q^3T^2}\right)\left(1+\frac{1}{q^{2N}T}\right)[N-1]
+ \Big([N-1]+q^{N-2}\Big)\Big([N-1]+q^{2-N}\Big)[N]
+ \Big(2qT+q^3T^2\Big)\Big(1+q^{2N}T\Big)[N-1]
$$
Both (\ref{KRborr_r}) and (\ref{KRborr_ur}) are in agreement with the results of
\cite{CM} for $N=2,3,4$ and $N=2,3$ respectively.

For $T=-1$  these expressions reduce to
\be
H^{6^3_2}_r = [N]^2-\left(q-\frac{1}{q}\right)^4\!\cdot[N+1][N-1],
\ \ \ \ \ \ \ \ \ \ H^{6^3_2}=[N]\cdot H^{6^3_2}_r
\ee
known from \cite{AENV} (where also generalizations of Borromean HOMFLY
to arbitrary triples of symmetric representations was found).

\subsubsection{Still another coloring: the 3-component link $6^3_3(v2)$ of \cite{CM})}

If instead of ${\cal R}_1{\cal R}_2{\cal R}_1{\cal R}_2{\cal R}_1{\cal R}_2$
we consider ${\cal R}_1{\cal R}_2^{-1}{\cal R}_1{\cal R}_2{\cal R}_1^{-1}{\cal R}_2$, then
the braid becomes
\be
\begin{array}{ccccccccccc}
\bullet && \bullet && \circ &\\
& \circ && \bullet && \bullet \\
\end{array}
\label{633braid}
\ee
and the KR complex
\be
v_{00} \stackrel{d_0}{\longrightarrow} 6v_{10}
\stackrel{d_1}{\longrightarrow} 6v_{20} + 9v_{21}
\stackrel{d_2}{\longrightarrow} 2v_{30} +  18v_{31}
\stackrel{d_3}{\longrightarrow} 9v_{41} + 6v_{42}
\stackrel{d_4}{\longrightarrow} 6v_{52}
\stackrel{d_5}{\longrightarrow} v_{63}
\ee
is changed for
\be
v_{21} \stackrel{\tilde d_0}{\longrightarrow} 2v_{10}+4v_{31}
\stackrel{\tilde d_1}{\longrightarrow} v_{00} + 4v_{20} + 4v_{21} +4v_{41}+2v_{42}
\stackrel{\tilde d_2}{\longrightarrow}         4v_{10} +2v_{30}+  10v_{31}+4v_{52}
\stackrel{\tilde d_3}{\longrightarrow}
\nn \\
\stackrel{\tilde d_3}{\longrightarrow} 2v_{20} +4v_{21} + 4v_{41} + 4v_{42} + v_{63}
\stackrel{\tilde d_4}{\longrightarrow} 4v_{31}+2v_{52}
\stackrel{\tilde d_5}{\longrightarrow} v_{41}
\ee
Note that the right endpoint of the complex is $v_{41}$ rather than $v_{42}$ --
this is the property of the coloring (\ref{633braid}).

The primary polynomial, associated with this complex, is
\be
&&\left(\frac{q^{4N-4}}{q^{2N}T^2}\right)^{-1}\cdot{\P}^{6^3_3(v2)}_r \ =\
{q^{2-2N}\cdot(qT)^2}\ +\ [N-1]\times \ \ \ \ \ \ \ \ \ \ \ \ \ \ \ \ \ \ \ \ \ \
\ \ \ \ \ \ \ \ \ \ \ \ \ \ \ \ \ \ \ \ \ \ \ \ \ \ \ \ \ \ \ \ \ \ \ \ \ \ \ \ \ \ \ \
\nn \\ \nn \\
&&\times{\footnotesize\arraycolsep=1mm\left|\begin{array}{c|ccccccccccc}
\hline && \\
\l[N-1] \times & 1 & 2q+4[2] & q^2+4q[2]+4+4[2]^2+2& 4q + 2q[2]^2+10[2]+4[2] & 2q[2]+4+4[2]^2+4+1 &
4[2]+2[2] & [2]^2 \\ && \\
\l[N-2]\times & 0&0& 2[2] & 4[2]^2 & 4[2]+[2](1+[2]^2) & 2[2]^2& 0 \\ &&\\
q^{-N}\ \times & 0&2q &2q^2+4q[2] &4q+2q[2]^2 & 2q[2] & 0 & 0 \\
&&\\ \hline \end{array}\right|}
\nn
\ee
The table can be rewritten as
\be
\left|\begin{array}{c|ccccccccccc}
\hline && \\
\l[N-1]  \times &  &&&&&& \underline{q^2}   &&&&(1)\\ & \\
\l[N-1]  \times & &&&2q^3& 6q^2 &  6q & 2  &&&&(0)\\ && \\
\l[N-1]  \times & &&9q^2&22q&19\cdot\underline{1}& \frac{6}{q} & \frac{1}{q^2}  &&&& (1) \\ && \\
\l[N-1]  \times & &6q& 18 & \frac{16}{q} & \frac{4}{q^2} &  & &&&& (0) \\ && \\
\l[N-1]  \times & 1 &\frac{4}{q}& 4\cdot\underline{\frac{1}{q^2}}&&  &  &  &&&& (1) \\ && \\ && \\
\l[N-2]  \times & & & &  &q^3&2\underline{q^2}& 0 &&&(-1) \\ && \\
\l[N-2]  \times & & & & 4q^2 &8q& 4& 0 &&&(0) \\ && \\
\l[N-2]  \times & & &2q & 8 &\frac{8}{q}&2\frac{2}{q^2}& &&&(0) \\ && \\
\l[N-2]  \times & & &\frac{2}{q} & 4\cdot\underline{\frac{1}{q^2}} &\frac{1}{q^3}&& 0 &&&(-1) \\ && \\
 && \\ && \\
q^{-N} \ \times & 0 & 2q & 6q^2+4  & 2q^3+8q + \frac{2}{q} &2q^2+2&0&0 &&&& (0)  \\ &&\\
\hline
\end{array}\right|
\nn
\ee
If we pick up only the five underlined terms -- the minimum allowed by Euler characteristics
in lines -- then the table reduces to
\be
\stackrel{?}{\sim}
\left|\begin{array}{c|cc|ccc|cccccc}
\hline &&&&&& \\
\l[N-1]  \times & 0&0& \frac{1}{q^2}&0& 1 &0& q^2 \\
&&&&\!\!\!\!\!\searrow\ \ \ \ \ &&\ \ \ \ \swarrow\!\!\!\!\!\\
\l[N-2] \times &0&0&0&\frac{1}{q^2} &0&q^2& 0 \\  &&&&&&  \\
\hline
\end{array}\right|
\sim \left|\begin{array}{ ccccccccc}
\hline && \\ 0 & 0 & q^{-N}  & 0 & [N-1]  & 0 & q^N \\
&& \\
\hline
\end{array}\right|
\label{promo633v2}
\ee
what would give rise to a very compact polynomial
\be
{\cal P}^{6^3_3(v2)}_r \stackrel{?}{=}
1+q^{N-2}[N-1] + q^{2N}[N-1]^2T^2 + q^{3N+2}[N-1]T^4
\label{wrongKR633v2}
\ee
Unfortunately this is {\it not} the right answer -- it does
not match the results of  \cite{CM}.
The real reason for this would come from analysis of morphisms --
which we postpone to a separate publication.
However, this failure is easy to expect, because formula(\ref{wrongKR633v2})
has two apparent irregularities: instead of the typical factor $[N-1][N-2]$
for 3-component links we got $[N-1]^2$, and the third term in the complex
which normally has simple, but non-trivial cohomology, vanishes.

\bigskip
The right answer arises if we keep {\it three} seemingly trivial pairs
in (\ref{promo633v2}),
coming from the lines with vanishing Euler characteristics:
\be
q^{2-2N}+[N-1]\cdot
\left|\begin{array}{c|cc|ccc|cccccc}
\hline &&&&&& \\
\l[N-1]  \times & 0&0& \frac{1}{q^2}+1
&\frac{1}{q}\!\!\!\!\!\!\!\!\!\!\!\!\!\!
& \!\!\!\!\!\!\!\!\!\!\!\!\!\!  1 &0& q^2 \\
&&&&\!\!\!\!\!\!\!\!\!\!\!\!\!\!\!\!\!\!\!\!\!\!\!\!\!\!\!\!\!\!\!\!\!\!\!\!
\searrow\ \searrow \ \ \ \ \ \ \ \ \ \ \ \ \  \searrow\!\!\!\!\!\!\!\!
\swarrow\!\!\!\!\!\!\!\!\!\!\!\!\!\!\!\!\!\!\!\!\!\!\!\!\!\!\!\!\!\!\! &
&\ \ \ \ \swarrow\!\!\!\!\!\\
\l[N-2] \times &0&0&0&\frac{1}{q^2}+1+q^2 &\frac{1}{q}+\boxed{q}&q^2& 0 \\  &&&&&&  \\
\hline
\end{array}\right|
\nn\\ \nn \\
q^{2-2N}+[N-1]\cdot
\left|\begin{array}{ cc|ccc|cccc}
\hline && \\ 0 & 0 & q^{-N}+q^{2-N}  & \underline{q^{1-N}} & \underline{q^{2-N}}+[N-2]  & 0 & q^N \\
&& \\
\hline
\end{array}\right|
\label{promo633v2a}
\ee
It remains to substitute the two underlined terms $\ [N-1]\Big(q^{1-N}\Longrightarrow q^{2-N}\Big)\ $
by $\ q^{2-N}\cdot q^{1-N} = q^{3-2N}\Longrightarrow q^{N-2}\cdot q^{2-N}=1\ $ in reduced case,
and $\ [N-1][N]\Big(q^{1-N}\Longrightarrow q^{2-N}\Big)\ $
by $\ [N-1]\Big(q^{1-N}\cdot q^{1-N} = q^{2-2N}\Longrightarrow q^{N-1}\cdot q^{2-N}=q\Big)\ $
in unreduced case.
This provides the right reduced and unreduced polynomials:
\be
{\cal P}^{6^3_3(v2)}_r = \frac{q^{2N-2}}{(qT)^2}\left\{\Big(q^{2-2N}+({q^2+1})q^{-N}[N-1]\Big)(qT)^2
+ q^{3-2N}(qT)^3 + \Big(1+ q[N-1][N-2]\Big)(qT)^4 +\right.\nn\\+ q^N[N-1](qT)^6\Big\} =
\nn\ee
\vspace{-0.5cm}
\be
= 1 + q^{N-1}[2][N-1] + q^2T + \Big(q^{2N}+q^{2N+1}[N-1][N-2]\Big)T^2 + q^{3N+2}[N-1]T^4,
\label{KR633v2}
\ee
\vspace{-0.2cm}
\be
{\cal P}^{6^3_3(v2)} = [N] + q^{N-1}[2][N][N-1] + q[N-1]T +
q^{2N+1}[N-1]\Big(1+[N][N-2]\Big)T^2 +\nn\\+ q^{3N+2}[N][N-1]T^4
\nn\ee
For $N=2,3$ this reproduces the answers from \cite{CM} in reduced case
\be
N=2: &&   2+q^2 + q^2T + q^4T^2+q^8T^4 \nn \\
N=3: &&    2+2q^2+q^4 + q^2T + (2q^6+q^8)T^2+ (q^{10}+q^{12})T^4
\ee
and in unreduced case
\be
N=2: &&   \frac{2}{q}+3q+q^3 + qT + q^5T^2+(q^7+q^9)T^4 \nn \\
N=3: &&    \frac{2}{q^2}+4+5q^2+3q^4 +q^6 + (1+q^2)T +
(q^4+3q^6+3q^8+q^{10})T^2+\nn\\&&+ (q^8+2q^{10}+2q^{12}+q^{14})T^4
\ee

\subsubsection{Knot $[3,4]$ (also known as $8_{19}$)
\label{torus34}}

\be
\dim_0=&v_{0,0}&=[N]^2\nn\\[1.5mm]
\dim_1=&8v_{1,0}&=8[N][N-1]\nn\\[1.5mm]
\dim_2=&12v_{2,0}+16v_{2,1}&=12[2][N][N-1]+16[N-1]^2\nn\\[1.5mm]
\dim_3=&8v_{3,0}+48v_{3,1}&=8[2]^2[N][N-1]+48[2][N-1]^2\nn\\[1.5mm]
\dim_4=&2v_{4,0}+48v_{4,1}+20v_{4,2}&=2[2]^3[N][N-1]+48[2]^2[N-1]^2+20[N-1]^2+20[2][N-1][N-2]
\nn\\[1.5mm]
\dim_5=&16v_{5,1}+40v_{5,2}&=16[2]^3[N-1]^2+40[2][N-1]^2+40[2]^2[N-1][N-2]\nn\\[1.5mm]
\dim_6=&20v_{6,2}+8v_{6,3}&=20[2]^2[N-1]^2+20[2]^3[N-1][N-2]+8[N-1]^2+\nn\\[1.5mm]
&&\phantom{=}+8([2]+[2]^3)[N-1][N-2]\nn\\[2mm]
\dim_7=&8v_{7,3}&=8[2][N-1]^2+8([2]^2+[2]^4)[N-1][N-2]\nn\\[1.5mm]
\dim_8=&v_{8,4}&=[N-1]^2+([2]+[2]^3+[2]^5)[N-1][N-2]
\nn\ee

\bigskip

\be
q^{8-8N}{\P}^{[3,4]}_r =
\boxed{q^{2-2N}}\ +\ [N-1]\times
 \ \ \ \ \ \ \ \ \ \ \ \ \ \ \ \
\ \ \ \ \ \ \ \ \ \ \ \ \ \ \ \ \ \ \ \ \ \ \ \
\ \ \ \ \ \ \ \ \ \ \ \ \ \ \ \ \ \ \ \ \ \ \ \
\nn \\ \nn \\
\times
\left.\begin{array}{|cc|ccccccccccccccccc }
\hline && \\
\l[N-2] &\times & &&&&&&&& \underline{q^5}       &&&&(1) \\ && \\
\l[N-2] & \times & &&&&&&& \underline{\underline{8q^4}} & 6q^3    &&&&(2)  \\ &&\\
\l[N-1] &\times &  &&&& 2q^4 & \underline{16q^3} & 20q^2 & 8q & 1   &&&&(-1)   \\
\l[N-2] & \times & &&&&&& \underline{\underline{28q^3}} &40q^2 &14q     &&&&(2)  \\ &&\\
\l[N-1] &\times & &&&8q^3 & 54q^2 &\underline{\underline{88q}} & 48 & \frac{8}{q} & &&&&(2)  \\
\l[N-2] & \times & &&&&& 40q^2 & \underline{\underline{92q}} & 64&\frac{14}{q} &&&&(-2)     \\ &&\\
\l[N-1] &\times &&&12q^2 & 64q & \underline{\underline{142}} &
\frac{88}{q} & \frac{20}{q^2} &\!\!\!\!\!\!\!\phantom{5_{5_{5_{5_{5_{5_{5_{5_5}}}}}}}}& &&&& (2)\\
\l[N-2] & \times &&&&&20q & \underline{\underline{80}} &
\frac{92}{q} & \frac{40}{q^2}& \frac{6}{q^3} &&&&(-2)
\\&& \\
\l[N-1] & \times & q^2 & 8q & 28 &\underline{\frac{56}{q}}
& \frac{50}{q^2} & \frac{16}{q^3}
&&& &&&& (-1) \\
\l[N-2] & \times &&&&&\underline{\frac{20}{q}}&\frac{40}{q^2} & \frac{28}{q^3} & \frac{8}{q^4}
& \frac{1}{q^5} &&&& (1)\\ &&\\
q^{-N} & \times & 2q^2 &8q & 12 & \frac{8}{q}& \frac{2}{q^2} &&&& &&&&(0) \\ &&\\
q^{-N} & \times &&&\underline{\underline{12 q^2}} & 16q & 6&&&& &&&&(2) \\ &&\\
q^{-N} &\times &&&& \underline{\underline{8q^3}} & 6q^2 &&&& &&&& (2)\\ &&\\
q^{-N}&\times &&&&&\underline{\underline{2q^4}} &&&& &&&&(2)
  &&\\ &&\\
\hline\hline &&\\
\l[N-2] &\times & &&&&&&&&  q^5     \\ && \\
\l[N-2] &\times & &&&&&&&2q^4&   \\ && \\
\l[N-2] & \times &&&&&&&q^3 \\ && \\
q^{-N} & \times &&&&&&q^5 \\ && \\
q^{-N} & \times &&&&&& 2q^3 \\ & \\
q^{-N} & \times &&&&& 2q^2 \\&&\\
q^{-N} & \times &&&& {q}\\&& \\ && \\
q^{-N} & \times &&& 2q^2 \\ && \\
q^{-N} & \times &&&& 2q^3 \\ && \\
q^{-N} &\times &&&&& 2q^4 \\ &&\\
\hline\hline
\end{array}\right|
\ee

\bigskip

The first three lines are transformed as follows:
\be
\begin{array}{cccccc} \\
q^3[N-1][N-2] \Lar q^4[N-1][N-2] \Lar q^5[N-1][N-2] \\ \\
\hline\hline  \\
q^{8-2N} \Lar [2]q^4 \Lar q^{2N} \\
\end{array}
\label{34n1n2}
\ee
and the last seven lines -- as
\be
q^{-N}[N-1]\times \left|\begin{array}{ccccccccc} \hline\\
&& 2q^2 & q \\
&&& 2q^3 & 2q^2\\
&&&& 2q^4 & 2q^3 \\
&&&&& q^5 &&& \\ \\
\hline\hline \\
&& q^2 & q^3 & [2]q^3 & [2]q^4
\\ \\
\hline \end{array}\right|
\ee
If we made the maximal possible cancelations {\it inside}
this table, only two terms would remain, $q^2$ and $q^5$,
but separated by a gap -- and the full contribution to the
KR polynomial would be $\sim q^{-N}[N-1]\Big( q^2(qT)^2 + q^5(qT)^5\Big)$
i.e. proportional to $N-1$.
Keeping more terms inside the table makes the full contribution
much smaller:
\be
\begin{array}{ccccccc}
\\
q^{-N}\times && q^2[N-1]& q^3[N-1] \\ \\
\l[2]q^{-N}\times &&&& q^3[N-1]& q^4[N-1] \\ \\
\hline\hline
\\
q^{-N}\times && q^{4-N} & q^{N+1} \\ \\
\l[2]q^{-N} \times &&&& q^{5-N} & q^{N+2} \\
\end{array}
\label{34qn1}
\ee
what leads to $\sim q^{-N}\Big( q^{4-N}(qT)^2 +q^{N+1}(qT)^3 +[2]q^{5-N}(qT)^4
+ [2]q^{N+2}(qT)^5\Big)$.
Of course, at $T=-1$ this coincides with
$q^{-N}[N-1]\Big( q^2(qT)^2 + q^5(qT)^5\Big)$.

Collecting all the contributions, we obtain:
\be
{\cal P}^{[3,4]}_r  =q^{8N-8}\Big\{ q^{2-2N} + q^{4-2N}(qT)^2 + q\cdot (qT)^3 +
\underline{\underline{[2]q^{5-2N}(qT)^4 +[2]q^2\cdot(qT)^5}}
+ \nn\\+\underline{q^{8-2N}(qT)^6 + [2]q^4\cdot(qT)^7 + q^{2N}(qT)^8}\Big\}
=
\nn\ee
\vspace{-0.5cm}
\be
\boxed{ = q^{6N-6}\Big\{ 1 + q^4T^2 +q^{2N+2}T^3 + [2]q^7T^4 + [2]q^{2N+5}T^5
+ q^{12}T^6 + [2]q^{2N+9}T^7 + q^{4N+6}T^8\Big\}}
\label{KRT35}
\ee
what is in perfect agreement with \cite{DGR} and \cite{DMMSS}:
\be
1+q^4T^2 + a^2q^2T^3 + [2]q^7T^4 + [2]q^5T^5a^2 + q^{12}T^6 + [2]q^9T^7a^2 +q^6T^8a^4
\ee

Eq.(\ref{KRT35}) is the answer at generic $N$.
For the special low value of $N=1$ it is substituted by Abelian answer ${\cal P}^{[3,4]}_r(N=1)=1$,
and it is also not directly applicable in the case of $N=2$.
The reason for this is that two substitutions, which we have made,
should be made differently, or should {\it not} be made at all, when $N=2$.
The first of them is (\ref{34n1n2}).
Indeed, here we substitute zero in the first line by non-vanishing second line --
and this should not be done.
Therefore the four terms, coming from this second line,  should be omitted when $N=2$:
they are underlined in (\ref{KRT35}).
The second substitution irrelevant at $N=2$ is in (\ref{34qn1}).
Here we encounter a map, proportional to the product of two quantum numbers: $[2][N-1]$.
Usually we substitute $[2][N-1] \Longrightarrow q[2][N-1]$ by $[2]q^{2-N}\Longrightarrow [2]q^{N-1}$,
eliminating the factor $[N-1]$, but when $N-1<2$ one should rather eliminate the bigger factor $[2]$,
substituting the original map by $q^{-1}[N-1]\Longrightarrow q^2[N-1]$.
This means that for $N=2$ the two double-underlined terms in (\ref{KRT35})
should be substituted by
$[N-1]q^{2-N}(qT)^4 +[N-1]q^{5-N}\cdot(qT)^5= q^4T^4 + q^8T^5$.
In result, for $N=2$ the answer (\ref{KRT35}) is changed for
\be
{\cal P}^{[3,4]}_r(N=2) = q^{6}\Big\{ 1 + q^4T^2 +q^{6}T^3 + q^6T^4 + q^{10}T^5\Big\}
\label{KRT35N2}
\ee
in accordance with \cite{katlas} (in fact the answers on this site itself are given only for
unreduced Khovanov polynomials, but running the attached program one can get the
reduced polynomials at $N=2$ as well).

\bigskip

In unreduced case we need appropriate modification of
(\ref{34n1n2}) and (\ref{34qn1}):
\be
\begin{array}{|ccccc|}
\hline&&&&\\
q^3[N][N-1][N-2] \Lar q^4[N][N-1][N-2] \Lar q^5[N][N-1][N-2] \\ &&&&\\
\hline\hline&&&& \\
q^{6-2N}[N-2] \Lar [2]q^4[N-2] \Lar q^{2N+2}[N-2]
\\&&&&\\\hline
\end{array}
\label{34N1N2}
\ee
and
\be
\begin{array}{|cccccc|}
\hline&&&&&\\
q^{-N}\times && q^2[N][N-1]& q^3[N][N-1]&& \\&&&&& \\
\l[2]q^{-N}\times &&&& q^3[N][N-1]& q^4[N][N-1] \\ &&&&&\\
\hline\hline&&&&&
\\
q^{-N}\times && q^{3-N}[N-1] & q^{N+2}[N-1]&& \\&&&&& \\
\l[2]q^{-N} \times &&&& q^{4-N}[N-1] & q^{N+3}[N-1] \\&&&&&\\\hline
\end{array}
\label{34qN1}
\ee
The latter is already familiar from the previous examples,
the former is new.

With the help of these relations we get:
$$
q^{8N-8}\Big\{ q^{2-2N}[N] +
q^{3-2N}[N-1]\,(qT)^2 + q^2[N-1]\, (qT)^3 +
[2]q^{4-2N}[N-1](qT)^4 +[2]q^3[N-1]\,(qT)^5 +
$$
$$
+ q^{6-2N}[N-2](qT)^6 + [2]q^4[N-2]\,(qT)^7 + q^{2N+2}[N-2]\,(qT)^8\Big\}
=
$$
\vspace{-0.5cm}
\be
 = q^{6N-6}\Big\{ [N] + q^3[N-1]T^2 +q^{2N+3}[N-1]T^3
+ [2]q^6[N-1]T^4 + [2]q^{2N+6}[N-1]\,T^5 + \nn \\
+ q^{10}[N-2]\,T^6 + [2]q^{2N+9}[N-2]\,T^7 + q^{4N+8}[N-2]\,T^8\Big\}
\label{KRT34urhypoth}
\ee
There is nothing to compare this formula with at generic values of $N$.
Unreduced superpolynomials for torus knots and links are {\it not}
provided by the evolution method of \cite{DMMSS,evo} and its relatives
of \cite{AgSh,Che}. On the other hand, the knot $[3,4]$ has $8$ crossings,
and as such it remained beyond reach of the matrix-factorization
calculations in \cite{CM}.
In the absence of both known results and explicitly constructed morphisms,
we need to rely upon other mind of arguments.
As we shall see in s.\ref{cofo} below it is most plausible that
(\ref{KRT34urhypoth}) is {\it over-reduced} -- all {\it known} answers
imply that unreduced KR polynomials for 3-strand knots should contain only
$[N-1]$ factors, not $[N-2]$.
Accordingly, we find eq.(\ref{KRT34ur}) from sec.\ref{cofo}
to be more plausible than (\ref{KRT34urhypoth}).

Instead, at $N=2$ eq.(\ref{KRT34urhypoth}) is exactly what needed
(while (\ref{KRT34ur}) will need further reduction in this case).
Indeed, both corrections, which led from (\ref{KRT35}) to (\ref{KRT35N2})
are unneeded in this case: the last four terms already enter (\ref{KRT34urhypoth})
with the coefficients $[N-2]$, which vanish at $N=2$,
and the simplified products are now $[2][N]$ rather than $[2][N-1]$,
so there is no need to switch from elimination of $[N]$ in generic case
to elimination of $[2]$ when $N=2$.
In full accordance with this prediction, it is enough to just
substitute $N=2$ into (\ref{KRT34urhypoth}) to reproduce the answer from \cite{katlas}.

\subsubsection{General case
\label{3strgencase}}

We omit the powers of $q$ because they depend in a clear way on position in the table.
Horizontal lines are inserted to simplify reading the table.

\centerline{{ \footnotesize
$
\begin{array}{llllllllllllllllll}
&&&&&&&&&\ldots \\ \\
\hline \\
&&&&&&& s_{70}
&\ldots & (2\delta_{n>7}+\delta_{n7}) \\
&&&&&&&&\ldots &(-\delta_{n>7}+\delta_{n6}+2\delta_{n5})\\ \\
&&&&&& s_{60} & 6s_{70} + s_{71}
&\ldots & (2\delta_{n>6}+2\delta_{n6}) \\
&&&&&&&
&\ldots & (2\delta_{n5}+\delta_{n4}) \\ \\
&&&&& s_{50} & 5s_{60}+s_{61} &
15s_{70}+5s_{71}+s_{72}
&\ldots & (2\delta_{n>5}+\delta_{n5}) \\
&&&&&& & s_{72}+s_{73}
&\ldots & (\delta_{n5}+2\delta_{n4})\\ \\
\hline \\
&&&&s_{40} & 4s_{50} +s_{51}& 10s_{60}+4s_{61}+s_{62}&
20s_{70}+10s_{71}+3s_{72}+s_{73}    \!\!\!\!\!\!\!\!\!\!\!\!
&\ldots & (2\delta_{n>4}-\delta_{n4} )\\
&&&&& & s_{62} + s_{63} & 4s_{72}+5s_{73}
&\ldots & (-\delta_{n>4}+2\delta_{n4}+\delta_{n3})\\ \\
&&& s_{30} & 3s_{40}+s_{41} &  6s_{50} +3s_{51}+s_{52}
&   10s_{60}+6s_{61}+2s_{62} +s_{63}&
 15s_{70}+10s_{71}+3s_{72}+s_{73}    \!\!\!\!\!\!\!\!\!\!\!\!
&\ldots & (2\delta_{n>3}-2\delta_{n3})\\
&&&& & s_{52} &3s_{62}+4s_{63}& 6s_{72}+8s_{73}
&\ldots & (-2\delta_{n>3}+2\delta_{n3})\\ \\
&& s_{20} &\!\!\! 2s_{30}+s_{31}  & \underline{\underline{3s_{40}+2s_{41} +s_{42}}}
& 4s_{50} + 3s_{51} +s_{52}  &
5s_{60}+4s_{61}+s_{62} & 6s_{70}+5s_{71}+s_{72}
&\ldots &  (2\delta_{n>2}-\delta_{n2}) \\
&&&& s_{42} &\underline{\underline{2s_{52}}} & 3s_{62} + 4s_{63}
& 4s_{72}+5s_{73}
&\ldots &  (-2\delta_{n>2}+\delta_{n2})\\ \\
1 & s_{10} & s_{20}+s_{21} & \underline{s_{30}+s_{31}} &s_{40}+s_{41}& s_{50}+s_{51}
& s_{60}+s_{61} & s_{70}+s_{71}
&\ldots
 & (-\delta_{n>1}= \delta_{n1}-1)\\
&&&& \underline{s_{42}} & s_{52} & s_{62}+s_{63}& s_{72}+s_{73}
&\ldots
& (\delta_{n>1}=1-\delta_{n1})
\\ \\ \hline \hline \\
2 & s_{10} &s_{20} & s_{30} &s_{40} & s_{50} & s_{60} & s_{70}  & \ldots
& (0) \\ \\
&& \underline{\underline{s_{20}}}& 2s_{30}& 3s_{40} & 4s_{50} & 5s_{60} &
6s_{70}   &\ldots
& (2\delta_{n>1})  \\ \\
&&& \underline{\underline{s_{30}}}& 3s_{40} & 6 s_{50} & 10s_{60} &
15s_{70}  & \ldots & (2\delta_{n>2}) \\ \\
&&&& \underline{\underline{s_{40}}}&   4 s_{50} & 10s_{60} & 20s_{70}  & \ldots & (2\delta_{n>3}) \\ \\
&&&&& \underline{\underline{s_{50}}}&   s_{60} & 15s_{70}  & \ldots & (2\delta_{n>4}) \\ \\
&&&&&& \underline{\underline{s_{60}}}& 6s_{70}  & \ldots & (2\delta_{n>5}) \\ \\
&&&&&&& \underline{\underline{s_{70}}}   & \ldots & (2\delta_{n>6}) \\ \\
&&&&&&& \ \ \ \ \ \ \ \ \ \ \ldots
\\ \\
\end{array}
$
}}

\bigskip

If the structure is not fully clear at this level, here are the
next three columns of the table:

\bigskip

\centerline{{\footnotesize
$\ \ \ \
\begin{array}{lllllllllllllll}
&&&&&&&&&\ldots \\ \\
&&&& s_{10,0} &\ldots & (2\delta_{n>10}-\delta_{n,10}) \\
&&&&&\ldots &(-\delta_{n>10} +2\delta_{n,10}+\delta_{n9}+2\delta_{n7}+\delta_{n6}) \\
\\ \\
&&& s_{90} & 9s_{10,0} + s_{10,1} &\ldots &(2\delta_{n>9}-2\delta_{n9}) \\
&&&&&\ldots &(-2\delta_{n>9}+2\delta_{n9}+\delta_{n7}+2\delta_{n6}) \\ \\
&& s_{80}& 8s_{90}+s_{91}   & 36s_{10,0}+8s_{10,1}+s_{10,2}
&\ldots & (2\delta_{n>8}-\delta_{n8}) \\
&&&&   s_{10,2}+s_{10,3}+s_{10,4} + s_{10,5}
&\ldots & (-2\delta_{n>8}+\delta_{n8}+2\delta_{n6}+\delta_{n5})\\ \\
\hline \\
 &   & 7s_{80}+s_{81}
  & 28s_{90}+7s_{91}+s_{92}  & 84s_{10,0}+28s_{10,1}+6s_{10,2}+s_{10,3}
&\ldots &(2\delta_{n>7}+\delta_{n7}) \\
&& & s_{92}+s_{93}+s_{94} & 7s_{10,2}+8s_{10,3}+8s_{10,4} + 8s_{10,5}
&\ldots & (-\delta_{n>7}+\delta_{n6}+2\delta_{n5})
\\ \\
 &    & 21s_{80} + 6s_{81}+s_{82}
  & 56s_{90}+21s_{91}+5s_{92}+s_{93 }  & 126s_{10,0}+56s_{10,1}+15s_{10,2}+4s_{10,3}+s_{10,4}
 \!\!\!\!\!\!\!\!\!\!\!\! \!\!\!\!\!\!\!\!\!\!\!\!   \!\!\!\!\!\!\!\!\!\!\!\!
&\ldots &  (2\delta_{n>6}+2\delta_{n6}) \\
 && s_{82} + s_{83}+s_{84}
 & 6s_{92}+7s_{93}+7s_{94} & 21s_{10,2}+26s_{10,3}+27s_{10,4} + 27s_{10,5}
&\ldots & (2\delta_{n5}+\delta_{n4}) \\ \\
  &     & 35 s_{80}+15s_{81}+4s_{82}+s_{83}
   & 70s_{90}+35s_{91}+10s_{92}+3s_{93}+s_{94}
   & 126s_{10,0}+70s_{10,1}+20s_{10,2}+6s_{10,3}+2s_{10,4}+s_{10,5}
   \!\!\!\!\!\!\!\!\!\!\!\!  \!\!\!\!\!\!\!\!\!\!\!\!  \!\!\!\!\!\!\!\!\!\!\!\!
    \!\!\!\!\!\!\!\!\!\!\!\!   \!\!\!\!\!\!\!\!\!\!\!\!
&  \ldots
& \ \ \ \ \ \ \ \ \ \ \ \        (2\delta_{n>5}+\delta_{n5}) \\
 &  & 5s_{82}+6s_{83}+6s_{84}
 & 15s_{92}+19s_{93}+20s_{94} & 35s_{10,2}+45s_{10,3}+48s_{10,4} + 49s_{10,5}
&\ldots & (\delta_{n5}+2\delta_{n4})\\ \\
\hline \\
&   &
  35s_{80}+20s_{81}+6s_{82} +2s_{83} +s_{84}
   & 56s_{90}+35s_{91}+10s_{92}+3s_{93}+s_{94}  & 84s_{10,0}+56s_{10,1}+15s_{10,2}+4s_{10,3}+s_{10,4}
 \!\!\!\!\!\!\!\!\!\!\!\!   \!\!\!\!\!\!\!\!\!\!\!\!  \!\!\!\!\!\!\!\!\!\!\!\!
&\ldots & (2\delta_{n>4}-\delta_{n4} )\\
 &  & 10s_{82}+13s_{83}+14s_{84}
 & 20s_{92}+26s_{93}+28s_{94} & 35s_{10,2}+45s_{10,3}+48s_{10,4} + 49s_{10,5}
&\ldots & (-\delta_{n>4}+2\delta_{n4}+\delta_{n3})\\ \\
&  &21s_{80}+15s_{81}+4s_{82}+s_{83}
 & 28s_{90}+21s_{91}+5s_{92}+s_{93}  & 36s_{10,0}+28s_{10,1}+6s_{10,2}+s_{10,3}
&\ldots & (2\delta_{n>3}-2\delta_{n3})\\
 &  & 10s_{82}+13s_{83}+14s_{84}
 & 15s_{92}+19s_{93}+20s_{94} & 21s_{10,2}+26s_{10,3}+27s_{10,4} + 27s_{10,5}
&\ldots & (-2\delta_{n>3}+2\delta_{n3})\\ \\
  &  & 7s_{80}+6s_{81}+s_{82}
   & 8s_{90}+7s_{91}+s_{92}  & 9s_{10,0}+8s_{10,1}+s_{10,2}
&\ldots & (2\delta_{n>2}-\delta_{n2}) \\
 &  & 5s_{82}+6s_{83}+6s_{84}
  & 6s_{92}+7s_{93}+7s_{94} & 7s_{10,2}+8s_{10,3}+8s_{10,4} + 8s_{10,5}
&\ldots &  (-2\delta_{n>2}+\delta_{n2})\\ \\
 &  & s_{80}+s_{81}
 & s_{90}+s_{91} & s_{10,0}+s_{10,1}
&\ldots & (-\delta_{n>1})\\
 &  & s_{82}+s_{83}+s_{84}
 & s_{92}+s_{93}+s_{94} & s_{10,2}+s_{10,3}+s_{10,4}+s_{10,5}
&\ldots & (\delta_{n>1})
\\ \\  \hline\hline \\
  &   & s_{80}& s_{90} & s_{10,0} & \ldots & (0) \\ \\
  &   &7s_{80}& 8s_{90} & 9s_{10,0}  &\ldots & (2\delta_{n>1})  \\ \\
  &   & 21s_{80} & 28s_{90} & 36s_{10,0} & \ldots & (2\delta_{n>2}) \\ \\
  &   & 35s_{80} & 56s_{90} & 84s_{10,0} & \ldots & (2\delta_{n>3}) \\ \\
  &   & 35s_{80} & 70s_{90} & 126s_{10,0} & \ldots & (2\delta_{n>4}) \\ \\
  &   & 21s_{80} & 56s_{90} & 126s_{10,0} & \ldots & (2\delta_{n>5}) \\ \\
  &   & 7s_{80} & 28s_{90} & 84s_{10,0} & \ldots & (2\delta_{n>6}) \\ \\
  &  & s_{80} & 8s_{90} & 36s_{10,0} & \ldots & (2\delta_{n>7}) \\ \\
  &  & & s_{90} & 36s_{10,0} & \ldots & (2\delta_{n>8}) \\ \\
  & & &  &  s_{10,0} & \ldots & (2\delta_{n>9}) \\ \\
  &&&&& \ldots \\ \\
 \end{array}
$
}}

\noindent
We remind that $s_{m,k}$ is non-vanishing when $\ m\leq n+k\ $ and $\ k\leq \frac{m}{2}\ $ and
therefore $\ m\leq 2n$.

\bigskip

Alternated sums are linear combinations of the identities like
\be
2 + \sum_{m=1}^{2n} (-)^m s_{m,0} = 0, \nn \\
\sum_{m=2k}^{2n} (-)^m s_{m,k} = \delta_{k,n}, \ \ \ k>0
\label{sumrules0}
\ee
Then, for example, in the lowest non-trivial line we have
\be
s_{42} - s_{52} + (s_{62}+s_{63}) - (s_{72}+s_{73}) + (s_{82}+s_{83}+s_{84}) - \ldots = \nn \\
= \Big(s_{42}-s_{52}+s_{62} -s_{72} + s_{82} - \ldots\Big)
+\Big(s_{63}-s_{73}+s_{83} - \ldots\Big) + \Big(s_{84}-\ldots\Big) + \ldots = \nn \\
=\delta_{n2} + \delta_{n3} + \delta_{n4} + \ldots = 1-\delta_{n1} = \delta_{n>1}
\ee
In general alternated sums in the lines with $[N-1]$ are evaluated with the help of the formulas
\be
\sum_{m=p}^{2n} (-)^{m+p} \frac{(m-1)!}{(p-1)!(m-p)!}\,s_{m0}
+ \sum_{k=1}^p\left(\sum_{m=p+k}^{2n} (-)^{m+p} \frac{(m-2k)!}{(p-k)!(m-p-k)!}\,s_{m,k}\right)
= 2\delta_{n>p} +\gamma_p\delta_{n,p}
\ee
where the deviation $\gamma_p$ is periodic in $p$, $\ \gamma_{p+6}=\gamma_p$,
and
\be
\gamma_1 = 1, \ \ \gamma_2=-1, \ \ \gamma_3 = -2, \ \ \gamma_4 = -1, \ \
\gamma_5 = +1, \ \ \gamma_6 = +2, \ \ \gamma_7 = +1, \ \ \ldots
\ee
Similar periodicity occurs in the lines with $[N-2]$:
the alternated sums are equal to
\be
\beta_{p} \delta_{n>p}
\ee
with $\beta_{p+6} = \beta_p$
\be
\beta_1 = -1, \ \ \beta_2 = -2, \ \ \beta_3 = -2, \ \ \beta_4 = -1, \ \ \beta_5 = 0, \ \ \beta_6 = 0,
\ \ \ldots
\ee
(in the two lowest line we actually write $\gamma_1\delta_{n>1}$ and $-\beta_1\delta_{n>1}$,
because they begins from a "wrong" place).
Here are a few more Euler characteristics in the $[N-2]$-lines:
\be
\begin{array}{lllllllll}
%
%
&&&&&\ldots &(-\delta_{n>16} +2\delta_{n,16}+\delta_{n,15}+2\delta_{n,13}+\delta_{n,12}
+2\delta_{n.10}+\delta_{n.9}) \\ \\
&&&&&\ldots &(-2\delta_{n>15}+2\delta_{n,15}+\delta_{n,13}+2\delta_{n,12}
+\delta_{n,10}+2\delta_{n9})\\ \\
&&&&&\ldots &(-2\delta_{n>14}+\delta_{n,14} +2\delta_{n,12}+\delta_{n,11}
+2\delta_{n9}+\delta_{n8}) \\ \\
&&&&&\ldots &(-\delta_{n>13} +\delta_{n,12}+2\delta_{n,11}+\delta_{n9}+2\delta_{n8}) \\ \\
&&&&&\ldots &(2\delta_{n,11}+\delta_{n,10}+2\delta_{n8}+\delta_{n7}) \\ \\
&&&&&\ldots &(\delta_{n,11}+2\delta_{n,10}+\delta_{n8}+2\delta_{n7}) \\ \\
\end{array}
\ee
The structure and values of Euler characteristics for arbitrary $n$ and $p$ is fully clear from these expressions.

\subsubsection {Knot $[3,5]$}

Before proceeding generically we analyze two more examples
-- already with the help of this general table --
to better see, how the things work.
For the case $n=5$ the table becomes:

\bigskip

\be
\!\!\!\!\!\!\!\!\!\!\!\!\!\!
q^{2-2N} + [N-1]\cdot
\left|
\begin{array}{cc|ccccccc|cccccccccc}
\hline
 &&&&&&&& \\
&&&&&&&&&&&& 0 &&&&(0)&\\
\l[N-2] &\times &&&&&&&&&&& q^7 &&&&(1)&  \\
&&&&&&&& \\
\hline
 &&&&&&&& \\
&&&&&&&&&&& 0 & &&&&(0)&\\
\l[N-2] &\times & &&&&&&&&& 2q^6& &&&&(2)& \\
 &&&&&&&&&&&&&& \\
&&&&&&&&&& 0 &&  &&&&(0)&\\
\l[N-2] &\times & &&&&&&&& 2q^5 && &&&&(2)&\\
 &&&&&&&&&&&&& \\
\l[N-1] &\times &&&&&&&&q^3 &&& &&&&(1)&\\
\l[N-2] &\times &&&&&&&& q^4 &&& &&&&(1)&\\
 &&&&&&&&&&&&& \\
\hline &&&&&&&&&&&&& \\
\l[N-1] &\times &&&&&&& \underline{2q^2} & &&&&(2)&\\
\l[N-2] &\times &&&&&&&&q^2 &&&&(-1)&\\&&&&&&&&&&&&& \\
\l[N-1] &\times &&&&&& 2q  &&&&&&(2)&\\
\l[N-2] &\times &&&&&&& 2q & &&&&(-2)&\\ &&&&&&&&&&&&&\\
\l[N-1] &\times &&&&& 2q^0 &&& &&&&(2)&\\
\l[N-2] &\times &&&&&&2q^0 &&&&&&(-2)&\\&&&&&&&&&&&&& \\
\l[N-1] &\times & &&& \frac{1}{q} &&&& &&&&(-1)& \\
\l[N-2] &\times & &&& & \frac{1}{q} &&& &&&&(1)& \\ &&&&&&&&&&&&&\\
\hline\hline &&&&&&&&&&&&& \\
q^{-N} & \times & 0 & 0 & 2q^2 &&&& &&&&(2)&\\
&&&&& 2q^3 &&&&&&&(2)&\\
&&&&&& 2q^4 &&&&&&(2)&\\
&&&&&&& 2q^5 & &&&&(2)&\\
&&&&&&&&&&&&& \\
\hline
\end{array}\right|
\label{35tab1}
\ee

\bigskip

\noindent
As usual we begin with the adjacent lines, multiplied by $[N-1]$ and $[N-2]$.
From above table it is already clear, that a reordering takes place in the
middle -- at the central cross, and adjacent items are naturally grouped in pairs
along differently directed diagonals.
Accordingly we need three kinds of substitutions:
\be
\l[N-1] \Longrightarrow [N-2]  \ \ \ \ {\rm by} \ \ \ \ q^{2-N} \Longrightarrow 0
\label{35tab11}
\ee
$$
q^2[N-1] \Longrightarrow q^4[N-2] \ \ \ \ {\rm by} \ \ \ \ q^{6-N}+q^{4-N} \Longrightarrow q^{N+1}
$$
and
$$
\l[N-2] \Longrightarrow q[N-2] \ \ \ \  {\rm by} \ \ \ \ q^{3-N} \Longrightarrow q^{N-2}
$$
in the left lower corner, the central cross  and the right upper corner respectively.
The underlined item $2q^2$ is split into two $q^2$, which are handled in two different ways.
Actually, in the right upper corner we have a choice between two sequences --
the pair of arrows in the last two of the following lines
could be either horizontal, or diagonal:
$$
\begin{array}{ccccccccccccccc}
\\
&&&&&&q^6[N-2] \Lar q^7[N-2] \\  \\
&&&&q^5[N-2] \Lar q^6[N-2] \\      \\
&& q^4[N-2] \Lar q^5[N-2] \\
&\nearrow \\
&& & \nearrow\\
q^2[N-1] \Lar q^3[N-1] \\
\end{array}
$$
If all the arrows are horizontal, this is cohomologically equivalent to
\be
\begin{array}{ccccccccccccccc}
&&&&&&q^{9-N} \Lar q^{N+4}\\
&&&&q^{8-N} \Lar q^{N+3} \\
& & q^{7-N} \Lar q^{N+2} \\
q^{4-N} \Lar q^{N+1} & \\
\end{array}
\label{35tab2}
\ee
If we took two diagonal arrows instead, we would rather get
$$
{\footnotesize
\begin{array}{ccccccccccccccc}
&&&&&&q^{9-N} \Lar q^{N+4}\\
&&&&q^{8-N} \Lar q^{N+3} \\
& & q^{5-N}+q^{7-N} \Lar q^{N+2} \\
q^{4-N}+q^{6-N} \Lar q^{N+1} & \\
\end{array}
}
$$
i.e. more terms would remain, thus the horizontal arrows are the right choice.
These substitutions (\ref{35tab2}) convert our table (\ref{35tab1}) into

\bigskip

$$
q^{2-2N} + [N-1]\cdot
\left|
\begin{array}{cc|ccccccc|cccccccccc}
\hline
 &&&&&&&&& \\
q^N  & \times &&&&&&&& q & q^2 & q^3 & q^4 \\ &&&&&&&&&&&&&& \\
q^{-N} & \times &&&&&&&q^4& q^7 & q^8 &  {q^9} &  \\
&&&&&&&&&&&&&& \\
q^{-N} & \times & & &  &\frac{1}{q}\cdot q^2&2q^2&2q\cdot q^2&q^2\cdot q^2&&&&&& \\
&&&&&&&&&&&&&& \\
\hline \hline && &&&&&&&&&& \\
q^{-N} & \times &0 &0 & 2q^2 & 2q^3&2q^4&2q^5& &&&&& \\
 &&&&&&&&&&&&& \\
\hline
\end{array}\right|
$$

\bigskip

\noindent
The first two lines come from (\ref{35tab2}), the third line -- from application
of (\ref{35tab11}).
Putting all the terms from the last two lines into one, we get:
$$
q^{2-2N} + [N-1]\cdot
\left|
\begin{array}{cc|ccccccc|cccccccccc}
\hline
 &&&&&&&&& \\
q^N  & \times &&&&&&&& q & q^2 & q^3 & q^4 \\
&&&&&&&&&&&&&& \\
q^{-N} & \times & & &  & & && q^4&q^7&q^8&q^9&& \\
&&&&&&&&&&&&&& \\
q^{-N} & \times &0 &0 & 2q^2 & 2q^3+q&2q^4+2q^2 &2q^5+2q^3& q^4  \\
 &&&&&&&&&&&&& \\
\hline
\end{array}\right|
$$
and eliminating the cohomologically trivial pairs in this new line,
\be
q^{2-2N} + [N-1]\cdot
\left|
\begin{array}{cc|ccccccc|cccccccccc}
\hline
 &&&&&&&&& \\
q^N  & \times &&&&&&&& q & \underline{q^2} & \underline{q^3} & \underline{q^4} \\
&&&&&&&&&&&&&& \\
q^{-N} & \times & & &  & & && q^4&\underline{q^7}&\underline{q^8}&\underline{q^9}&& \\
&&&&&&&&&&&&&& \\
q^{-N} & \times &0 &0 & q^2 & q^3&q^3[2] &q^4[2]&   \\
 &&&&&&&&&&&&& \\
\hline
\end{array}\right|
\label{35tab3}
\ee

\bigskip

\noindent
Now we have three increasing sequences and
time comes to multiplication by $[N-1]$ and further
elimination of cohomologically trivial parts.
This is already a standard procedure in this paper,
the only new point is that in the middle sequence we have a jump
in power by three between $q^4$ and $q^7$,
and $\ q^4[N-1]\Longrightarrow q^7[N-1]\ $ is substituted by
$\ q^{8-N}+q^{6-N}=q^{7-N}[2] \Longrightarrow q^{N+5} +q^{N+3}=q^{N+4}[2]\ $.
Similarly, in unreduced case -- which splits at this step --
we substitute $\ q^4[N]\Longrightarrow q^7[N]\ $  by
$\ q^{7-N}+q^{5-N} =q^{6-N}[2]\Longrightarrow q^{N+6} +q^{N+4}=q^{N+5}[2]$.

Continuing with the unreduced case, we obtain
\be
\!\!\!\!\!\!\!\!\!\!\!\! \!\!\!\!\!\!\!\!\!
q^{2-2N} +
\left|
\begin{array}{cc|ccccccc|cccccccccc}
\hline
 &&&&&&&&& \\
 q^{2N} & \times &&&&&&&& & q^2\cdot q^{-2} && q^4\cdot q^{-2} \\
 &&&&&&&&& \\
 1 & \times &&&&&&&& q\cdot q^{2} && q^3\cdot q^{2} \\
 &&&&&&&&& \\   &&&&&&&&& \\
 1 & \times &&&&q^3\cdot q^{-2}&&q^4[2]\cdot q^{-2}&&q^4[2]&& q^7   \\
 &&&&&&&&& \\
 q^{-2N} & \times &0&0 &q^2\cdot q^2 &&q^3[2]\cdot q^2 &&q^{7}[2]&&q^{10} \\
 &&&&&&&&&\\
 \hline
\end{array}\right|
\label{35tab4}
\ee

\bigskip

This table is the answer:
$$
{\cal P}^{[3,5]}_r = q^{10(N-1)}\Big( q^{2-2N} + q^{4-2N}(qT)^2 + q\cdot (qT)^3
+ \underline{\underline{q^{5-2N}[2](qT)^4 + q^2[2](qT)^5}} +
$$
\vspace{-0.7cm}
\be
 + \underline{\underline{q^{7-2N}[2](qT)^6 + (q^4[2]}}+\underline{q^3)(qT)^7 + (q^{2N} + q^{10-2N})(qT)^8
+ q^6[2](qT)^9 + q^{2+2N}(qT)^{10}}\Big) =
\label{KRT35ra}
\ee
\vspace{-0.2cm}
$$
\boxed{\ = q^{8(N-1)}\Big(
1+\Big(q^4T^2 +q^2T^3a^2\Big)\Big(1 + q^3[2]T^2+q^7[2]T^4\Big)
+  q^8 T^7a^2 + (q^{16}+q^6a^4)T^8 +q^{13}[2]T^9a^2+q^{10}T^{10}a^4\Big)\ }
$$
with $a=q^N$, what is in perfect agreement with (159) of \cite{DMMSS}.

As explained in sec.\ref{torus34} this formula requires substantial modification
for the special low value of $N=2$ (for Khovanov polynomial):
all underlined terms, coming from the piece of $\P$, which was originally proportional to $[N-2]$
(for clarity they are also underlined in the earlier table (\ref{35tab3})),
should be omitted, and in the double underlined terms, where $[N-1]$ was eliminated for generic $N$,
at $N=2$ one should rather eliminate a bigger factor $[2]$.
Thus
\be
{\cal P}^{[3,5]}_r(N=2) =q^8\Big\{1 +
 q^4T^2 + q^6T^3 +  q^{6} T^4 + q^{10} T^5 + q^{10}T^6+q^{14}T^7\Big\}
\ee
in perfect agreement with \cite{katlas}.

\bigskip

In unreduced case (\ref{35tab3}) is multiplied by $[N]$, and in the further transformations
we use $[N]$ instead of $[N-1]$, so that (\ref{35tab4}) is changed for
$$
q^{2-2N}[N] + [N-1]\cdot
\left|
\begin{array}{cc|ccccccc|cccccccccc}
\hline
 &&&&&&&&& \\
 q^{2N} & \times &&&&&&&& & q^2\cdot q^{-1} && q^4\cdot q^{-1} \\
 &&&&&&&&& \\
 1 & \times &&&&&&&& q\cdot q && q^3\cdot q \\
 &&&&&&&&& \\   &&&&&&&&& \\
 1 & \times &&&&q^3\cdot q^{-1}&&q^4[2]\cdot q^{-1}&&q^5[2]&& q^8   \\
 &&&&&&&&& \\
 q^{-2N} & \times &0&0 &q^2\cdot q &&q^3[2]\cdot q &&q^{6}[2]&&q^{9} \\
 &&&&&&&&&\\
 \hline
\end{array}\right|
\label{35tab4a}
$$
and unreduced KR polynomial is
\be
{\cal P}^{[3,5]} =
q^{10(N-1)}\Big\{ q^{2-2N}[N] + \Big(q^{3-2N}(qT)^2 + q^2 (qT)^3 + q^{4-2N}[2](qT)^4
+ q^3[2](qT)^5 + q^{6-2N}[2](qT)^6 + \nn \\
+ (q^2+q^4+q^6)(qT)^7 + (q^{2N+1} + q^{9-2N})(qT)^8
+ (q^4+q^8)(qT)^9 + q^{3+2N}(qT)^{10}\Big)\cdot[N-1]\Big\}
\label{KRT35ura}
\ee
Like in the $[3,4]$ case, there is nothing to compare this formula with --
it is new.
However, eq.(\ref{KRT35ura}) will be confirmed in sec.\ref{cofo}.
To get superpolynomial, one makes the substitutions
\be
\boxed{
q^N=a, \ \ \ \ \ \ \l[N-p\,] = \frac{q^{2p}+a^2T}{q^{p-1}(1-q^2)a}
}
\label{KRtosup}
\ee
(assuming that quantum numbers $[N-p\,]$ are independent variables --
not made out of $q^N$).

\subsubsection{The link $[3,6]$}

This time the table is

\bigskip

\be
\!\!\!\!\!\!\!\!\!\!\!\!\!\!\!\!\!\!\!\!\!\!
q^{2-2N} + [N-1]\cdot
\left|
\begin{array}{cc|ccccccc|cccccccccc}
\hline
 &&&&&&&& \\
 &&&&&&&&&&&&  &&0&&(0)&\\
\l[N-2] &\times &&&&&&&&&&&&& q^9 &&(1)&  \\
&&&&&&&& \\
&&&&&&&&&&&&& 0 &&&(0)&\\
\l[N-2] &\times &&&&&&&&&&&& 2q^8 &&&(2)&  \\
&&&&&&&& \\
&&&&&&&&&&&& 0 &&&&(0)&\\
\l[N-2] &\times &&&&&&&&&&& 2q^7 &&&&(2)&  \\
&&&&&&&& \\
\hline
 &&&&&&&& \\
&&&&&&&&&&& 0 & &&&&(0)&\\
\l[N-2] &\times & &&&&&&&&& q^6& &&&&(1)& \\
 &&&&&&&&&&&&&& \\
\l[N-1]&\times &&&&&&&
&& 2q^4 &&  &&&&(2)&\\
 & & &&&&&&&& 0 && &&&&(0)&\\
 &&&&&&&&&&&&& \\
\l[N-1] &\times &&&&&&
&&2q^3 &&& &&&&(2)&\\
 & &&&&&&&& 0 &&& &&&&(0)&\\
 &&&&&&&&&&&&& \\
\hline &&&&&&&&&&&&& \\
\l[N-1] &\times &&&&&&& {2q^2} & &&&&(2)&\\
\l[N-2] &\times &&&&&&&&q^2 &&&&(-1)&\\&&&&&&&&&&&&& \\
\l[N-1] &\times &&&&&& 2q  &&&&&&(2)&\\
\l[N-2] &\times &&&&&&& 2q & &&&&(-2)&\\ &&&&&&&&&&&&&\\
\l[N-1] &\times &&&&& 2q^0 &&& &&&&(2)&\\
\l[N-2] &\times &&&&&&2q^0 &&&&&&(-2)&\\&&&&&&&&&&&&& \\
\l[N-1] &\times & &&& \frac{1}{q} &&&& &&&&(-1)& \\
\l[N-2] &\times & &&& & \frac{1}{q} &&& &&&&(1)& \\ &&&&&&&&&&&&&\\
\hline\hline &&&&&&&&&&&&& \\
q^{-N} & \times & 0 & 0 & 2q^2 &&&& &&&&(2)&\\
&&&&& 2q^3 &&&&&&&(2)&\\
&&&&&& 2q^4 &&&&&&(2)&\\
&&&&&&& 2q^5 & &&&&(2)&\\
&&&&&&&& 2q^6 &&&&(2)&\\
&&&&&&&&&&&&& \\
\hline
\end{array}\right|
\label{36tab1}
\ee

\bigskip

\noindent
We shall see a little later that a few more cohomologically trivial combinations
should be kept in this table in order to  reproduce the known superpolynomial,
but for a while we can continue with the table in its present form.

The next iteration:
$$
q^{2-2N} + [N-1]\cdot
\left|
\begin{array}{cc|ccccccc|cccccccccc}
\hline
 &&&&&&&&& \\
 \l[N-2] &\times &&&&&&&&&&&&& q^9 \\
 &&&&&&&&& \\
\hline
 &&&&&&&&& \\
q^N  & \times &&&&&&&& q & q^2 & q^3 & & 2q^5 \\ &&&&&&&&&&&&&& \\
q^{-N} & \times &&&&&&&q^4& \overline{q^5} & q^6+q^8 & & 2q^{10} &  \\
&&&&&&&&&&&&&& \\
q^{-N} & \times & & &  &\frac{1}{q}\cdot q^2&2q^2&2q\cdot q^2&q^2\cdot q^2&&&&&& \\
&&&&&&&&&&&&&& \\
\hline \hline && &&&&&&&&&& \\
q^{-N} & \times &0 &0 & 2q^2 & 2q^3&2q^4&2q^5&\overline{2}q^6 &&&&& \\
 &&&&&&&&&&&&& \\
\hline
\end{array}\right|
$$

\bigskip

\noindent
and further, eliminating one overlined pair and {\it inserting} three underlined ones,

\bigskip

\centerline{
$
q^{2-2N} + [N-1]\cdot
\left|
\begin{array}{cc|ccccccc|cccccccccc}
\hline
 &&&&&&&&& \\
 \l[N-2] &\times &&&&&&&&&&&&& q^9 \\
 &&&&&&&&& \\
\hline
 &&&&&&&&& \\
q^N  & \times &&&&&&&& q & q^2 & q^3
& \underline{\underline{\underline{q^4}}} & 2q^5+\underline{\underline{\underline{q^3}}} \\
&&&&&&&&&&&&&& \\
q^{-N} & \times &0 &0 &  q^2 & q^3  &q^3[2]&q^4[2]& q^5[2] & 0+\underline{q^6[2]}
&\underbrace{q^7[2]+\underline{q^5[2]}}_{q^8+(2q^2+1)\cdot q^4}
&0+\underline{\underline{q^9}} &2q^{10}+\underline{\underline{q^8}}& \\
 &&&&&&&&&&&&& \\
\hline
\end{array}\right|  \sim
$
}

\bigskip

\centerline{
$
\sim q^{2-2N} +
\left|
\begin{array}{cc|ccccccc|cccccccccc}
\hline
 &&&&&&&&& \\
 \l[N-1][N-2] &\times &&&&&&&&&&&&& q^9 \\
 &&&&&&&&& \\
 q^N[N-1]  &\times &&&&&&&&&&&&(2q^2+1)\cdot q^3 & \\
  &&&&&&&&& \\
\hline
 &&&&&&&&& \\
q^{2N}  & \times &&&&&&&& & 1 &  & {q^2} &   \\
&&&&&&&&&&&&&& \\
1 &\times &&&&&&&&  q^3 &&q^5 &&& \\
&&&&&&&&&&&&&& \\&&&&&&&&&&&&&& \\
1 &\times &&&& q &&q^2[2]&&q^4[2]&&q^7&&& \\
&&&&&&&&&&&&&& \\
q^{-2N} & \times &0 &0 &  q^4 &   &q^5[2]&&q^7[2]&&q^{10}  \\
 &&&&&&&&&&&&& \\
\hline
 &&&&&&&&&&&&& \\
q^{-N}[N-1] & \times &&&&&&&  & 0 & (2q^2+1)\cdot q^4 &&(2q^{2}+1)\cdot q^8&\\
&&&&&&&&&&&&& \\
\hline
\end{array}\right|
$
}

\bigskip

\noindent
Insertions are {\it allowed}, because there were many terms of this kind
at these places in original table from sec.\ref{3strgencase}.
Insertions are {\it needed} to provide the right answer for the superpolynomial.

Indeed, from the last table we read:
\be
{\cal P}^{[3,6]}_r = q^{12N-12}\Big\{ q^{2-2N} + q^{4-2N}(qT)^2 + q\cdot(qT)^3 + q^{5-2N}[2](qT)^4
+ q^2[2](qT)^5 + q^{7-2N}[2](qT)^6 + \nn \\
+ (q^4[2]+q^3)(qT)^7 + (q^{2N}+q^{10-2N})(qT)^8 +q^6[2](qT)^9 + q^{2N+2}(qT)^{10} + \nn \\
+ (2q^2+1)\Big(q^{4-N}(qT)^8 + q^{8-N}(qT)^{10}+q^{N+3}(qT)^{11}\Big)[N-1] +
q^9[N-1][N-2](qT)^{12}\Big\}
\label{KR36r}
\ee
Using the rules (\ref{KRtosup}) we obtain $q^{10(N-1)}$ times:
$$
1+q^4T^2+q^2a^2T^3+q^7[2]T^4+q^5a^2T^5 + q^{11}[2]T^6+(q^9[2]+q^8)a^2T^7 +
\ \   (q^{16}+q^6a^4)T^8 + q^{13}[2]a^2T^9 + q^{10}a^4T^{10} +
$$
\vspace{-0.6cm}
\be
+ (2q^2+1)\Big(q^{10}T^8+q^{16}T^{10}+q^{12}a^2T^{11}\Big) \,
\underbrace{\frac{q^2+a^2T}{1-q^2}}_{q^N[N-1]}
+ q^{18}T^{12} \,\underbrace{\frac{(q^2+a^2T)(q^4+a^2T)}{(1-q^2)^2}}_{q^{2N+1}[N-1][N-2]}
\ee
what is the right answer from eq.(148) of \cite{DMMSS}.
Like for the knots $[3,4]$ and $[3,5]$ the answer is not directly applicable for Khovanov polynomial
at $N=2$: in this case it should be further reduced by the rules, explained in sec.\ref{torus34}.

\bigskip

As usual, only the terms in the first two lines of (\ref{KR36r}) are seriously affected
in the switch to the unreduced KR polynomial, while the last line is just multiplied by $[N]$:
\be
{\cal P}^{[3,6]} = q^{10N-10}\Big\{ [N] + \Big(q^3T^2 + q^{2N+3}T^3 + q^{6}[2]T^4
+ q^{2N+6}[2]T^5 + q^{10}[2]T^6 + \nn \\
+ \underbrace{(q^{10}[2]+q^7)q^{2N}}_{q^{2N+9}[3]} T^7
+ \underbrace{(q^{4N+7}+q^{15})}_{\!\!\!\!\!\!\!\!\!\!\!\!\!\!\!\!\!\!\!\!\!
q^{2N+11}(q^{2N-4}+q^{4-2N})
\!\!\!\!\!\!\!\!\!\!\!\!\!\!\!\!\!\!\!\!\!}\,T^8
+\underbrace{(q^{11}+q^{15})q^{2N}}_{q^{2N+13}(q^2+q^{-2})}T^9 + q^{4N+11}T^{10}
\Big)[N-1] + \nn \\
+ (2q^2+1)\Big(q^{N+10}T^8 + q^{N+16}T^{10}+q^{3N+12}T^{11}\Big)[N][N-1] +
q^{2N+19}T^{12}[N][N-1][N-2]\Big\}
\label{KR36}
\ee

\bigskip

We return to generic consideration of generic $[3,n]$ knots and links in section \ref{cofo},
where a more powerful technique will be applied.

\subsection{Some other families of  3-strand knots and links}

In what follows we use the general notation $(a_1,b_1|a_2,b_2|a_3,b_3|\ldots)$ for the
$3$-strand braid
${\cal R}_1^{a_1}{\cal R}_2^{b_1}{\cal R}_1^{a_2}{\cal R}_2^{b_2}
{\cal R}_1^{a_3}{\cal R}_2^{b_3}\ldots$\
In this notation the torus braid is $[3,n] = (1,1)^n$, while the figure-eight
knot $4_1$ is associated with $(1,-1)^2$.

\subsubsection{${\cal R}_1^n$\ : back to two-strands}

In this case the third strand is actually unlinked with the other two,
so the answer should be a product of $2$-strand knot/link and an unknot $[N]$.
Another obvious feature is that already the very first differential $d_0$ is made
only of the morphism $\alpha_1$, with no $\alpha_2$ -- as we know this
implies that the kernel of $d_0$ is $N$- rather than 1-component.
It is instructive to see how this works in the general context of
$3$-strand links.

In this case all the spaces at hypercube vertices are of the type
$$
v_{k0} =
[2]^{k-1}[N][N-1] = q[2]^{k-1}\Big([N-1] + q^{-N}\Big)[N-1]
$$
In particular, no $[N-2]$ factors show up.
The  primary polynomial is
\be
q^{-n(N-1)}\cdot \P_r^{(n,0)} =
q^{2-2N} + [N-1]\cdot\left|\begin{array}{cc|ccccccccccccccccccccccc}
\hline && \\
&& 1 & qT &\ldots & (qT)^k &\ldots &(qT)^n \\
&&\\
\hline &&\\
\l[N-1]&\times & q^2 & nq & \ldots & q[2]^{k-1}C^k_n & \ldots & q[2]^{n-1} \\ && \\
q^{-N}& \times & 2\underline{q^2} &nq &\ldots & q[2]^{k-1}C^k_n &\ldots & q[2]^{n-1} \\
&&\\ \hline
\end{array}\right|
\ee
Underlined term (one unit out of two) will contribute to the zeroth cohomology.
To understand what are the other contributions, consider some particular example,
say, $n=5$:

$$
q^{2-2N} + [N-1]\cdot\left|\begin{array}{cc|ccccccccccccccccccccccc}
\hline &&\\
&& && && & \underline{q^5} &&&& (1)\\
&& \\
&& && && 5\underline{q^4} & 4q^3 &&&& (1)\\
&& \\
&& && & 10\underline{q^3} &15q^2 & 6q &&&&(1) \\
&& \\
&& && 10\underline{q^2} & 20q& 15& \frac{4}{q} &&&&(1) \\
&&\\
\l[N-1]&\times & q^2 & 5q &  10 & \frac{10}{q} & \frac{5}{q^2} & \frac{1}{q^3} &&&&(0)\\ && \\ && \\
q^{-N}& \times & 2\underline{q^2} &5q & 10 & \frac{10}{q} &\frac{5}{q^2} & \frac{1}{q^3}
&&&&(1) \\ && \\
&&&& 10\underline{q^2}& 20q & 15&\frac{4}{q} &&&&(1) \\ && \\
&&&&&10\underline{q^3} & 15q^2 & 6q &&&& (1) \\ && \\
&&&&&& 5\underline{q^4} & 4q^3 &&&&(1)\\ && \\
&&&&&&&\underline{q^5} &&&&(1) \\
&&\\ \hline
\end{array}\right|
$$

\noindent
Contributing to cohomologies are underlined terms, which form our familiar
increasing pairs, $q^k[N-1] \Longrightarrow q^{k+1}[N-1]$, which we substitute by
$q^{k+2-N} \Longrightarrow q^{N+k-1}$.
In unreduced case the similar substitution will be
$q^k[N] \Longrightarrow q^{k+1}[N]$ by $q^{k+1-N} \Longrightarrow q^{N+k}$.
Thus the answer is:
\be
q^{-n(N-1)}\cdot {\cal P}_r^{(n,0)} =
\underbrace{q^{2-2N}+ q^{-N}[N-1]\cdot q^2}_{q^{2-N}([N-1]+q^{-N})
= q^{1-N}[N]} \ + \nn \\
+ [N-1]\cdot\Big(\underbrace{[N-1]+q^{-N}}_{\frac{1}{q}[N]}\Big)\Big(q^{4-N}(qT)^2
+ q^{N+1}(qT)^3 +q^{6-N}(qT)^4 + q^{N+3}(qT)^5 + \ldots \Big)
\ee
i.e.
$$
{\cal P}_r^{(n,0)} =
q^{(n-1)(N-1)}[N] \left\{ 1 \ + \ q^{N-2}\Big( q^{4-N}(qT)^2 + q^{6-N}(qT)^4 + \ldots\Big)
+ q^{N-2}\Big( q^{N+1}(qT)^3 + q^{N+3}(qT)^5 + \ldots\Big)\ +
\phantom{5^{5^{5^{5^{5^{5^{5^{5^5}}}}}}}}\right.
$$
\vspace{-0.6cm}
\be
\ \ \ \ \ \ \ \ \ \ \ \ \ \ \ \ \ \
\left. +\ \underbrace{q^n\cdot (qT)^n [N-1]}_{{\rm for\ even}\ n}\right\}
\ee
where we easily recognize (\ref{KR2strlinkur}).
Note that this time the factor $[N]$ remains and is not traded for $[N-1]$
in the process of eliminating of cohomologically trivial part --
we proceed in exactly the same way as with the other 3-strand knots
and do not pay attention to unification of $[N-1]$ and $q^{-N}$ into $\frac{1}{q}[N]$
at intermediate stages.
the remaining overall factor $[N]$ describes the independent unknot component
of the 3-strand link ${\cal R}_1^n$.

\bigskip

For unreduced KR polynomial we get instead:
$$
{\cal P}^{(n,0)} = q^{(n-1)(N-1)}[N] \left\{ [N] \ + \
q^{N-2}[N-1]\Big( q^{3-N}(qT)^2 + q^{5-N}(qT)^4 + \ldots\Big) +
\phantom{5^{5^{5^{5^{5^{5^{5^{5^5}}}}}}}}\right.
$$
\vspace{-0.6cm}
\be
\left.
+ q^{N-2}[N-1]\Big( q^{N+2}(qT)^3 + q^{N+4}(qT)^5 + \ldots\Big)\
 +\ \underbrace{q^n\cdot (qT)^n [N][N-1]}_{{\rm for\ even}\ n}\right\}
\ee
-- and this is in accordance with (\ref{P2odd}) and (\ref{P2even}).

\subsubsection{${\cal R}_1^n{\cal R}_2^m$\,: still no $[N-2]$ -- composites
\label{compos0} }

In this family all spaces are of the types $v_{k0} = [2]^{k-1}[N][N-1]$
and $v_{k1}= [2]^{k-2}[N-1]^2$ -- this is the most general
case, when no
contributions with $[N-2]$ factors appear in expressions for KR polynomials.
Actually this family consists of composite links and knots, but it
includes a number of  examples
which we already examined in other braid representations,
therefore its analysis can also serve as an illustration of topological
invariance of our formalism.

From now on we use the following pictorial representation of the braid:
$$
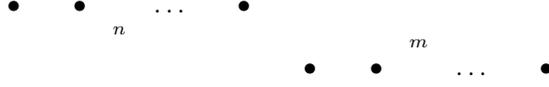
\begin{array}{ccccccccccccccccccccc}
\bullet && \bullet && \ldots &&\bullet \\ &&& ^n &&&&&&&&  _m \\
&&&&&&&&     \bullet && \bullet && \ldots &&\bullet\\
\end{array}
$$
Black and white dots show relative positions of matrices ${\cal R}_1^{\pm 1}$
and ${\cal R}_2^{\pm 1}$ in the first and the second lines respectively.
In our current examples there are no white dots, and there are $n$ black ones
in the first line and $m$ black ones in the second line.

The primary polynomial for this family has the following form:
\be
q^{-(n+m)(N-1)}\cdot \P_r^{(n,m)} =
q^{2-2N} + \ \ \ \ \ \ \ \ \ \ \ \ \ \ \ \ \ \ \ \ \ \ \
\ \ \ \ \ \ \ \ \ \ \ \ \ \ \ \ \ \ \ \ \ \ \
\ \ \ \ \ \ \ \ \ \ \ \ \ \ \ \ \ \ \ \ \ \ \  \nn \\ \nn \\
\!\!\!\!\!\!\!\!\!\!\!\!\!\!
+ [N-1]\cdot\left|\begin{array}{cc|ccccccccccccccccccccccc}
\hline && \\
&& 1 & qT &\ldots & (qT)^k &\ldots &(qT)^{n+m} \\
&&\\
\hline &&\\
&&&&& (C^k_n+C^k_m)(v_{k0}-v_{k1}) +C^k_{n+m}v_{k1} \\ &&&&& || \\
\l[N-1]&\times & q^2 & (n+m)q & \ldots &
[2]^{k-2}\Big(q^2(C^k_n+C^k_m) + C^k_{n+m}\Big)
& \ldots & [2]^{n+m-2} \\ && \\
q^{-N}& \times & 2 {q^2} &(n+m)q &\ldots & q[2]^{k-1}(C^k_n+C^k_m) &\ldots & 0 \\
&&\\ \hline
\end{array}\right|
\ee

\bigskip

\noindent
As usual, to understand the structure of reduction to KR polynomials,
we consider examples.

\bigskip

\underline{${\cal R}_1{\cal R}_2$ -- unknot (torus knot $[3,1]$)}


\be
\begin{array}{ccccc}
\bullet\ \\ \\ & \ \bullet  \\ \\ \\ \\ \\
\end{array} \ \ \ \ \ \ \ \
q^{-2(N-1)}\cdot \P_r^{(1,1)} =
q^{2-2N} + [N-1]\cdot\left|\begin{array}{cc|ccccccccccccccccccccccc}
\hline && \\
\l[N-1]&\times & q^2 & 2q & 1 \\ && \\
q^{-N}& \times & 2 q^2 &2q & 0 \\
&&\\ \hline
\end{array}\right| \sim q^{2-2N}
\ee

\bigskip

\noindent
since the table contains two cohomologically trivial sequences.
Thus ${\cal P}^{(1,1)}_r = 1$ and ${\cal P}^{(1,1)} = [N]$.

\bigskip

\underline{${\cal R}_1^2{\cal R}_2$ -- Hopf link}


\be
\begin{array}{ccccc}
\bullet&& \bullet \\ \\ & &&\ \bullet \\ \\ \\ \\ \\
\end{array} \ \ \ \ \ \ \ \
q^{-3(N-1)}\cdot \P_r^{(2,1)} =
q^{2-2N} + [N-1]\cdot\left|\begin{array}{cc|ccccccccccccccccccccccc}
\hline && \\
&& 1& qT&(qT)^2 & (qT)^3 \\ && \\
\hline && \\
\l[N-1]&\times & q^2 & 3q & 2+q[2] & [2] \\ && \\
q^{-N}& \times & 2q^2 &3q & \underbrace{q[2]}_{1+\boxed{q^2}} & 0 \\
&&\\ \hline
\end{array}\right| \nn \\
\sim q^{2-2N} + q^{-N}[N-1]\cdot q^2\cdot (qT)^2
\ \ \ \ \ \ \ \ \ \ \ \ \ \ \ \ \ \ \ \ \ \ \ \ \ \ \ \ \ \ \ \ \ \
\ \ \ \ \ \ \ \ \ \ \ \ \ \ \ \ \ \ \ \ \ \ \ \ \ \ \ \ \ \ \ \ \ \
\ee

\bigskip

\noindent
The only element inside the table, which contributes to cohomologies is boxed.\\
Thus $\ {\cal P}^{(2,1)}_r = q^{N-1}\Big(1+q^{N+2}T^2\,[N-1]\Big)\ $ and
$\ {\cal P}^{(2,1)} = [N]{\cal P}^{(2,1)}_r$

\bigskip

\underline{${\cal R}_1^3{\cal R}_2$ -- trefoil}
%
\be
\ \ \ \ \ \
\begin{array}{ccccccc}
\bullet&& \bullet&&\bullet \\ \\ & &&&&\ \bullet \\ \\ \\ \\ \\ \\ \\ \\
\end{array} \!\!\!\!\!\!\!\!\!\!\!
\!\!\!\!\!\!\!\!\!\!\!\!\!\!\!\!\!\!\!\!\!\! \!\!\!\!\!\!\!\!\!\!\!\!\!\!\!\!\!\!
\!\!\!\!\!\!\!\!\!\!\!\!\!\!\!\!\!\!
q^{-4(N-1)}\cdot \P_r^{(3,1)} =
q^{2-2N} + [N-1]\cdot\left|\begin{array}{cc|ccccccccccccccccccccccc}
\hline && \\
\l[N-1]&\times & q^2 & 4q & 3+3q[2]& q[2]^2+3[2] &[2]^2 \\ && \\
q^{-N}& \times & 2q^2 &4q & 3q[2] &q[2]^2 &0 \\
&&\\ \hline
\end{array}\right| \nn \\
\sim q^{2-2N} + [N-1]\cdot\left|\begin{array}{cc|ccccccccccccccccccccccc}
\hline &&\\
&& 1 & qT &(qT)^2 & (qT)^3 &(qT)^4 \\
&&\\
\hline && \\
q^{-N}& \times & 0 &0 & q^2 &q^3 &0 \\
&&\\ \hline
\end{array}\right|
\sim q^{2-2N} + q^{-N}\Big(q^{4-N}(qT)^2 + q^{N+1}(qT)^3\Big)
\nn
\ee

\bigskip

\noindent
since the first line in the first table is cohomologically trivial. \\
Thus $\ {\cal P}^{(3,1)}_r = q^{2N-2}\Big(1 +q^4T^2+q^{2N+2}T^3\Big)\ $ and
$\ {\cal P}^{(3,1)} = q^{2N-2}\Big([N] + (q^3T^2 +q^{2N+3}T^3)[N-1]\Big)$.

\bigskip

\underline{${\cal R}_1^4{\cal R}_2$ -- the torus link $[2,4]$ or $4^2_1(v2)$ of \cite{CM}}
%
%
\be
\ \ \ \ \ \
\begin{array}{cccccccccccccccc}
\bullet&& \bullet&&\bullet&&\bullet \\ \\ & &&&&&&\ \bullet \\ \\ \\ \\ \\ \\ \\ \\
\end{array} \!\!\!\!\!\!\!\!\!\!\!
\!\!\!\!\!\!\!\!\!\!\!\!\!\!\!\!\!\!\!\!\!\! \!\!\!\!\!\!\!\!\!\!\!\!\!\!\!\!\!\!
\!\!\!\!\!\!\!\!\!\!\!\!\!\!\!\!\!\!
q^{-5(N-1)}\cdot \P_r^{(4,1)} =
q^{2-2N} + [N-1]\cdot\left|\begin{array}{cc|ccccccccccccccccccccccc}
\hline && \\
\l[N-1]&\times & q^2 & 5q & 4+6q[2] & 6[2]+4q[2]^2 & 4[2]^2+q[2]^3 & [2]^3 \\ && \\
q^{-N}& \times & 2q^2 &5q & 6q[2] & 4q[2]^2 & q[2]^3 & 0 \\
&&\\ \hline
\end{array}\right| \nn \\ \nn \\
\sim q^{2-2N} + [N-1]\cdot\left|\begin{array}{cc|ccccccccccccccccccccccc}
\hline && \\
q^{-N}& \times & 0 &0 & q^2 & q^3 & \underline{q^4} & 0 \\
&&\\ \hline
\end{array}\right|
\sim q^{2-2N} + q^{-N}\Big(q^{4-N}(qT)^2 + q^{N+1}(qT)^2 + q^4[N-1](qT)^4\Big)
\nn
\ee

\bigskip

\noindent
where we left unpaired the last element in the second table -- as usual.
Thus
\be
{\cal P}^{(4,1)}_r = q^{3N-3}\Big(1+q^4T^2 +q^{2N+2}T^3+q^{N+6}T^4[N-1]\Big),\nn \\
{\cal P}^{(4,1)} = q^{3N-3}\Big([N]+(q^3T^2 +q^{2N+3}T^3)[N-1]+q^{N+6}T^4[N][N-1]\Big)
\label{KR421}
\ee

\bigskip

\underline{\underline{${\cal R}_1^n{\cal R}_2$ -- 2-strand torus knots and links $[2,n]$}}

\bigskip

From above examples it is already clear what the entire series $(n,1)$ is going to be:
\be
\ \ \ \ \ \
\begin{array}{cccccccccc}
\bullet&& \bullet&&\ldots &&\bullet \\ &&& ^n \\ & &&&&&&\ \bullet \\ \\ \\ \\ \\ \\ \\ \\
\end{array} \!\!\!\!\!\!\!\!\!\!\!  \!\!\!\!\!\!\!\!\!\!\! \!\!\!\!\!\!\!\!\!\!\!
\!\!\!\!\!\!\!\!\!\!\!\!\!\!\!\!\!\!\!\!\!\! \!\!\!\!\!\!\!\!\!\!\!\!\!\!\!\!\!\!
\!\!\!\!\!\!\!\!\!\!\!\!\!\!\!\!\!\!
q^{-(n+1)(N-1)}\cdot \P_r^{(n,1)} \sim
q^{2-2N} + [N-1]\cdot\left|\begin{array}{cc|ccccccccccccccccccccccc}
\hline && \\
&& 1 & qT & (qT)^2&(qT)^3& \ldots &(qT)^n&(qT)^{n+1}\\
&&\\ \hline && \\
q^{-N}& \times & 0 &0 & q^2 &q^3 &\ldots &q^n & 0 \\
&&\\ \hline
\end{array}\right|
\label{2stragentab}
\ee
\noindent
Now the answer depends on whether $n$ is odd (knots) or even (links):
in the former case all the items in the table come in pairs and
\be
{\cal P}^{(2k+1,1)}_r = q^{2k(N-1)}\Big( 1 + q^4T^2 +q^{2N+2}T^3 +
q^8T^4+q^{2N+6}T^5 + \ldots + q^{4k}T^{2k} + q^{2N+4k-2}T^{2k+1} \Big), \nn \\
{\cal P}^{(2k+1,1)} = q^{2k(N-1)}\left\{ [N] + \Big(q^3T^2 +q^{2N+3}T^3 +
q^7T^4+q^{2N+7}T^5 + \ldots + q^{4k-1}T^{2k} + q^{2N+4k-1}T^{2k+1} \Big)[N-1]\right\}=\nn \\
= q^{2k(N-1)}\left\{ [N] + \frac{1}{q}\Big(q^4T^2 + (q^4T^2)^2 + \ldots +(q^4T^2)^k\Big)
(1+q^{2NT})[N-1]\right\}
\nn
\ee
while in the latter case the last term remains unpaired and
\be
\!\!\!\!\!\!\!\!\!\!
{\cal P}^{(2k,1)}_r = q^{(2k-1)(N-1)}\Big( 1 + q^4T^2 +q^{2N+2}T^3 +
q^8T^4+q^{2N+6}T^5 + \ldots + q^{4k-4}T^{2k-2} + q^{2N+4k-6}T^{2k-1}+\nn\\
 + q^{N+4k-2}T^{2k}[N-1]\Big) \
= \ q^{N-1}\cdot {\cal P}^{(2k-1,1)}_r + q^{2kN +2k-1}T^{2k}[N-1],
 \nn \\ \nn\\
{\cal P}^{(2k,1)} = q^{(2k-1)(N-1)}\left\{ [N] + \Big(q^3T^2 +q^{2N+3}T^3 +
q^7T^4+q^{2N+7}T^5 + \ldots + q^{4k-5}T^{2k-1} + q^{2N+4k-5}T^{2k-1} \Big)[N-1]+\right.\nn\\
\left. + q^{N+4k-2}T^{2k}[N][N-1]\right\}
= q^{N-1}\cdot {\cal P}^{(2k-1,1)} + q^{2kN +2k-1}T^{2k}[N][N-1]
\nn
\ee

This family is of course equivalent to $2$-strand torus knots $[2,n]$,
since the first Reidemeister move is sufficient to eliminate the single $R_2$.
Thus above calculation -- and agreement with (\ref{P2odd}) and (\ref{P2even}) --
is actually a check of invariance under the first
Reidemeister move (within particular family).

\bigskip

In particular, for the knot $5_1 = [2,5]$
\be
{\cal P}^{5_1}_r = q^{4(N-1)}\Big( 1 + q^4T^2 +q^{2N+2}T^3 +
q^8T^4+q^{2N+6}T^5\Big), \nn \\
{\cal P}^{5_1} = q^{4(N-1)}\left\{[N] + \Big(q^3T^2 +q^{2N+3}T^3 +
q^7T^4+q^{2N+7}T^5\Big)[N-1]\right\}
\label{KR51}
\ee
and for the link $[2,6] = 6^2_1(v1)$ of \cite{CM}
\be
{\cal P}^{[2,6]}_r =q^{5(N-1)}\Big( 1 + q^4T^2 +q^{2N+2}T^3 +
q^8T^4+q^{2N+6}T^5 + q^{N+10}T^{6}[N-1]\Big), \nn \\
{\cal P}^{[2,6]} = q^{5(N-1)}\left\{ [N] + \Big(q^3T^2 +q^{2N+3}T^3 +
q^7T^4+q^{2N+7}T^5   \Big)[N-1]+  q^{N+10}T^{6}[N][N-1]\right\}
\label{KR[26]}
\ee
-- what reproduces the answers from \cite{CM} (for unreduced ${\cal P}^{[6.2]}$
it was found there only for $n=2,3,4$).

\bigskip

\underline{${\cal R}_1^2{\cal R}_2^2$ -- composite of two Hopf links}
%
%
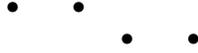
\be
\begin{array}{ccccccc}
\bullet&& \bullet \\  & && \ \bullet &&\bullet \\ \\ \\ \\ \\ \\ \\ \\
\end{array}  \!\!\!\!\!\!\!\!\!\!\!\!\!\!\!\!\!\!\!\!\!\! \!\!\!\!\!\!\!\!\!\!\!\!\!\!\!\!\!\!
\!\!\!\!\!\!\!\!\!\!\!\!\!\!\!\!\!\!
q^{-4(N-1)}\cdot \P_r^{(2,2)} =
q^{2-2N} + [N-1]\cdot\left|\begin{array}{cc|ccccccccccccccccccccccc}
\hline && \\
\l[N-1]&\times & q^2 & 4q & 6+2q^2 & 4[2] & [2]^2 \\ && \\
q^{-N}& \times & 2q^2 &4q & 2q[2] & 0 & 0 \\
&&\\ \hline
\end{array}\right| \ \ \ \ \ \ \ \ \ \ \nn \\
\sim q^{2-2N}  + [N-1]\cdot\left|\begin{array}{cc|ccccccccccccccccccccccc}
\hline && \\
\l[N-1]&\times & 0 & 0 & 0 & 0 & q^2 \\ && \\
q^{-N}& \times & 0 & 0 & 2q^2 & 0 & 0 \\
&&\\ \hline
\end{array}\right| \ \sim
q^{2-2N}+ [N-1]\Big(2q^{2-N}(qT)^2+q^2[N-1](qT)^4\Big)
\nn
\ee
and reduced KR polynomial
\be
\!\!\!\!\!\!\!
{\cal P}^{(2,2)}_r = q^{2N-2}\Big(1 +2q^{N+2}T^2[N-1] +q^{2N+4}T^4[N-1]^2\Big)
= \left\{q^{N-1}\Big(1+q^{N+2}T^2[N-1]\Big)\right\}^2 = \Big\{{\cal P}^{(2,1)}_r\Big\}^2
\ee
is the square of reduced KR polynomial for the Hopf link --
as it should be for reduced knot/link polynomials.

Unreduced polynomial factorizes into a product of reduced and unreduced ones:
\be
{\cal P}^{(2,2)} = [N]\cdot{\cal P}^{(2,2)}_r
= {\cal P}^{(2,1)}\cdot{\cal P}^{(2,1)}_r
\ee

\bigskip

\underline{Generic ${\cal R}_1^n{\cal R}_2^m$ -- a composite of two 2-strand knots/links}

\bigskip

Now it is clear that the same will be true in general.
Like HOMFLY, the reduced KR polynomial for a composition just factorizes
into a product of reduced polynomials for the constituents:
\be
\boxed{
{\cal P}^{(n,m)}_r = {\cal P}^{(n,1)}_r\cdot {\cal P}^{(m,1)}_r =
{\cal P}^{[2,n]}_r\cdot{\cal P}^{[2,m]}_r
}
\label{KRnmr}
\ee

In unreduced case things can be more complicated.
Even for HOMFLY the product of two unreduced polynomials exceeds
the polynomial for a composite by a factor of $[N]$.
Within the $(n,m)$ family a natural possibility would be
\be
{\cal P}^{(n,m)}
\stackrel{?}{=} {\cal P}^{[2,n]}\cdot{\cal P}^{[2,m]}_r   \ \ \ \ n\geq m
\label{KRnmur}
\ee
In such a relation unreduced is preferably the polynomial with $n\geq m$ --
clearly this product is smaller than its alternative if unreduced
is the factor with $n<m$ (since the bigger is $n$, the more is the gain
in diminishing unreduced polynomial from $[N]$ times the unreduced one).
However, as we shall see, there are few chances for this formula
to hold even within the $(n,m)$ family.
Moreover, if morphisms are not fully specified,
sometimes it seems possible to have even a smaller polynomial
-- and the right unreduced KR polynomial can lie in between that minimal
one and (\ref{KRnmur}).

To illustrate the situation, we consider two more examples.

\bigskip

\underline{${\cal R}_1^4{\cal R}_2^2$ }
%
%
\be
\ \ \ \ \ \ \ \ \ \ \
\begin{array}{ccccccc}
\bullet & \bullet & \bullet & \bullet \\  & &&&  \ \bullet &\bullet
\\ \\ \\ \\ \\ \\ \\ \\ \\ \\
\end{array}  \!\!\!\!\!\!\!\!\!\!\!\!\!\!\!\!\!\!\!\!\!\!
\!\!\!\!\!\!\!\!\!\!\!\!\!\!\!\!\!\! \!\!\!\!\!\!\!\!\!\!\!\!\!\!\!\!\!\!
\!\!\!\!\!\!\!\!\!\!\!\!
q^{-6(N-1)}\cdot \P_r^{(4,2)} =
q^{2-2N} + [N-1]\cdot\left|\begin{array}{cc|ccccccccccccccccccccccc}
\hline && \\
&&&&&&&&\boxed{q^4} &&&& (1) \\ && \\
&&&&&& q^4 & 6\underline{q^3} & 4q^2 &&&& (-1) \\ && \\
&&&&& 4q^3 & 17\underline{q^2} & 18 q & 6 &&&& (-1) \\ && \\
&&&& 7q^2 &24q & 31& \frac{18}{q} & \frac{4}{q^2} &&&& (0) \\ && \\
\l[N-1]&\times & q^2 & 6q & 15 & \frac{20}{q} & \frac{15}{q^2} & \frac{6}{q^3} &\frac{1}{q^4}
&&&& (0)\\
&& \\ &&\\
q^{-N}& \times & 2q^2 &6q & 7 & \frac{4}{q} & \frac{1}{q^2} &&&& (0) \\ &&\\
&&&& 7\underline{\underline{q^2}} & 8q & 3 &&&& (2)\\ &&\\
&&&&&4\underline{q^2} & 3q^2 &&&& (1)\\ && \\
&&&&&&\underline{q^4} &&&& (1) \\
&&\\ \hline
\end{array}\right|
\nn
\ee

\bigskip

\noindent
Underlined are the terms that we choose to contribute to cohomologies,
and the boxed element goes directly to the KR polynomial.
The  underlined pair in the upper part of the table
\be
q^2[N-1] \Longrightarrow q^3[N-1]\ \ \ {\rm is\ substituted\  by} \ \ \
\ q^{4-N} \Longrightarrow q^{N+1},
\label{diminnm}
\ee
so that the next iteration is
\be
q^{2-2N} + q^4[N-1]^2\cdot(qT)^6 + q^{N+1}[N-1](qT)^5 + q^{-N}[N-1]
\cdot \left|\begin{array}{cccccccccc}
\hline && \\
0 & 0 & 2\underline{q^2} & \underline{q^3} & 2q^4 & 0 & 0 \\
&&\\ \hline
\end{array}\right|
\ee
Underlined are the terms which we are going to reduce by the rule
(\ref{diminnm}), and the result is
\be
{\cal P}^{(4,2)}_r = q^{4N-4} \left\{ 1 + q^4T^2 + q^{2N+2}T^3 +
\Big(q^{N+2}T^2 + 2q^{N+6}T^4 + q^{3N+4}T^5\Big)[N-1] +
q^{2N+8}T^6[N-1]^2\right\} = \nn \\
= q^{N-1}\Big(1+q^{N+2}T^2[N-1]\Big)\cdot
q^{3N-3}\Big(1+q^4T^2+q^{2N+2}T^3+q^{N+6}T^4[N-1]\Big)
= {\cal P}^{(2,1)}_r\cdot{\cal P}^{(4,1)}_r \ \ \ \ \ \ \ \ \
\ee

\bigskip

What we need to modify in unreduced case is the rule (\ref{diminnm}),
which was used in two places.
Therefore

\be
{\cal P}^{(4,2)} = q^{4N-4} \left\{ [N] + \Big(q^3T^2 + q^{2N+3}T^3\Big)[N-1]
+ \Big(q^{N+5}T^4+q^{3N+5}T^5\Big)[N-1]^2 + \right.\nn \\
\left. + \Big(q^{N+2}T^2 + q^{N+6}T^4\Big)[N][N-1]
q^{2N+8}T^6[N][N-1]^2\right\} =
\nn
\ee
\vspace{-0.5cm}
\be
= q^{3N-3}\left\{[N]+\Big(q^3T^2+q^{2N+3}T^3\Big)[N-1]+q^{N+6}T^4[N][N-1]\Big)
\cdot q^{N-1}\Big(1+q^{N+2}T^2[N-1]\Big)\right\}
={\cal P}^{(4,1)}\cdot {\cal P}^{(2,1)}_r \ \ \ \ \ \ \ \ \
\nn
\ee
This equation confirms the rule (\ref{KRnmur}).

Alternative product $\ {\cal P}^{(2,1)} \cdot{\cal P}^{(4,1)}_r$,
where unreduced is the polynomial with the smaller $m<n$,
would in this case imply that unreduced polynomial is just $[N]\cdot{\cal P}^{(4,2)}_r$
and is clearly bigger than the true unreduced ${\cal P}^{(4,2)}$.

\bigskip

\underline{${\cal R}_1^4{\cal R}_2^3$ -- one more example }
%
%
%
\be
\begin{array}{ccccccccc}
\bullet & \bullet & \bullet & \bullet \\  & &&& \ \bullet &\bullet &\bullet
\\ \\ \\ \\ \\ \\ \\ \\ \\ \\
\end{array}  \!\!\!\!\!\!\!\!\!\!\!\!\!\!\!\!\!\!\!\!\!\! \!\!\!\!\!\!\!\!\!\!\!\!\!\!\!\!\!\!
\!\!\!\!\!\!\!\!\!\!\!\!\!\!\!\!\!\! \!\!\!\!\!\!\!\!\!\!\!
q^{-7(N-1)}\cdot \P_r^{(4,3)} =
q^{2-2N} + [N-1]\cdot\left|\begin{array}{cc|ccccccccccccccccccccccc}
\hline && \\
&&1 & qT & (qT)^2 & (qT)^3 &(qT)^4 & (qT)^5 & (qT)^6 & (qT)^7 \\
&&\\ \hline && \\
&&&&&&&&&\underline{q^5} & (1) \\ && \\
&&&&&&&&7\underline{\underline{q^4}} &5q^3& (2) \\ && \\
&&&&&& q^4 & 21\underline{\underline{q^3}} & 28q^2 &10q& (-2) \\ && \\
&&&&& 5q^3 & 37\underline{q^2} & 63 q & 42 & \frac{10}{q}& (-1) \\ && \\
&&&& 9q^2 &40q & 71& \frac{63}{q} & \frac{28}{q^2} & \frac{5}{q^3}& (0) \\ && \\
\l[N-1]&\times & q^2 & 7q & 21 & \frac{35}{q} & \frac{35}{q^2} & \frac{21}{q^3}
&\frac{7}{q^4} & \frac{1}{q^5}
& (0)\\
&& \\ &&\\
q^{-N}& \times & 2q^2 &7q & 9 & \frac{5}{q} & \frac{1}{q^2} &&&& (0) \\ &&\\
&&&& 9\underline{\underline{q^2}} & 10q & 3 &&&& (2)\\ &&\\
&&&&&5\underline{\underline{q^3}} & 3q^2 &&&& (2)\\ && \\
&&&&&&\underline{q^4} &&&& (1) \\
&&\\ \hline
\end{array}\right|
\nn
\ee
This time four terms in the upper part of the table come in two pairs
and we substitute
$$
[N-1]\cdot\Big(q^2\Longrightarrow 2q^3 \Longrightarrow 2q^4 \Longrightarrow q^5\Big)
$$
by
$$
q^{4-N} \Longrightarrow \underline{q^{N+1}} + q^{5-N} \Longrightarrow \underline{q^{N+2}} + q^{6-N}
\Longrightarrow q^{N+3}
$$
Repeating this once again for the two freshly underlined terms, we substitute
$\ q^N[N-1]\cdot\Big(q \Longrightarrow q^2\Big)\ $ by $\ q^3 \Longrightarrow q^{2N}\ $,
and the next iteration is
\be
q^{-7(N-1)}\cdot \P_r^{(4,3)} \ \sim \
q^{2-2N}  + q^3\cdot(qT)^5 + q^{2N}\cdot (qT)^6 + q^{N+3}[N-1]\cdot(qT)^7 + \nn \\
+ q^{-N}[N-1]
\cdot \left|\begin{array}{ccccccccccccc}
\hline && \\
0 & 0 & 2{q^2} & 2{q^3} & 2\,\boxed{q^4} & q^5 & \boxed{q^6} & 0 &&&& (2) \\
&&\\ \hline
\end{array}\right|
\ee
In this small table we need to pick up to terms, which contribute to cohomologies
straightforwardly -- as usual these should stand as far to the right
as only possible, i.e. these are the two items in boxes.
Other terms in the table can be further simplified with the help of
(\ref{diminnm}) to provide $q^{-N}\Big(2q^{4-N} \Longrightarrow 2q^{N+1}
\Longrightarrow q^{6-N} \Longrightarrow q^{N+3}\Big)$, and finally
\be
{\cal P}^{(4,3)}_r
= q^{5N-5}\Big\{ 1 + 2q^4T^2 + 2q^{2N+2}T^3 +
q^8T^4 + q^{2N+6}T^5 + q^{N-2}\Big(q^4(qT)^4 + q^6(qT)^6\Big)[N-1]+\nn \\
+ q^{2N-2}\Big(q^3\cdot(qT)^5 + q^{2N}\cdot (qT)^6 + q^{N+3}[N-1]\cdot(qT)^7\Big)\Big\} = \nn \\
= q^{5N-5}\Big\{ 1 + 2q^4T^2 + 2q^{2N+2}T^3 + q^8T^4 + 2q^{2N+6}T^5 +q^{4N+4}T^6 +
\Big( q^{N+6}T^4 + q^{N+10}T^6 + q^{3N+8}T^7\Big) [N-1]\Big\} = \nn \\
= q^{2N-2}\Big(1+q^4T^2+q^{2N+2}T^3\Big)\cdot
q^{3N-3}\Big(1+q^4T^2+q^{2N+2}T^3+q^{N+6}T^4[N-1]\Big)
= {\cal P}^{(3,1)}_r\cdot{\cal P}^{(4,1)}_r \ \ \ \ \ \ \ \ \
\ee

\bigskip

In unreduced case we get instead
\be
q^{7-7N}\P^{(4,3)} \sim q^{2-2N}[N]
+ [N][N-1]^2\Big(\ 0 \ \ 0 \ \ 0 \ \ 0 \ \ q^2 \ \ 2q^3 \ \ 2q^4 \ \ q^5\ \Big)
+ q^{-N}[N][N-1]\Big(\ 0 \ \ 0 \ \ \underline{2q^2} \ \ \underline{2q^3} \ \
q^4 \ \ 0 \ \ 0 \ \ 0 \ \Big)
\nn
\ee
In the second bracket we reduce the underlined pair to
\be
\Big(\ 0 \ \ 0 \ \  \boxed{2q^{3-2N}[N-1]} \ \ \boxed{2q^2[N-1]} \ \
\underline{q^{4-N}[N][N-1]} \ \ 0 \ \ 0 \ \ 0 \ \Big)
\label{secbra1}
\ee
As to the first bracket,
there are different reduction options now
(if we do not specify the morphisms from general principles).

The minimal option is to first get rid of $[N]$ (rather than of $[N-1]$) whenever possible.
Then the next iterations of the first bracket are:
\be
\l[N-1]^2\left(\begin{array}{cccccccccccc}
&&&&0& \underline{q^{N+2}} & \underline{q^{N+3}} &\boxed{q^{N+4}} \\
0&0&0&0\\
&&&&\boxed{q^{3-N}}&q^{4-N} & \boxed{q^{5-N}} & 0 \end{array}\right)
\sim \nn \\ \nn \\ \nn \\
\sim \left(\begin{array}{cccccccccccc}
&&&& 0 & \boxed{q^{4}[N-1]} & \boxed{q^{2N+1}[N-1]} &\boxed{q^{N+4}[N-1]^2} \\
0&0&0&0\\
&&&&\boxed{q^{3-N}[N-1]^2}&\underline{q^{4-N}[N-1]^2} & \boxed{q^{5-N}[N-1]^2} & 0 \end{array}\right)
\label{firstbra1}
\ee
The underlined items in (\ref{secbra1}) and (\ref{firstbra1}) are transformed from
$\ q^{4-N}[N][N-1] \Longrightarrow q^{4-N}[N-1]^2\ $ \\
to $\ q^{5-2N}[N-1] \Longrightarrow 0\ $
and we get the hypothetical unreduced polynomial
\be
\tilde{\cal P}^{(4,3)}  \stackrel{?}{=}
q^{5N-5}\Big\{
[N] + (2q^3T^2+q^7T^4)(1+q^{2N}T)[N-1] +  q^{4N+5}[N-1]T^6
+ q^{N+5}T^4[N-1]^2\Big(1+q^4T^2+q^{2N+4}T^3\Big)
\Big\}
\nn
\ee
This expression gets proportional to $[N]$ at $T=-1$, but clearly it
is too ugly to be a right answer (implied by the true morphisms,
once they will be specified).

Thus we sacrifice extreme minimality and proceed instead in our usual way --
postponing the reduction with the help of $[N]$ (rather than of $[N-1]$)
to the very last step.
This means that the first bracket is rather converted into:
\be
\l[N][N-1]\left(\begin{array}{cccccccccccc}
&&&&0& \underline{q^{N+1}} & \underline{q^{N+2}} &\boxed{q^{N+3}} \\
0&0&0&0\\
&&&&\boxed{q^{4-N}}&q^{5-N} & \boxed{q^{6-N}} & 0 \end{array}\right)
\sim \nn \\ \nn \\ \nn \\
\sim \left(\begin{array}{cccccccccccc}
&&&& 0 & \boxed{q^{2}[N-1]} & \boxed{q^{2N+1}[N-1]} &\boxed{q^{N+3}[N][N-1]} \\
0&0&0&0\\
&&&&\boxed{q^{4-N}[N][N-1]}&\underline{q^{5-N}[N][N-1]} & \boxed{q^{6-N}[N][N-1]}
& 0 \end{array}\right)
\label{firstbra2}
\ee
Now the underlined items   (\ref{secbra1}) and (\ref{firstbra2}) are transformed from
$\ q^{4-N}[N][N-1] \Longrightarrow q^{5-N}[N][N-1]\ $ \\
to $\ q^{5-2N}[N-1] \Longrightarrow q^4[N-1]\ $
and
\be
{\cal P}^{(4,3)} \stackrel{?}{=}
q^{5N-5}\Big\{ [N] + \Big(2q^3T^2 + q^7T^4+q^{2N+5}T^5)(1+q^{2N}T)[N-1]
+q^{N+6}T^4\Big(1 + q^{4}T^6 + q^{2N+2}T^3\Big) [N][N-1]\Big\}
\nn
\ee
This version looks somewhat more plausible, despite it is bigger than the previous one.
Finally, the product (\ref{KRnmur}) is even bigger:
\be
{\cal P}^{(4,1)}\cdot{\cal P}^{(3,1)}_r
= q^{3N-3}\left\{[N]+\Big(q^3T^2+q^{2N+3}T^3\Big)[N-1]+q^{N+6}T^4[N][N-1]\right\}
\cdot q^{2N-2}\Big(1+q^4T^2+q^{2N+2}T^3\Big)
\nn
\ee
In particular, the terms $q^3T^2$ here will enter with the coefficient
$q[N]+[N-1] =2[N-1]+q^N > 2[N-1]$.

\bigskip

Thus we confirmed once again the general formula (\ref{KRnmr}),
describing reduced KR polynomials  for the family ${\cal R}_1^n{\cal R}_2^m$
of {\it composite} knots and links, made from a pair of the 2-strand ones.
As to unreduced KR polynomials for composites, the question remains open:
though (\ref{KRnmur}) holds in some examples, it is hardly true even within
the $(n,m)$ family and has much less chances to work in general,
where there is no natural ordering between the constituents of the composite.
Most probably there is {\it no} universal rule for building unreduced KR
polynomials for composites from those of their constituents.

\subsection{Five intersections}

\subsubsection{The 2-component link ${\cal R}_1{\cal R}_2{\cal R}_1{\cal R}_2{\cal R}_1$
(the torus $[2,4]$ or $4^2_1(v2)$ of \cite{CM})}

The primary polynomial, associated with the braid
$$
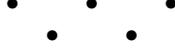
\begin{array}{ccccccccc}
\bullet && \bullet && \bullet \\
& \bullet && \bullet
\end{array}
$$
is described by the following table:

$$
q^{2-2N} +[N-1]\cdot\left|\begin{array}{cc|ccccccccccccc}
\hline &&\\
&& v_{0,0} \lar 5v_{10} \lar 4v_{20}+6v_{21} \lar v_{30}+9v_{31}\lar 3v_{41}+2v_{42} \lar v_{52}\\
&&\\ \hline && \\
\l[N-1] & \times & && && && q^3 && 3q^2 && \!\!q \\
\l[N-2] &\times & &&&&&& &&&& \ q^2 \\ && \\
\l[N-1] & \times & && && 4q^2 && 11q && 8 && \!\!\!\!\frac{1}{q} \\
\l[N-2] &\times & && && && && 2q && \ 2 \\ && \\
\l[N-1] & \times & q^2 && 5q && 10 && \frac{10}{q} && \!\!\frac{3}{q^2} &&  \\
\l[N-2] &\times & &&&&&& && \ \ \ \frac{2}{q} && \frac{1}{q^2} \\ && \\
q^{-N} & \times & 2q^2 && 5q && 4 && \frac{1}{q} \\ && \\
q^{-N} & \times& && && 4q^2 && 2q \\&&\\
q^{-N} & \times& && &&  && q^3 \\
&&\\
\hline
\end{array}\right|
$$

\bigskip

$$
\sim\  q^{2-2N} +[N-1]\cdot\left|\begin{array}{cc|ccccccccccccc}
\hline &&\\
\l[N-1] & \times & && && &&  && q^2 &&   \\
\l[N-2] &\times & &&&&&& &&&& q^2 \\ && \\
\l[N-1] & \times & && && && \frac{1}{q} &&  &&   \\
\l[N-2] &\times & &&&&&& &&\frac{1}{q} &&  \\ && \\
q^{-N} & \times& 0 &&0 && 2q^2 && q^3  \\&&\\
\hline
\end{array}\right|
$$

\bigskip

$$
\ \sim \ q^{2-2N}+q^{-N}[N-1]\cdot\left|\begin{array}{ccccccccccc}\hline \\
0 && 0 && 2q^2 && q^3+\frac{1}{q}\cdot q^2 && q^2\cdot q^2 && 0 \\ \\
\hline\end{array}\right|
$$

\bigskip

$$
\ \sim \
q^{2-2N}+q^{-N}[N-1]\cdot\left|\begin{array}{ccccccccccc}\hline \\
0 && 0 && \underline{q^2} && \underline{q^3}  && q^4 && 0 \\ \\
\hline\end{array}\right|
$$

The two underlined items are handled according to the usual rule:
$q^2[N-1] \Longrightarrow q^3[N-1]$ is substituted by $q^4\Longrightarrow q$,
and we finally obtain the reduced KR polynomial
\be
{\cal P}_r^{4^2_1(v2)} =
q^{5N-5}\Big(q^{2-2N} + q^{4-2N}(qT)^2 + q\cdot (qT)^3 +q^{4-N}(qT)^4[N-1]\Big)
= \nn \\
= q^{3N-3}\Big(1 + q^4T^2 + q^{2N+2}T^3 + q^{N+6}T^4 [N-1]\Big)
\label{KR421v2r}
\ee
and it indeed coincides with the answer for $4^2_1(v2)$ in \cite{CM}.

In unreduced case we get instead
\be
{\cal P}^{4^2_1(v2)} =
q^{3N-3}\Big([N] + q^3T^2[N-1] + q^{2N+3}T^3[N-1] + q^{N+6}T^4 [N][N-1]\Big)
\label{KR421v2ur}
\ee
-- again in accord with \cite{CM}.

These two expressions are of course the same as (\ref{KR421}), which was obtained in another
braid representation of the same link.

\subsubsection{Another initial coloring:
the 2-component   link ${\cal R}_1{\cal R}_2{\cal R}_1^{-1}{\cal R}_2{\cal R}_1$
(the torus $[2,4]$ with inverse orientation or $4^2_1(v1)$ of \cite{CM})}

The primary polynomial, associated with the braid
$$
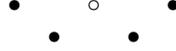
\begin{array}{ccccccccc}
\bullet && \circ && \bullet \\
& \bullet && \bullet
\end{array}
$$
is described by the following table:

$$
\left(\frac{q^{4N-4}}{q^NT}\right)^{-1}{\cal P}^{4^2_1(v1)}_r
= q^{2-2N}\cdot (qT) +[N-1]\cdot
\ \ \ \ \ \ \ \ \ \ \ \ \ \ \ \ \ \ \ \ \ \ \ \ \ \
\ \ \ \ \ \ \ \ \ \ \ \ \ \ \ \ \ \ \ \ \ \ \ \ \ \
\ \ \ \ \ \ \ \ \ \ \ \ \ \ \ \ \ \ \ \ \ \ \ \ \ \
\ \ \ \ \ \ \ \ \ \ \ \ \ \ \ \ \ \ \ \ \ \ \ \ \ \
$$
\centerline{
$
\cdot\left|\begin{array}{c|ccccccccccccc}
\hline &&\\
& v_{10} \lar v_{00}+2v_{20}+2v_{21} \lar 4v_{10}+v_{30}+5v_{31}
\lar 2v_{20}+4v_{21}+ \lar 4v_{31}+v_{52} \lar v_{41}\\
&&&&&&& +2v_{41}+2v_{42}
&&\\ \hline && \\
\l[N-1] \times & q && q^2+2q[2]+2 && 4q+q[2]^2+5[2]
&& 2q[2] + 4 + 2[q]^2+2 && 4[2]+[2] && [2]^2 \\
&&\\
\l[N-2] \times & &&&&&&2[2] && [2]^2 && \\ &&\\
q^{-N}\ \times & q && 2q^2+2q[2] && 4q+q[2]^2 && 2q[2] \\ && \\
\hline
\end{array}\right|
$}

\bigskip

In more detail the table is
$$
\left|\begin{array}{c|ccccccccccccccccccccc}
\hline && \\
\l[N-1] \times & && && &&  &&  && \underline{q^2} &&&& (-1) \\
\l[N-2] \times & &&&&&& &&  \underline{q^2} && &&(1) \\ && \\
\l[N-1] \times & && && q^3 && 4q^2 && 5q && 2 &&&&(0) \\ && \\
\l[N-1] \times &  && 3q^2 && 11q && 12 && 5q^{-1} && q^{-2} &&&& (0) \\
\l[N-2] \times & && && &&  2q &&  2 && && (0) \\ && \\
\l[N-1] \times & q && 4 && 6\underline{q^{-1}} && 2q^{-2} &&  && &&&& (1) \\
\l[N-2] \times & && && && 2\underline{q^{-1}} && q^{-2} &&&&(-1) \\ && \\
q^{-N} \ \times &  &&  && q^3 && 2\underline{q^2} && && &&&&(-1) \\ && \\
q^{-N} \ \times& && 4q^2 && 6q && 2 && && &&&& (0)\\&&\\
q^{-N} \ \times& q&& 2&& q^{-1} && && && &&&&(0) \\
&&\\
\hline
\end{array}\right|
$$

Contributing to cohomologies are two underlined pairs in the upper part
and one underlined element in the lower part of the table:

$$
\left|\begin{array}{c|ccccccccccccccccccccc}
\hline && \\
\l[N-1] \times & 0&0& q^{-1} &&&q^2 \\ & && & \searrow \ \  & \ \ \swarrow \\
\l[N-2] \times & &&&q^{-1}  & q^2 \\ &\\
q^{-N} \ \times & &&& q^2 \\ &\\
\hline
\end{array}\right|
\sim \left|\begin{array}{c|ccccccccccccccccccccc}
\hline && \\
q^N \ \times &0 &0&0&0&0& 1 \\ \\
q^{-N} \ \times &0&0&q&q^2&0&0 \\ \\
\hline\end{array}\right|
$$

After multiplication by $[N-1]$ or $[N][N-1]$ (in the case of reduced
and unreduced polynomials respectively) the two terms in the second line
can be further reduced: $\ [N-1]\Big(q\Longrightarrow q^2\Big)\ $
gets substituted by $\ q^{3-N}\Longrightarrow q^N\ $
and $\ [N][N-1]\Big(q\Longrightarrow q^2\Big)\ $
-- by $ \ [N-1]\Big(q^{2-N}\Longrightarrow q^{N+1}\Big)\ $.
Therefore the KR polynomials are:
\be
{\cal P}^{4^2_1(v1)}_r = \frac{q^{3N-3}}{(qT)}\Big\{
q^{2-2N}\cdot (qT) + q^{-N}\Big(q^{3-N}(qT)^2 + q^N(qT)^3\Big)+ q^N[N-1](qT)^5\Big\}
= \nn \\
= q^{N-1}\Big( 1 + q^2T+ q^{2N}T^2+q^{3N+2}[N-1]T^4\Big), \nn \\ \nn \\
{\cal P}^{4^2_1(v1)} = q^{N-1}\Big( [N] + (qT+ q^{2N+1}T^2)[N-1]+q^{3N+2}[N][N-1]T^4\Big)
\label{KR421v1}
\ee
-- in full accordance with the results of \cite{CM}.

\subsubsection{One more initial coloring:
the 2-component whitehead link ${\cal R}_1{\cal R}_2^{-1}{\cal R}_1{\cal R}_2^{-1}{\cal R}_1$
(the $L5a1$ of \cite{katlas} or $5^2_1$ of \cite{CM})}

Now the braid becomes
$$
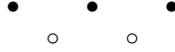
\begin{array}{ccccccccc}
\bullet && \bullet && \bullet \\
& \circ && \circ
\end{array}
 \ \ \ \ \ \ \ \ \ \ \ \ \
$$
and the primary polynomial changes for

$$
q^{2-2N}\cdot(qT)^2 +[N-1]\cdot \ \ \ \ \ \ \ \ \ \ \ \ \ \ \ \ \ \ \ \ \ \ \ \
\ \ \ \ \ \ \ \ \ \ \ \ \ \ \ \ \ \ \ \ \ \ \ \ \ \ \ \ \ \ \ \ \ \ \ \ \ \ \ \ \ \ \ \ \ \ \ \
\ \ \ \ \ \ \ \ \ \ \ \ \ \ \ \ \ \ \ \ \ \ \ \ \ \ \ \ \ \ \ \ \ \ \ \ \ \ \ \ \ \ \ \ \ \ \ \
$$

$$
\cdot\left|\begin{array}{cc|ccccccccccccc}
\hline &&\\
&& v_{20}\lar 2v_{10}+3v_{31}\lar v_{00}+6v_{21}
\lar 3v_{10}+6v_{31}+ v_{52} \lar 3v_{20}+2v_{41} \lar v_{30}\\
&& && && +v_{41}+2v_{42} \\
&& v_{00} \lar 5v_{10} \lar 4v_{20}+6v_{21} \lar v_{30}+9v_{31}\lar 3v_{41}+2v_{42} \lar v_{52}\\
&&\\ \hline && \\
\l[N-1] & \times & && && q^2 && \!\!q && 2q^2 && q^3 \\
\l[N-2] &\times & &&&&&& \ q^2 &&&&  \\ && \\
\l[N-1] & \times &q^2 && 3q && 4 && 6q+ \frac{1}{q} && 3q^2 +4 && 2q \\
\l[N-2] &\times & && &&  2q && 2 && &&  \\ && \\
\l[N-1] & \times & 1 && 2q+ \frac{3}{q} && q^2 + 6 + \frac{1}{q^2}&& 3q+ \frac{6}{q}
&& 3 + \frac{2}{q^2} && \frac{1}{q}  \\
\l[N-2] &\times & &&&&\ \frac{2}{q} && \frac{1}{q^2}&&  &&  \\ && \\
q^{-N} & \times & 1 && 2q && 2q^2 && 3q && 3 &&\frac{1}{q} \\ && \\
q^{-N} & \times& q^2 && &&  &&  && 3q^2 && 2q\\&&\\
q^{-N} & \times&  && &&  &&  && && q^3\\
&&\\
\hline
\end{array}\right|
$$

\bigskip

$$
\sim\  q^{2-2N}\cdot(qT)^2 +[N-1]\cdot\left|\begin{array}{cc|ccccccccccccc}
\hline &&\\
\l[N-1] & \times & && && &&  && 2q^2 && q^3   \\
\l[N-2] &\times & &&&&&& q^2 &&&&  \\ && \\
\l[N-1] & \times &  && && 1 && 2q && q^2 && \\ && \\
\l[N-1] & \times & && \frac{1}{q} && &&   &&  &&   \\
\l[N-2] &\times & &&&&\frac{1}{q} && && &&  \\ && \\
q^{-N} & \times& 1 &&q && q^2 && 0 && q^2 && q^3  \\&&\\
\hline
\end{array}\right|
$$

\bigskip

\bigskip

\noindent
Applying four standard substitutions, \\
$[N-1]\Big(q^2\Longrightarrow  q^3\Big)\ $ by $\ q^{4-N}\Longrightarrow q^{N+1}$, \\
$q^2[N-2] \Longrightarrow q^2[N-1]\ $ by $\ 0\Longrightarrow q^N$, \\
$[N-1]\Big(1\Longrightarrow 2q\Longrightarrow q^2\Big)\ $ by
$\ q^{2-N} \Longrightarrow q^{N-1}+q^{3-N} \Longrightarrow q^N$ \\
and $\ \frac{1}{q}[N-1] \Longrightarrow \frac{1}{q}[N-2]\ $ by $\ q^{1-N}\Longrightarrow 0$,
we convert this into

\be
q^{2-2N}\cdot (qT)^2+[N-1]\cdot\left|\begin{array}{cc|ccccccccccc}\hline &&\\
q^N &\times &    && && && \frac{1}{q}&&2 && q           \\ && \\
q^{-N}&\times & 1 &&2q && 2\,\boxed{q^2} && \underline{q^3}
&& \underline{q^2}+\underline{\underline{q^4}} && \underline{\underline{q^3}} \\ && \\
 \hline\end{array}\right|
\nn \\
\sim q^{2-N}[N-1]\cdot (qT)^2 + \left|\begin{array}{cc|ccccccccccc}\hline &&\\
q^{2N} &\times &    && && && &&\frac{1}{q^2} && \frac{1}{q}           \\ && \\
1 &\times & && && && q &&q^2 &&  \\ &&\\ && \\
1 &\times & && \frac{1}{q} &&1  && && &&  \\ &&\\
q^{-2N}&\times & q^2 &&\underline{q^3} && \underline{q^2} &&   &&  &&   \\ && \\
 \hline\end{array}\right|
\label{P521a}
\ee

\bigskip

\noindent
The underlined items in the first of these tables form two cohomologically trivial parts
and are eliminated.
In the second table we eliminated the $[N-1]$ factor -- just one (boxed) term remains,
which is taken away from the table --
and inserted the $q^{2-2N}\cdot (qT)^2$ term inside the table,
where it is underlined and can be eliminated together with another underlined term
in the last line.

In result we get:
$$
{\cal P}^{5^2_1}_r = \frac{q^{3(N-1)}}{(q^NT)^2}\left(q^{2-N}[N-1]\cdot (qT)^2
+ q^{2-2N} + \frac{1}{q}(qT)+ (qT)^2 + q\cdot(qT)^3 + (q^2+q^{2N-2})(qT)^4 + q^{2N-1}(qT)^5
\right)=
$$
\vspace{-0.5cm}
\be
=  \frac{1}{q^{N+1}T^2}+ \frac{q^{N-3}}{T} + \Big(q[N-1]+q^{N-1}\Big) + q^{N+1}T +
\Big(q^{N+3}+q^{3N-1}\Big)T^2 + q^{3N+1}T^3
\label{KR521r}
\ee

\bigskip

In unreduced case we have instead of (\ref{P521a}):
$$
q^{2-2N}\underbrace{[N]}_{\!\!\!\!\!\!\!\!\!\!\!\!\!\!\!\frac{1}{q}[N-1]+q^{N-1}
\!\!\!\!\!\!\!\!\!\!\!\!\!\!\!\!\!\!}
\cdot (qT)^2+[N][N-1]\cdot\left|\begin{array}{cc|ccccccccccc}\hline &&\\
q^N &\times &    && && && \frac{1}{q}&&2 && q           \\ && \\
q^{-N}&\times & 1 &&2q && 2\,\boxed{q^2} && \underline{q^3}
&& \underline{q^2}+\underline{\underline{q^4}} && \underline{\underline{q^3}} \\ && \\
 \hline\end{array}\right|
$$

$$
\sim q^{2-N}[N][N-1]\cdot (qT)^2 \ +\  [N-1]\cdot\left|\begin{array}{cc|ccccccccccc}\hline &&\\
q^{2N} &\times &    && && && &&\frac{1}{q} && 1           \\ && \\
1 &\times & && && && 1 &&q &&  \\ &&\\ && \\
1 &\times & && 1 &&q  && && &&  \\ &&\\
q^{-2N}&\times & q &&\underline{q^2} && \underline{q} &&   &&  &&   \\ && \\
 \hline\end{array}\right|
 + q^{1-N}\cdot (qT)^2
$$

\bigskip

\noindent
where the usual shifts of powers occur in the second table,
because we now eliminate the factor $[N]$ rather than $[N-1]$.
Another difference is that the term $q^{2-2N}[N](qT)^2$ can not be directly
inserted inside the table, because the table remains multiplied by $[N-1]$.
Thus $[N]$ in this term should be split in two pieces: the one proportional to $[N-1]$
is inserted and enters a cohomologically trivial pair (underlined), which can be eliminated;
the second piece remains and will be later absorbed in the inverse transformation
of $[N-1]$ into $[N]$ in one of the terms in the answer.
Keeping all this in mind, we obtain:
$$
{\cal P}^{5^2_1} = \frac{q^{3(N-1)}}{(q^NT)^2}\, \Big\{q^{2-N}[N][N-1]\cdot (qT)^2 +
$$
$$
\left.
+ \Big(q^{1-2N} +  (qT)+ q\cdot(qT)^2 +  (qT)^3 + (q+q^{2N-1})(qT)^4 + q^{2N}(qT)^5\Big)[N-1]
+ q^{1-N}\cdot (qT)^2
\right\}=
$$
\vspace{-0.5cm}
\be
= \Big(q[N-1] + q^{N-1}\Big)[N]
+ \left\{\frac{1}{q^{N+2}T^2}+ \frac{q^{N-2}}{T} + q^NT +
\Big(q^{N+2}+q^{3N}\Big)T^2 + q^{3N+2}T^3\right\}[N-1]
\label{KR521ur}
\ee

Expressions (\ref{KR521r}) and (\ref{KR521ur}) coincide with the answers from \cite{CM}.

\subsection{Six intersections}

\subsubsection{The knot ${\cal R}_1^3{\cal R}_2{\cal R}_1{\cal R}_2 $
(torus $[2,5]$ or $5_1$)}

The KR complex for the braid
\ \ \ \ \ \ \ \ \ \ \ \ \ \ \ \ \ \ \ \ \ \ \ \ \ \ \ \ \ \ \ \ \ \
\ \ \ \ \ \ \ \ \ \ \ \ \ \ \ \
is
\vspace{-0.7cm}
$$
\begin{array}{cccccccccccc}
\bullet & \bullet & \bullet & & \bullet \\
&&& \bullet && \bullet    \\
\end{array}
$$
\be
v_{00} \stackrel{d_0}{\longrightarrow} 6v_{10}
\stackrel{d_1}{\longrightarrow} 7v_{20} +8v_{21}
\stackrel{d_2}{\longrightarrow} 4v_{30} + 16v_{31}
\stackrel{d_3}{\longrightarrow} v_{40} + 11v_{41} + 3v_{42}
\stackrel{d_4}{\longrightarrow} 3v_{51}+3v_{52}
\stackrel{d_5}{\longrightarrow} v_{62}
\label{51complex}
\ee
and the primary polynomial:
\be
q^{6-6N}\cdot\P^{5_1}_r = q^{2-2N} + [N-1]\cdot
\left|\begin{array}{cc|cccccccccccccccccccccccccccccc}
\hline && \\
\l[N-1] & \times &&&&&q^4&3q^3&q^2 \\
\l[N-2] & \times &&&&&&& q^3\\ && \\
\l[N-1] & \times &&&&4q^3&14q^2&12q&2 \\
\l[N-2] & \times &&&&&&3q^2& 3q\\ && \\
\l[N-1] & \times &&&7q^2&24q&28&\frac{12}{q}&\frac{1}{q^2} \\
\l[N-2] & \times &&&&&3q&6& \ \frac{3}{q} \\ && \\
\l[N-1] & \times &q^2&6q&15&\frac{20}{q}&\frac{12}{q^2}&\frac{3}{q^3}& \\
\l[N-2] & \times &&&&&\ \frac{3}{q}&\ \frac{3}{q^2}&\ \frac{1}{q^3} \\ && \\ && \\
q^{-N}  & \times &2q^2&6q&7&\frac{4}{q}&\frac{1}{q^2}&& \\ && \\
q^{-N}  & \times &&&7q^2&8q&3&& \\ && \\
q^{-N}  & \times &&&&4q^3&3q^2&& \\ && \\
q^{-N}  & \times &&&&&q^4&& \\
&&\\ \hline
\end{array}\right|
\label{51tab} \\ \nn \\
\sim
q^{2-2N}  + [N-1]\cdot
\left|\begin{array}{cc|ccccccccccc }
\hline && \\
\l[N-1] &\times &&&& \frac{1}{q} &&q^3 \\
\l[N-2] & \times &&&&&\frac{1}{q} && q^3 \\ && \\
q^{-N} &\times &0&0&2q^2&q^3&q^4 \\
&&\\ \hline
\end{array}\right| \nn \\ \nn \\
\sim
q^{2-2N} + q^{-N}[N-1]\cdot
\left|\begin{array}{cccccccccccccc }
\hline && \\
0&0&q^2&q^3&q^4&q^5& 0 \\
&&\\ \hline
\end{array}\right|
\nn
\ee
Thus we arrive at pattern (\ref{2stragentab}), which is characteristic for 2-strand knots and links $[2,n]$
-- this time with $n=5$, i.e. we obtain the KR polynomial for the knot $5_1=[2,5]$:
\be
{\cal P}^{5_1}_r = q^{4N-4}\Big(1+q^4T^2+q^{2N+2}T^3+q^8T^4+q^{2N+6}T^5\Big), \nn \\
{\cal P}^{5_1} = q^{4N-4}\Big([N]+(q^3T^2 +q^7T^4)(1+q^{2N}T)[N-1]\Big)
\label{KR51a}
\ee

\bigskip

Just the same hypercube describes a whole set of knots, associated with different
colorings of the same knot diagram -- at hypercube level this corresponds to
different choices of initial vertex, leading to different KR complexes:
the differentials are made from morphisms, which point away from the
initial vertex.
In the particular example, considered in this subsection, the hypercube has
$2^5=64$ vertices, and the set of emerging knots includes:
\be
\begin{array}{|c|c|c|c|}
\hline
\mbox{Name}&w&\mbox{Braid word}\\
\hline
6_2&2&[1, 1, 1, -2, 1, -2]\\
\hline
5_2&4&[1, 1, 1, 2, -1, 2]\\
\hline
5_1&6&[1, 1, 1, 2, 1, 2]\\
\hline
4_1&0&[1, -1, -1, 2, -1, 2]\\
\hline
3_1&4&[1, 1, 1, 2, 1, -2]\\
&&[1, 1, -1, 2, 1, 2]\\
&&[1, 1, 1, -2, 1, 2]\\
\cline{2-3}
&2&[1, 1, 1, 2, -1, -2]\\
&&[1, 1, 1, -2, -1, 2]\\
\cline{2-3}
&-4&[1, -1, -1, -2, -1, -2]\\
\hline
\mbox{unknot}&0&[1, 1, 1, -2, -1, -2]\\
&&[1, 1, -1, 2, -1, -2]\\
&&[1, -1, -1, -2, 1, 2]\\
\cline{2-3}
&2&[1, 1, -1, 2, 1, -2]\\
&&[1, 1, -1, 2, -1, 2]\\
&&[1, -1, -1, 2, 1, 2]\\
\cline{2-3}
&-2&[1, -1, -1, 2, -1, -2]\\
&&[1, -1, -1, -2, 1, -2]\\
&&[1, -1, -1, -2, -1, 2]\\
\hline
\end{array}
\nn\ee
Since simpler knots have already appeared in our considerations
and a their evaluation would just once again validate the topological invariance
of our construction from \cite{DM3},
we now analyze only the new cases of $5_2$ and $6_2$.

\subsubsection{Another initial vertex: the knot ${\cal R}_1^3{\cal R}_2{\cal R}_1^{-1}{\cal R}_2 $
(the twisted knot $5_2$)}

The KR complex for
\ \ \ \ \ \ \ \ \ \ \ \ \ \ \ \ \ \ \ \ \ \ \ \ \ \ \ \ \ \ \ \ \ \
\ \ \ \ \ \ \ \ \ \ \ \ \ \ \ \
is
\vspace{-0.7cm}
$$
\begin{array}{cccccccccccc}
\bullet & \bullet & \bullet & & \circ \\
&&& \bullet && \bullet    \\
\end{array}
 \ \ \ \ \ \ \ \ \ \ \ \ \ \ \ \ \ \ \ \ \ \ \ \ \ \ \ \ \ \ \ \ \ \
$$

\centerline{
{\footnotesize
$
v_{10} \stackrel{d_0}{\longrightarrow} v_{00}+3v_{20}+2v_{21}
\stackrel{d_1}{\longrightarrow} 5v_{10} + 3v_{30}+7v_{31}
\stackrel{d_2}{\longrightarrow} 4v_{20}+6v_{21}+v_{40}+6v_{41}+3v_{42}
\stackrel{d_3}{\longrightarrow} v_{30}+9v_{31}+2v_{51}+3v_{52}
\stackrel{d_4}{\longrightarrow} 5v_{41}+v_{62}
\stackrel{d_5}{\longrightarrow} v_{51}
$
}}

\bigskip

\noindent
The primary polynomial is obtained by reordering the columns in (\ref{51tab}):

$$
\left(\frac{q^{5N-5}}{q^{N}T}\right)^{-1}\cdot
\P^{5_2}_r = q^{2-2N}\cdot(qT)\  +
\ \ \ \ \ \ \ \ \ \ \ \ \ \ \ \ \ \ \ \ \ \ \ \ \ \ \ \ \ \ \ \ \ \ \ \
\ \ \ \ \ \ \ \ \ \ \ \ \ \ \ \ \ \ \ \ \ \ \ \ \ \ \ \ \ \ \ \ \ \ \ \
\ \ \ \ \ \ \ \ \ \ \ \ \ \ \ \ \ \ \ \ \ \ \ \ \ \ \ \ \ \ \ \ \ \ \ \
$$
\be
+ [N-1]\cdot
\left|\begin{array}{cc|cccccccccccccccccccccccccccccc}
\hline && \\
\l[N-1] & \times &&&&q^4&2q^3&q^2&q^3 \\
\l[N-2] & \times &&&&&&q^3& \\ && \\
\l[N-1] & \times &&&3q^3&9q^2&q^3+9q&5q^2+2&3q \\
\l[N-2] & \times &&&&&3q^2&3q& \\ && \\
\l[N-1] & \times &&3q^2&13q&4q^2+18&11q+\frac{9}{q}&10+\frac{1}{q^2}&\frac{3}{q} \\
\l[N-2] & \times &&&&3q&6&\frac{3}{q}& \\ && \\
\l[N-1] & \times &q&q^2+5&5q+\frac{10}{q}&10+\frac{7}{q^2}
&\frac{10}{q}+\frac{2}{q^3}&\frac{5}{q^2}&\frac{1}{q^3} \\
\l[N-2] & \times &&&&\frac{3}{q}&\frac{3}{q^2}&\frac{1}{q^3}& \\ && \\ && \\
q^{-N}  & \times &q&2q^2+3&5q+\frac{3}{q}&4+\frac{1}{q^2}&\frac{1}{q}&& \\ && \\
q^{-N}  & \times &&3q^2&6q&4q^2+3&2q&& \\ && \\
q^{-N}  & \times &&&3q^3&3q^2&q^3&& \\ && \\
q^{-N}  & \times &&&&q^4&&& \\ && \\
&&\\ \hline
\end{array}\right|
 \nn
 \ee
 \be
\sim
q^{2-2N}\cdot(qT)\  + [N-1]\cdot
\left|\begin{array}{cc|ccccccccccc }
\hline && \\
\l[N-1] &\times &&&&&q&2q^2&q^3 \\&& \\
\l[N-1] &\times &&&\frac{1}{q}&& q^3 && \\
\l[N-2] & \times &&&&\frac{1}{q}&& q^3& \\ && \\
q^{-N} &\times &0&0&0&2q^2&0&0&0 \\
&&\\ \hline
\end{array}\right|
\nn \ee
\be
\sim
q^{2-2N}\cdot(qT)\ +  [N-1]\cdot
\left|\begin{array}{cc|cccccccccccc }
\hline && \\
q^N &\times & 0&0&0 & 0 & 0 & 1 & q \\ && \\
q^{-N} &\times & 0 & 0 & q & 2q^2 & q^3 & 0 & 0 \\
&&\\ \hline
\end{array}\right|
 \nn
 \ee
 \be
\sim
q^{2-2N}\cdot(qT)\ +
\left|\begin{array}{cccccccccccc}
\hline \\
&&&& & q & q^{2N} \\ && \\
\hline && \\
 0&0&0 & 0 & 0 & q^2 & q^{2N-1} \\ && \\
 0 & 0 & q^{3-2N} & 1+q^{4-2N} & q & 0 & 0 \\
&&   \\ \hline && \\
&& {q^{2-2N}} & q+q^{3-2N} & q^2\\ \\
\hline
\end{array}\right|
\nn
\ee

\bigskip

\noindent
Therefore
\be
{\cal P}^{5_2}_r = \frac{q^{5N-5}}{q^NT}\Big( q^{2-2N}(qT) +q^{3-2N}(qT)^2
+ (1+q^{4-2N})(qT)^3 +q\cdot (qT)^4 + q^2(qT)^5+q^{2N-1}(qT)^6\Big) = \nn \\
=q^{2N-2}\Big( 1 + q^2T + (q^4+q^{2N})T^2 + q^{2N+2}T^3 + q^{2N+4}T^4+q^{4N+2}T^5\Big)
= q^{2N}T + (1+q^{2N}T^2)\cdot {\cal P}^{3_1}_r, \nn \\ \nn \\
{\cal P}^{5_2} = q^{2N-2}[N] + \Big(q^{2N-1}T + (q^{2N+1}+q^{4N-1})T^2
+q^{4N+1}(T^3+T^4) +q^{6N+1}T^5\Big)[N-1]
\label{KR52}
\ee
--in accordance with \cite{CM} (unreduced polynomial was found there for $N=2,3,4$).

\subsubsection{One more initial vertex:
${\cal R}_1^3{\cal R}_2^{-1}{\cal R}_1{\cal R}_2^{-1} $
(the knot $6_2$)}

The KR complex for
\ \ \ \ \ \ \ \ \ \ \ \ \ \ \ \ \ \ \ \ \ \ \ \ \ \ \ \ \ \ \ \ \ \
\ \ \ \ \ \ \ \ \ \ \ \ \ \ \ \
is
\vspace{-0.7cm}
$$
\begin{array}{cccccccccccc}
\bullet & \bullet & \bullet & & \bullet \\
&&& \circ && \circ    \\
\end{array}
 \ \ \ \ \ \ \ \ \ \ \ \ \ \ \ \ \ \ \ \ \ \ \ \ \ \ \ \ \ \ \ \ \ \
$$

\centerline{
{\footnotesize
$
v_{20} \stackrel{d_0}{\longrightarrow} 2v_{10}+4v_{31}
\stackrel{d_1}{\longrightarrow} v_{00}+8v_{21} +3v_{41}+3v_{42}
\stackrel{d_2}{\longrightarrow} 4v_{10} + 12v_{31}+v_{51}+3v_{52}
\stackrel{d_3}{\longrightarrow} 6v_{20} + 8v_{41} + v_{62}
\stackrel{d_4}{\longrightarrow} 4v_{30}+2v_{51}
\stackrel{d_5}{\longrightarrow} v_{40}
$
}}

\bigskip

\noindent
The primary polynomial:

\centerline{
$
\left(\frac{q^{4N-4}}{q^{2N}T^2}\right)^{-1}\cdot\P^{6_2}_r = q^{2-2N}(qT)^2 +
\l[N-1]\cdot
\left|\begin{array}{cc|cccccccccccccccccccccccccccccc}
\hline && \\
\l[N-1] & \times &&&&&&&\underline{q^4} &&&& (1) \\ &&\\
\l[N-1] & \times& &&&&& \underline{\underline{\underline{6q^3}}} & 3q^2 &&&& (-3) \\&&\\
\l[N-1] & \times&&&& \underline{q^3} & \underline{\underline{\underline{\underline{15q^2}}}}
&14q & 3 &&&&(3=4-1)\\
\l[N-2] & \times &&&&&\underline{q^3}&& &&& (1)\\ && \\
\l[N-1] & \times& && 4q^2 & \underline{\underline{\underline{22q}}}
& 24 & \frac{10}{q} & \frac{1}{q^2} &&&& (-3) \\&&\\
\l[N-1] & \times& q^2 & 6q & \underline{17} &\frac{18}{q}
& \frac{9}{q^2} & \frac{2}{q^3}& &&&&(1)\\&&\\
\l[N-1] & \times & 1 &\frac{4}{q} & \frac{3}{q^2} & \underline{\frac{1}{q^3}} & && &&&& (-1) \\
\l[N-2] & \times &&&\frac{3}{q}& \frac{3}{q^2}& \underline{\frac{1}{q^3}} & &&&&(1)
\\ && \\ && \\&&\\
q^{-N}  & \times &1&2q&2q^2&&&& && (1) \\ && \\
q^{-N}  & \times &&&&4q&6&\frac{4}{q}&\frac{1}{q^2} && (-1) \\ &&\\
q^{-N}  & \times &q^2&&&&6q^2&8q&3 && (1+1)\\ && \\
q^{-N}  & \times &&&&&&4q^3&3q^2 && (-1) \\ && \\
q^{-N}  & \times &&&&&&&q^4 &&(1) \\ && \\
&&\\ \hline
\end{array}\right|
$
}

\bigskip

\noindent
Leaving only the underlined  items in the upper part of the table,
we can substitute it by
\be
\sim
q^{2-2N}\cdot(qT)^2\  + [N-1]\cdot
\left|\begin{array}{cc|ccccccccccc }
\hline && \\
\l[N-1] &\times &0&0&1&3q&4q^2&3q^3&q^4 &&&(0) \\&& \\
\l[N-1] & \times &&&& q^3+\frac{1}{q^3} & &  \\
\l[N-2] & \times &&&&  &q^3+\frac{1}{q^3}&&&&  \\ && \\
 && \\
q^{-N} &\times &1&q&2q^2&0&0&0&0  &&&(2) \\ &&\\
q^{-N} &\times &0&0&0&q&q^2&q^3&q^4 &&& (0) \\
&&\\ \hline
\end{array}\right|
\nn
\ee
\be
\sim
q^{2-2N}\cdot(qT)^2\  + [N-1]\cdot
\left|\begin{array}{cc|cccccccccccc }
\hline && \\
q^N& \times &&& &\frac{1}{q} & 2 & 2q & q^2       \\ && \\ && \\
q^{-N}& \times &&& q^2 &2q^3 & 2q^4 & q^5      \\ && \\
q^{-N}& \times &&&&\frac{1}{q}+q^5       \\ && \\
q^{-N} &\times &1&q&2q^2&0&0&0&0  &&& \\ &&\\
q^{-N} &\times &0&0&0&q&q^2&q^3&q^4 &&&  \\
&&\\ \hline
\end{array}\right|
\nn
\ee

\be
\sim
q^{2-2N}\cdot(qT)^2\  + [N-1]\cdot
\left|\begin{array}{cc|cccccccccc }
\hline && \\
q^N& \times &0&0&0 &q^{-1} & 2 & 2q & q^2       \\ && \\
 q^{-N} &\times &1&q&2q^2&\frac{1}{q}+q^3&0&0&0  &&& \\ &&\\
  \hline
\end{array}\right|
\nn
\ee
\be
\sim
\left|\begin{array}{cc|cccccccccc }
\hline && \\
q^{2N} & \times &0&0&0 &0 & q^{-2} & q^{-1} & 1       \\ && \\
1 & \times &0&0&0 &q & q^2 &q^3 \\ && \\
1&\times & 0 &\frac{1}{q}& 1& q & \\ && \\
 q^{-2N} &\times &q^2&0&q^4+\underline{q^2}&\underline{q}&0&0&0  &&& \\ &&\\
  \hline
\end{array}\right|
\nn
\ee
At the last stage we absorbed the item from outside the table (underlined),
and it makes a cohomologically trivial pair with another underlined item --
and they both can be eliminated.
From this we read:
$$
{\cal P}^{6_2}_r = \frac{q^{2N-4}}{T^2}\left( q^{2-2N} + \frac{1}{q}\,(qT)
+ (1+q^{4-2N})(qT)^2 + 2q\,(qT)^3
+ (q^{2N-2}+q^2)(qT)^4
+ (q^{2N-1}+q^3)(qT)^5 + q^{2N}(qT)^6\right) =
$$
\vspace{-0.5cm}
\be
= \frac{1}{q^2T^2}+\frac{q^{2N-4}}{T} + (q^2+q^{2N-2}) + 2q^{2N}T +
(q^{4N-2}+\underbrace{q^{2N+2}}_{
\!\!\!\!\!\!\!\!\!\!\!\!\!\!{\rm instead\ of}\ q^{N+2}\!\!\!\!\!\!\!\!\!\!\!})T^2
+ (q^{4N} + q^{2N+4})T^3 +q^{4N+2}T^4
\label{KR62r}
\ee
which coincides with the answer from \cite{CM}
(up to a misprint in the underbraced term).

\bigskip

In unreduced case the two last tables are
\be
q^{2-2N}[N](qT)^2\  + [N][N-1]\cdot
\left|\begin{array}{cc|cccccccccc }
\hline && \\
q^N& \times &0&0&0 &q^{-1} & 2 & 2q & q^2       \\ && \\
 q^{-N} &\times &1&q&2q^2&\frac{1}{q}+q^3&0&0&0  &&& \\ &&\\
  \hline
\end{array}\right|
\nn
\ee
\be
\sim
q^{1-N}(qT)^2\  +
[N-1]\cdot \left|\begin{array}{cc|cccccccccc }
\hline && \\
q^{2N} & \times &0&0&0 &0 & q^{-1} & 1 & q       \\ && \\
1 & \times &0&0&0 &1 & q &q^2 \\ && \\
1&\times & 0 &1& q& q^2 & \\ && \\
 q^{-2N} &\times &q&0&q^3+\underline{q}&\underline{1}&0&0&0  &&& \\ &&\\
  \hline
\end{array}\right|
\nn
\ee
where in the item outside the table we substituted $\ [N]=\frac{1}{q}[N-1] + q^{N-1}$,
absorbed the first piece inside the table from where it can be eliminated
together with another underlined term.
Thus
\be
{\cal P}^{6_2} = q^{N-1}+
\label{KR62ur}
\ee
\vspace{-0.4cm}
$$
\!\!\!\!\!\!\!\!
+\frac{q^{2N-4}}{T^2}[N-1]\Big( q^{1-2N} + (qT)
+ (q+q^{3-2N})(qT)^2 + q[2](qT)^3
+ (q^{2N-1}+q)(qT)^4
+ (q^{2N}+q^2)(qT)^5 + q^{2N+1}(qT)^6\Big) =
$$
\vspace{-0.5cm}
\be
\!\!\!\!\!\!\!\!
= q^{N-1}+[N-1]\Big(\frac{1}{q^3T^2}+\frac{q^{2N-3}}{T} + (q^{2N-1}+q)
+ q^{2N}[2] T + (q^{4N-1}+ {q^{2N+1}})T^2 + (q^{4N+1} + q^{2N+3})T^3 +q^{4N+3}T^4\Big)
\nn
\ee
again in accordance  with   \cite{CM} (where it was calculated for $N=2,3,4$).

\subsubsection{The braid
${\cal R}_1^2{\cal R}_2{\cal R}_1{\cal R}_2^{2} $
(the knot $5_1$)}

The KR complex for
\ \ \ \ \ \ \ \ \ \ \ \ \ \ \ \ \ \ \ \ \ \ \ \ \ \ \ \ \ \ \ \ \ \
\ \ \ \ \ \ \ \ \ \ \ \ \ \ \ \
is
\vspace{-0.7cm}
$$
\begin{array}{cccccccccccc}
\bullet & \bullet & & \bullet & & \\
&& \bullet && \bullet & \bullet    \\
\end{array}
 \ \ \ \ \ \ \ \ \ \ \ \ \ \ \ \ \ \ \ \ \ \ \ \ \ \ \ \ \ \ \ \ \ \
$$
\be
v_{00} \stackrel{d_0}{\longrightarrow} 6v_{10}
\stackrel{d_1}{\longrightarrow} 6v_{20} + 9v_{21}
\stackrel{d_2}{\longrightarrow} 2v_{30} + 18v_{31}
\stackrel{d_3}{\longrightarrow} 11v_{41} + 4v_{42}
\stackrel{d_4}{\longrightarrow} 2v_{51}+4v_{52}
\stackrel{d_5}{\longrightarrow} v_{62}
\ee

\bigskip

\noindent
The primary polynomial:

\bigskip

\centerline{
$
q^{6-6N}\P^{}_r = q^{2-2N} + [N-1]
\cdot\left|\begin{array}{c|cccccccccccccccccccccccccccccc}
\hline && \\
\l[N-1]\times & q^2 & 6q & 6q[2]+9 & 2q[2]^2 + 18[2] & 11[2]^2+4  & 2[2]^3+4[2] & [2]^2    \\ &&\\
\l[N-2] \times &&&&& 4[2]& 4[2]^2 & [2]^3            \\ && \\
q^{-N}\ \times & 2q^2 & 6q &  6q[2] & 2q[2]^2 &        \\
&&\\ \hline
\end{array}\right|
$
}

\centerline{
$
= q^{2-2N} + [N-1]
\cdot\left|\begin{array}{c|lllllllllllllccccccccc}
\hline && \\
\l[N-1]\times &&&& &  & 2\underline{q^3} & q^2 &&&&(-1) \\ && \\
\l[N-1]\times &&&&2q^3 & 11\underline{q^2} &10q & 2 &&&& (1) \\ && \\
\l[N-1]\times & && 6q^2 & 22q & 26\cdot\underline{1} & 10q^{-1} &q^{-2}&&&&(1) \\ &&\\
\l[N-1]\times & q^2 & 6q & 15 & 20\underline{q^{-1}}& 11q^{-2} &  2q^{-3} &  &&&&(-1)\\
&&\\ &&\\
\l[N-2] \times &&&&&&&\underline{q^3} &&&&(1)\\ && \\
\l[N-2] \times &&&&&&4\underline{q^2}&3q&&&&(-1) \\ && \\
\l[N-2] \times &&&&& 4q&8\cdot\underline{1}&3q^{-1}&&&&(-1) \\ && \\
\l[N-2] \times &&&&& 4\underline{q^{-1}}&4q^{-2}&q^{-3}&&&& (1)            \\ && \\ && \\
q^{-N}\ \times & 2q^2 & 6q &  6q[2] & 2q[2]^2 & &&&&        \\
&&\\ \hline
\end{array}\right|
$
}

\bigskip

\noindent
All the four underlined terms in the $[N-1]$ lines have counterparts in the
$[N-2]$ lines, and after substitution of $\ [N-1] \Longrightarrow [N-2]\ $
by $\ q^{2-N}\Longrightarrow 0$, we remain only with $q^{-N}$ terms:

\be
q^{2-2N}+q^{-N}[N-1]\cdot\left|\begin{array}{cccccccccccccccc} \hline &&\\
&&&&&q^5&0 \\
&&&&q^4\\
&&&&q^2 \\
&&&q \\ \\
&&&2q^3\\
&&6q^2&4q \\
2q^2&6q & 6 & 2q^{-1} \\
\\ \hline
\end{array}\right|
\sim
q^{2-2N}+q^{-N}[N-1]\cdot\left|\begin{array}{cccccccccccccccc} \hline &&\\
&&&&&q^5&0 \\
&&&&q^4\\
&&&q^3\\
0&0&q^2 \\
\\ \hline
\end{array}\right|
\nn
\ee

\bigskip

\noindent
It is easy to recognize the familiar structure (\ref{2stragentab})
-- thus we indeed deal with the torus knot $[2,5]$, alternatively named $5_1$.
The final step -- elimination of the factor $[N-1]$, it is slightly
different in reduced and unreduced cases, and
\be
{\cal P}^{5_1}_r = q^{6N-6}\Big(q^{2-2N} + q^{4-2N}(qT)^2 + q\cdot(qT)^3 + q^{6-2N}(qT)^4
+q^{3}\cdot(qT)^5\Big) = \nn \\
= q^{4N-4}\Big(1+q^4T^2 +q^{2N+2}T^3+q^8T^4 + q^{2N+6}T^5\Big),  \\
{\cal P}^{5_1} = q^{6N-6}\left\{q^{2-2N}[N] + \Big(q^{3-2N}(qT)^2 + q^2\cdot(qT)^3 + q^{5-2N}(qT)^4
+q^{4}\cdot(qT)^5\Big)[N-1]\right\} =
\nn
\ee
\vspace{-0.4cm}
$$
= q^{4N-4}\left\{[N]+\Big(q^3T^2 +q^{2N+3}T^3+q^7T^4 + q^{2N+7}T^5\Big)[N-1]\right\}
= q^{4N-4}\left([N]+q^3T^2(1+q^4T^2)(1+q^{2N}T)[N-1]\right)
$$

\subsubsection{Another initial vertex: the braid
${\cal R}_1^2{\cal R}_2^{-1}{\cal R}_1{\cal R}_2^{-2} $
(the knot $6_3$)}

The KR complex for
\ \ \ \ \ \ \ \ \ \ \ \ \ \ \ \ \ \ \ \ \ \ \ \ \ \ \ \ \ \ \ \ \ \
\ \ \ \ \ \ \ \ \ \ \ \ \ \ \ \
is
\vspace{-0.7cm}
$$
\begin{array}{cccccccccccc}
\bullet & \bullet & & \bullet & & \\
&& \circ && \circ & \circ    \\
\end{array}
 \ \ \ \ \ \ \ \ \ \ \ \ \ \ \ \ \ \ \ \ \ \ \ \ \ \ \ \ \ \ \ \ \ \
$$

\centerline{
{\footnotesize
$
v_{30} \stackrel{d_0}{\longrightarrow} 3v_{20}+3v_{41}
\stackrel{d_1}{\longrightarrow} 3v_{10} + 9v_{31} + v_{51} +2v_{52}
\stackrel{d_2}{\longrightarrow} v_{00} + 9v_{21} + 5v_{41}+4v_{42}+ v_{62}
\stackrel{d_3}{\longrightarrow} 3v_{10} + 9v_{31} + v_{51} +2v_{52}
\stackrel{d_4}{\longrightarrow} 3v_{20}+3v_{41}
\stackrel{d_5}{\longrightarrow} v_{30}
$
}}

\bigskip

\noindent
The primary polynomial:

$$
\left(\frac{q^{3N-3}}{q^{3N}T^3}\right)^{-1}\cdot\P^{6_3}_r = q^{2-2N}(qT)^3\ \ \ +\ \ \
\l[N-1]\cdot \ \ \ \ \ \ \ \ \ \ \ \ \ \ \ \ \ \ \ \ \ \ \ \ \ \
\ \ \ \ \ \ \ \ \ \ \ \ \ \ \ \ \ \ \ \ \ \ \ \ \ \
\ \ \ \ \ \ \ \ \ \ \ \ \ \ \ \ \ \ \ \ \ \ \ \ \ \
$$

\centerline{
$
\cdot\left|\begin{array}{c|cccccccccccccccccccccccccccccc}
\hline && \\
\l[N-1]\times & q[2]^2 & 3q[2]+3[2]^2 & 3q+9[2]+[2]^3+2[2] &
q^2+9+5[2]^2+4+[2]^2 & 3q+9[2]+[2]^3+2[2] & 3q[2]+3[2]^2 & q[2]^2 \\ &&\\
\l[N-2] \times &&& 2[2]^2 & 4[2]+[2]^3 & 2[2]^2 \\ && \\
q^{-N}\ \times & q[2]^2 & 3q[2]&3q & 2q^2 &3q &3q[2] & q[2]^2 \\
&&\\ \hline
\end{array}\right|
$
}

\bigskip

\noindent
Rewriting the table in a more convenient form, we get:

\bigskip

$
\left|\begin{array}{c|ccccccccccccccc}
\hline && \\
\l[N-1]\times  &&&&&&& \underline{q^3} &&& (1)   \\ && \\
\l[N-1]\times  &&&&&q^3&6\underline{\underline{\underline{q^2}}}& 2q &&&(-3) \\ && \\
\l[N-1]\times  &&&q^3&7\underline{q^2}
&17\underline{\underline{\underline{\underline{q}}}}&9& q^{-1} &&&(3=-1+4)  \\ && \\
\l[N-1]\times  &q^3&6q^2&17\underline{q}&25\cdot\underline{\underline{\underline{1}}}
&14q^{-1}&3q^{-2}& &&&(-2=1-3)   \\ && \\
\l[N-1]\times  &2q&9&14\underline{\underline{q^{-1}}}&6q^{-2}&q^{-3}&& &&&(2)  \\ && \\
\l[N-1]\times  &q^{-1}&3\underline{q^{-2}}&q^{-3}&&&& &&&(-1)  \\ && \\
&&\\
\l[N-2]\times & &&&q^3&2\underline{q^2}&&& &(1)  \\ && \\
\l[N-2]\times & &&2q^2&7\underline{q}&4&&& &(-1)  \\ && \\
\l[N-2]\times & &&4&7\underline{q^{-1}}&2q^{-2}&&& &(-1)   \\ && \\
\l[N-2]\times & &&2\underline{q^{-2}}&q^{-3}&&&& &(1)  \\ && \\
&&\\
q^{-N}\ \times & &&&&&&q^3   \\ && \\
q^{-N}\ \times & &&&&&3q^2&2q   \\ && \\
q^{-N}\ \times & &&&2q^2&3q&3&q^{-1}   \\ && \\
q^{-N}\ \times & q^3&3q^2&3q&&&&   \\ && \\
q^{-N}\ \times & 2q&3&&&&&   \\ && \\
q^{-N}\ \times & q^{-1}&&&&&&   \\
&&\\ \hline
\end{array}\right|
$

\bigskip

\noindent
As usual cohomologically non-trivial part of $\P$ is concentrated around inverse
diagonal -- in underlined terms.
In order to match two of the four  underlined terms in the $[N-2]$ lines,
we need to add an extra pair in the $[N-1]$ lines as well.

After that the upper part of the table is reduced, and we obtain:
\be
\sim
q^{2-2N}\cdot(qT)^3\  + [N-1]\cdot
\left|\begin{array}{cc|ccccccccccc}
\hline && \\
\l[N-1] &\times &0&0&q^{-1}&3&4q&3q^2&q^3   \\&& \\
\l[N-1] & \times &&q^{-2}&q^{-1}+q& q^2 & &  \\
\l[N-2] & \times &&&q^{-2}&q^{-1}+q& q^2&  \\ && \\
 && \\
q^{-N} &\times & q^{-1}&1&q&q^2&0&q^2&q^3&   \\
&&\\ \hline
\end{array}\right|
\nn
\ee

\bigskip

\noindent
Reducing now the $[N-1]$ and $[N-2]$ factors inside the table, we get:
\be
q^{2-2N}\cdot(qT)^3\  + [N-1]\cdot
\left|\begin{array}{cc|ccccccccccc}
\hline && \\
q^N &\times & 0&0&0&q^{-2}&2q^{-1} & 2 & q \\ &&\\
q^{-N} &\times &0&0&q&2q^{2}&2q^3&q^2& 0   \\&& \\
q^{-N} & \times &0&1&[2]q^2& q^4 &0 &0  & 0   \\ && \\
&&\\
q^{-N} &\times & q^{-1}&1&q&q^2&0&q^2&q^3    \\
&&\\ \hline
\end{array}\right|
\nn
\ee

\bigskip

\noindent
Collecting all the terms, proportional to $q^{-N}$,
$$
\Big( q^{-1}\ \ 2 \ \ 3q+q^3 \ \ 3q^2+q^4 \ \ 2q^3 \ \ 2q^2 \ \ q^3\Big)
\sim \Big( q^{-1} \ \ 2 \ \ 3q \ \ 2q^2\ \ 0 \ \ 0 \ \ 0\Big)
$$
we get:

\be
\left(\frac{q^{3N-3}}{q^{3N}T^3}\right)^{-1}\cdot\P^{6_3}_r \sim
q^{2-2N}\cdot(qT)^3\  + [N-1]\cdot
\left|\begin{array}{cc|ccccccccccc}
\hline && \\
q^N &\times & 0&0&0&q^{-2}&2q^{-1} & 2 & q \\ &&\\
q^{-N} &\times &q^{-1}&2&3q&2q^{2}&0&0& 0   \\
&&\\ \hline
\end{array}\right|
\nn
\ee
In reduced case this is cohomologically equivalent to
\be
\left|\begin{array}{c|ccccccccccc}
\hline && \\
q^{2N}\times &0&0&0&0&q^{-3} & q^{-2} & q^{-1} \\ && \\
1\ \times & 0 & 0 & 0 & 1 &q & q^2 & \\ && \\ && \\
1\ \times & 0&q^{-2}&q^{-1} & 2 & \\ && \\
q^{-2N}\times &q & q^2 & 2\underline{q^3} &\underline{q^2} \\
&&\\ \hline
\end{array}\right|
\nn
\ee
The second underlined term in the last line comes from the item outside the table
and is eliminated together with another underlined term,
so that
$$
\!\!\!\!\!\!\!\!\!\!\!
{\cal P}^{6_3}_r = \frac{1}{q^3T^3}\Big(q^{1-2N}+(q^{2-2N}+q^{-2})(qT) + (q^{3-2N}+q^{-1})(qT)^2
+ 3(qT)^3 + (q^{2N-3}+q)(qT)^4+(q^{2N-2}+q^2)(qT)^5 + q^{2N-1}(qT)^6\Big) =
$$
\vspace{-0.4cm}
\be
= \frac{1}{q^{2N+2}T^3} + 3 + q^{2N+2}T^3\ +\
(q^{N-2}+q^{2-N})\left(\frac{1}{q^{N+2}T^2} +\frac{1}{q^NT}+q^NT  + q^{N+2}T^2\right)
\label{KR63r}
\ee

\bigskip

In unreduced case we have instead
\be
\left(\frac{q^{3N-3}}{q^{3N}T^3}\right)^{-1}\cdot\P^{6_3} \sim
q^{2-2N}[N](qT)^3\  + [N][N-1]\cdot
\left|\begin{array}{cc|ccccccccccc}
\hline && \\
q^N &\times & 0&0&0&q^{-2}&2q^{-1} & 2 & q \\ &&\\
q^{-N} &\times &q^{-1}&2&3q&2q^{2}&0&0& 0   \\
&&\\ \hline
\end{array}\right|
\nn\\ \nn \\
\sim q^{1-N}\cdot(qT)^3 + [N-1]\cdot \left|\begin{array}{c|ccccccccccc}
\hline && \\
q^{2N}\times &0&0&0&0&q^{-2} & q^{-1} & 1 \\ && \\
1\ \times & 0 & 0 & 0 & q^{-1} &1 & q & \\ && \\ && \\
1\ \times & 0&q^{-1}&1 & 2q & \\ && \\
q^{-2N}\times &1 & q & 2\underline{q^2} &\underline{q} \\
&&\\ \hline
\end{array}\right|
\ \ \ \ \ \ \ \ \ \ \ \ \ \ \ \
\nn
\ee
and
\be
{\cal P}^{6_3} = q^{1-N} \ +\ \frac{[N-1]}{q^3T^3}\Big(q^{-2N}+(q^{1-2N}+q^{-1})(qT) + \nn \\
(q^{2-2N}+1)(qT)^2
+ (2q+q^{-1})(qT)^3 + (q^{2N-2}+1)(qT)^4+(q^{2N-1}+q)(qT)^5 + q^{2N}(qT)^6\Big) =
\nn \\
= [N]\ + \ [N-1]\left(\frac{1}{q^{2N+3}T^3} + [2] + q^{2N+3}T^3 \ +\
(q^{N-1}+q^{1-N})\Big(\frac{1}{q^{N+2}T^2} +\frac{1}{q^NT}+q^NT  + q^{N+2}T^2\Big)\right)
= \nn
\ee
\vspace{-0.6cm}
\be
= [N]\ + \ [N-1]\left(\frac{1}{q^{2N+3}T^3} + [2] + q^{2N+3}T^3\right)
\ + \ [2N-2]\left(\frac{1}{q^{N+2}T^2} +\frac{1}{q^NT}+q^NT  + q^{N+2}T^2 \right)
\label{KR63ur}
\ee

\noindent
Both (\ref{KR63r}) and (\ref{KR63ur}) coincide with the answers from \cite{CM}
(available only for $N=2$ and $N=3$ in unreduced case).

\bigskip

\noindent

\subsubsection{The 3-component link $\Big({\cal R}_1^2{\cal R}_2\Big)^2 $
(torus link $[3,3]$ or $6^3_3(v1)$ of \cite{CM})}

The KR complex for the braid
\ \ \ \ \ \ \ \ \ \ \ \ \ \ \ \ \ \ \ \ \ \ \ \ \ \ \ \ \ \ \ \ \ \
\ \ \ \ \ \ \ \ \ \ \ \ \ \ \ \
\vspace{-0.7cm}
is
$$
\begin{array}{cccccccccccc}
\bullet & \bullet && \bullet &  \bullet \\
&& \bullet &&& \bullet    \\
\end{array}
$$
\be
v_{00} \stackrel{d_0}{\longrightarrow} 6v_{10}
\stackrel{d_1}{\longrightarrow} 7v_{20} +8v_{21}
\stackrel{d_2}{\longrightarrow} 4v_{30} + 16v_{31}
\stackrel{d_3}{\longrightarrow} v_{40} + \underline{10v_{41} + 4v_{42}}
\stackrel{d_4}{\longrightarrow} \underline{2v_{51}+4v_{52}}
\stackrel{d_5}{\longrightarrow} v_{62}
\label{33complex1}
\ee
It differs from (\ref{51complex}) only by the coefficients in the underlined terms
-- but now this is a 3-component link and the KR polynomial will be
essentially different.

The primary polynomial is
\be
q^{6-6N}\cdot\P^{6^3_3(v1)}_r =\ \ \ \ \ \ \ \ \ \ \ \ \ \ \ \ \ \ \ \ \ \ \ \ \ \ \ \ \ \
\ \ \ \ \ \ \ \ \ \ \ \ \ \ \ \ \ \ \ \ \ \ \ \ \ \ \ \ \ \ \ \ \ \ \ \ \
\ \ \ \ \ \ \ \ \ \ \ \ \ \ \ \ \ \ \ \ \ \ \ \ \ \ \ \ \ \ \ \ \ \ \ \ \
\ \ \ \ \ \ \ \ \ \ \ \ \ \ \ \
\nn \\
= q^{2-2N} + [N-1]\cdot
\left|\begin{array}{c|cccccccccccccccccccccccccccccc}
\hline && \\
\l[N-1]\times & q^2 & 6q &7q[2]+8 & 4q[2]^2+16[2] & q[2]^3+10[2]^2+4 &
2[2]^3+4[2] & [2]^2 \\ && \\
\l[N-2]\times & &&&& 4[2] & 4[2]^2 & [2]^3 \\ && \\
q^{-N}\ \times & 2q^2 & 6q & 7q[2] & 4q[2]^2 & q[2]^3\\ && \\
\hline
\end{array}\right|
\nn \\ \nn \\
= q^{2-2N} + [N-1]\cdot
\left|\begin{array}{cc|cccccccccccccccccccccccccccccc}
\hline && \\
\l[N-1] & \times &&&&&q^4&2q^3&q^2&&&&(0) \\
\l[N-2] & \times &&&&&&& \underline{q^3}&&&(1)\\ && \\
\l[N-1] & \times &&&&4q^3&13\underline{q^2}&10q&2 &&&&(1) \\
\l[N-2] & \times &&&&&&4\underline{q^2}& 3q &&&(-1)\\ && \\
\l[N-1] & \times &&&7q^2&24q&27\cdot\underline{1}&\frac{16}{q}&\frac{1}{q^2} &&&& (1) \\
\l[N-2] & \times &&&&&4q&8\cdot\underline{1}& \ \frac{3}{q} &&& (-1) \\ && \\
\l[N-1] & \times &q^2&6q&15&20\cdot\underline{\frac{1}{q}}&\frac{11}{q^2}&\frac{2}{q^3}&  &&&&(-1) \\
\l[N-2] & \times &&&&&\ 4\cdot\underline{\frac{1}{q}}
&\ \frac{4}{q^2}&\ \frac{1}{q^3} &&& (1)\\ && \\ && \\
q^{-N}  & \times &2q^2&6q&7&\frac{4}{q}&\frac{1}{q^2}&& \\ && \\
q^{-N}  & \times &&&7\underline{\underline{q^2}}&8q&3&& \\ && \\
q^{-N}  & \times &&&&4\underline{q^3}&3q^2&& \\ && \\
q^{-N}  & \times &&&&&\underline{q^4}&& \\
&&\\ \hline
\end{array}\right|
\nn
\ee
Note that, in variance from (\ref{dimstor33}) which describes the same torus link $[3,3]$,
this time the space $v_{51}$ is involved, still its contribution to the very first
line of above table belongs to cohomologically trivial combination, and the
main contribution comes from the underlined term in the second line,
which is proportional to $[N-1][N-2]$. rather than to $[N-1]^2$.

All the rest is already rather standard. The underlined pairs in the upper part
of the table are of the type $\ [N-1] \Longrightarrow [N-2]\ $ and get substituted
by $\ q^{2-N} \Longrightarrow 0$, thus at the next stage we obtain
\be
q^{2-2N} + q^3[N-1][N-2](qT)^6 +
q^{-N}[N-1] \cdot \left|\begin{array}{cccccccccccccc}
\hline \\
0 & 0 & 2\underline{q^2} & \overline{q^3} &q^4 & 0 & 0 \\ \\
 &  &  & \underline{q} & \overline{q^2}+q^4 & \\ \\
\hline\end{array}\right|
\ee
where the first line comes from the lower part of the previous table,
while the second line -- from reduction of pairs in its upper part.
Underlined and overlined are the two pairs, which look cohomologically trivial --
however, we know well, that the right answers require that something
remains in the order $T^3$ (this should not be too difficult to see from
analysis of morphisms -- at level $T^3$ their cohomologies are still
rather simple to study explicitly).
If we leave the overlined pair intact and eliminate only the underlined one,
then the right answer emerges:
$$
q^{-N}[N-1] \cdot \left|\begin{array}{cccccccccccccc}
\hline \\
0 & 0 &  {q^2} & q^3 &q^2+2q^4 & 0 & 0 \\ \\
\hline\end{array}\right|  \
\sim q^{-N}(q^2+2q^4)[N-1](qT)^4 + q^{-N}
\Big(q^2\cdot q^{2-N}\cdot(qT)^2 +
q^3\cdot q^{N-2} \cdot (qT)^3\Big)
$$
in reduced case,
{\footnotesize
$$
q^{-N}[N-1][N] \cdot \left|\begin{array}{cccccccccccccc}
\hline \\
0 & 0 &  {q^2} & q^3 &q^2+2q^4 & 0 & 0 \\ \\
\hline\end{array}\right| \
\sim q^{-N}(q^2+2q^4)[N-1][N](qT)^4 + q^{-N}
\Big(q^2\cdot q^{1-N}\cdot(qT)^2 +
q^3\cdot q^{N-1} \cdot (qT)^3\Big)[N-1]
$$
}
in unreduced case, and
$$
{\cal P}^{6^3_3(v1)}_r = q^{6N-6}\Big(q^{2-2N}
+  q^{4-2N}(qT)^2 +
q \cdot (qT)^3\Big)+ q^{-N}(q^2+2q^4)[N-1](qT)^4 +q^3[N-1][N-2](qT)^6\Big) =
$$
\vspace{-0.5cm}
\be
= q^{4N-4}\Big(1 + q^4T^2 + q^{2N+2}T^3 + (q^{2N+4}+q^{2N+6})[N-1]T^4
+ q^{2N+7}[N-1][N-2]T^6\Big)
\label{KR33true1}
\ee
$$
{\cal P}^{6^3_3(v1)} =
q^{4N-4}\Big([N] + q^3T^2(1+q^{2N}T)[N-1] + (q^{2N+4}+q^{2N+6})[N][N-1]T^4
+ q^{2N+7}[N][N-1][N-2]T^6\Big)
$$
in agreement with (\ref{KR33true}) and \cite{CM} (for $N=2,3$).

\subsubsection{Another coloring: the 3-component link
$\Big({\cal R}_1^2{\cal R}_2^{-1}\Big)^2 $
($6^3_1(v2)$ of \cite{CM})}

The KR complex for the braid
\ \ \ \ \ \ \ \ \ \ \ \ \ \ \ \ \ \ \ \ \ \ \ \ \ \ \ \ \ \ \ \ \ \
\ \ \ \ \ \ \ \ \ \ \ \ \ \ \ \
\vspace{-0.7cm}
is obtained by reordering of (\ref{33complex1}):
$$
\begin{array}{cccccccccccc}
\bullet & \bullet && \bullet &  \bullet \\
&& \circ &&& \circ    \\
\end{array}
$$
\be
v_{20} \stackrel{d_0}{\longrightarrow} 2v_{10}+4v_{31}
\stackrel{d_1}{\longrightarrow} v_{00} +8v_{21} + 2v_{41} + 4v_{42}
\stackrel{d_2}{\longrightarrow} 4v_{10} + 12v_{31}+4v_{52}
\stackrel{d_3}{\longrightarrow} 6v_{20} + 8v_{41} + 4v_{62}
\stackrel{d_4}{\longrightarrow} 4v_{30} + 2v_{51}
\stackrel{d_5}{\longrightarrow} v_{40}
\nn
\ee

The primary polynomial is
\be
q^{4-2N}T^2\cdot\P^{6^3_1(v2)}_r =q^{2-2N}\cdot(qT)^2 + [N-1]\cdot
\ \ \ \ \ \ \ \ \ \ \ \ \ \ \ \ \ \ \ \ \ \ \ \ \ \ \ \ \ \ \ \ \ \ \ \ \
\ \ \ \ \ \ \ \ \ \ \ \ \ \ \ \ \ \ \ \ \ \ \ \ \ \ \ \ \ \ \ \ \ \ \ \ \
\ \ \ \ \ \ \ \ \ \ \ \ \ \ \ \
\nn \\
\cdot\left|\begin{array}{c|cccccccccccccccccccccccccccccc}
\hline && \\
\l[N-1]\times & q[2] & 2q+4[2] &q^2+8+2[2]^2+4 & 4q+12[2]+4[2] & 6q[2]+8[2]^2+[2]^2 &
4q[2]^2+2[2]^3 & q[2]^3 \\ && \\
\l[N-2]\times & &&4[2]&4[2]^2& [2]^3 &  &  \\ && \\
q^{-N}\ \times & q[2] & 2q & 2q^2 & 4q & 6q[2]& 4q[2]^2 & q[2]^3\\ && \\
\hline
\end{array}\right|
\nn \\ \nn \\
= q^{2-2N}\cdot(qT)^2 + [N-1]\cdot
\left|\begin{array}{cc|cccccccccccccccccccccccccccccc}
\hline && \\
\l[N-1] & \times &&&&& & &q^4&&&&(1) \\ && \\
\l[N-1] & \times &&&&& &6q^3 &3q^2&&&&(-3) \\ && \\
\l[N-1] & \times &&&&&15q^2&14q  &3&&&&(4) \\ && \\
\l[N-1] & \times &&&3q^2&20q&24 &10q^{-1} &q^{-2}&&&&(-2) \\ && \\
\l[N-1] & \times &q^2&6q&16&10q^{-1}&9q^{-2} &2q^{-3} &&&&&(2) \\ && \\
\l[N-1] & \times &1&4q^{-1}&2q^{-2}&& & &&&&&(-1) \\ && \\
&& \\
\l[N-2] & \times &&&&&q^3&& &&&(1)\\ && \\
\l[N-2] & \times &&&&4q^2&3q&& &&&(-1)\\ && \\
\l[N-2] & \times &&&4q&8&3q^{-1}&& &&&(-1)\\ && \\
\l[N-2] & \times &&&4q^{-1}&4q^{-2}&q^{-3}&& &&&(1)\\ && \\
&&\\
q^{-N}  & \times &&&&&&& {q^4}&&&&  \\ && \\
q^{-N}  & \times &&&&& &4q^3&3q^2 \\ && \\
q^{-N}  & \times &&&&&6q^2&8q&3 \\ && \\
q^{-N}  & \times &&&2q^2&4q&6&\frac{4}{q}&\frac{1}{q^2} \\ && \\
q^{-N}  & \times &q^2&2q& \\ && \\
q^{-N}  & \times &1&& \\
&&\\ \hline
\end{array}\right|
\nn
\ee

\be
\sim q^{2-2N}(qT)^2 + [N-1]\cdot\left|\begin{array}{c|ccccccccccccccccccccc}
\hline && \\
\l[N-1]\times & 0 & q^{-1}\!\!\!\! & 2\!\!\!\!& 2q & \!\!\!\!4q^2 &3q^3 & q^4 \\ & \\
\l[N-2]\times & 0 & 0 & q^{-1} & 1+q^2 & {q^3} & 0 & 0 \\ & \\
\hline && \\
q^{-N} \ \times & 1 &q &q^2&& q^2 &q^3 & q^4 \\
&&\\ \hline
\end{array}\right|
\nn
\ee
The last one is the minimal version of reduction -- only the
"must" items are kept (and selected from the usual places along the diagonals).
However, the right answer requires to keep a couple of the seemingly trivial
pairs in the $[N-2]$ lines, we put them in additional line:
\be
\sim q^{2-2N}(qT)^2 + [N-1]\cdot\left|\begin{array}{c|ccccccccccccccccccccc}
\hline && \\
\l[N-1]\times & 0 & q^{-1}\!\!\!\! & 2\!\!\!\!& 2q & \!\!\!\!4q^2 &3q^3 & q^4 \\ & \\
\l[N-2]\times & 0 & 0 & q^{-1} & 1+q^2 & {q^3} & 0 & 0 \\ & \\
\l[N-2]\times & 0 & 0 & q & 1+q^2 & {q} & 0 & 0 \\ & \\
\hline && \\
q^{-N} \ \times & 1 &q &q^2&& q^2 &q^3 & q^4 \\
&&\\ \hline
\end{array}\right|
\nn \\ \nn \\
= q^{2-2N}(qT)^2 + [N-1]\cdot\left|\begin{array}{c|ccccccccccccccccccccc}
\hline && \\
\l[N-1]\times & 0 & q^{-1}\!\!\!\! & 2\!\!\!\!& 2q & \!\!\!\!4q^2 &3q^3 & q^4 \\
& &&\!\!\!\!\! \searrow \ \ \ \ \ \ \ \ \
&\!\!\!\!\!\!\!\!\!\searrow\!\!\!\!\!\swarrow \ \ \ \ \ \ \ \ \swarrow\!\!\!\!\!&
&\!\!\!\!\!\!\!\swarrow \ \ \ \ \   \\
\l[N-2]\times & 0 & 0 & q^{-1}+q & 2+2q^2 & \boxed{q}+q^3 & 0 & 0 \\ & \\
 \hline && \\
q^{-N} \ \times & 1 &q &q^2&& q^2 &q^3 & q^4 \\
&&\\ \hline
\end{array}\right|
\nn \\ \nn \\
\sim q^{2-2N}(qT)^2 + q[N-1][N-2] (qT)^4 +
[N-1]\cdot\left|\begin{array}{c|ccccccccccccccccccccc}
\hline && \\
\l[N-1] \times & 0 &0 & 0 & q& 2q^2 & 2q^3 & q^4 \\ & \\
q^N \ \times & & & & q^{-1}&2& q \\ & \\
q^{-N} \ \times &  &q &2q^2&&  & &  \\ & \\
\hline && \\
q^{-N} \ \times & 1 &q &q^2&& q^2 &q^3 & q^4 \\ & \\
\hline
\end{array}\right|
\nn \\ \nn \\
\sim q^{2-2N}(qT)^2 + q[N-1][N-2] (qT)^4 +
[N-1]\cdot\left|\begin{array}{c|ccccccccccccccccccccc}
\hline && \\
 q^N \ \times & 0 &0 &0 & 0 &1& q & \boxed{q^2} \\ & \\
q^{-N} \ \times & & & & \underline{q^3}  &\underline{q^4} &\underline{q^5}&   \\ & \\
\hline && \\
q^{-N} \ \times & 1 &q &q^2&& \underline{q^2} &\underline{q^3} & \underline{q^4} \\ & \\
\hline && \\
q^N \ \times & & & & q^{-1}&2& q \\ & \\
q^{-N} \ \times &  &q &\boxed{2q^2}&&  & &  \\ & \\
\hline
\end{array}\right|
\nn
\ee

Now, we eliminate the underlined items as belonging to cohomologically trivial pairs,
take the boxed items outside the table and perform the usual reduction of the
$[N-1]$ factor:

\be
\sim q^{2-2N}(qT)^2 + q[N-1][N-2] (qT)^4 + 2q^{2-N}[N-1](qT)^2 + q^{2+N}[N-1](qT)^6 +
\nn \\
+\l[N-1]\cdot\left|\begin{array}{c|ccccccccccccccccccccc}
\hline && \\
 q^N \ \times & 0 &0 &0 & q^{-1} &3& 2q & 0 \\ & \\
q^{-N} \ \times & 1 &2q &q^2&0&0  &0 &0  \\ & \\
\hline
\end{array}\right|
\label{631v2a} \\ \nn \\
\sim  q[N-1][N-2] (qT)^4 + 2q^{2-N}[N-1](qT)^2 + q^{2+N}[N-1](qT)^6 +
\nn \\
+\left|\begin{array}{c|ccccccccccccccccccccc}
\hline && \\
 q^{2N} \ \times & 0 &0 &0 & 0 &q^{-2}& 2q^{-1} & 0 \\
1\ \times &0 & 0 & 0& q &2q^2 & 0 & 0 \\ & \\
1\ \times &0&q^{-1} &1 &0& 0 & 0 & 0 \\
q^{-2N} \ \times & q^2 &\underline{q^3} &\underline{q^2}&0& 0 &0 &0  \\ & \\
\hline
\end{array}\right|
\label{631v2b}
\ee
where the first item is moved from outside to inside the table (underlined)
and is eliminated together with another underlined term.
In result the reduced KR polynomial is
\be
{\cal P}_r^{6^3_1(v2)} = \frac{q^{2N-2}}{(qT)^2}\Big\{
q^{2-2N} +q^{-1}(qT) +\Big(1+2q^{2-N}[N-1]\Big)(qT)^2 +q\cdot(qT)^3 + \nn \\
\Big(q^{2N-2}+2q^2 +q[N-1][N-2]\Big)(qT)^4 + 2q^{2N-1}(qT)^5 + q^{2+N}[N-1](qT)^6
\Big\}
= \nn \\
=\frac{1}{q^2T^2} + \frac{q^{2N-4}}{T} + \Big(q^{2N-2} + 2q^N[N-1]\Big)
+q^{2N}T + \nn \\
+\Big(q^{4N-2} + 2q^{2N+2} + q^{2N+1}[N-1][N-2]\Big)T^2
+ 2q^{4N}T^3 + q^{3N+4}[N-1]T^4
\label{KR631r}
\ee
This answer is in nice accordance with the result of  \cite{CM}:
\be
N=2:& \frac{1}{q^2T^2} + \frac{1}{T} + 3q^2 + q^4T + 3q^6T^2 + 2q^8T^3 + q^{10}T^4, \nn \\
N=3:& \frac{1}{q^2T^2} + \frac{q^2}{T} + (2q^2+3q^4) + q^6T +
(q^6+3q^8+q^{10})T^2 + 2q^{12}T^3 + q^{13}[2]T^4
\nn
\ee

\bigskip

In unreduced case the difference appears in transition from (\ref{631v2a})
to (\ref{631v2b}):
\be
\sim q^{2-2N}\underbrace{[N]}_{\!\!\!\!\!\!\!\!\!\! \frac{1}{q}[N-1] + q^{N-1}
\!\!\!\!\!\!\!\!\!\!}(qT)^2 + q[N][N-1][N-2] (qT)^4 + 2q^{2-N}[N][N-1](qT)^2 + q^{2+N}[N][N-1](qT)^4 +
\nn \\
+\l[N][N-1]\cdot\left|\begin{array}{c|ccccccccccccccccccccc}
\hline && \\
 q^N \ \times & 0 &0 &0 & q^{-1} &3& 2q & 0 \\ & \\
q^{-N} \ \times & 1 &2q &q^2&0&0  &0 &0  \\ & \\
\hline
\end{array}\right|
\label{631v2ura} \\ \nn \\ \nn \\
\sim  q[N][N-1][N-2] (qT)^4 + 2q^{2-N}[N][N-1](qT)^2 + q^{2+N}[N][N-1](qT)^4 +
\nn \\
+[N-1]\cdot\left|\begin{array}{c|ccccccccccccccccccccc}
\hline && \\
 q^{2N} \ \times & 0 &0 &0 & 0 &q^{-1}& 2 & 0 \\
1\ \times &0 & 0 & 0& 1 &2q & 0 & 0 \\ & \\
1\ \times &0&1 &q &0& 0 & 0 & 0 \\
q^{-2N} \ \times & q &\underline{q^2} &\underline{q}&0& 0 &0 &0  \\ & \\
\hline
\end{array}\right| + q^{1-N}\cdot (qT)^2
\label{631v2urb}
\ee
so that
\be
{\cal P}^{6^3_1(v2)} = \frac{q^{2N-2}}{(qT)^2}[N-1]\Big\{q^{1-2N}
 +(qT) +\Big(\underline{q}+2q^{2-N}[N]\Big)(qT)^2 +(qT)^3 + \nn \\
+\Big(q^{2N-1}+2q +q[N][N-2]\Big)(qT)^4 + 2q^{2N}(qT)^5 + q^{2+N}[N](qT)^6
\Big\} + \underline{q^{N-1}}
= \nn \\
= \Big(\frac{1}{q^3T^2} + \frac{q^{2N-3}}{T}\Big)[N-1]
 + \Big(\underline{q^{2N-2}[N]} + 2q^N[N][N-1]\Big)
+q^{2N-1}[N-1]T + \nn \\
+\Big((q^{4N-1} + 2q^{2N+1})[N-1] + q^{2N+1}[N][N-1][N-2]\Big)T^2
+ 2q^{4N+1}[N-1]T^3 + q^{3N+4}[N][N-1]T^4
\label{KR631ur}
\ee
(the two underlined terms at the l.h.s. combine to form the underlined
item at the r.h.s.).
This answer
coincides with that of \cite{CM} for $N=2$ and $N=3$.


\subsubsection{The 2-component link $6_2^2$ ($L6a2$) in the 3-strand representation}

With the braid ${\cal R}_1^4{\cal R}_2^{2}{\cal R}_1^2{\cal R}_2:$
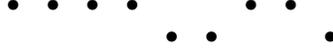
\vspace{-0.6cm}
$$
\begin{array}{ccccccccc}
\bullet&\bullet &\bullet&\bullet &&&\bullet&\bullet& \\
&&&&\bullet&\bullet&&&\bullet
\end{array}
$$
one associates a KR complex
\be
v_{00} \longrightarrow 9v_{10} \longrightarrow 18v_{20}+18v_{21} \longrightarrow
21v_{30}+63v_{31} \longrightarrow 15v_{40}+95v_{41}+16v_{42} \longrightarrow \nn\\
\longrightarrow 6v_{50}+80v_{51}+40v_{52} \longrightarrow
v_{60}+39v_{61}+44v_{62} \longrightarrow 10v_{71}+26v_{72}
\longrightarrow v_{81}+8v_{82} \longrightarrow v_{92}
\ee
This 2-component link has a very simple Jones polynomial $q^7+q^{11}$,
still it is a really $9$-intersection link $9^2_{49}=L9n15$.

\bigskip

However, the change of coloring to the braid
${\cal R}_1^4{\cal R}_2^{-2}{\cal R}_1^2{\cal R}_2:$
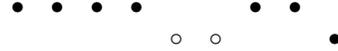
\vspace{-0.8cm}
$$
\ \ \ \ \ \ \ \ \ \ \ \ \ \ \ \ \ \ \ \ \ \ \ \ \ \ \ \ \ \ \ \ \ \ \ \ \ \ \ \ \ \
\ \ \ \ \ \ \ \ \ \ \ \ \ \ \ \ \ \ \ \ \ \ \ \ \ \ \ \
\begin{array}{ccccccccc}
\bullet&\bullet &\bullet&\bullet &&&\bullet&\bullet& \\
&&&&\circ&\circ&&&\bullet
\end{array}
$$
provides a $6$-intersection 2-component link $6^2_2$.
Associated with this braid is the KR complex
\be
v_{20} \longrightarrow 2v_{10}+v_{30}+6v_{31} \longrightarrow
v_{00}+2v_{20}+12v_{21} + 6v_{41}+15v_{41} \longrightarrow \nn\\
\longrightarrow 7v_{10}+42v_{31}+27v_{51}+8v_{52}\longrightarrow
15v_{20}+6v_{21}+54v_{41}+16v_{42}+19v_{61} +16v_{62} \longrightarrow \nn \\
\longrightarrow 20v_{30}+15v_{31}+38v_{51}+32v_{52}+7v_{71}+14v_{72}
\longrightarrow 15v_{40}+20v_{41}+14v_{61}+28v_{62}+v_{81}+6v_{82}
\longrightarrow \nn \\ \longrightarrow
6v_{50}+15v_{51}+2v_{71}+12v_{72}+v_{92} \longrightarrow
v_{60}+6v_{61}+2v_{82} \longrightarrow v_{71}
\ee
Reduced primary polynomial in this case is:
$$
\left(\frac{q^{7N-7}}{(q^NT)^2}\right)^{-1}\P^{6_2}_r = q^{2-2N}(qT)^2 + [N-1]\cdot
\ \ \ \  \ \ \ \ \ \ \ \ \ \ \ \ \ \ \ \ \ \ \ \ \ \ \ \ \ \ \ \ \ \ \ \
$$

\centerline{
$
\left|\begin{array}{c||c|c|c|c|c}\hline &&&&& \\
& 1 & (qT) & (qT)^2 & (qT)^3 & (qT)^4 \\ &&&&& \\
\hline &&&&& \\
\l[N-1]\times & q[2] & 2q+q[2]^2+6[2]& q^2+2q[2]+12+21[2]^2& 7q+42[2]+27[2]^3+8[2]&
15q[2]+6+54[2]^2+16+19[2]^4+16[2]^2  \\ &&&&&  \\
\l[N-2]\times& &&&8[2]^2 &16[2]+16[2]^3  \\ &&&&&  \\
q^{-N}\ \times &q[2]&2q+q[2]^2&2q^2+2q[2]&7q &15q[2] \\&&&&& \\
\hline
\end{array}\right.
$
}

\centerline{
$
\left|
\begin{array}{c||c|c|c|c|c}\hline &&&&& \\
& (qT)^5 & (qT)^6 & (qT)^7 & (qT)^8 & (qT)^9 \\ &&&&& \\
\hline &&&&& \\
\l[N-1]\times & 20q[2]^2+15[2]+38[2]^3+ &
15q[2]^3+20[2]^2+14[2]^4+ &
6q[2]^4+15[2]^3+2[2]^5+& q[2]^5+6[2]^4+2[2]^4 & [2]^5  \\
& +32[2]+7[2]^5+14[2]^3 &  + 28[2]^2+[2]^6+6[2]^4 & +12[2]^3+[2]^5 &&\\
&&&&& \\
\l[N-2]\times& 32[2]^2+14[2]^4&28[2]^3+6[2]^5 & 12[2]^4+[2]^6 & 2[2]^5& \\ &&&&& \\
q^{-N}\ \times &  20q[2]^2 & 15q[2]^3 &6q[2]^4 & q[2]^5 & \\&&&&& \\
\hline
\end{array}\right|
$
}

\bigskip

Now, before
drawing and manipulating entire tables
we calculate the generating function of alternated sums:
\be
u(q)=
q[2] - q\cdot (2q+q[2]^2+6[2]) +\ldots = q^8-q^6+q^4-2q^2+1
\ee
where the entry in column $k=0,\ldots,9$ is taken with the weights $(-q)^k$.
The answer is a polynomial $\sum_i m_iq^i$ and its coefficients $m_i$
are exactly the alternated sums in the $[N-1]$ lines, presented in the
last column of the detailed table below
(the smaller $i$ the lower is the line).
Similarly for the $[N-2]$ lines:
\be
v(q)=-q^3\cdot 8[2]^3+q^4\cdot(16[2]+16[2]^3) - \ldots =
q^{13}-2q^{11}+q^9-q^5+2q^3-q
\ee
and for the $q^{-N}$ lines:
\be
w(q)=q[2] - q\cdot(2q+q[2]^2) + \ldots =  q^{14}-q^{12}+q^{10}-q^8+q^6+q^4-q^2
\ee
The linear combination
$$q^{7-2\cdot 2}(u(q)+q^{-2}w(q)+q^{-2}\cdot q^2)
$$
gives the Jones polynomial, while
$$\frac{A^{7-2}}{q^7}
\left( u(q)\frac{\{A/q\}}{\{q\}} + v(q)\frac{\{A/q\}}{\{q\}} +w(q) A^{-1}\right)\frac{\{A/q\}}{\{q\}}
+ \frac{A^{5-2}}{q^5}
$$
is the HOMFLY polynomial.

The detailed table looks like

\bigskip

$$
\left| \begin{array}{c|ccccccccccccc}
\hline &&&&&&&&&&&&\\ &1&(qT)&(qT)^2&(qT)^3&(qT)^4&(qT)^5&(qT)^6&(qT)^7&(qT)^8&(qT)^9\\
&&&&&&&&&&&\\ \hline
&&&&&&&&&&&&\\
\l[N-1]\times&&&&&&&&&q^6&q^5&&(0)\\
\l[N-1]\times&&&&&&&q^6&9q^5&13q^4&5q^3&&(0)\\
\l[N-1]\times&&&&&&7q^5&41q^4&66q^3&42q^2&10q&&(0)\\
&&&&&&&&&&&&\\
\l[N-1]\times&&&&&19q^4&107q^3&188\underline{q^2}&147q&58&10q^{-1}&&(1)\\
\l[N-1]\times&&&&27q^3&161q^2&313\underline{q}&281&135q^{-1}&37q^{-2}&5q^{-3}&&(-1)\\
&&&&&&&&&&&&\\
\l[N-1]\times&&q^3&24q^2&138q&291\cdot\underline{1}&293q^{-1}&158q^{-2}&48q^{-3}&9q^{-4}&q^{-5}&&(1)\\
\l[N-1]\times&q^2&10q&56&131\cdot\underline{\underline{q^{-1}}}&136q^{-2}&87q^{-3}
&36q^{-4}&3q^{-5}&&&&(-2)\\
\l[N-1]\times &1&7q^{-1}&21\cdot\underline{q^{-2}}&27q^{-3}&19q^{-4}&7q^{-5}&q^{-6}&&&&&(1)\\
&&&&&&&&&&&&\\
\hline &&&&&&&&&&&&\\
\l[N-2]\times&&&&&&&&q^6&2\underline{q^5}&&(1)\\
\l[N-2]\times&&&&&&&6q^5&18\underline{\underline{q^4}}&10q^3&&(-2)\\
\l[N-2]\times&&&&&&14q^4&58\underline{q^3}&63q^2&20q&&(1)\\
&&&&&&&&&&&&\\
\l[N-2]\times&&&&&16q^3&88q^2&144q&92&20q^{-1}&&(0)\\
&&&&&&&&&&&&\\
\l[N-2]\times&&&&8q^2&64q&148\cdot\underline{1}&144q^{-1}&63q^{-2}&10q^{-3}&&(1)\\
\l[N-2]\times&&&&16&64\cdot\underline{\underline{q^{-1}}}&88q^{-2}&58q^{-3}&18q^{-4}&q^{-5}&&(-2)\\
\l[N-2]\times&&&&8\cdot\underline{q^{-2}}&16q^{-3}&14q^{-4}&6q^{-5}&q^{-6}&&&(1)\\
&&&&&&&&&&&&\\ \hline &&&&&&&&&&&&\\
q^{-N}\ \times&&&&&&&&&\underline{q^6}&&&(1)\\
q^{-N}\ \times&&&&&&&&6\underline{q^5}&5q^4&&&(-1)\\
q^{-N}\ \times&&&&&&&15\underline{q^4}&24q^3&10q^2&&&(1)\\
q^{-N}\ \times&&&&&&20\underline{q^3}&45q^2&36q&10&&&(-1)\\
q^{-N}\ \times&&&&&15\underline{q^2}&40q&45&24q^{-1}&5q^{-2}&&&(1)\\
q^{-N}\ \times&&q^3&4\underline{q^2}&7q&15&20q^{-1}&15q^{-2}&6q^{-3}&q^{-4}&&&(1)\\
q^{-N}\ \times&q^2&4\underline{q}&2&&&&&&&&&(-1)\\
q^{-N}\ \times&1&q^{-1}&&&&&&&&&&(0)\\
&&&&&&&&&&&&\\ \hline
\end{array}\right|
$$

\bigskip

\noindent

Picking up the underlined items  from the table, we get:
\be
\left|\begin{array}{c|ccccccccccc}
\hline & \\
\l[N-1]\times&  0 & 0 & q^{-2} &2q^{-1} & 1&\underline{q} & \underline{q^2} & 0 & 0 & 0     \\ &\\
\l[N-2]\times& 0 & 0 & 0 & q^{-2} & 2q^{-1}& 1 & \underline{q^3}
& \underline{\underline{2q^4}} & \underline{q^5}& 0 \\ & \\
q^{-N}\ \times &0 &q&q^2&0&q^2&q^3&q^4&q^5& q^6 & 0 \\
&\\ \hline
\end{array}\right|
\nn \\ \nn \\
\sim
\left|\begin{array}{c|ccccccccccc}
\hline & \\
1\ \times&  0 & 0 & 0 &0 & 0&q^{3-N} & q^N & 0 & 0 & 0     \\ &\\
1\ \times& 0 & 0 & 0 & 0 & 0& 0 & q^{6-N} & q^{N+1}+q^{7-N} & q^{N+2}& 0 \\ & \\
q^{-N}\ \times &0 &q&1+q^2&2q&2q^2&q^3 &q^4&q^5& q^6 & 0 \\
&\\ \hline
\end{array}\right|
\nn \\ \nn \\
\sim \left|\begin{array}{c|ccccccccccc}
\hline & \\
q^N \ \times&  0 & 0 & 0 &0 & 0&0 &  \underline{1} & \underline{q} & q^2 & 0     \\ &\\
q^{-N}\ \times &0 &0&0&q&2q^2&2q^3&q^4+q^6&q^5+q^7& q^6 & 0 \\
&\\ \hline
\end{array}\right|
\nn \\ \nn \\
\sim \left|\begin{array}{c|ccccccccccc}
\hline & \\
q^N \ \times&  0 & 0 & 0 &0 & 0&0  & {1} & {q} & \boxed{q^2} & 0     \\ &\\
q^{-N}\ \times &0 &0&0&q&2q^2&2q^3&q^4&0 & 0 & 0 \\
&\\ \hline
\end{array}\right|
\nn
\ee
Now it remains to multiply by $[N-1]$ and further reduce the lines,
except for the boxed element.
Then we get
\be
\left(\frac{q^{7N-7}}{(q^NT)^2}\right)^{-1}\P^{6_2}_r \sim q^{2-2N}(qT)^2 +
\l[N-1]\cdot
\left|\begin{array}{c|ccccccccccc}
\hline & \\
q^N \ \times&  0 & 0 & 0 &0 & 0&0  & {1} & {q} & \boxed{q^2} & 0     \\ &\\
q^{-N}\ \times &0 &0&0&q&2q^2&2q^3&q^4&0 & 0 & 0 \\
&\\ \hline
\end{array}\right|
\nn \\ \nn \\
\!\!\!\!\! \!\!\!\!\! \!\!\!\!\! \!\!\!\!\! \sim q^{2-2N}(qT)^2 + q^{N+2}[N-1](qT)^8 +
\left|\begin{array}{cccccccccccc}
\hline & \\
0&0&0& q^{3-2N} & 1+q^{4-2N} & q+q^{5-2N} & q^2+q^2& q^{2N-1}& 0 & 0\\
&\\ \hline
\end{array}\right|
\label{P622r}
\ee
so that
\be
\!\!\!\!\!
{\cal P}^{6^2_2}_r =
q^{3N-3}\Big(1+ q^2T + (q^{2N}+q^4)T^2 + (q^{2N+2}+q^6)T^3 +2q^{2N+4}T^4
+ q^{4N+2}T^5 + q^{3N+6}[N-1]T^6\Big)
\label{KR622r}
\ee

In unreduced case modified is just the last step: instead of (\ref{P622r}),
\be
\left(\frac{q^{7N-7}}{(q^NT)^2}\right)^{-1}\P^{6_2} \sim q^{2-2N}(qT)^2[N] +
\l[N][N-1]\cdot
\left|\begin{array}{c|ccccccccccc}
\hline & \\
q^N \ \times&  0 & 0 & 0 &0 & 0&0  & {1} & {q} & \boxed{q^2} & 0     \\ &\\
q^{-N}\ \times &0 &0&0&q&2q^2&2q^3&q^4&0 & 0 & 0 \\
&\\ \hline
\end{array}\right|
\nn \\ \nn \\
\!\!\!\!\!\!\!
\sim q^{2-2N}(qT)^2[N] + q^{N+2}[N][N-1](qT)^8 + [N-1]\cdot
\left|\begin{array}{cccccccccccc}
\hline & \\
0&0&0& q^{2-2N} & q+q^{3-2N} & q^2+q^{4-2N} & q+q^3& q^{2N}& 0 & 0\\
&\\ \hline
\end{array}\right|
\nn
\ee
so that
\be
{\cal P}^{6^2_2} =
q^{3N-3}\Big([N]+ q[N-1]T + (q^{2N+1}+q^3)[N-1]T^2 + (q^{2N+3}+q^5)[N-1]T^3
+ \nn \\
+q^{2N+4}[2][N-1]T^4 + q^{4N+3}[N-1]T^5 + q^{3N+6}[N][N-1]T^6\Big)
\label{KR622ur}
\ee

These results are in accord with \cite{CM} (note that there is   no
orientation dependence for $6^2_2$ and both versions of \cite{CM} are in fact
equivalent -- related by the usual change $q,t\longrightarrow q^{-1},t^{-1}$).

\section{The $4$-strand examples
\label{4stra}}

\setcounter{equation}{0}

\subsection{Decomposition of ${\cal R}$-matrices}

A product of the fundamental representations decomposes with the help of
the representation tree \cite{MMMkn1}:
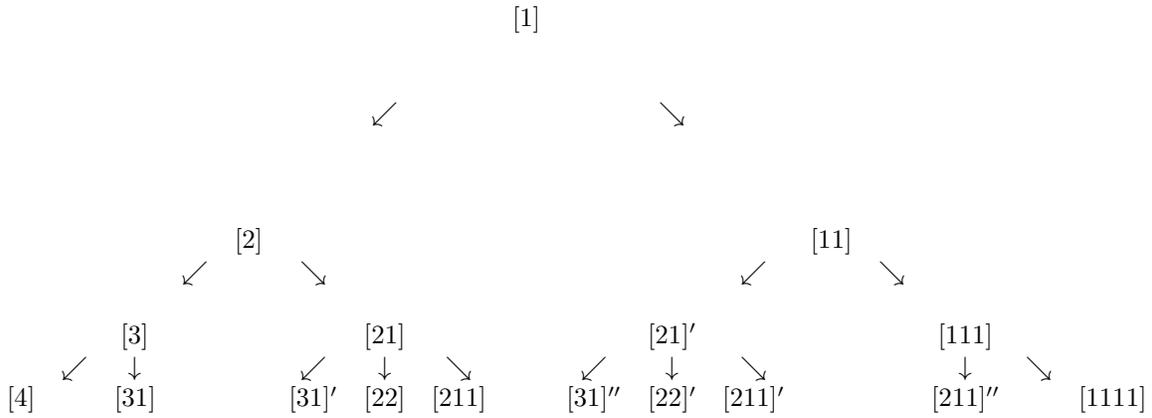
$$
\begin{array}{ccccccccccccccccccccccccccccccccccccc}
\\
&&&&&&&&[1]&&&&&&&&\\    \\ \\
&&&&&& \swarrow &&&& \searrow &&&&&& \\ \\  \\ \\
&&&&[2] &&&& &&&& [11]&&&& \\
&&&\swarrow &&\searrow &&& &&&\swarrow && \searrow &&&  \\  \\
&&[3] &&   &&[21]&& && [21]'&& &&[111] && \\
&\swarrow &\downarrow&  &  &\swarrow & \downarrow & \searrow&
&\swarrow & \downarrow & \searrow& &  &\downarrow & \searrow& \\
\l[4] && [31] &&          & [31]' & [22] & [211] && [31]''&[22]'& [211]' & && [211]'' && [1111] \\
\end{array}
$$

\bigskip

\underline{2-strand case:}

In the beginning of this paper, we decomposed the ${\cal R}$ matrix, acting in the space
$[1]\otimes [1] = [2] +[11]$ as
\be
{\cal R} \sim q^{-1}P_{[2]} - qP_{[11]} \sim I - q[2]P_{11}
\label{deco2str}
\ee
and associated $I$ with the black and $P_{11}$ with the white points in the hypercube vertices.
Then the contribution of $n$ white points,
\be
v_n = \frac{1}{[N]}{\rm Tr}_q \Big([2]P_{11}\Big)^n = [2]^n \frac{D_{11}}{[N]} = [2]^{n-1}[N-1]
\ee
was identified with the dimension of the vector space at the corresponding
hypercube vertex.

\bigskip

\underline{3-strand case:}

In this case we have two ${\cal R}$ matrices, acting on
the first two and on the last two of representations in the product $[1]\otimes [1]\otimes [1]$.
They related by the mixing (Racah) matrix  \cite{MMMkn1}:
\be
{\cal R}_2 = U{\cal R}_1U^\dagger
\ee
From the look at representation tree it is clear that the decomposition (\ref{deco2str})
implies that
\be
{\cal R}_1 \sim I - q[2](P_{[21]'}+P_{111}) = I - q[2]\pi_1
\ee
Therefore
\be
{\cal R}_2 = I - q[2]\pi_2, \ \ \ \ \ \pi_2= U\pi_1U^\dagger,
\ \ \ \ \ \ \
U = \left(\begin{array}{cc} c & s \\ -s & c \end{array}\right) =
\left(\begin{array}{cc} \frac{1}{[2]} & \frac{\sqrt{[3]}}{[2]} \\
-\frac{\sqrt{[3]}}{[2]} & \frac{1}{[2]} \end{array}\right)
\ee
and the quantities of interest are now
\be
v_{n,k} = [2]^n \cdot \frac{1}{[N]}{\rm Tr}_q \Big(\pi_1\pi_2\Big)^k =
[2]^n \frac{ D_{111} + c^{2k}D_{21}}{[N]}
= [2]^{n-1}[N-1] \, \frac{[N-2] + [2]^{1-2k}[N+1]}{[3]}
\ee
where $c=[2]^{-1}$ is the non-trivial diagonal element of the mixing matrix
in the $[21]$ sector, it appears because
\be
\underbrace{\left(\begin{array}{cc}  0 & 0 \\ 0 & 1 \end{array}\right)}_{\pi_1}\
\underbrace{\left(\begin{array}{cc}  c & s \\ -s & c \end{array}\right)
\left(\begin{array}{cc}  0 & 0 \\ 0 & 1 \end{array}\right)
\left(\begin{array}{cc}  c & -s \\ s & c \end{array}\right)}_{\pi_2} =
\left(\begin{array}{cc}  0 & 0 \\ 0 & 1 \end{array}\right)
\left(\begin{array}{cc}  s^2 & cs \\ cs & c^2 \end{array}\right) =
\underbrace{\left(\begin{array}{cc}  0 & 0 \\ cs & c^2 \end{array}\right)}_{\pi_1\pi_2}
\ee
The crucial next feature was the fact that
\be
\frac{[N-2] + [2]^{1-2k}[N+1]}{[3]} =
[2]^{1-2k}\left([N-1] + [2][N-2]\sum_{i=0}^{k-2}\ [2]^{i}\right)
\ee
-- this provided the formulas (\ref{dim3strands}) that we used in the study
of the 3-strand examples:
\be
v_{n,k} = [2]^{n-2k}[N-1]\left([N-1] + [2][N-2]\sum_{i=0}^{k-2}\ [2]^{i}\right)
\ee
One can formulate the phenomenon as follows: after elimination of denominator
($[3]$) dimensions become bilinear combinations of $[N-1]$ and $[N-2]$
(in particular, $[N+1]$ disappeared).
As usual, dependence on the total number $n$ of white points is only in the overall
factor $[2]^n$.

\bigskip

\underline{4-strand case:}

This time there are three ${\cal R}$-matrices and thus two types of mixing matrices
\be
{\cal R}_2 = U_{21}{\cal R}_1U_{21}^\dagger, \ \ \ \ \ \ {\cal R}_3 = U_{31}{\cal R}_1U_{31}^\dagger,
\ee
explicitly evaluated in \cite{MMMkn1}.
Now ${\cal R}_i \sim I - q[2]\pi_i$
where the 4-strand avatar of the projection operator $P_{11}$ is now
\be
\pi_1 = P_{[31]''} + P_{[22]'} +P_{[211]'}+P_{[211]''}+P_{[1111]},
\ \ \ \ \ \pi_2 = U_{21}\pi_1U_{21}^\dagger, \ \ \ \ \pi_3 = U_{31}\pi_1U_{31}^\dagger
\ee
According to \cite{MMMkn1} the mixing matrices are:

\bigskip

\centerline{
$
U_{21} = \left(\begin{array}{c|ccc|cc|ccc|cccccc}
1 &&&&&&&&&\\
\hline
&1 &&&&&&&&\\
&& c_2 & s_2&&&&& \\
&& -s_2 & c_2&&&&& \\
\hline
&&&& c_2 & s_2&&&& \\
&&&& -s_2 & c_2&&&& \\
\hline
&&&&&&1&& \\
&&&&&&& c_2 & s_2& \\
&&&&&&& -s_2 & c_2& \\
\hline
&&&&&&&&&1
\end{array}\right)
\ \ \ \ \
V = \left(\begin{array}{c|ccc|cc|ccc|ccccc}
1 &&&&&&&&&\\
\hline
&c_3 & s_3 &&&&&&&\\
&-s_3 & c_3&&&&&&& \\
&&& 1&&&&&\\
\hline
&&&& cc_2 & ss_2&&& \\
&&&& -ss_2 & cc_2&&& \\
\hline
&&&&&& c_3 & s_3 &\\
&&&&&& -s_3 & c_3 &\\
&&&&&&&&1& \\
\hline
&&&&&&&&&1
\end{array}\right)
$}

\be
U_{31} = U_{21}VU_{21},
\ee
where the five blocks correspond to the sectors $[4]$, $[31]$, $[22]$, $[211]$ and $[1111]$
and the size of the block is the multiplicity with which this representation appears
in the product $[1]^{\otimes 4}$.
The cosines and sines of the mixing angles are
\be
c_k = \frac{1}{[k]}, \ \ \ s_k = \sqrt{1-c_k^2} = \frac{\sqrt{[k+1][k-1]}}{[k]}, \ \  \ \ \
cc_2 = 2c_2^2-1 = -\frac{[4]}{[2]^3},\ \ \
ss_2 = -\sqrt{1-cc_2^2} = -\frac{2\sqrt{[3]}}{[2]^2}
\ee

\bigskip

Since $\pi_i^2 = \pi_i$, the dimensions $\ v_{n|\vec k}\ $ do not depend on powers of  projectors
-- and thus are the same for many of hypercube vertices.
They rather depend on the
words, made from the letters $k_i=1,2,3$ without repetitions:
\be
v^{(n)}_{\vec k} = \frac{[2]^n}{[N]}{\rm Tr}_q \Big(\prod_i \pi_{k_i}\Big)
= [2]^n\Tr \Big(D \ \prod_i \pi_{k_i}\Big)
\ee
where quantum trace is substituted by the ordinary trace with additionally inserted diagonal
matrix of reduced quantum dimensions
\be
D = {\rm diag}\left(\frac{D_4}{[N]},\ \frac{D_{31}}{[N]}(1,1,1),\ \frac{D_{22}}{[N]}(1,1),
\frac{D_{211}}{[N]}(1,1,1),\ \frac{D_{1111}}{[N]}\right)
\ee
As usual, the dependence on the number $n$ of white points is in the simple
common factor $[2]^n$.

The lowest examples are:
\be
v_{0} = [N]\cdot v_{00} = [N]^3, \nn \\ \nn \\
v_{n} = [N] \cdot v_{n0} = [2]^{n-1}[N]^2[N-1], \nn \\ \nn \\
v^{(n)}_{(12)^k} = v^{(n)}_{(23)^k}   = [N]\cdot v_{n,k} =
[2]^{n-2k}[N][N-1]\left([N-1] + [2][N-2]\sum_{i=0}^{k-2}\ [2]^{i}\right), \nn \\
v^{(n)}_{(13)^k} = v^{(n)}_{13}=  \frac{[2]^n}{[N]}D_{11}^2 = [2]^{n-2}[N][N-1]^2, \nn \\ \nn \\
v^{(n)}_{123} = \frac{[2]^{n-2}}{[N]}\Big([2]^2D_{1111}+q^2D_{22} +2q^2D_{211}\Big)
= [2]^{n-3}[N-1]^3, \nn \\
v^{(n)}_{1231} = v^{(n)}_{1232} =v^{(n)}_{123} =[2]^{n-3}[N-1]^3,\nn\\
v^{(n)}_{(12)^23} = [2]^{n-5}[N-1]^2\Big([N-1]+[2][N-2]\Big),\nn\\
v^{(n)}_{(123)^2} = [2]^{n-6}[N-1]\Big([2][N-1]^2 + 2[N-1][N-2]+[N-2][N-3]\Big), \nn \\
v^{(n)}_{(123)^21} = v^{(n)}_{(123)^23} = [2]^{n-6}[N-1]\Big([2][N-1]^2+2[N-1][N-2]+[N-2][N-3]\Big), \nn \\
v^{(n)}_{(123)^22} = [2]^{n-7}[N-1]\Big( [N-1]^2+3[2][N-1][N-2]+[2][N-2][N-3]\Big),\nn \\
\ldots
\ee
For one particular series of dimensions we have:
\be
v^{(n)}_{(123)^3} = [2]^{n-8}[N-1]\Big([2][N-1]^2 + 6[N-1][N-2]+(q^2+5+q^{-2})[N-2][N-3]\Big), \nn \\
v^{(n)}_{(123)^4} = [2]^{n-10}[N-1]\Big([2][N-1]^2 + 14[N-1][N-2]
+(q^4+7q^2+19+7q^{-2}+q^{-4})[N-2][N-3]\Big), \nn \\
\ldots \nn \\
\  \ \ \ \ \ \ \ \
v^{(n)}_{(123)^k} = [2]^{n-2k-2}[N-1]\Big([2][N-1]^2 + 2(2^{k-1}-1)[N-1][N-2]+\xi_k[N-2][N-3]\Big)
\ee
where the first several coefficients $\xi_k$ are:
\be
k=1 & 0 \nn\\
k=2 & 1 \nn\\
k=3 & [3]+4 \nn \\
k=4 & [5]+6[3]+12 \nn \\
k=5 & [7] + 8[5] + 25[3]+33 \nn \\
k=6 & [9] +10[7] + 42[5] + 91[3] + 89 \nn \\
k=7 & [11] + 12[9] + 63[7]+185[5]+313[3]+243
\nn \\
\ldots \nn \\
\ee

We see that like $v^{(n)}_{n,k}$ in the 3-strand case were bilinear combinations
of $[N-1]$ and $[N-2]$, these 4-strand space dimensions are
trilinear combinations of $[N-1]$, $[N-2]$, $[N-3]$.
Some factors $[N]$ also appear when {\it not} all the four strands are
involved into intersections -- then we substitute them by
$[N] = q[N-1]+q^{1-N}$ -- thus also some terms with $q^{-N}$ factors appear.
Note, that there is no need to reduce them further to $[N-2]$ or $[N-3]$.

\subsection{Separation of {\it partly-sterile} contributions}

Primary polynomial in the 4-strand case is decomposed as
\be
\left( \frac{q^{n_\bullet(N-1)}}{(q^NT)^{n_\circ}}\right)^{-1}\P^{\cal L}
= q^{3-3N} + A[N-1]^3 + \frac{[N-1][N-2]}{[2]}\Big(B[N-1]  + C[N-3]\Big)
\label{decomp1}
\ee
with $N$-independent polynomials $A,B,C$, with integer coefficients.
However, these coefficients are not obligatory positive and there is also
a factor $[2]$ in {\it denominator} in the last two terms.
The same phenomenon was already present in the 3-strand case --
and this what the origin of the $q^{-N}$ lines in our tables.
As we know from that experience, contributions to such lines
arise from the factors $[N]$, which appear when some of the strands
are "sterile" -- not participate in any intersection.
This means that such contributions (hypercube vertices) should be handled
separately -- and only the rest, i.e. {\it really $m=4$-strand}
contributions should be decomposed like in (\ref{decomp1}).

\subsection{Unknot as a torus $[4,1]$}

For the KR complex
\be
\boxed{
v_0 \longrightarrow \underbrace{3[N]v_{10}}_{v^{(1)}_{1}+v^{(1)}_{2}+v^{(1)}_{3}}
\longrightarrow \underbrace{3[N]v_{21}}_{v^{(2)}_{12}+v^{(2)}_{23}+v^{(2)}_{13}} }
\longrightarrow v^{(3)}_{123}
\ee
we have
\be
q^{3-3N}\P^{[4,1]}_r = [N]^3 + 3[N]^2[N-1]\cdot(qT) + 3[N][N-1]^2\cdot(qT)^2
+ [N-1]^3\cdot (qT)^3 = \nn \\
= q^{3-3N} + [N-1]\Big(q^3[N-1]^2(1+T)^3 + 3q^{3-N}[N-1](1+T)^2 + 3q^{3-2N}(1+T)\Big)
\label{PPt41}
\ee
where we substituted $[N]$ by $q[N-1]+q^{1-N}$.
Note that this was done in all the three {\it partly-sterile} contributions,
which we put in the box (in this particular case the majority of spaces are boxed,
but for complicated knots and links the majority will be non-sterile).
From this formula it is obvious that reduced and non-reduced KR polynomials are
\be
{\cal P}^{[4,1]}_r = q^{3N-3}\cdot q^{3-3N}=1, \ \ \ \ \ \  {\cal P}^{[4,1]} = [N]
\ee
as they should be for the unknot.

\bigskip

If we did not handle the {\it partly-sterile} contributions separately,
we would get decomposition (\ref{decomp1}) with
\be
qA= 3+3q^2+q^4+3q^2[2]^2T +3q^3[2]T^2+  q^4T^3 =(T+1)(q^4T^2+2q^4T+3q^2T+q^4+3q^2+3)   , \nn \\
qB=-3\Big(2q^2[2]^2+(2+3q^2+2q^4)T+q^3[2]T^2\Big) = -3(T+1)(q^3[2]T+[2]^2) , \nn \\
qC=3q^2(T+1)
\label{ABCt41}
\ee
These formulas are, of course, consistent with (\ref{PPt41}); they are obtained from it
by the substitutions
\be
q^{2-N} = [N-1]-q[N-2],\nn\\
q^{4-2N} = [N-1]^2 - \frac{q^2+2}{[2]}[N-1][N-2] +\frac{q^2}{[2]}[N-2][N-3]
\ee
However,  (\ref{ABCt41}) is of little use for construction of KR polynomials
-- it is (\ref{PPt41}) that should be used.

\subsection{The 2-component torus link $T[4,2]$ (the $4^2_1$ link of \cite{CM})}

\bigskip

First of all, with the 4-strand braid $[4,2]$
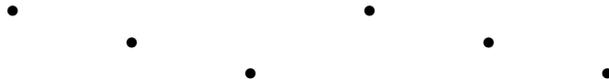
\vspace{-0.9cm}
$$ \ \ \ \ \ \ \ \ \ \ \ \ \ \ \ \ \ \ \ \ \ \ \ \ \ \ \ \ \ \ \ \ \ \ \ \
\ \ \ \ \ \ \ \ \ \ \ \ \ \ \ \ \ \ \
\begin{array}{cccccccccccccccccccccccccccccccccccccccccccccccccccccc}
\bullet &&&&&&&&&&&&\bullet \\
&&&&\bullet &&&&&&&&&&&&\bullet \\
&&&&&&&& \bullet &&&&&&&&&&&& \bullet
\end{array}
$$

\bigskip

\noindent
we naturally associate the product
\be
\Tr\Big((1+\tau\pi_1)(1+\tau\pi_2)(1+\tau\pi_3)(1+\tau\pi_1)(1+\tau\pi_2)(1+\tau\pi_3)\Big)
= \nn \\ =
\Tr\Big( \ \boxed{I +   (6\tau + 3\tau^2) \pi +   ( 8\tau^2+8\tau^3 ) \pi_1\pi_2
 +  (4\tau^2+4\tau^3+\tau^4  ) \pi_1\pi_3
+ 4\tau^4 (\pi_1\pi_2)^2} + \nn\\
+(8\tau^3+9\tau^4+2\tau^5)\pi_1\pi_2\pi_3
    +  \tau^4 \pi_1\pi_2\pi_3\pi_2
+ 4\tau^5 (\pi_1\pi_2)^2\pi_3  + \tau^6 (\pi_1\pi_2\pi_3)^2\Big)
\ee
where symbol $\ \Tr\ $ implies the possibility of cyclic permutations
in the product and identifications of, say, $\pi_1\pi_2$ with $\pi_2\pi_3$
(but not with $\pi_1\pi_3$) and we temporarily introduced a parameter $\tau = qT[2]$.
The first line at the r.h.s. contains {\it partly-sterile} items,
which are actually $0,1,2,3$-strand expressions times powers of $[N]$,
while the second line consists of the {\it non-sterile}, i.e. essentially
4-strand contributions, to be expanded into $A$, $B$, $C$ polynomials.

\bigskip

Collecting terms with the same power of $\tau$ we can read from this the KR complex:
\be
\boxed{v_0 \longrightarrow \underbrace{6[N]v_{10}}_{2v^{(1)}_{1}+2v^{(1)}_{2}+2v^{(1)}_{3}}
\longrightarrow
\underbrace{3[N]v_{20}}_{v^{(2)}_{1}+v^{(2)}_{2}+v^{(2)}_{3}}
+ \underbrace{8[N]v_{21} }_{v^{(2)}_{12}\ {\rm or}\ v^{(2)}_{23}} +
4v^{(2)}_{13} }
\longrightarrow
\boxed{8[N]v_{31} + 4v^{(3)}_{13}} +  8 v^{(3)}_{123} \longrightarrow\nn\\
\longrightarrow
\boxed{v^{(4)}_{13} + 2[N]v_{42}} + 10v^{(4)}_{123}  + 2v^{(4)}_{1232}
\longrightarrow
2v^{(5)}_{123}+4v^{(5)}_{(12)^23}
\longrightarrow v^{(6)}_{123123}
\ee
and the primary polynomial
\be
q^{6-6N}\P^{[4,2]}_r = \boxed{[N]^3\ +\ 6[N]^2[N-1](qT)\ +\ [N]\Big(3[2][N][N-1]+(8+4)[N-1]^2\Big)(qT)^2}
+ \nn \\ +\left(\,\boxed{(8+4)[2][N] [N-1]^2  } + 8[N-1]^3\right)(qT)^3
+ \nn \\
+ \left(\,\boxed{[N]\Big([2]^2[N-1]^2+2[N-1]\big([N-1]+[2][N-2]\big)\Big)} + 12[2][N-1]^3  \right)(qT)^4+\nn\\
+ \Big(2[2]^2[N-1]^3 + 4[N-1]^2\big([N-1]+[2][N-2]\big)\Big)(qT)^5 + \nn \\
+[N-1]\Big([2][N-1]^2+2[N-1][N-2]+[N-2][N-3]\Big)(qT)^6
\label{pri42}
\ee
Now, substituting $[N] = q[N-1]+q^{1-N}$ and re-expanding into $A,B,C$ polynomials plus $q^{-N}$
terms, we get:
\be
q^{6-6N}\P^{[4,2]}_r = q^{3-3N} +
q^{-2N}[N-1]\Big(3q^3 + 6q^3T+3q^4[2]T^2 \Big)
+ q^{-N}[N-1][N-2]\cdot 2q^5[2]T^4  + \nn \\
+q^{-N}[N-1]^2\Big(3q^3 +12q^3T+6q^4[2]T^2 + 12q^3T^2+12q^4[2]T^3 +q^5([2]^2+2)T^4\Big) + \nn \\
+[N-1]^3\Big(3q^3+ 6q^3T +3q^4[2]T^2+ 12q^3T^2+12q^4[2]T^3+8q^3T^3 +\nn\\
+ q^5([2]^2+2)T^4 +12q^4[2]T^4 +q^5(2[2]^2+4)T^5+[2]q^6T^6\Big)
+\nn \\
+[N-1]^2[N-2]\Big(2q^5[2]T^4 + 8q^5[2]T^5 +2q^6T^6\Big)
+[N-1][N-2][N-3]\Big(q^6T^6\Big)
\ee
From this we can now read our usual tables:
\be
q^{6-6N}\P^{[4,2]}_r = q^{3-3N} +[N-1]\cdot
\ \ \ \ \ \ \ \ \ \ \ \ \ \ \ \ \ \ \ \ \ \ \ \ \ \ \ \ \ \ \ \ \ \
\ \ \ \ \ \ \ \ \ \ \ \ \ \ \ \ \ \ \ \ \ \ \ \ \ \ \ \ \ \ \ \ \ \
\ \ \ \ \ \ \ \ \ \ \ \ \ \ \ \ \ \ \ \ \ \ \ \ \ \ \ \ \ \ \ \ \ \
 \\
\left|\begin{array}{c|cccccccc}
\hline&&&&&&&&\\
&1 & (qT)& (qT)^2 & (qT)^3 & (qT)^4 & (qT)^5 & (qT)^6 \\
&&&&&&&&\\
\hline&&&&&&&&\\
\l[N-1]^2\times & 3q^3 & 6q^2 & 3q^2[2]+12q& 12q[2]+8 & q([2]^2+2)+ 12[2] & 2[2]^2+4 & [2] \\
&&&\\
\l[N-1][N-2]\times &&&&&2q[2]& 8[2]& 2 \\
&&&\\
\l[N-2][N-3]\times &&&&&&& 1 \\
&&&\\
q^{-N}[N-1]\times & 3q^3 & 12q^2 & 6q^2[2] + 12q & 12q[2] & q([2]^2+2) &&& \\
&&&\\
q^{-N}[N-2]\times & &&&&2q[2] \\
&&&\\
q^{-2N}\ \times &3q^3& 6q^2&3q^2[2]\\
&&&&&&&&\\ \hline
\end{array}\right|
\nn\ee

\bigskip

\noindent
Now we could follow our standard procedure,
but instead we prefer to abbreviate and formalize it.

\section{Concise formulation of the procedure and more complicated examples
\label{cofo}}

\setcounter{equation}{0}

We are now ready to formulate the shortened sequence of steps for building
the KR polynomials.
Within our mnemonic approach it does not actually require
explicit construction of the hypercube (which would be necessary for explicit
construction of morphisms).
We can directly start from the spaces in KR {\it complex}
(direct sums of spaces at the given distance from initial vertex)
and formulate mnemonic rules for construction of differentials.
Surprisingly or not, the rules appear very simple and arbitrariness,
though persists, is satisfactorily low.

As one could observe in the previous examples, complexity of our construction
is increased as we increase the number $m$ of strands in quite a systematic way.
For $m=2$ the entire dependence of $\P$ on $N$
was concentrated in two structures -- $q^{-N}$ and $[N-1]$: they
arose in the primary polynomials $\P$
with the coefficients, which are positive polynomials of $q$ and $T$,
but do not depend on $N$.
For $m=3$ there were three structures instead of two,
$q^{-2N}$, $q^{-N}[N-1]$ and $[N-1][N-2]$;
for $m=4$ there were four of them,
$q^{-3N}$, $q^{-2N}[N-1]$, $q^{-N}[N-1][N-2]$ and $[N-1][N-2][N-3]$;
generalization to higher $m$ is obvious.
We emphasize that this rule depends only on the number of strands $m$ --
in above sense
the $N$-dependence for complicated $m$-strand knots and links is just the same
as for the simple (say, torus) ones.
This decomposition of $\P$ resembles the conjecture
of \cite{DMMSS} and \cite{MMMkn1} that superpolynomials for the $m$-strand braids
can be decomposed into linear combinations of appropriate  Mac Donald characters.
However, the basis that we use here is somewhat different -- and thus
different is the entire formalism.

Of course, in KR polynomials for {\it knots} only the $N$-dependent powers of $q$
can survive, while quantum numbers contain $(q-q^{-1})$ in denominators --
thus {\it all} the elements of original basis could be naturally present only
for the $m$-component links, where the "superpolynomial" (actually, "superseries")
is allowed to have $(q-q^{-1})^{m-1}$ in denominator.
Thus the next steps of our procedure form recursive transition to another -- monomial -- basis.
First, for knots it is guaranteed that the structure $[N-1][N-2]\ldots[N+1-m]$ is cohomologically
equivalent to a combination of $q^{-N}[N-1]\ldots[N+2-m]$ and $q^{N}[N-1]\ldots[N+2-m]$.
Next, these two are substituted by combinations of
$\big(q^{-2N}\oplus 1 \oplus q^{2N}\big)[N-1]\ldots [N+3-m]$ and all the way down to
$q^{-(m-1)N}\oplus \ldots \oplus q^{(m-1)N}$.
For $l$-component links the $l$ lowest elements of the original basis can also survive --
but with considerably simplified coefficients.

Technically, the procedure is remarkably simple.
Decomposition of primary polynomial $\P$ in our new basis is
unique and straightforward: one just iteratively evaluate $\P$ at $N=1$,
$N=2$, \ldots, $N=m$ and read the coefficient functions -- and the "miracle"
is that after that the answers for $N>m$ are provided automatically.
This knowledge and the use of such distinguished decomposition
simplifies the entire procedure and in fact by itself it fixes many of the ambiguities
which we encountered in above examples -- there we used more ambiguous decompositions
with {\it a priori} wider basises (for example, for $m=3$ our original basis,
which was inspired by dimensions $v_{m,k}$ of the graded vector spaces, included
separately $[N-2][N-1]$,\ $[N-1]^2$,\ $q^{-N}[N-1]$,\ $q^{-N}[N-2]$,\ $q^{-2N}$,
though only three of these structures are linearly independent).

In result this  procedure suffers from relatively small ambiguities, which can often be fixed
by minor additional knowledge: for example, one can ask for certain regularity
within a given family of knots/links (like, torus, or twist -- in the spirit of the
evolution method of \cite{DMMSS} and \cite{evo}),
or compare with the known from  \cite{katlas}
Khovanov polynomials at $N=2$.
One should only remember that Khovanov polynomials
and, more generally, KR polynomials at $N<m$ are often further reduced as compared
to generic answer -- but this is very easy to take into account, as we already saw in
sec.\ref{torus34}, and this further reduction is performed within the chosen resolution
of ambiguity, so it can still be used to fix it, at least partly.

In the following subsection we try to use our experience from the previous sections to
better formulate the building procedure for KR polynomial for a link diagram ${\cal L}$,
which is an $m$-strand braid.
In the next subsections we apply it first to
a few examples, that we already examined in a less systematic approach,
and then proceed to new more complicated examples.

\subsection{The algorithm (still ambiguous, but not too much)}

\subsection{The algorithm}

\noindent

\phantom{.}

1) Calculate the primary polynomial $\P^{\cal L} = [N]\cdot\P^{\cal L}_r$
by the standard technique of HOMFLY calculus,
exhaustively described in \cite{MMMkn1,Anopaths,AnoMcabling}
and already illustrated by numerous examples in the present text.

\bigskip

2) Decompose these primary polynomials as functions of $N$:
\be
\boxed{
\P_r^{\cal L} = \frac{q^{n_\bullet(N-1)}}{(q^NT)^{n_\circ}}\
\sum_{k=0}^{m-1} A_k^{\cal L}(q,T)\cdot \left(q^{(k+1-m)N}\prod_{i=1}^k [N-i]\right), }  \nn \\
\P^{\cal L} = [N]\cdot \P_r = \frac{q^{n_\bullet(N-1)}}{(q^NT)^{n_\circ}}\
\sum_{k=0}^{m-1} A_k^{\cal L}(q,T)\cdot \left(q^{(k+1-m)N}\prod_{i=0}^k [N-i]\right)
\label{expaPP}
\ee
where $m$ is the number of strands, and emerging functions $A_0,\ldots,A_{m-1}$ no longer depend on $N$ -- only on $q$ and $T$.

\bigskip

3) At $T=-1$ the functions $A_k$ for knots  should be proportional to $\{q\}^k = (q-q^{-1})^k$,
while for $l$-component links restricted are the functions with $k\geq l$:\
$A_{k\geq l}\sim \{q\}^{k+1-l}$.
As functions of $T$ they should be cohomologically equivalent to multiples of
$(1+q^2T)^{k+1-l}$, i.e.
\be
A_k = (1+q^2T)^{k+1-l}\bar A_k \ +\ (1+T)\times {\rm positive\ polynomial}
\label{divis}
\ee

\bigskip

4) Each of {\it non-trivial} functions $A_1,\ldots,A_{l-1}$ is cohomologically equivalent to a
combination of just a few "diagonal" functions:
\be
A_k(q,T) = \sum_{\{nu\}} q^{\mu_k}T^{\nu_k} \Big(1+q^2T+q^4T^2 + \ldots + (q^2T)^{s_k}\Big)
\ + (1+T)\times {\rm positive\ polynomial}
\label{diags}
\ee
The choice of $\nu$ defines $\mu$ and $s$.

\bigskip

5) In most cases there is some finite freedom in selecting $\nu$ -- and this is what
is going to be fixed by the study of morphisms, beyond this paper.
However, for relatively simple knots and links the choice is not too big
and, most important, it does {\it not} depend on $N$
-- thus it can be easily fixed by comparison with existing knowledge,
including the Jones-Khovanov polynomials for $N=2$, available at \cite{katlas}.

\bigskip

6) Diagonal functions are further reduced with the help of identities like
\be
\l[N-1](1+q^2T) = q^{2-N} + q^NT  + q[N-2](1+T)\ \sim\  q^{2-N} + q^NT
\label{redur}
\ee
and
\be
\l[N][N-1](1+q^2T) = (q^{1-N}+q^{N+1}T)[N-1]+ q[N-1]^2(1+T)\ \sim\  q^{1-N} + q^{N+1}T
\label{reduur}
\ee
applied to reduced and unreduced polynomials respectively.
When $s_k$ is odd, this is a uniquely defined procedure.
For $s_k$ even one term remains unpaired and there is an additional freedom
to choose which -- say, at one or another end of the diagonal.

To handle $A_k$ with the property (\ref{divis}), one may need to apply
(\ref{redur}) several times, e.g.
\be
\l[N-1][N-2](1+q^2T)^2 = \Big(q^{2-N} + q^NT  + q[N-2](1+T)\Big)
\Big(q^{3-N} + q^{N-1}T  + q[N-3](1+T)\Big) \sim \nn\\
\sim \Big(q^{2-N}+q^NT\Big)\Big(q^{3-N}+q^{N-1}T\Big)=q^{5-2N} + q^2[2]T +q^{2N-1}T^2
\label{redur2}
\ee
and
\be
\l[N-1][N-2][N-3](1+q^2T)^3 \sim
\Big(q^{2-N}+q^NT\Big)\Big(q^{3-N}+q^{N-1}T\Big)\Big(q^{4-N}+q^{N-2}T\Big)= \nn \\
= q^{9-3N} + q^{5-N}[3]T+q^{N+1}[3]T^2+q^{3N-3}T^3
\label{redur3}
\ee
In unreduced case instead of (\ref{reduur}) we have:
\be
\l[N][N-1][N-2](1+q^2T)^2 \sim \Big(q^{1-N}-q^{N+1}T\Big)\Big(q^{3-N}+q^{N-1}T)[N-1]=\nn\\
=\Big(q^{4-2N} + (q^4+1)T +q^{2N}T^2\Big)[N-1]
\label{reduur2}
\ee
and
\be
\l[N][N-1][N-2][N-3](1+q^2T)^3 \sim
\Big(q^{1-N}+q^{N+1}T\Big)\Big(q^{3-N}+q^{N-1}T\Big)\Big(q^{4-N}+q^{N-2}T\Big)[N-1]= \nn \\
= \Big(q^{8-3N} + q^{-N}(q^8+q^4+q^2)T+q^{N}(q^4+q^2+q^{-2})T^2+q^{3N-2}T^3\Big)[N-1]
\label{reduur3}
\ee
Note that reduction is only to $[N-1]$, though in (\ref{reduur2}) it {\it could be}
to $[N-2]$ and in (\ref{reduur3}) -- to $[N-3]$,
providing $\Big(q^{1-N}-q^{N+1}T\Big)\Big(q^{2-N}+q^{N}T\Big)[N-2]$
and $\Big(q^{1-N}+q^{N+1}T\Big)\Big(q^{2-N}+q^{N}T\Big)\Big(q^{3-N}+q^{N-1}T\Big)[N-3]$
respectively.
This, however, would cause discrepancy from the answers of \cite{CM} and would thus
correspond to some other choice of morphisms.

Also, additional consideration is required for low values of $N<m$,
when application of the rules (\ref{redur}) and (\ref{reduur}) can complicate rather than
simplify the polynomial -- then such substitutions should {\it not} be done, and the answer
gets different from the one for generic $N$ (i.e. is not obligatory obtained just by substituting
the low value of $N$ into the general formula),
see sec.\ref{torus34} above for a representative example.

\bigskip

7) The remaining cohomologically non-trivial terms form the KR polynomial.

\bigskip

One can easily recognize these steps in our above examples -- and we saw
that they indeed can easily reproduce the known answers.
In particular, this procedure does not necessarily eliminate {\it everything},
proportional to $T+1$ -- and here the role of diagonal anzatz (\ref{diags}) seems important.
Clearly this diagonal structure reflects some basic property of morphisms
and it will be in the focus of further studies of this story.

\bigskip

Now we provide a few more complicated examples, which are technically difficult to handle
without the concise formulation of the present section.
Drastic simplification is provided by the properties (\ref{divis}) and (\ref{diags})
of expansion (\ref{expaPP}), which we now use to the full extent.
In practice,  we begin every example below from the list of possible
diagonal functions for $A_0, \ldots, A_{m-1}$, deduced from the knowledge of $\P$.
We no longer construct $\P$ from the space dimensions $v$, as we did so far,
but simply apply the technique of \cite{MMMkn1,Anopaths,AnoMcabling}
(which is now available as computer programs) --
this is enough for the purposes of the present paper.
Still, explicit knowledge of spaces, which form the KR complex, -- which we overlook in
such approach -- remains important for future discussion of morphisms.
From this point of view the more detailed considerations in the previous sections
of the present paper have their own advantages.

\subsection{The 2-component torus link $[4,2]$ revisited}

In this case the concise procedure implies that we {\it begin} from eq.(\ref{pri42}).
Still, making the story a little slower, we calculate the trace of a product of
$10\times 10$ matrices,
\be
\P_r^{[4,2]} = q^{6N-6}\Tr_{10\times 10} \Big\{D(1+\tau\pi_1)(1+\tau\pi_2)(1+\tau\pi_3)
(1+\tau\pi_1)(1+\tau\pi_2)(1+\tau\pi_3)\Big\}
\ee
with $\tau = [2]qT$ and expand it according to (\ref{expaPP}):
\be
q^{6-6N}\P^{[4,2]}_r = A_0q^{-3N} + A_1\cdot q^{-2N}[N-1] + A_2\cdot q^{-N}[N-1][N-2]
+ A_3\cdot [N-1][N-2][N-3]
\ee
This is an easy computation, because $A_0$ is just the value of the trace at $N=1$,
and after that $A_1,A_2,A_3$ are extracted from its values at $N=2,3,4$.
One easily checks in this way that $A_2$ and $A_3$ are proportional to $T+1$,
and can be neglected in the further analysis.
$A_0$ is just a monomial (as usual),
\be
A_0^{[4,2]}=q^3
\ee
The only non-trivial piece is $A_1$. We represent it as a table
(keeping in mind more complicated examples, we change positions of lines and columns).
\be
A_1^{[4,2]} =
\left|\begin{array}{c|ccccccccccccc}
\hline &&\\
&  q^3 & q^5 & q^7 & q^9 & q^{11} \\
&&\\ \hline && \\
1 &  3 & 3 & 1&& \\
T &  6 & 12& 6&&\\
T^2 &3 & 21& 21&3& \\
T^3 && 12 & 32&12&\\
T^4 && 1& 17&17&1 \\
T^5 &&   &2 &8& 2 \\
T^6 &&   & &1&1\\
&&\\ \hline && \\
& (0) &(1)&(1)&(1)&(0) \\
&&\\ \hline
\end{array}\right|
\ \sim \
\left|\begin{array}{c|ccccccccccccc}
\hline &&\\
&  q^3 & q^5 & q^7 & q^9 & q^{11} \\
&&\\ \hline && \\
1 &   & \times & && \\
T &   & & \times&&\\
T^2 & & \otimes& &\times& \\
T^3 &&  & \otimes&&\\
T^4 && \times& &\boxed{\otimes}& \\
T^5 &&    &\times&&  \\
T^6 &&    &&\times& \\
&&\\ \hline && \\
& (0) &(1)&(1)&(1)&(0) \\
&&\\ \hline
\end{array}\right|
\ee
The first table just means that $A_1^{[4,2]} = (3q^3+3q^5+q^7)+T(6q^3+12q^6+6q^7)+\ldots$
The second table shows only the items from the first one, which can contribute to the
cohomology.
To construct the second table from the first one, we need to calculate
alternated sums (Euler characteristics) along columns: they are given in the last line
and there are just three non-vanishing entries, all equal to one.
Thus, according to the rule (\ref{diags}), for the second table
we should pick up one of the diagonals
with three entries, marked by crosses.
Clearly there are three possible choices,
the right one turns out to be the middle diagonal, with crosses circled.
Moreover, since there are three entries on this selected diagonal,
only two of them can be further reduced with the help of (\ref{redur}) and (\ref{reduur}), and
we should select the one, which remains intact -- the right choice is boxed.
These selections provide
\be
\P^{[4,2]}_r = q^{6N-6}\Big(\underbrace{q^{3-3N}}_{q^{-3N}A_0} +
\underbrace{q^{-2N}\cdot q^5T^2\Big(q^{2-N} + q^NT\Big)}_{q^5T^2\cdot
q^{-2N}\Big([N-1](1+q^2T) - q[N-2](T+1)\Big)}+
q^9T^4 q^{-2N}[N-1]\Big)= \nn \\
=q^{3-3N}\Big(1 + q^4T^2+q^{2N+2}T^3+q^{N+6}[N-1]T^4\Big), \nn \\ \nn \\
\P^{[4,2]} =
q^{6N-6}\Big(\underbrace{q^{3-3N}}_{q^{-3N}A_0}[N] +
\underbrace{q^{-2N}\cdot q^5[N-1]T^2\Big(q^{1-N} + q^{N+1}T\Big)}_{q^5T^2\cdot
q^{-2N}\Big([N][N-1](1+q^2T) - q[N-1]^2(T+1)\Big)}+
q^9T^4 q^{-2N}[N][N-1]\Big)= \nn \\
= q^{3-3N}\Big([N] + q^3T^2(1+q^{2N}T)[N-1]+q^{N+6}[N][N-1]T^4\Big)
\ee
--in accordance with \cite{CM} and our 2- and 3-strand calculations in
(\ref{KRo24}) and (\ref{KR421}), (\ref{KR421v2r}), (\ref{KR421v2ur})
respectively. This confirms topological invariance of our procedure.

\subsection{The 2-strand knots and links, once again}

As a further demonstration of concise procedure, we once again
consider the 2-strand case.
Here
\be
q^{-n(N-1)}\P^{[2,n]}_r =  \Tr_{2\times 2} \Big\{D\cdot(I+
\tau \pi)^n\Big\}
= \underbrace{q^{1-N}}_{A_0q^{-N}} +\underbrace{\left(q+
\sum_{k=1}^n [2]^{k-1}C^k_n(qT)^k\right)}_{A_1}[N-1]
\ee
Then
\be
A_1^{[2,n]} = \left|\begin{array}{c|ccccccccccccc}
\hline &&\\
& q & q^3 & q^5 & q^7 & \ldots & q^{2i-1} & \ldots & q^{2k-1} & \ldots & q^{2n-1} \\
&&\\ \hline && \\
1 & 1    \\
T & n    \\
T^2 & C^2_n & C^2_n   \\
T^3 & C^3_n & 2C^3_n & C^3_n  \\
T^4 & C^4_n & 3 C^4_n & 3C^4_n & C^4_n \\
\ldots & \\
T^k & C^k_n  & (k-1)C^k_n & C^2_{k-1}C^k_n & C^3_{k-1}C^k_n & \ldots & C^{i-1}_{k-1}C^k_n &\ldots& 1 \\
\ldots & \\
T^n &  1 & n-1& C^2_{n-1} & C^3_{n-1} & \ldots & C^{i-1}_{n-1} & \ldots & C^{k-1}_{n-1} & \ldots &  1   \\
&&\\ \hline && \\
& (0) & (1) & (1) & (1) & \ldots &(1) & \ldots & (1) \ldots & (1)\\
&&\\ \hline
\end{array}\right|
\ee
In this case there is just one suitable diagonal, and the only arbitrariness is the choice of the
boxed item for even $n$ (for odd $n$ the box should be removed):
\be
A_1^{[2,n]} \sim \left|\begin{array}{c|ccccccccccccc}
\hline &&\\
& q & q^3 & q^5 & q^7 &   \ldots & q^{2k-1} & \ldots & q^{2n-1} \\
&&\\ \hline && \\
1 & \\
T & \\
T^2 &&\otimes \\
T^3 &&& \otimes \\
T^4 &&&& \otimes \\
\ldots &\\
T^k &&&&&&\otimes \\
\ldots &\\
T^n &&&&& &&&\boxed{\otimes} \\
&&\\ \hline && \\
& (0) & (1) & (1) & (1) & \ldots &(1) & \ldots & (1)  \\
&&\\ \hline
\end{array}\right|
\ee
Application of (\ref{redur}) and (\ref{reduur}) converts this into the standard answers:
\be
{\cal P}^{[2,n]}_r = q^{(n-1)(N-1)}\left( 1 +q^4T^2 + q^{2N+2}T^3 + \ldots + q^{4i}T^{2i} +
q^{2N+4i+2}T^{2i+1} + \ldots + \underbrace{q^{N+2n-2}[N-1]T^N}_{{\rm if}\ n\ {\rm is\ even}}\right),\nn\\
\\
{\cal P}^{[2,n]}_r = q^{(n-1)(N-1)}\left\{ [N] +
\Big(q^3T^2 + \ldots + q^{4i-i}T^{2i} + \ldots\Big)\Big(1+q^{2N}T\Big)[N-1]
+ \underbrace{q^{N+2n-2}[N][N-1]T^N}_{{\rm if}\ n\ {\rm is\ even}}\right\}
\nn
\ee
The last term is present only   for even $n$, i.e. for the 2-strand links.



\subsection{$3$-strand torus knots and links
\label{3stratorus2}}

This time
\be
q^{-2n(N-1)}\P^{[3,n]}_r =  \Tr_{4\times 4} \left\{D\cdot
\Big((I+\tau \pi_1)(I+ \tau \pi_2)\Big)^n\right\}
= \underbrace{q^{2-2N}}_{q^{-2N}A_0^{[3,N]}} +
A_1^{[3,n]}\cdot q^{-N}[N-1]+A_2^{[3,n]}[N-1][N-2]
\nn
\ee
The matrices are $4\times 4$ and act as numbers on representations $[3]$ and $[111]$,
while as $2\times 2$ matrices on the space of two representations $[21]$.

To understand the general structure it is enough to look at the first few examples.

\subsubsection{The unknot ($n=1$)}

Both $A_2^{[3,1]}$ and $A_1^{[3,1]}$ are cohomologically trivial,
thus
\be
{\cal P}^{[3,1]}_r = q^{2N-2}\cdot q^{2-2N}=1,\nn \\
{\cal P}^{[3,1]} = q^{2N-2}\cdot q^{2-2N}[N]=[N]
\ee

\subsubsection{The trefoil ($n=2$)}

The function $A_2^{[3,2]} = q^3(1+T)^2(2q^2T^2+T^2+2T+1)$
is cohomologically trivial,
\be
A_1^{[3,2]}=
\left|\begin{array}{c|ccccc}
\hline && \\
& q^2 &q^4 &q^6 \\
&\\ \hline && \\
1 & 2&1\\
T& 4&4 \\
T^2&2&8&2 \\
T^3&&4&4\\
T^4&&&1\\
&&\\ \hline && \\
&(0)&(1)&(-1) \\
&&\\
\hline
\end{array}\right|
\ \sim \
\left|\begin{array}{c|ccccc}
\hline && \\
& q^2 &q^4 &q^6 \\
&\\ \hline && \\
1 & \\
T&   \\
T^2&&\otimes  \\
T^3&&&\otimes \\
T^4& \\
&&\\ \hline && \\
&(0)&(1)&(-1) \\
&&\\
\hline
\end{array}\right|
\ee
Then
$$q^4T^2(1+q^2T)\cdot q^{-N}[N-1] \ \stackrel{(\ref{redur})}{\sim}\ q^{4-N}T^2(q^{2-N}+q^NT)$$
in reduced, and
$$q^4T^2(1+q^2T)\cdot q^{-N}[N][N-1] \ \stackrel{(\ref{reduur})}{\sim}\
q^{4-N}T^2(q^{1-N}+q^{N+1}T)[N-1]$$
in unreduced case,
so that
\be
{\cal P}^{[3,2]}_r = q^{4N-4}\Big(q^{2-2N} + q^{4-N}T^2(q^{2-N}+q^NT)\Big)=
q^{2N-2}\Big(1+q^4T^2+q^{2N+2}T^3\Big),\nn \\
{\cal P}^{[3,2]}_r = q^{4N-4}\Big(q^{2-2N}[N] + q^{4-N}T^2(q^{1-N}+q^{N+1}T)[N-1]\Big)=
q^{2N-2}\Big([N]+q^3T^2(1+q^{2N}T)[N-1]\Big)
\ee

\subsubsection{A link ($n=3$)}

Since the link is $3$-component,
and $N^3 = N(N-1)(N-2)+3N(N-1)+N$, our functions $A_2$ and $A_1$
should contain at least one and three unbalanced cohomologically non-trivial terms.
Indeed,
\be
A_2^{[3,3]}=
\left|\begin{array}{c|ccccc}
\hline && \\
& q^3 &q^5 &q^7 & q^9 \\
&\\ \hline && \\
1 & 1 \\
T& 6  \\
T^2&15&6  \\
T^3&20&22&2 \\
T^4&15&30&9 \\
T^5&6&18&12   \\
T^6&1&4&5&1     \\
&&\\ \hline && \\
&(0)&(0)&(0)&(1) \\
&&\\
\hline
\end{array}\right|
\ \sim \
\left|\begin{array}{c|ccccc}
\hline && \\
& q^3 &q^5 &q^7 & q^9 \\
&\\ \hline && \\
1 &  \\
T&   \\
T^2&  \\
T^3& \\
T^4& \\
T^5&   \\
T^6&&&&\boxed{\otimes}     \\
&&\\ \hline && \\
&(0)&(0)&(0)&(1) \\
&&\\
\hline
\end{array}\right|
\ee
and
\be
A_1^{[3,3]}=
\left|\begin{array}{c|ccccc}
\hline && \\
& q^2 &q^4 &q^6 & q^8 \\
&\\ \hline && \\
1 & 2&1\\
T&  6&6\\
T^2&6&21&6 \\
T^3&2&24&24&2 \\
T^4&&9&24&9 \\
T^5&&&6&6  \\
T^6&&&&1  \\
&&\\ \hline && \\
&(0)&(1)&(0)&(2) \\
&&\\
\hline
\end{array}\right|
\ \sim \
\left|\begin{array}{c|ccccc}
\hline && \\
& q^2 &q^4 &q^6 & q^8 \\
&\\ \hline && \\
1 & \\
T&  \\
T^2& &\otimes&\\
T^3& &&\otimes \\
T^4& &&\boxed{\otimes}&\boxed{2\otimes}\\
T^5&  \\
T^6&  \\
&&\\ \hline && \\
&(0)&(1)&(0)&(2) \\
&&\\
\hline
\end{array}\right|
\ee
where we selected the items in $A_1$ at allowed
(by the values of partial Euler characteristics)
places on the main diagonal.
This is not a minimal choice, but it satisfies three additional mnemonic rules:
$(i)$ there are no gap along the diagonal; $(ii)$ there is a contribution in the $T^3$ line,
and $(iii)$ contributions with additional factors $[N-1]$ are concentrated in one power in $T$, and those with $[N-1][N-2]$ -- in another power.
Anyhow, this is the choice which leads to the right answer (and which should
be justified by the future analysis of morphisms).
The two un-boxed crosses are reduced by the rule (\ref{redur}) in reduced case:
$$q^4T^2(1+q^2T)\cdot q^{-N}[N-1] \sim q^{4-N}T^2(q^{2-N}+q^NT)$$
and by the rule (\ref{reduur}) in unreduced case:
$$q^4T^2(1+q^2T)\cdot q^{-N}[N][N-1] \sim q^{4-N}T^2(q^{1-N}+q^{N+1}T)[N-1]$$
so that
\be
{\cal P}^{[3,3]}_r =
q^{6N-6}\Big(q^{2-2N} + q^{4-N}T^2(q^{2-N}+q^NT) + (q^6T^4  + 2q^8T^4)q^{-N}[N-1]
+ q^{9}T^6[N-1][N-2]\Big) = \nn \\
= q^{4N-4}\Big(1 +  q^4T^2+q^{2N+2}T^3+ (q^{N+4}+2q^{N+6})[N-1]T^4
+q^{2N+7}[N-1][N-2]T^6\Big)\nn \\
\label{KR33n} \\
{\cal P}^{[3,3]} = q^{4N-4}\Big([N] +  q^3T^2(1+q^{2N}T)[N-1]+
(q^{N+4}+2q^{N+6})[N][N-1]T^4 +q^{2N+7}[N][N-1][N-2]T^6\Big)
\nn
\ee
in agreement with (\ref{KR33true}) and \cite{CM},
and (in reduced case) also with \cite{DGR} and \cite{DMMSS,Che}.

\subsubsection{Knot $8_{19}$ ($n=4$)}

The function
\be
A_2^{[3,4]} =
\left|\begin{array}{c|ccccccc}
\hline && \\
& q^3 &q^5 &q^7 & q^9 &q^{11} & q^{13} \\
&\\ \hline && \\
1 &  1& \\
T&   8& \\
T^2&28&12&  \\
T^3&56&64&8& \\
T^4&70&142&54&2 \\
T^5&56&168&128&16   \\
T^6&28&112& 140&48     \\
T^7&8&40& 72&48&8   \\
T^8&1&6&14&15&6&1   \\
   \\
&&\\ \hline && \\
&(0)&(0)&(0)&(1)&(-2)&(1) \\
&&\\
\hline
\end{array}\right|
\ \sim \
\left|\begin{array}{c|ccccccc}
\hline && \\
& q^3 &q^5 &q^7 & q^9 &q^{11} & q^{13} \\
&\\ \hline && \\
1 &  \\
T&   \\
T^2&  \\
T^3& \\
T^4& \\
T^5&   \\
T^6&&&&\otimes   \\
T^7&&&&&2\otimes  \\
T^8&&&&&&\otimes  \\
&&\\ \hline && \\
&(0)&(0)&(0)&(1)&(-2)&(1) \\
&&\\
\hline
\end{array}\right|
\ee
contributes
$$q^9T^6(1+q^2T)^2[N-1][N-2]\ \stackrel{(\ref{redur2})}{\sim}
q^9T^6(q^{2-N}+q^NT)(q^{3-N}+q^{N-1}T)$$
to reduced polynomial.
Similarly,

\be
A_1^{[3,4]} =
\left|\begin{array}{c|ccccc}
\hline && \\
& q^2 &q^4 &q^6 & q^8 & q^{10} \\
&\\ \hline && \\
1 & 2&  1& \\
T&  8&  8&\\
T^2&12&40&12& \\
T^3& 8&72&72&  8&  \\
T^4& 2&56&128&56&2 \\
T^5&  &16&88& 88&16 \\
T^6&  &  &20& 48&20\\
T^7&  &   & &  8&8 \\
T^8&  &   & &  & 1\\
&&\\ \hline && \\
&(0)&(1)&(0)&(0)&(-1) \\
&&\\
\hline
\end{array}\right|
\ \sim \
\left|\begin{array}{c|ccccc}
\hline && \\
& q^2 &q^4 &q^6 & q^8 & q^{10} \\
&\\ \hline && \\
1 & \\
T&  \\
T^2& &\otimes \\
T^3& &&\otimes  \\
T^4& &&\otimes&\otimes \\
T^5& &&&\otimes&\otimes \\
T^6&  \\
T^7&  \\
T^8&  \\
&\\ \hline && \\
&(0)&(1)&(0)&(0)&(-1) \\
&&\\
\hline
\end{array}\right|
\ee
contributes
$$
q^4T^2(1+q^2T^2+q^4T^2)(1+q^2T)\cdot q^{-N}[N-1]\ \stackrel{(\ref{redur})}{\sim}\
q^{4-N}T^2(1+q^2T^2+q^4T^2)\Big(q^{2-N}+q^NT\Big)
$$

Putting all together we get:
\be
{\cal P}^{[3,4]}_r =
q^{8N-8}\Big(q^{2-2N} + q^{4-N}T^2(1+q^3[2]T^2 )\Big(q^{2-N}+q^NT\Big)
+ q^9T^6(q^{2-N}+q^NT)(q^{3-N}+q^{N-1}T)\Big) = \nn \\
\boxed{=q^{6N-6}\Big(1 + q^4T^2+q^{2N+2}T^3 +q^7[2]T^4+q^{2N+5}[2]T^5+q^{12}T^6
+ q^{2N+9}[2]T^7 +q^{4N+6}T^8\Big)}\label{KRT34}
\ee
in agreement with (\ref{KRT35}) and thus with \cite{DGR} and \cite{DMMSS}.
We remind that the cases $N=1$ and $N=2$ should be considered separately:
see the paragraph before eq.(\ref{KRT35N2}).

In unreduced case we have instead
$$
\!\!\!\!\!\!\!\!\!\!{\cal P}^{[3,4]} =
q^{8N-8}\left\{q^{2-2N}[N] + \Big(q^{4-N}T^2(1+q^3[2]T^2) (q^{1-N}+q^{N+1}T )
+ q^9T^6(q^{1-N}+q^{N+1}T)(q^{3-N}+q^{N-1}T)\Big)[N-1]
\right\} =
$$
\vspace{-0.5cm}
\be
\boxed{= q^{6N-6}\left\{[N] + \Big(q^3T^2 +q^6[2]T^4 +q^{11}T^6+q^{2N+7}T^7 \Big)
(1+q^{2N}T)[N-1]\right\}}
\label{KRT34ur}
\ee
This is {\it different} from our (\ref{KRT34urhypoth}),
\be
q^{6N-6}\Big\{ [N] + q^3[N-1]T^2 +q^{2N+3}[N-1]T^3
+ [2]q^6[N-1]T^4 + [2]q^{2N+6}[N-1]\,T^5 + \nn \\
+ q^{10}[N-2]\,T^6 + [2]q^{2N+9}[N-2]\,T^7 + q^{4N+8}[N-2]\,T^8\Big\}
\label{KRT34urhypoth1}
\ee
which, we think, is {\it over}reduced -- down to $[N-2]$,
while, as we suggest in the present section \ref{cofo}, the 3-strand
knot should not be reduced more than to $[N-1]$.
Another argument is that with the present prescription the case $n=4$
nicely suites into general formulas valid for all $n$ --
in the spirit of \cite{DMMSS} and \cite{evo}, and we avoid
"accidental" cancelations, possible in this particular example.
Still, rigorous way to choose between these possibilities requires
the study of morphisms.

As already mentioned at the end of sec.\ref{torus34},
eq.(\ref{KRT34ur}) should be modified at the special value of $N=2$,
just as in reduced case.
As explained in that section, one should throw away all the terms,
coming from $A_2^{[3,4]}$, because they were originally proportional to
vanishing $[N-2]$.
The second modification, when the factor $[2]$ is eliminated instead of $[N-1]$,
is unneeded in unreduced case for 3-strand knots, because actually eliminated is
$[N]$, not $[N-1]$, and it is not smaller than $[2]$.
Thus instead of (\ref{KRT34ur}),  we get for $N=2$
\be
{\cal P}^{[3,4]}(N=2) =
q^{6N-6}\left\{[N] + \Big(q^3T^2 +q^6[2]T^4  \Big)
(1+q^{2N}T)[N-1]\right\}
\ee
and this is just the same as substitution of $N=2$ into generically wrong
(\ref{KRT34urhypoth1}).

\subsubsection{Knot $10_{124}$ ($n=5$)}

The knowledge of alternated sums in this case
requires the reduction of the two functions to be
\be
A_2^{[3,5]} \sim
\left|\begin{array}{c|ccccccccccc}
\hline && \\
& q^3 &q^5 &q^7&q^{9}&q^{11}&q^{13}&q^{15}&q^{17}   \\
&\\ \hline && \\
1 & \\
T&   \\
T^2&  \\
T^3& \\
T^4& \\
T^5 &\\
T^6& &&&\otimes\\
T^7& &&&&2\otimes\\
T^8& &&&&&2\otimes\\
T^9& &&&&&&2\otimes\\
T^{10}& &&&&&&&\otimes  \\
&&\\ \hline && \\
&(0)&(0)&(0)&(1)&(-2)&(2)&(-2)&(1) \\
&&\\
\hline
\end{array}\right|
\ \ \ \ \
A_1^{[3,5]}\sim
\left|\begin{array}{c|ccccccccc}
\hline && \\
& q^2 &q^4 &q^6&q^8&q^{10}&q^{12}   \\
&\\ \hline && \\
1 & \\
T&   \\
T^2& &\otimes \\
T^3& &&\otimes\\
T^4& &&\otimes&\otimes\\
T^5 &&&&\otimes&\otimes\\
T^6 &&&&&\otimes& \\
T^7 &&&&&&\otimes \\
T^8&\\
T^9&\\
T^{10}  \\
&&\\ \hline && \\
&(0)&(1)&(0)&(0)&(0)&(-1) \\
&&\\
\hline
\end{array}\right|
\nn
\ee
what further implies
\be
q^{10(N-1)}\Big(q^9T^6(1+q^4T^2)(1+q^2T)^2[N-1][N-2] +
q^4T^2  (1+q^3[2]T^2+q^6T^4) (1+q^2T)\cdot q^{-N}[N-1] + q^{2-2N}\Big)\nn \\
\ \stackrel{(\ref{redur})\&(\ref{redur2})}{\sim}\
q^{8N-8}\Big(1 + q^{N+2}T^2(1+q^3[2]T^2+q^6T^4)(q^{2-N}+q^NT) +
q^{2N+7}T^6(1+q^4T^2)(q^{2-N}+q^NT)(q^{3-N}+q^{N-1}T)\Big)
\nn
\ee
Therefore
\be
{\cal P}^{[3,5]}_r =q^{8N-8}\Big(1+q^4T^2+q^{2N+2}T^3+q^7[2]T^4+q^{2N+5}[2]T^5
+q^{11}[2]T^6 +\nn\\
+(2q^{2N+8}+q^{2N+10})T^7
+(q^{16}+q^{4N+6})T^8+q^{2N+13}[2]T^9+q^{4N+10}T^{10}\Big)=
\label{KRT35r}
\ee
\vspace{-0.5cm}
$$
\boxed{
=q^{8N-8}\Big\{ 1 + q^4T^2\Big(\underline{1+q^3[2]T^2+q^6T^4}\Big)(1+q^{2N-2}T)
+ q^{12}T^6\underline{\Big(1+q^4T^2\Big)} (1+q^{2N-2}T)(1+q^{2N-4}T)\Big\}
}
$$
and
\be
\!\!\!\!
\boxed{
{\cal P}^{[3,5]} =q^{8N-8}\Big\{[N]+\Big(q^3T^2\underline{(1 +q^3[2]T^2
+q^{6}T^4)}
+ q^{11}T^6\underline{(1+q^4T^2)}(1+q^{2N-4}T)\Big)(1+q^{2N}T)[N-1]\Big\}
}
\label{KRT35ur}
\ee
These formulas are in accordance with (\ref{KRT35ur}),
which is in turn consistent with \cite{DMMSS}, -- and with
(\ref{KRT35ura}).
Moreover, now we wrote them in the form, allowing generalization
to arbitrary $n$: it is clear that the only things that depend on $n$
are the two underlined functions -- literally read from diagonals
in the tables for $A_2$ and $A_1$.
Moreover, like in the first function, coming from $A_1$,
the overall coefficient $q^4T^2$ does not depend on $n$,
in the second function, coming from $A_2$, intact is the last coefficient:
it depends on $n$, but in an obvious way: $q^3(q^2T)^{2n-3}=q^{4n-3}T^{2n}$.

\subsubsection{A $3$-component link ($n=6$)}

Like for $[n,3]$, there will be one and three non-compensated
non-trivial contributions, proportional to $[N-1][N-2]$ and to $[N-1]$
respectively, but this time they will be considerably more sophisticated.
Moreover we will need to make two iterations to formulate what are the
proper reductions of $A_2$ and $A_1$.
Justification of these choices is provided by the answer from \cite{DMMSS}
-- and by the general structure of the formulas for all $n$ at once.

The structure of reduction is crucially restricted by the values of alternated
sums in columns of our tables, i.e. by decomposition of Euler characteristic
(HOMFLY polynomial):
\be
HOMFLY = q^{12N-12}\Big(E_2(q)\cdot[N-1][N-2]\ +\ E_1(q)\cdot q^{-N}[N-1]\ +\ q^{2-2N}\Big)
\ee
The coefficient $E_2(q)$, which  controls reduction of the function $A_2$,
in the case of $[3,6]$ is quite tricky:
after the term $q^{21}$, associated with the double box,
is subtracted it is not divisible by $(1-q^2)^2$.
Instead the best possible decomposition is
\be
E_2^{[3,6]} = q^9-2q^{11}+2q^{13}-q^{15}+2q^{17}-2q^{19}+q^{21} =
q^9 (1-q^2)^2 + \boxed{q^{13}(1+2q^4)(1-q^2)} + \boxed{\boxed{q^{21}}}
\ee
or even
\be
\!\!\!\!\! E_2^{[3,6]} = q^9-2q^{11}+2q^{13}-q^{15}+2q^{17}-2q^{19}+q^{21} =
q^9(1+q^4) (1-q^2)^2 + \boxed{q^{15}(1+q^4)(1-q^2)} + \boxed{\boxed{q^{21}}}
\ee
For the first choice the reduction of $A_2$ would be represented as
\be
A_2^{[3,6]} \ \stackrel{?}{\sim}\
\left|\begin{array}{c|ccccccccccc}
\hline && \\
& q^3 &q^5 &q^7&q^{9}&q^{11}&q^{13}&q^{15}&q^{17}&q^{19}&q^{21}   \\
&\\ \hline && \\
1 & \\
T&   \\
T^2&  \\
T^3& \\
T^4& \\
T^5 &\\
T^6& &&&\otimes\\
T^7& &&&&2\otimes\\
T^8& &&&&&2\boxed{\otimes}\\
T^9& &&&&&&\boxed{\otimes}\\
T^{10}& &&&&&&&\boxed{2\otimes}  \\
T^{11}& &&&&&&&&\boxed{2\otimes} \\
T^{12}& &&&&&&&&&\boxed{\boxed{\otimes}}\\
&&\\ \hline && \\
&(0)&(0)&(0)&(1)&(-2)&(2)&(-1)&(2)&(-2)&(1) \\
&&\\
\hline
\end{array}\right|
\ee
The second choice is not minimal, but instead it shifts boxes
further to the right -- what is always "good" for torus knots and links
(and perhaps even more generally, for link diagrams with all vertices black).
Whatever the reason, this choice is the one which provides the right answer
and dictates the proper form of reduced table:
\be
A_2^{[3,6]} \ {\sim}\
\left|\begin{array}{c|ccccccccccc}
\hline && \\
& q^3 &q^5 &q^7&q^{9}&q^{11}&q^{13}&q^{15}&q^{17}&q^{19}&q^{21}   \\
&\\ \hline && \\
1 & \\
T&   \\
T^2&  \\
T^3& \\
T^4& \\
T^5 &\\
T^6& &&&\otimes\\
T^7& &&&&2\otimes\\
T^8& &&&&&2{\otimes}\\
T^9& &&&&&&2{\otimes}\\
T^{10}& &&&&&&\boxed{\otimes}&(1+\boxed{2)\otimes}  \\
T^{11}& &&&&&&&\boxed{\otimes}&\boxed{2\otimes} \\
T^{12}& &&&&&&&&&\boxed{\boxed{\otimes}}\\
&&\\ \hline && \\
&(0)&(0)&(0)&(1)&(-2)&(2)&(-1)&(2)&(-2)&(1) \\
&&\\
\hline
\end{array}\right|
\ee
As a corollary, the contribution of $A_2(q,T)$ to reduced polynomial is
\be
\boxed{\boxed{q^{21}T^{12}[N-1][N-2]}} + \boxed{q^{15}T^{10}(1+2q^2)(1+q^2T)\cdot[N-1][N-2]}
+ q^9T^6(1+q^4T^2)(1+q^2T)^2[N-1][N-2]\nn \\
\sim q^{21}T^{12}[N-1][N-2]  + \boxed{q^{15}T^{10}(1+2q^2)(q^{3-N}+q^{N-1}T)[N-1]}
 + q^9T^6(1+q^4T^2)\Big(q^{2-N}+q^NT\Big)(q^{3-N}+q^{N-1}T)
\nn
\ee

\bigskip

The Euler characteristic in the $A_1$ sector is much simpler:
\be
E_1^{[3,6]}=q^4+2q^{14}
\ee
but this implies that diagonals in $A_1$ are rather complicated:
they should have no gaps, despite there is a huge gap in $E_1^{[3,6]}$,
and at the same time, modulo $3$  boxed term, the contribution  should be
{\it positively} divisible by $1+q^2T$.
An option, which keeps boxed terms as far to the right as only possible, is
\be
A_1^{[3,6]} \ \stackrel{?}{\sim}\
\left|\begin{array}{c|ccccccccc}
\hline && \\
& q^2 &q^4 &q^6&q^8&q^{10}&q^{12}&q^{14}   \\
&\\ \hline && \\
1 & \\
T&   \\
T^2& &\otimes \\
T^3& &&\otimes\\
T^4& &&\otimes&\otimes\\
T^5 &&&&\otimes&\otimes\\
T^6 &&&&&\otimes&\otimes \\
T^7 &&&&&&\otimes&\otimes \\
T^8&&&&&&&\boxed{3\otimes}\\
T^9&\\
T^{10}&  \\
T^{11} &  \\
T^{12} &  \\
&&\\ \hline && \\
&(0)&(1)&(0)&(0)&(0)&(0)&(2) \\
&&\\
\hline
\end{array}\right|
\nn
\ee
However, the right option turns to be different:
\be
A_1^{[3,6]}\sim
\left|\begin{array}{c|ccccccccc}
\hline && \\
& q^2 &q^4 &q^6&q^8&q^{10}&q^{12}&q^{14}   \\
&\\ \hline && \\
1 & \\
T&   \\
T^2& &\otimes \\
T^3& &&\otimes\\
T^4& &&\otimes&\otimes\\
T^5 &&&&\otimes&\otimes\\
T^6 &&&&&\otimes& \\
T^7 &&&&&&\otimes& \\
T^8&&&&&&\boxed{\otimes}&\boxed{2\otimes}\\
T^9&\\
T^{10}&  \\
T^{11} &  \\
T^{12} &  \\
&&\\ \hline && \\
&(0)&(1)&(0)&(0)&(0)&(0)&(2) \\
&&\\
\hline
\end{array}\right|
\nn
\ee
what contributes
\be
\boxed{(q^{12}T^8+2q^{14}T^8)\cdot q^{-N}[N-1]} +
q^4T^2(1+ q^3[2]T^2+q^6T^4) (1+q^2T)\cdot q^{-N}[N-1] \nn \\
\ \stackrel{(\ref{redur})}{\sim}\
(1+2q^2)q^{12-N}T^8[N-1]+ q^{4-N}T^2(1+ q^3[2]T^2+q^6T^4)(q^{2-N}+q^NT)
\ee

\bigskip

Combining the contributions from $A_2$ and $A_1$, we get:
\be
{\cal P}^{[3,6]}_r =
q^{12N-12}\Big\{q^{2-2N}
+ \boxed{(1+2q^2)q^{12-N}T^8[N-1]}+ q^{4-N}T^2(1+ q^3[2]T^2+q^6T^4)(q^{2-N}+q^NT)
+ \label{KR36ra} \\
\!\!\!\!\!\!\!+q^{21}T^{12}[N-1][N-2]  + \boxed{q^{15}T^{10}(1+2q^2)(q^{3-N}+q^{N-1}T)[N-1]}+\nn\\
 + q^9T^6(1+q^4T^2)\Big(q^{2-N}+q^NT\Big)(q^{3-N}+q^{N-1}T)
\Big\}
 = \nn \\ \nn\\
= q^{12N-12}\Big\{ q^{21}T^{12}[N-1][N-2]
+ (1+2q^2)\Big(q^{12-N}T^8+q^{18-N}T^{10}+q^{N+14}T^{11}\Big)[N-1]+ \nn \\
+q^{2-2N}\Big(1+q^{4}T^2+q^{2N+2}T^3+q^{7}[2]T^4+q^{2N+5}[2]T^5+q^{10}T^6+q^{2N+8}T^7+\nn \\
+ q^{12}T^6+q^{2N+9}[2]T^7+q^{4N+6}T^8
+ q^{16}T^8+q^{2N+13}[2]T^9+q^{4N+10}T^{10}\Big)\Big\}
\nn\ee
This reproduces (\ref{KR36r}) and thus the answer from \cite{DGR} and \cite{DMMSS} --
see discussion after (\ref{KR36r}).

\bigskip

In unreduced case we should just change factors $(q^{2-N}+q^NT)$ wherever they appear
to $(q^{1-N}+q^{N+1}T)[N-1]$ and introduce a factor of $[N]$ in all other terms:
\be
{\cal P}^{[3,6]} =
q^{12N-12}\Big\{q^{2-2N}[N]
+ \boxed{(1+2q^2)q^{12-N}T^8[N][N-1]}+ \phantom{q^{4-N}T^2(1+ q^3[2]T^2+q^6T^4)(q^{1-N}+q^{N+1}T)}
\label{KR36ura} \\
+q^{4-N}T^2(1+ q^3[2]T^2+q^6T^4)(q^{1-N}+q^{N+1}T)[N-1]+q^{21}T^{12}[N][N-1][N-2]  + \nn\\
+\boxed{q^{15}T^{10}(1+2q^2)(q^{3-N}+q^{N-1}T)[N][N-1]}
 + \nn\\+q^9T^6(1+q^4T^2)\Big(q^{1-N}+q^{N+1}T\Big)(q^{3-N}+q^{N-1}T)[N-1]
\Big\}
 = \nn \\ \nn\\
= q^{12N-12}\Big\{ q^{21}T^{12}[N][N-1][N-2]
+ (1+2q^2)\Big(q^{12-N}T^8+q^{18-N}T^{10}+q^{N+14}T^{11}\Big)[N][N-1]
+\nn \\+q^{2-2N}[N]
+q^{2-2N}\Big(q^{3}T^2+q^{2N+3}T^3+q^{6}[2]T^4+q^{2N+6}[2]T^5+q^{9}T^6+q^{2N+9}T^7+\nn\\
+ q^{11}T^6+(q^{2N+7}+q^{2N+11})T^7+q^{4N+7}T^8
+ q^{15}T^8+(q^{2N+11}+q^{2N+15})T^9+q^{4N+11}T^{10}\Big)[N-1]\Big\}
\nn\ee
This reproduces (\ref{KR36}).
The last two lines can also be rearranged so that conversion of the factor $[N-1]$ into
$[N]$ at $T=-1$ gets obvious:
$$
{\cal P}^{[3,6]}= q^{12N-12}\Big\{ q^{21}T^{12}[N][N-1][N-2]
+ (1+2q^2)\Big(q^{12-N}T^8+q^{18-N}T^{10}+q^{N+14}T^{11}\Big)[N][N-1]
+q^{2-2N}[N] +
$$
\vspace{-0.5cm}
\be
q^{2-2N}\Big(q^3T^2+q^6[2]T^5+q^{10}[2]T^6+q^{2N+7}T^7+q^{15}T^8+q^{2N+11}T^9\Big)(1+q^{2N}T)[N-1]
\ee

\subsubsection{Generic $n$}

After these examples, the general structure of the answer for
3-strand torus knots should be partly clear:
the matrix $A_2^{[3,n]}$ is reduced to the vicinity of the
main diagonal, which ends at $T^{2n}$,
while $A_1^{[3,n]}$ -- to that of diagonals, which begins at $T^2$.
"Vicinity" can consist of additional shorter diagonals under
the main ones.

As usual, the structure of reductions is dictated by decomposed
Euler characteristic, with additional requirement that diagonals
should have no gaps -- and to fill the gap one can need to add
more diagonals.
One more restriction is that contributions of diagonals
(minus certain boxed terms for links)
should be divisible by $(1+q^2T)$ -- once in the case of $A_1$
and twice in the case of $A_2$.

Anyhow, the starting point are Euler characteristics.
Especially simple is
\be
E_1^{[3,n]}= q^4+(\omega^n+\bar\omega^n)\cdot q^{2n+2}
\ee
where $\omega=e^{2\pi i/3}$ and $\bar\omega=e^{-2\pi i/3}$, i.e.
the coefficient in the second term is $-1$ for knots and $+2$
for links.
In both cases we have a huge gap in $E_1$,
and this means that there will be at least two diagonals --
but sometime two will be not enough, as we already saw in the
case of $n=6$.

As to $E_2$, it is less trivial.
The simplest thing is to draw: for $n=3k+1$
\be
\begin{array}{c|cccccccccccccccccccccc}
& q^9T^6 & q^{11}T^7 & q^{13}T^8 &\ldots \\
&&&\\
\hline &&\\
n=4& 1&-2&1 \\
n=7& 1&-2&2&-1&0&-1&2&-2&1 \\
n=10& 1&-2&2&-1&0&0&1&-2&1&0&0&-1&2&-2&1\\
n=13& 1&-2&2&-1&0&0&1&-2&2&-1&0&-1&2&-2&1&0&0&-1&2&-2&1 \\
\ldots & \\
\end{array}
\nn
\ee
We added also powers of $T$ which would appear at the main diagonal
of reduced $A_2$.
Diagonal ends at  $q^{4n-3}T^{2n}$.
There is clearly a periodicity in $n$ with period $6$, not just $3$.
and we need to consider the six cases separately.
These cases are similar, and the present paper we briefly describe just one of these cases.

\subsubsection{Knots from the series $n=6s+1$}

Here we have pictorially:
\be
E_2 = &\underbrace{(1221)00(1221)00\ldots 00(1221)}_{s \ {\rm times}}
0\underbrace{\overline{(1221)}00 \ldots 00\overline{(1221)}00\overline{(1221)}}_{s \ {\rm times}}
\nn\\
\nn\\
I&(1222)22(2222)22\ldots 22(2222)2(2222)22\ldots 22(2222)22(2221) \nn\\
II&(0001)22(1001)22\ldots 22(1\underline{001)2(100}1)22\ldots 22(1001)22(1000)\nn\\
\nn \\
III&(0001)22(2222)22\ldots 22(2222)2(2222)22\ldots 22(2222)22(1000)\nn\\
IV&(0000)00(1221)00\ldots 00(1221)0(1221)00\ldots 00(1221)00(0000)\nn\\
\ee
Here $(1221)$ and $\overline{(1221)}$ denote respectively the sequences
$(1,-2,2,-1)$ and $(-1,2,-1,2)$, after that the choice signs in the next lines is obvious.
To make these formulas readable we keep brackets in all lines, though they do not
carry any other information.

The first line (I) shows, how the gaps are filled in the main diagonal -- then
to restore the right values of alternated sums we would need to add the next diagonal
with the entries (II) -- with inverted signs w.r.t. (I).
This diagonal is already shorter -- has zeroes at the ends,
but it solves the problem only for $s=1$ ($n=7$):
\be
\begin{array}{ccccccccccccccccc}
E_2 &(1221)0(1221) \nn \\
\label{E2s1} \\
I & (1222)2(2221)\nn \\
II&(0001)2(1000)\nn\\
\end{array}
\ee

For higher $s>1$ diagonal II still has gaps
(zero entries, not located at the ends).
Thus we need to change II for III and add a compensating next diagonal IV.
This IV has always has gaps, and can not be the final step --
instead it is exactly the same as original $E_2$, only shorter, as if $s$ was
changed to $s-1$. One more step would give the right answer for $s=2$:
\be
\begin{array}{cccccccccccccccccccccccccccccc}
E_2& (1221)00(1221)0(1221)00(1221)\nn \\
\label{E2s2} \\
I& (1222)22(2222)2(2222)22(2221)\nn \\
III& (0001)22(2222)2(2222)22(1000)\nn \\
V&(0000)00(1222)2(2221)00(0000)\nn \\
VI&(0000)00(0001)2(1000)00(0000) \nn\\
\end{array}
\ee
Thus the reduced function $A_2^{[3,6s+1]}$ has the following pattern:
it consists of $2s$ diagonals, each shorter by $6$ than the previous one,
and the last is of length $3$:

\centerline{
{\footnotesize
$
\left|\begin{array}{c|ccccccccccccccccccccccccc}
\hline
&\\
n=6s+1& q^3\ &q^5\ & q^7\ &q^9\ &q^{11}& q^{13} &q^{15}&q^{17}&q^{19}&q^{21}&q^{23}&q^{25}\!\!
& \!\!\ldots\!\!
& \!\!q^{4n-15}\!\!\!\!\!\! & q^{4n-13}\!\!\!\!\!\! &q^{4n-11}\!\!\!\!\!\!&
q^{4n-9}\!\!\!\!\!& q^{4n-7}\!\!\!\!\! & q^{4n-5}\!\!\!\!\!& q^{4n-3} \\
&&&\\
\hline
&&&\\
1&\\
T&\\
T^2 &&  \\
T^3 &&&  \\
T^4&&&\\
T^5&&&\\
\hline
T^6 &&&&\otimes \\
T^7 &&&&&2\otimes  \\
T^8 &&&&&&2\otimes \\
T^9 &&&&&&&2\otimes \\
\hline
T^{10} &&&&&&&\otimes&2\otimes\\
T^{11}   &&&&&&&&2\otimes&2\otimes \\
T^{12}&&&&&&&&&2\otimes&2\otimes\\
T^{13}&&&&&&&&&&2\otimes&2\otimes \\
\hline
T^{14}&&&&&&&&&&\otimes&2\otimes&2\otimes\\
 &\ldots &&&&&&&&&&2\otimes&2\otimes\\
T^{2n-6}& &&&&&&&&&&&2\otimes&\ldots&2\otimes\\
T^{2n-5}& &&&&&&&&&&&&&2\otimes&2\otimes\\
T^{2n-4}& &&&&&&&&&&&&&\otimes&2\otimes&2\otimes\\
T^{2n-3}& &&&&&&&&&&&&&&&2\otimes&2\otimes\\
\hline
T^{2n-2}& &&&&&&&&&&&&&&&&\otimes&2\otimes\\
T^{2n-1}& &&&&&&&&&&&&&&&&&&2\otimes\\
T^{2n}  & &&&&&&&&&&&&&&&&&&&\otimes\\
&\\
\hline
\end{array}\right|
$
}}

\bigskip

As to $E_1$, in the case of $n=6s+1$ there are many short diagonals,
which form a simple ordered structure.
The first has length $4$ and spreads from $q^4T^2$ to $q^{10}T^5$.
The next diagonals have length $6$ and spreads from $q^4T^2$ to $q^{10}T^5$,
then from $q^6T^4$ to $q^{16}T^9$, then
from $q^{12}T^8$ to $q^{22}T^{13}$ and so on,
from $q^{6k}T^{4k}$ to $q^{6k+10}T^{4k+5}$,
until the last ones are cut
by the right margin of the table at $q^{2n+2}=q^{12s+4}$,
so that the last diagonal of the length $6$ spreads from
$q^{12s-6}T^{8s-4}$ to $q^{12s+4}T^{8s+1}$
and below it is a small diagonal of the length $2$,
consisting of just two points $q^{12s}T^{8s}$ and $q^{12s+2}T^{8s+1}$:

\bigskip

\be
\!\!\!\!\!\!\!\!\!\!\!\!\!\!\!\!\!\!\!\!\!\!\!\!\!\!\!\!\!\!
A_1^{[3,6s+1]} \sim \left|\begin{array}{c|cccccccccccccccccccccccc}
\hline
&\\
n=6s+1& q^2\ &q^4\ & q^6\ &q^8\ &q^{10}&q^{12}&q^{14}&q^{16}&q^{18}&q^{20}&q^{22}&q^{24} &\ldots
& \!\! q^{2n-8}\!\!&\!\!  q^{2n-6}\!\! &\!\!  q^{2n-4}\!\!\! &\!\!\!  q^{2n-2}\!\!  &\!\!  q^{2n}\!\!
& \!\! q^{2n+2}\!\!  \\
&&&\\
\hline
&&&\\
1&\\
T&\\
T^2 &&\boxed{\otimes} \\
T^3 &&&\boxed{\otimes} \\
\hline
T^4 &&&\boxed{\otimes}&\boxed{\otimes}\\
T^5&&&&\boxed{\otimes}&\boxed{\otimes} \\
T^6&&&&&\otimes \\
T^7&&&&&&\otimes \\
\hline
T^8&&&&&&\otimes&\otimes \\
T^9   &&&&&&&\otimes&\otimes \\
T^{10}&&&&&&&&\otimes\\
T^{11}&&&&&&&&&\otimes\\
\hline
T^{12}&&&&&&&&&\otimes&\otimes\\
T^{13}&&&&&&&&&&\otimes&\otimes\\
T^{14}&&&&&&&&&&&\otimes\\
T^{15}&&&&&&&&&&&&\otimes\\
&&&&&&&&&&&&&&\ldots\\
&\ldots\\
T^{8s-4}&&&&&&&&&&&&&&\otimes&\otimes\\
T^{8s-3}&&&&&&&&&&&&&&&\otimes&\otimes\\
T^{8s-2}&&&&&&&&&&&&&&&&\otimes\\
T^{8s-1}&&&&&&&&&&&&&&&&&\otimes\\
\hline
T^{8s} &&&&&&&&&&&&&&&&&\otimes&\otimes\\
T^{8s+1}&&&&&&&&&&&&&&&&&&\otimes&\otimes \\
T^{8s+2} &&&   \\
&\ldots \\
T^{2n}& \\
&\\
\hline
\end{array}\right|
\nn
\ee

\bigskip

The case $n=6s+1$ is distinguished because there are deformations of the
clear structure: for example, $A_1$ contains integer number of
the $6$-point configuration, marked by boxes in the table
and represented by the product $(1+q^2T)(1+[2]q^3T^2)$ in the formula below.
However, for other values of $n$ the deviations from this structure
occur only at the right margin of the tables and are easy to understand
from our previous examples.

From these tables we can easily read expressions for KR polynomials:
\be
{\cal P}^{[3,6s+1]}_r \sim
q^{12s(N-1)}\left\{
1 \ +\  q^4T^2(1+[2]q^3T^2)\Big(1+q^6T^4+ \ldots +(q^6T^4)^{2s-1}\Big)
\underbrace{q^{2N-2}(1+q^2T)\cdot q^{-N}[N-1]}_{\!\!\!\sim 1+q^{2N-2}T\!\!\!} +
\right.\nn \\
+ \left(q^9T^6\Big(1+q^4T^2+\ldots + (q^4T^2)^{6s-3}  \Big)
\ +\  q^{15}T^{10}\Big(1+q^4T^2+\ldots + (q^4T^2)^{6s-6}\Big) +
 \ \ \ldots \phantom{5^{5^{5^{5^5}}}}\!\!\!\!\!\!\!\!\!\!\!
+q^{12s+3}T^{8s+2}\right)
\cdot \nn \\  \left.
\cdot\underbrace{q^{2N-2}(1+q^2T)^2[N-1][N-2]}_{
\!\!\!\!\!\!\sim q^{3}+q^{2N}[2]T+q^{4N-3}T^2\!\!\!\!\!\!}
\right\}
\nn
\ee
and similarly for the unreduced KR polynomial.

It is straightforward, but still instructive to repeat the analysis of sec.\ref{2strator},
examine relation to MacDonald dimensions and observe the appearance
of $\gamma$-factors.




\subsection{The twisted knot $6_1$ }

As any twisted knot this one can be represented in terms of two antiparallel
strands and additional lock element
(see \cite{evo} for details of the twisted knot calculus)
-- in this representation it was analyzed
in \cite{DM3}.
Now we do this in the representation by an ordinary braid  --which
in this case is 4-strand:
\be
(1,1,2,-1,-3,2,-3):\ \ \
{\cal R}_1^2{\cal R}_2{\cal R}_1^{-1}{\cal R}_3^{-1}{\cal R}_2{\cal R}_3^{-1}
\ \ \ \ \ \ \begin{array}{ccccccccccccc}
\bullet && \bullet &&&&\circ \\
&&&&\bullet &&&&\bullet \\
&&&&&&\circ &&&&\circ
\end{array}
\ee
(we remind that ${\cal R}_1$ and ${\cal R}_3$, as well as projectors $\pi_1$ and $\pi_3$,
commute, thus their mutual ordering is non-essential).

In this case three of the vertices in knot diagram are white, i.e. three inverse
${\cal R}$-matrices are present, and
\be
\left(\frac{q^{4N-4}}{q^{3N}T^3}\right)^{-1}\P^{6_1}_r =
\ \ \ \ \ \ \ \ \ \ \ \ \ \ \ \ \ \ \ \ \ \ \ \ \ \ \ \ \ \ \ \ \ \ \ \ \ \ \ \ \ \ \ \ \ \nn \\
= \Tr_{10\times 10}
\left\{D\cdot\Big(I+ [2]qT \pi_1\Big)^2\Big(I+ [2]qT \pi_2\Big)\Big([2]\pi_1+ qT\cdot I\Big)
\Big([2]\pi_3+ qT\cdot I\Big)\Big(I+ [2]qT \pi_2\Big) \Big([2]\pi_3+ qT\cdot I\Big)\right\} = \nn
\ee
\vspace{-0.4cm}
\be
= A_0^{6_1}q^{-3N} + A_1^{6_1}q^{-2N}[N-1] + A_2^{6_1}q^{-N}[N-1][N-2] + A_3^{6_1}[N-1][N-2][N-3]
\ee
i.e., all the four functions $A_k(q,T)$ are cohomologically non-trivial:

\be
A_3^{6_1} = \left|\begin{array}{c|ccccccccccccc}
\hline &&\\
& q^3 & q^5 & q^7 & q^9 & q^{11}& q^{13} \\
&&\\ \hline && \\
1 &   1& 1 & &\\
T &   4& 9 & 3&\\
T^2 & 6& 25& 18& 1&\\
T^3 & 4& 30& 42& 8&\\
T^4 & 1& 16& 46&22&1&\\
T^5 &  & 3 & 21&28&4&\\
T^6 &  & & 3& 11&8& \\
T^7 &  & & & 1 &2& 1\\
&&\\ \hline & \\
& (0)&(0)& (1) & (-3)&(3)&(-1) \\
&\\ \hline
\end{array}\right| \ \sim \
\left|\begin{array}{c|ccccccccccccc}
\hline &&\\
& q^3 & q^5 & q^7 & q^9 & q^{11}& q^{13} \\
&&\\ \hline && \\
1 &   &  & &\\
T &   &  & &\\
T^2 & & & & &\\
T^3 & & & & &\\
T^4 & & &\otimes &&&\\
T^5 &  &  & &3\otimes&&\\
T^6 &  & & & &3\otimes&& \\
T^7 &  & & &  && \otimes\\
&&\\ \hline & \\
& (0)&(0)& (1) & (-3)&(3)&(-1) \\
&\\ \hline
\end{array}\right|
\ee
Clearly, this matrix -- and thus its relevant diagonal -- satisfies (\ref{divis}).
Therefore its contribution to reduced KR polynomial is
\be
q^7T^4[N-1][N-2][N-3](1+q^2T)^3 \ \stackrel{(\ref{redur3})}{\sim}\
q^7T^4\Big(q^{2-N}+q^NT\Big)\Big(q^{3-N}+q^{N-1}T\Big)\Big(q^{4-N}+q^{N-2}T\Big)
\ee

\be
A_2^{6_1} = \left|\begin{array}{c|ccccccccccccc}
\hline &&\\
& q & q^3 & q^5 & q^7 & q^9 & q^{11}& q^{13} & q^{15} \\
&&\\ \hline && \\
1 &   1& 3 & 3& 1  &\\
T &   2& 16 & 26&15& 3&\\
T^2 & 1& 22& 72&62 &20&1 &\\
T^3 & & 8 & 79& 123&58&8 &\\
T^4 & & & 33&120   &92&24&1 &\\
T^5 &  &  & 7&53   &78&36&4 &\\
T^6 &  & & 1& 12   &29&26&8 & \\
T^7 &  & & & 1     &4 & 6&4 & 1\\
&&\\ \hline & \\
& (0)&(1)& (-3) & (3)&(-2)&(1)&(1)&(-1) \\
&\\ \hline
\end{array}\right| \ \sim \
\left|\begin{array}{c|ccccccccccccc}
\hline &&\\
& q & q^3 & q^5 & q^7 & q^9 & q^{11}& q^{13} & q^{15} \\
&&\\ \hline && \\
1 &   &  & &   &\\
T &   &  & && &\\
T^2 & & \otimes& & && &\\
T^3 & &  &  3\otimes& &\\
T^4 & & & &   3\otimes&& &\\
T^5 &  &  & &   &2\otimes&\otimes &\\
T^6 &  & & &    &&2\otimes&2\otimes&  \\
T^7 &  & & &      & & & \otimes & \otimes\\
&&\\ \hline & \\
& (0)&(1)& (-3) & (3)&(-2)&(1)&(1)&(-1) \\
&\\ \hline
\end{array}\right|
\nn
\ee
This time the signs of Euler characteristics in columns are not quite alternating
-- this means that there should be at least two diagonals contributing.
Additionally, each diagonal should satisfy (\ref{divis}) -- this dictates the
minimal choice as above.
From (\ref{redur2}) we now get the contribution of $A_2$ to reduced KR polynomial:
\be
\Big(q^3T^2(1+q^2T+q^6T^3)  \ +\ q^{11}T^5\Big)\cdot q^{-N}[N-1][N-2](1+q^2T)^2 \sim \nn \\
\sim  \Big(q^3T^2+q^5T^3+q^{10}[2]T^5\Big)\cdot q^{-N}(q^{2-N}+q^NT)(q^{3-N}+q^{N-1}T)
\ee

\be
A_1^{6_1} = \left|\begin{array}{c|ccccccccccccc}
\hline &&\\
& q^2 & q^4 & q^6 & q^8 & q^{10} & q^{12} & q^{14} \\
&&\\ \hline && \\
1 &   1& 2& 1&  \\
T &   3&13&13& 3& \\
T^2 & 1&19&41&19& 1&  \\
T^3 &  & 7&51&51& 8&\\
T^4 &  & 1&24&60&24& 1& \\
T^5 &  &  & 4&29&29& 4&\\
T^6 &  &  &  & 7&14& 7&\\
T^7 &  &  &  & 1& 3& 3& 1\\
&&\\ \hline && \\
&(-1) & (2) & (-2) & (2)& (-1) & (1)& (-1) \\
&&\\ \hline
\end{array}\right|
\ \sim \
\left|\begin{array}{c|ccccccccccccc}
\hline &&\\
& q^2 & q^4 & q^6 & q^8 & q^{10} & q^{12} & q^{14} \\
&&\\ \hline && \\
1 &     \\
T &   \otimes  \\
T^2 & &2\otimes  \\
T^3 & &&2\otimes  \\
T^4 & &&&2\otimes  \\
T^5 & &&&&\otimes  \\
T^6 & &&&&&\otimes  \\
T^7 & &&&&&& \otimes\\
&&\\ \hline && \\
&(-1) & (2) & (-2) & (2)& (-1) & (1)& (-1) \\
&&\\ \hline
\end{array}\right|
\nn
\ee
With the help of (\ref{redur}) we read the contribution to reduced KR polynomial:
\be
q^2T\Big(1+q^2T+q^4T^2+q^6T^3+ q^{10}T^5\Big)\cdot q^{-2N}[N-1](1+q^2T) \sim \nn \\
\ \sim \
q^2T\Big(1+q^2T+q^4T^2+q^6T^3+ q^{10}T^5\Big)\cdot q^{-2N}(q^{2-N}+q^NT)
\ee
Finally,
\be
A_0^{6_1} = q^6T^3
\ee
Collecting all the four contributions, we obtain:
\be
q^{N-4}T^{-3}\Big\{
q^7T^4\Big(q^{2-N}+q^NT\Big)\Big(q^{3-N}+q^{N-1}T\Big)\Big(q^{4-N}+q^{N-2}T\Big) + \nn \\
+\Big(q^3T^2+q^5T^3+q^{10}[2]T^5\Big)\cdot q^{-N}(q^{2-N}+q^NT)(q^{3-N}+q^{N-1}T) + \nn \\
+ q^2T\Big(1+q^2T+q^4T^2+q^6T^3+ q^{10}T^5\Big)\cdot q^{-2N}(q^{2-N}+q^NT) + q^6T^3\cdot q^{-3N}
\Big\} = \nn \\ \nn \\
=   \frac{q^{-2N}}{T^2} + \frac{1}{T}\Big(\underline{q^{2-2N}+q^{4-2N}}+q^{-2}\Big)
+ \Big( \underline{q^{2-2N}+q^{4-2N}} +\underline{\underline{q^{6-2N}+q^2}}+ 2 \Big) + \nn \\
+T\Big(\underline{\underline{q^{6-2N}}}+\overline{\overline{q^{12-2N}}}
+2\underline{\underline{q^2}}+\overline{\overline{q^4}} + q^{2N-2}\Big)
+T^2\Big(\overline{q^{10-2N}}+\overline{\overline{q^{12-2N}+q^4}}
+\overline{q^6+q^8+q^{10}} + q^{2N}\Big) + \nn \\
+T^3\Big(\overline{q^{10-2N} +q^6+2\underline{q^8}+q^{10}}+ q^{2N+2} + \underline{q^{2N+4}+q^{2N+6}}\Big)
+T^4\Big(\underline{q^8 +q^{2N+4}+q^{2N+6}}+ q^{4N}\Big)
\ee
After elimination of underlined and overlined cohomologically trivial pairs,
this finally provides
\be
\boxed{
{\cal P}^{6_1}_r = \frac{1}{q^{2N}T^2} + \frac{1}{q^2T} + 2
+(q^2+q^{2N-2})T+q^{2N}T^2+q^{2N+2}T^3+q^{4N}T^4
}
\label{KR61r}
\ee
in agreement with \cite{CM} and also with \cite{evo},
where (sec.5.2.6) the fundamental polynomial for the twisted knot $6_1$ was suggested to
be $\ 1+\Big(1+A^2q/t\Big)\{Aq\}\{A/t\}$, what in our current notation, see (\ref{KR41r}),
is equal to
\be
{\cal P}^{6_1}_r =
1 + \Big(1+q^{2N}T^{2}\Big)\left( \frac{1}{q^{2N}T^2}+\frac{1}{q^2T} + q^2T + q^{2N}T^2\right)
\ee
what is exactly (\ref{KR61r}).

\bigskip

In unreduced case we should use reduction formulas (\ref{reduur}),(\ref{reduur2}) and (\ref{reduur3})
to get:
\be
q^{N-4}T^{-3}\cdot q^6T^3\cdot q^{-3N} \underbrace{[N]}_{
\!\!\!\!\!\!\!\!\!\!\!\!\frac{1}{q}[N-1]+q^{N-1}\!\!\!\!\!\!\!\!\!} +
q^{N-4}T^{-3}\Big\{
q^7T^4\Big(q^{1-N}+q^{N+1}T\Big)\Big(q^{3-N}+q^{N-1}T\Big)\Big(q^{4-N}+q^{N-2}T\Big) + \nn \\
+\Big(q^3T^2+q^5T^3+q^{10}[2]T^5\Big)\cdot q^{-N}(q^{1-N}+q^{N+1}T)(q^{3-N}+q^{N-1}T) + \nn \\
+ q^2T\Big(1+q^2T+q^4T^2+q^6T^3+ q^{10}T^5\Big)\cdot q^{-2N}(q^{1-N}+q^{N+1}T)\Big\}[N-1]
= \nn \\ \nn \\
=   q^{1-N}+\Big\{ \frac{q^{-2N-1}}{T^2} + \frac{1}{T}\Big(\underline{q^{1-2N}+q^{3-2N}}
+\boxed{q^{-1}}\Big)
+ \Big( \underline{q^{1-2N}+q^{3-2N}} +\underline{\underline{q^{5-2N}+q^3}}+ q
+\boxed{q^{-1}} \Big) + \nn \\
+T\Big(\underline{\underline{q^{5-2N}}}+\overline{\overline{q^{11-2N}}}
+q+\underline{\underline{q^3}}
+\overline{\overline{q^5}} + q^{2N-1}\Big)
+T^2\Big(\overline{q^{9-2N}}+\overline{\overline{q^{11-2N}+q^5}}
+\overline{q^5+q^7+q^{11}} + \boxed{q^{2N+1}}\Big) + \nn \\
+T^3\Big(\overline{q^{9-2N} +q^5+q^7} +\underline{q^9}+\overline{q^9+q^{11}}+ \boxed{q^{2N+1}}
+ \underline{q^{2N+5}+q^{2N+7}}\Big)
+T^4\Big(\underline{q^9 +q^{2N+5}+q^{2N+7}}+ q^{4N+1}\Big)\Big\}[N-1]
\nn
\ee
so that
\be
{\cal P}^{6_1} = \overbrace{[N]}^{\!\!\!\!\!\!\!\!\!q^{1-N}+q[N-1]\!\!\!\!\!\!\!\!}
+\Big\{\frac{1}{q^{2N+1}T^2} + \frac{1}{qT} + \frac{1}{q}
+(q+q^{2N-1})T+q^{2N+1}T^2+q^{2N+1}T^3+q^{4N+1}T^4\Big\}[N-1] = \nn \\
\boxed{=[N] + \Big(\frac{1}{q^{2N+1}T^2}+ \frac{1}{q} + qT+ q^{2N+1}T^3\Big)(1+q^{2N}T)[N-1]}
\label{KR61ur}
\ee
Note two pairs of boxed terms -- they seem to form cohomologically trivial pairs,
but they did {\it not} do so in reduced polynomial, therefore we keep them in the
unreduced case as well.
Note also that if we performed a "deeper" reduction to $[N-2]$ and $[N-3]$ in
(\ref{reduur2}) and (\ref{reduur3}), the answer (\ref{KR61ur}) would also change.
However, it is (\ref{KR61ur}) that is consistent with the result in \cite{CM},
available there for $N=2$ and $N=3$.

\subsection{The 2-component link $L6a1$, one of the two orientations ($6^2_3(v1)$ of \cite{CM})}

As usual, the link diagram depends on mutual orientation of knots in the link.
In this case both are represented by 4-strand braids.
We begin with the simpler one with $8$ intersections:
$$
{\cal R}_1{\cal R}_2^{-1}{\cal R}_3{\cal R}_2^{-1}{\cal R}_1{\cal R}_2^{-1}
{\cal R}_3^{-1}{\cal R}_2^{-1}:
\ \ \ \
\begin{array}{cccccccccccccccc}
\bullet &&&&&&&&\bullet \\
&& \circ &&&& \circ &&&& \circ &&&& \circ \\
&&&&\bullet &&&&&&&&\circ
\end{array}
$$

This example is distinguished in the present paper, because
this was the only case, where we originally disagreed with the
previously published results.
Now the misprints there are already corrected and there is now
a complete agreement.
Still, when this paper was being written, the issue was still
unclear and we used the chance
to demonstrate in detail the
trial-and-error process, which leads to what we think is the right answer --
even if there is nothing to compare it at the end.
Since this may be instructive,
we decided to keep this subsection as it was originally written --
thus the story will be a little longer than usual.

The four $A_k$ functions in this case are all cohomologically non-trivial.
We begin with their most naive reduction:
\be
A_3^{6^2_3(v1)} = \left|\begin{array}{c|ccccccccccccc}
\hline &&\\
& q & q^3 & q^5 & q^7 & q^9 & q^{11} & q^{13} \\
&&\\ \hline && \\
1 &   1&3&3&1&  \\
T &   3&15&22&11&1&  \\
T^2 & 3&27&62&45&7&  \\
T^3 & 1&20& 84&96& 24& \\
T^4 & & 5& 50& 115& 51 \\
T^5 & &  & 10& 60 & 69&4  \\
T^6 & &  &   & 10 & 35&16& \\
T^7 & &  &   &    &  5&9&2\\
T^8 & &  &   &    &   &1&1 \\
&&\\ \hline && \\
& (0)&(0)&(-1)&(4)&(-6)&(4)&(-1) \\
&& \\ \hline
\end{array}\right|
\ \sim \
\left|\begin{array}{c|ccccccccccccc}
\hline &&\\
& q & q^3 & q^5 & q^7 & q^9 & q^{11} & q^{13} \\
&&\\ \hline && \\
1 &     \\
T &      \\
T^2 &   \\
T^3 & && \otimes  \\
T^4 & &&& 4\otimes  \\
T^5 & &&&&6\otimes   \\
T^6 & &&&&&4\otimes  \\
T^7 & &&&&&& \otimes \\
T^8 &   \\
&&\\ \hline && \\
& (0)&(0)&(-1)&(4)&(-6)&(4)&(-1) \\
&& \\ \hline
\end{array}\right|
\nn
\ee
The contribution to reduced KR polynomial is
\be
q^5T^3(1+q^2T)^4[N-1][N-2][N-3] \ \stackrel{(\ref{redur3})}{\sim}\
q^5T^3(1+q^2T)\Big(q^{2-N}+q^NT\Big)\Big(q^{3-N}+q^{N-1}T\Big)\Big(q^{4-N}+q^{N-2}T\Big)
\nn
\ee

\be
A_2^{6^2_3(v1)} = \left|\begin{array}{c|ccccccccccccc}
\hline &&\\
& q^{-1} & q & q^3 & q^5 & q^7 & q^9 & q^{11} & q^{13}& q^{15}\\
&&\\ \hline && \\
1 & 1& 5&10& 10&  5& 1 \\
T & 1&15&49& 68& 45& 13& 1&     \\
T^2 &&10&79&173&156& 59& 7&   \\
T^3 && 1&41&203&282&143&24&  \\
T^4 &&  & 5& 96&281&212&51& \\
T^5 &&  &  & 14&132&193&77& 4&  \\
T^6 &&  &  &   & 23& 89&67&16&  \\
T^7 &&  &  &   &  1& 16&26&13&2 \\
T^8 &&  &  &   &   &  1& 3& 3&1  \\
&&\\ \hline && \\
& (0) & (-1) & (4) & (-6) & (5) & (-3) & (0) & (2)& (-1) \\
&& \\ \hline
\end{array}\right|
\nn \\
\stackrel{?}{\sim} \
\left|\begin{array}{c|ccccccccccccc}
\hline &&\\
& q^{-1} & q & q^3 & q^5 & q^7 & q^9 & q^{11} & q^{13}& q^{15}\\
&&\\ \hline && \\
1 &     \\
T & &\otimes    \\
T^2 &&&4\otimes   \\
T^3 &&\times&&6\otimes   \\
T^4 &&&4\times&&5\otimes   \\
T^5 &&&&6\times&&3\otimes&\otimes   \\
T^6 &&&&&5\times&&\otimes&2\otimes  \\
T^7 &&&&&&3\times&&&\otimes \\
T^8 &&&&&&&\times \\
&&\\ \hline && \\
& (0) & (-1) & (4) & (-6) & (5) & (-3) & (0) & (2)& (-1) \\
&& \\ \hline
\end{array}\right|
\ee
Note that additional restriction on the choice of diagonals is imposed
by divisibility over $(1+q^2T)^2$.
The contribution to KR polynomial of encycled crosses is
\be
qT\Big(1+2q^2T+q^4T^2+q^6T^3+q^{10}T^4 \Big)(1+q^2T)^2\cdot q^{-N}[N-1][N-2] \nn \\
\ \stackrel{(\ref{redur2})}{\sim}\
q^{1-N}T\Big(1+2q^2T+q^4T^2+q^6T^3+q^{10}T^4 \Big)\Big(q^{2-N}+q^NT\Big)\Big(q^{3-N}+q^{N-1}T\Big)
\nn
\ee

\be
\!\!\!\!\!\!\!\!\!\!\!\!\!\!\!\!\!\!
A_1^{6^2_3(v1)}= \left|\begin{array}{c|ccccccccccccc}
\hline &&\\
& 1 & q^2 & q^4 & q^6 & q^8 & q^{10} & q^{12} & q^{14}   \\
&&\\ \hline && \\
1 & 1&  4& 6&  4&  1  \\
T & 1& 12&31&31&12&1   \\
T^2 &&  7&49&84&49&7 \\
T^3 &&   &23&102&102&23    \\
T^4 &&   &  &46& 116& 46 \\
T^5 &&   &  & 3&  64& 64&4&\\
T^6 &&   &  &  &  16& 44&16& \\
T^7 &&   &  &  &   2& 12&12&2\\
T^8 &&   &  &  &    &  1&2 &1   \\
&&\\ \hline && \\
&(0)& (-1)&(1)&(-2)&(2)&(-2)&(2)&(-1) \\
&& \\ \hline
\end{array}\right|
\ \stackrel{?}{\sim} \
\left|\begin{array}{c|ccccccccccccc}
\hline &&\\
& 1 & q^2 & q^4 & q^6 & q^8 & q^{10} & q^{12} & q^{14}   \\
&&\\ \hline && \\
1 &     \\
T &  &\boxed{\otimes}   \\
T^2 &&&\otimes   \\
T^3 & &&&2\otimes  \\
T^4 & &&&&2\otimes  \\
T^5 & &&&&&2\otimes  \\
T^6 & &&&&&&2\otimes  \\
T^7 & &&&&&&&{\otimes}\\
T^8 & \\
&&\\ \hline && \\
&(0)& (-1)&(1)&(-2)&(2)&(-2)&(2)&(-1) \\
&& \\ \hline
\end{array}\right|
\nn
\ee
Position of the boxed item in the upper corner is preferable,
because then the remaining diagonal is not split into two
{\it independent} multiples of $(1+q^2T)$.
The contribution to primary polynomial is
\be
q^4T^2\Big(1+q^2T+q^4T^2+q^6T^3+q^8T^4\Big)\cdot q^{-2N}(1+q^2T)[N-1]
+ q^{2}T\cdot q^{-2N}[N-1] \nn \\ \stackrel{(\ref{redur})}{\sim}\
q^{4-2N}T^2\Big(1+q^2T+q^4T^2+q^6T^3+q^8T^4\Big)\Big(q^{2-N}+q^NT\Big) +
q^{2-2N}T^7[N-1]
\nn
\ee
and, finally,
\be
A_0^{6^2_3(v1)} = q^8T^5
\nn
\ee
Putting all together, we obtain:
\be
\frac{q^{3N-3}}{q^{5N}T^5}\Big\{q^{2-2N}T[N-1] +
q^5T^3(1+q^2T)\Big(q^{2-N}+q^NT\Big)\Big(q^{3-N}+q^{N-1}T\Big)\Big(q^{4-N}+q^{N-2}T\Big) +\nn\\
+ q^{1-N}T\Big(1+2q^2T+q^4T^2+q^6T^3+q^{10}T^4 \Big)\Big(q^{2-N}+q^NT\Big)\Big(q^{3-N}+q^{N-1}T\Big)
+\nn \\ + q^{4-2N}T^2\Big(1+q^2T+q^4T^2+q^6T^3+q^8T^4\Big)\Big(q^{2-N}+q^NT\Big)+q^{8-3N}T^5\Big\}
= \nn \\ \nn \\
=\frac{[N-1]}{q^{4N+1}T^4} \ +\  \frac{\overline{q^{3-5N}}}{T^4}
+ \frac{1}{T^3}\Big(\overline{q^{3-5N}}+q^{5-5N}+\underline{q^{5-5N}+q^{1-3N}}+q^{-3N-1}\Big) +\nn\\
+ \frac{1}{T^2}\Big(q^{11-5N} + \underline{\underline{q^{7-5N}}}+\underline{q^{5-5N}+q^{1-3N}}
+2q^{1-3N} + \underline{\underline{2q^{3-3N}}}+q^{-N-3}\Big) + \nn \\
+ \frac{1}{T}\Big(\underline{\underline{q^{7-5N}}}+\overline{q^{9-5N}+q^{13-5N}+
2q^{5-3N}+q^{7-3N}+q^{9-3N}}+\underline{\underline{2q^{3-3N}}}+2q^{-N-1}\Big) + \nn \\
+\Big(q^{5-5N}+\overline{q^{9-5N}+q^{13-5N}+
2q^{5-3N}+q^{7-3N}+q^{9-3N}}+\underline{q^{7-3N}+q^{11-3N}+q^{5-N}+q^{3-N}}+2q^{1-N}\Big) + \nn \\
+T\Big(q^{11-5N}+\overline{\overline{q^{9-3N}+q^{7-N}}} + \underline{q^{7-3N}+q^{11-3N}+q^{5-N}+q^{3-N}}
+ q^{3-N}+q^{N-1}\Big)
+ \nn \\
 + T^2\Big(\overline{\overline{q^{9-3N}+q^{7-N}}}+q^{N+1}\Big)
 \ \ \ \ \ \ \ \ \ \ \ \ \ \ \ \ \ \ \ \ \ \ \ \ \ \
\label{623v1promor}
\ee
Eliminating the underlined and overlined pairs with adjacent powers of $T$, we finally get:
\be
{\cal P}_r^{6^2_3(v1)}\ \stackrel{?}{=}
\frac{[N-1]}{q^{4N+1}T^4} +
\frac{1}{T^3}\Big(\widetilde{q^{5-5N}}+q^{-1-3N}\Big)
+ \frac{1}{T^2}\Big(\widetilde{q^{11-5N}} + 2q^{1-3N}+q^{-N-3}\Big)
+ \frac{2}{q^{N+1}T}  + \nn\\
+ \Big(\widetilde{q^{5-5N}}+2q^{1-N}\Big)
+ \Big(\widetilde{q^{11-5N}} +q^{3-N}+q^{N-1}\Big)T +  q^{N+1}T^2
\label{KR623v1r0}
\ee
Of course, at $T=-1$ this reproduces the right Jones and HOMFLY polynomials
\be
\frac{1}{q^9}-\frac{1}{q^7}+\frac{3}{q^5}-\frac{2}{q^3}+\frac{2}{q}-2q+q^3
\ee
and
\be
\frac{[N-1]}{q^{4N+1}} +  q^{-3N} \left(2q-\frac{1}{q}\right)
+ q^{-N}\left(-q^3+2q-\frac{2}{q}+\frac{1}{q^3}\right)   + q^N\left(q-\frac{1}{q}\right)
\ee
Still,   expression (\ref{KR623v1r}) looks somewhat suspicious:
it seems natural to get rid of the four tilded
terms there (which cancel and drop away in reduction to HOMFLY polynomial at $T=-1$).
Note, however, that it is not obligatory easy to eliminate these terms
by picking up some other diagonals in our tables, because in (\ref{623v1promor})
each term with $q^{-5N}$ is rigidly linked to at least one term with $q^{-3N}$
--eliminating one, we would unavoidably create another.
The only hope is that additionally created terms could form cohomologically
trivial combinations.
Before demonstrating that this is what actually happens,
let us consider the implication of our naive answer for unreduced polynomial --
this can help, because unreduced Jones-Khovanov polynomial (at $N=2$)
is available at \cite{katlas,indiana}, and there will be at least something
to compare with.

\bigskip

In unreduced case (\ref{623v1promor}) is changed for
\be
\frac{q^{3N-3}}{q^{5N}T^5}\Big\{q^{2-2N}T[N][N-1] +
q^5T^3(1+q^2T)\Big(q^{1-N}+q^{N+1}T\Big)\Big(q^{3-N}+q^{N-1}T\Big)\Big(q^{4-N}+q^{N-2}T\Big)[N-1]
+\nn\\
+ q^{1-N}T\Big(1+2q^2T+q^4T^2+q^6T^3+q^{10}T^4 \Big)
\Big(q^{1-N}+q^{N+1}T\Big)\Big(q^{3-N}+q^{N-1}T\Big)[N-1]
+\nn \\
+ q^{4-2N}T^2\Big(1+q^2T+q^4T^2+q^6T^3+q^8T^4\Big)\Big(q^{1-N}+q^{N+1}T\Big)[N-1]
+q^{8-3N}[N]
T^5\Big\}
= \nn \\
=\frac{[N][N-1]}{q^{4N+1}T^4} \ +\ q^{5-5N}[N] \ +\
\Big\{\frac{\overline{q^{2-5N}}}{T^4}+
\frac{1}{T^3}\Big(\overline{q^{2-5N}}+\widetilde{q^{4-5N}}
+\underline{q^{4-5N}+q^{2-3N}}+q^{-3N-2}\Big) +\nn\\
+\frac{1}{T^2}\Big(\widetilde{q^{10-5N}} + \underline{\underline{q^{6-5N}}}+\underline{q^{4-5N}+q^{2-3N}}
+2q^{-3N} + q^{4-3N}+\underline{\underline{q^{4-3N}}}+q^{-N-2}\Big) + \nn \\
+ \frac{1}{T}\Big(\underline{\underline{q^{6-5N}}}+\overline{q^{8-5N}+q^{12-5N}+
2q^{6-3N}+\boxed{q^{4-3N}}}+\widetilde{\widetilde{q^{10-3N}}}
+{q^{2-3N}}+\underline{\underline{q^{4-3N}}}+2q^{-N}\Big) + \nn \\
+\Big(
\overline{q^{8-5N}+q^{12-5N}+
2q^{6-3N}+q^{4-3N} }+\underline{2q^{8-3N}+q^{12-3N}+q^{6-N}+q^{4-N}}+q^{2-N}+q^{-N}\Big) + \nn \\
+T\Big(\widetilde{q^{10-5N}}+\overline{\overline{q^{8-N}}}
+ \underline{2q^{8-3N}+q^{12-3N}+q^{6-N}+q^{4-N}}
+ q^{2-N}+q^{N}\Big) + \nn \\
+ T^2\Big(\widetilde{\widetilde{q^{10-3N}}}+\overline{\overline{q^{8-N}}}+q^{N+2}\Big)\Big\}[N-1]
\ \ \ \ \ \ \ \ \ \ \ \ \ \ \ \ \ \ \ \ \ \ \ \ \ \
\label{623v1promour}
\ee
The reason why the boxed term is paired with the one in lower rather than in the upper adjacent
line is that it did so in reduced case.
Thus
\be
{\cal P}^{6^2_3(v1)}\stackrel{?}{=}
\frac{[N][N-1]}{q^{4N+1}T^4} \ +\ \widetilde{\widetilde{ q^{5-5N}[N]}} \ +\
 \Big\{
\frac{1}{T^3}(\widetilde{q^{4-5N}}+q^{-3N-2})
+ \frac{1}{T^2}(\widetilde{q^{10-5N}}+2q^{-3N})
+ \nn \\
+ \frac{1}{q^{3N-2}T} +q^{-N} +(\widetilde{q^{10-5N}}+q^{2-N})T
\Big\}(1+q^{2N}T)[N-1]
\label{623v1promour2}
\ee
Now we can compare this expression at$N=2$ with the unreduced Jones-Khovanov polynomial
from \cite{katlas,indiana}, which for this purpose should be appropriately regrouped:
\be
\frac{[2]}{q^9T^4} + \frac{1}{q^8T^3} +\frac{2q^{-6}+q^{-4}}{T^2}+\frac{2}{q^2T}+(2q^{-2}+1)
+(1+q^2)T+q^4T^2 = \nn \\
= \frac{[2]}{q^9T^4} + q^{-1}[2] +\left(\frac{1}{q^8T^3}+\frac{2}{q^6T^2}
+\frac{1}{q^2}+T\right)(1+q^4T)
\ee
From this formula there can be no doubt that one should somehow eliminate
the tilded terms in (\ref{623v1promour2}), simultaneously  modifying the double-tilded one,
and that the right answer is:
\be
\boxed{
{\cal P}^{6^2_3(v1)} = \frac{[N][N-1]}{q^{4N+1}T^4} \ +\ q^{1-N}[N] \ +\
\left(\frac{1}{q^{3N+2}T^3} + \frac{2}{q^{3N}T^2}+q^{-N} +q^{2-N}T\right)(1+q^{2N}T)[N-1]
}
\label{KR623v1ur}
\ee
In reduced case this would imply
\be
{\cal P}^{6^2_3(v1)}_r =
\frac{[N-1]}{q^{4N+1}T^4} \ +\ q^{1-N} \ +\
\left(\frac{1}{q^{3N+1}T^3} + \frac{2}{q^{3N-1}T^2}+q^{1-N}
+q^{3-N}T\right)(1+q^{2N-2}T)=\nn \\
\boxed{
= \frac{[N-1]}{q^{4N+1}T^4}+ \frac{1}{q^{3N+1}T^3} +
\left(\frac{2}{q^{3N-1}}+\frac{1}{q^{N+3}}\right)\frac{1}{T^2}+
\frac{2}{q^{N+1}T} +2q^{1-N} +\Big(q^{3-N}+q^{N-1}\Big)T +q^{N+1}T^2
}
\label{KR623v1r}
\ee
Note that this is exactly (\ref{KR623v1r0}) with the tilded terms omitted
(as we suspected from the very beginning) -- but now we have a better
representation in the first line, which provides a tool for modifying original
tables.

As an intermediate step we can make another table: representing the coefficient in
front of $(1+q^{2N-2}T)$ inside braces in (\ref{623v1promor}):
\be
\begin{array}{c||cc||ccccc|cc||c|cc}
&&&&&&&&&&&\\
&  A_3 && & A_2 &&&&A_2^{aux}&&A_1&A_1^{aux} \\
&&&&&&&&&&&\\
\hline
&&&&&&&&&&&\\
T & && q^{4-2N}&&&&&&&&     \\
T^2 & && \boxed{1} & 2q^{6-2N} &&&&&&q^{4-2N} & \\
T^3 & q^{12-2N} & && \boxed{2q^2} & q^{8-2N}&&&&&q^{6-2N}&q^{6-2N}  \\
T^4 & q^6+q^8 & q^{14-2N} && &q^4 & q^{10-2N} && q^{12-2N} && q^{8-2N} & q^{6-2N} \\
T^5 & \boxed{q^{2N+2}} & q^{8}+q^{10}&&&& q^6 &q^{14-2N} & q^8&q^{12-2N}   & q^{10-2N} \\
T^6 &&\boxed{q^{2N+4}} &&&&&q^{10} &&  q^8 & q^{12-2N} \\
&&&&&&&&&&&&\\
\end{array}
\nn
\ee
In the columns from $A_3$ and $A_2$ the verticals are rigid:
they are defined by the factors $(q^{3-N}+q^{N-1}T)$ and $(q^{4-N}+q^{N-1}T)$
In the case of $A_1$ all items should be in separate columns,
we put them in one just to save space.
Thus the freedom is to add columns.
The boxed elements are the ones from the answer -- from the first line in
(\ref{KR623v1r}), -- and thus the corresponding columns should be contributing.
At the same time the non-boxed elements from these {\it must} columns
should drop away from the answer (\ref{KR623v1r}), i.e. should enter some
cohomologically trivial pairs.
Such pairs are formed by identical entries at some two {\it adjacent} lines.
Clearly, the content of the tree sets of columns $A_3$, $A_2$, $A_1$ is not sufficient
to provide all the needed cancelations.
But adding additional columns in $A_2^{aux}$ and $A_1^{aux}$ solves the problem
and reproduces (\ref{KR623v1r}).
However, addition of extra columns means that their counterparts should be
added into our original reduced (diagonal) tables:

\bigskip

\centerline{
$
\!\!\!\!\!\!
A_2 \sim
\left|\begin{array}{c|ccccccccccccc}
\hline &&\\
& q^{-1} & q & q^3 & q^5 & q^7 & q^9 & q^{11} & q^{13}& q^{15}\\
&&\\ \hline && \\
1 &     \\
T & &\otimes    \\
T^2 &&&4\otimes   \\
T^3 && &&6\otimes   \\
T^4 &&& &&5\otimes   \\
T^5 &&&& &&3\otimes&\otimes   \\
T^6 &&&&& &&\otimes&2\otimes  \\
T^7 &&&&&& &&&\otimes \\
T^8 &&&&&&&  \\
&&\\ \hline && \\
& (0) & (-1) & (4) & (-6) & (5) & (-3) & (0) & (2)& (-1) \\
&& \\ \hline
\end{array}\right|
\longrightarrow
\left|\begin{array}{c|ccccccccccccc}
\hline &&\\
& q^{-1} & q & q^3 & q^5 & q^7 & q^9 & q^{11} & q^{13}& q^{15}\\
&&\\ \hline && \\
1 &     \\
T & &\otimes    \\
T^2 &&&4\otimes   \\
T^3 && &&6\otimes   \\
T^4 &&& &&5\otimes &\otimes  \\
T^5 &&&& &&4\otimes&3\otimes   \\
T^6 &&&&& &&3\otimes&3\otimes  \\
T^7 &&&&&& &&\otimes&\otimes \\
T^8 &&&&&&& &&  \\
&&\\ \hline && \\
& (0) & (-1) & (4) & (-6) & (5) & (-3) & (0) & (2)& (-1) \\
&& \\ \hline
\end{array}\right|
$
}

\bigskip

\centerline{
$
A_1=
\left|\begin{array}{c|ccccccccccccc}
\hline &&\\
& 1 & q^2 & q^4 & q^6 & q^8 & q^{10} & q^{12} & q^{14}   \\
&&\\ \hline && \\
1 &     \\
T &  &\boxed{\otimes}   \\
T^2 &&&\otimes   \\
T^3 & &&&2\otimes  \\
T^4 & &&&&2\otimes  \\
T^5 & &&&&&2\otimes  \\
T^6 & &&&&&&2\otimes  \\
T^7 & &&&&&&&{\otimes}\\
T^8 & \\
&&\\ \hline && \\
&(0)& (-1)&(1)&(-2)&(2)&(-2)&(2)&(-1) \\
&& \\ \hline
\end{array}\right|
\ \longrightarrow \
\left|\begin{array}{c|ccccccccccccc}
\hline &&\\
& 1 & q^2 & q^4 & q^6 & q^8 & q^{10} & q^{12} & q^{14}   \\
&&\\ \hline && \\
1 &     \\
T &  &\boxed{\otimes}   \\
T^2 &&&\otimes   \\
T^3 & &&&3\otimes  \\
T^4 & &&&\otimes&3\otimes&&\times  \\
T^5 & &&&&\otimes&2\otimes&&\times  \\
T^6 & &&&&&&2\otimes  \\
T^7 & &&&&&&&{\otimes}\\
T^8 & \\
&&\\ \hline && \\
&(0)& (-1)&(1)&(-2)&(2)&(-2)&(2)&(-1) \\
&& \\ \hline
\end{array}\right|
$
}

\bigskip

\noindent
At the level of these diagonals one could also consider shifting
a couple of elements in the $A_1$ table to positions of crosses --
but this is forbidden, because in the full table $A_1$ there was
nothing at these positions -- and reduction assumes that one can
not produce negative contributions neither in the remnant nor in the ratio.

With this choice we get instead of (\ref{623v1promor}):
\be
\frac{q^{3N-3}}{q^{5N}T^5}\Big\{q^{2-2N}T[N-1] +
q^5T^3(1+q^2T)\Big(q^{2-N}+q^NT\Big)\Big(q^{3-N}+q^{N-1}T\Big)\Big(q^{4-N}+q^{N-2}T\Big) +\nn\\
+ q^{1-N}T\Big(1+2q^2T+q^4T^2+q^7[2]T^3+q^9[2]T^4 \Big)
\Big(q^{2-N}+q^NT\Big)\Big(q^{3-N}+q^{N-1}T\Big)
+\nn \\ + q^{4-2N}T^2\Big(1+2q^2T+q^3[2]T^2+q^6T^3+q^8T^4\Big)\Big(q^{2-N}+q^NT\Big)+q^{8-3N}T^5\Big\}
= \nn \\ \nn \\
=\frac{[N-1]}{q^{4N+1}T^4} \ +\  \frac{\overline{q^{3-5N}}}{T^4}
+ \frac{1}{T^3}\Big(\overline{q^{3-5N}}+\underline{2q^{5-5N}+q^{1-3N}}+q^{-3N-1}\Big) +\nn\\
+ \frac{1}{T^2}\Big(\underline{2q^{5-5N}+q^{1-3N}}
+ \underline{\underline{q^{11-5N} + q^{7-5N}+3q^{3-3N}}}
+2q^{1-3N} +  q^{-N-3}\Big) + \nn \\
+ \frac{1}{T}\Big(\underline{\underline{q^{11-5N}+q^{7-5N}+3q^{3-3N}}}
+\overline{q^{13-5N}+q^{9-5N}+q^{5-5N}+ q^{9-3N}+q^{7-3N}+2q^{5-3N}}+2q^{-N-1}\Big) + \nn \\
\!\!\!\!\!\!\!\!\!\!\!\!\!\!\!\!\!\!\!\!\!\!\!\!\!\!\!\!\!
+\Big(\overline{q^{13-5N}+q^{9-5N}+q^{5-5N}+q^{9-3N}+q^{7-3N}+2q^{5-3N}}
+\underline{q^{11-5N}+q^{11-3N}+q^{9-3N}+2q^{7-3N}+q^{5-N}+q^{3-N}}+2q^{1-N}\Big) + \nn \\
+T\Big(\underline{q^{11-5N}+q^{11-3N}+q^{9-3N}+2q^{7-3N}+q^{5-N}+q^{3-N}}
+\overline{\overline{q^{9-3N}+q^{7-N}+q^{5-N}}} + q^{3-N}+q^{N-1}\Big)
+ \nn
\ee
\vspace{-0.5cm}
\be
+ T^2\Big(\overline{\overline{q^{9-3N}+q^{7-N}+q^{5-N}}}+q^{N+1}\Big)
 \ \ \ \ \ \ \ \ \ \ \ \ \ \ \ \ \ \ \ \ \ \ \ \ \ \
\label{623v1promor1}
\ee
and elimination of underlined and overlined terms leads to (\ref{KR623v1r}):
\be
\!\!\!\!\!\!\!\! \!\!\!\!\!\!\!\! \boxed{
{\cal P}^{6^2_3(v1)}_r = \frac{[N-1]}{q^{4N+1}T^4}+ \frac{1}{q^{3N+1}T^3} +
\left(\frac{2}{q^{3N-1}}+\frac{1}{q^{N+3}}\right)\frac{1}{T^2}+
\frac{2}{q^{N+1}T} +2q^{1-N} +\Big(q^{3-N}+q^{N-1}\Big)T +q^{N+1}T^2
}
\label{KR623v1r1}
\ee

Unreduced version is:
\be
\frac{q^{3N-3}}{q^{5N}T^5}\Big\{q^{2-2N}T[N][N-1] +
q^5T^3(1+q^2T)\Big(q^{1-N}+q^{N+1}T\Big)\Big(q^{3-N}+q^{N-1}T\Big)\Big(q^{4-N}+q^{N-2}T\Big)[N-1]
+\nn\\
+ q^{1-N}T\Big(1+2q^2T+q^4T^2+q^7[2]T^3+q^9[2]T^4 \Big)
\Big(q^{1-N}+q^{N+1}T\Big)\Big(q^{3-N}+q^{N-1}T\Big)[N-1]
+\nn \\
+ q^{4-2N}T^2\Big(1+2q^2T+q^3[2]T^2+q^6T^3+q^8T^4\Big)\Big(q^{1-N}+q^{N+1}T\Big)[N-1]
+q^{8-3N}T^5[N]\Big\}
= \nn \\
\nn \\
=\frac{[N][N-1]}{q^{4N+1}T^4} \ +\ \Big\{ \frac{\overline{q^{2-5N}}}{T^4}
+ \frac{1}{T^3}\Big(\overline{q^{2-5N}}+\underline{2q^{4-5N}}+q^{-3N-2}\Big)
+ \frac{1}{T^2}\Big(\underline{2q^{4-5N}}
+ \underline{\underline{q^{10-5N} + q^{6-5N}}} +2q^{-3N} \Big) + \nn \\
+ \frac{1}{T}\Big(\underline{\underline{q^{10-5N}+q^{6-5N}}}
+\overline{q^{12-5N}+q^{8-5N}+q^{6-3N}+ q^{4-3N}}+\boxed{q^{4-5N}+q^{2-3N}}\Big) + \nn \\
+\Big(\overline{q^{12-5N}+q^{8-5N}+q^{6-3N}+q^{4-3N}}
+\underline{q^{10-5N}+ q^{8-3N}+q^{6-3N}} +q^{-N}\Big) + \nn
\ee
\vspace{-0.5cm}
\be
+T\Big(\underline{q^{10-5N} +q^{8-3N}+q^{6-3N}}  + q^{2-N} \Big)
\Big\}(1+q^{2N}T)[N-1]
 + \boxed{q^{5-5N}[N]} \ \ \ \ \ \ \ \ \ \ \ \ \ \ \ \ \ \ \ \ \
\label{623v1promour1}
\ee
To convert the terms in boxes we need an additional manipulation:
\be
(q^{2-3N}+q^{4-5N})\frac{1}{T}(1+q^{2N}T)[N-1]+\ q^{5-5N}\!
\underbrace{[N]}_{\!\!\!\!\!\!\!\!\!\!\!\frac{1}{q}[N-1]+q^{N-1}\!\!\!\!\!\!\!\!\!\!\!\!} =\nn\\
=\frac{q^{2-3N}}{T}(1+q^{2N}T)[N-1]+\left(\overline{\frac{q^{4-5N}}{T}}+q^{4-3N}
+ \overline{q^{4-5N}}\right)[N-1] + q^{4-4N} \sim
\nn \\
\!\!\!\!\!
\sim \frac{q^{2-3N}}{T}(1+q^{2N}T)[N-1] + q^{3-3N}
\underbrace{
\overbrace{[N]}^{\!\!\!\!\!\!\!\!\!\!\!q[N-1]+q^{1-N}\!\!\!\!\!\!\!\!\!\!\!\!}
}_{\!\!\!\!\!\!\!\!\!\!\!\frac{1}{q}[N-1]+q^{N-1}\!\!\!\!\!\!\!\!\!\!\!\!} =
\left( \overline{\frac{q^{2-3N}}{T}}+q^{2-N}+\overline{q^{2-3N}}\right)[N-1]+q^{2-2N}
\sim q^{1-N}\overbrace{[N]}^{\!\!\!\!\!\!\!\!\!\!\!q[N-1]+q^{1-N}\!\!\!\!\!\!\!\!\!\!\!\!}
\ee
and finally we arrive at (\ref{KR623v1ur}):
\be
\boxed{
{\cal P}^{6^2_3(v1)} = \frac{[N][N-1]}{q^{4N+1}T^4} \ +\ q^{1-N}[N] \ +\
\left(\frac{1}{q^{3N+2}T^3} + \frac{2}{q^{3N}T^2}+q^{-N} +q^{2-N}T\right)(1+q^{2N}T)[N-1]
}
\label{KR623v1ur1}
\ee
Thus we obtain what seems to be the right answers for reduced and unreduced KR
polynomials for $6^2_3(v2)$.
These answers are consistent with the known unreduced Jones-Khovanov polynomial,
but inconsistent with the answers from \cite{CM} (which, however, are in any case
incorrect for this particular link).
Not only we got the presumably correct formulas, we also reconstructed the
relevant differentials, whose cohomologies provide these answers --
in the future it remains only to compose these differentials from the
morphisms, suggested in \cite{DM3}.

In fact, eqs.(\ref{KR623v1r1}) and (\ref{KR623v1ur1}) appear to coincide with
the corresponding answers in \cite{CM}.

\subsection{The 2-component link $L6a1$, another orientation ($6^2_3(v2)$ of \cite{CM})}


In this orientation the braid is still 4-strand, but  has $10$ intersections:
$$
{\cal R}_1{\cal R}_2{\cal R}_3{\cal R}_2^{2}{\cal R}_1^{-1}{\cal R}_2^{2}{\cal R}_3^{-1}{\cal R}_2^:
\ \ \ \
\begin{array}{ccccccccccccccccccc}
\bullet &&&&&&&&&&\circ \\
&& \bullet &&&& \bullet&&\bullet &&&& \bullet && \bullet &&&& \bullet \\
&&&&\bullet &&&&&&&&&&&& \circ
\end{array}
$$
The function $A_3^{6^2_3(v2)}$ is cohomologically trivial,
\be
A_2^{6^2_3(v2)} = \left|\begin{array}{c|ccccccccccccc}
\hline &&\\
&  q^2 & q^4 & q^6 & q^8 & q^{10} & q^{12} & q^{14} & q^{16}& q^{18} & q^{20} \\
&&\\ \hline && \\
1 &   1&2&1  \\
T &   2&22&16&4  \\
T^2 & 1&76&126&53&4  \\
T^3 & &139&435&350&91  \\
T^4 & &159&782&1083&551&91  \\
T^5 & &119&826&1764&1484&525&63  \\
T^6 & &57&539&1642&2086&1217&320&29  \\
T^7 & &16&213&899&1613&1406&608&121&8\\
T^8 & &2&46&278&683&833&534&174&25&1 \\
T^9 & &&4&   42&144&238&212&102&24&2\\
T^{10} &&&&   2& 11& 25&30&20&7&1 \\
&&\\ \hline && \\
&(0)&(0)&(0)&(-1)&(3)&(-3)&(1)&(0)&(0)&(0) \\
&& \\ \hline
\end{array}\right|
\nn \\ \sim \
\left|\begin{array}{c|ccccccccccccc}
\hline &&\\
&  q^2 & q^4 & q^6 & q^8 & q^{10} & q^{12} & q^{14} & q^{16}& q^{18} & q^{20} \\
&&\\ \hline && \\
1 &     \\
T &     \\
T^2 &   \\
T^3 & &&&\times  \\
T^4 & &&&&3\times  \\
T^5 & &&&\otimes&&3\times  \\
T^6 & &&&&3\otimes&&\times  \\
T^7 & &&&\times&&3\otimes  \\
T^8 & &&&&3\times&&\otimes\\
T^9 & &&&&&3\times\\
T^{10} &&&&&&&\times \\
&&\\ \hline && \\
& (0)&(0)&(0)&(-1)&(3)&(-3)&(1)&(0)&(0)&(0) \\
&& \\ \hline
\end{array}\right|
\ee
Contribution of the middle diagonal to reduced polynomial is
\be
q^8T^5(1+q^2T)^3\cdot q^{-N}[N-1][N-2] \ \stackrel{(\ref{redur2})}{\sim}\
q^{8-N}T^5(1+q^2T)\Big(q^{2-N}+q^NT\Big)\Big(q^{3-N}+q^{N-1}T\Big)
\ee
\be
A_1^{6^2_3(v2)} = \left|\begin{array}{c|ccccccccccccc}
\hline &&\\
&  q^3 & q^5 & q^7 & q^9 & q^{11} & q^{13} & q^{15}& q^{17} & q^{19} \\
&&\\ \hline && \\
1 &   1& 1  \\
T &   4& 14&4 \\
T^2 & 3& 53& 53&4 \\
T^3 &&91&211&91&   \\
T^4 &&91&392&392&91   \\
T^5 &&63&406&721&406&63   \\
T^6 &&29&263&707&707&263&29   \\
T^7 &&8&105&403&612&403&105&8 \\
T^8 &&1&23&129&279&279&129&23&1 \\
T^9 &&&2&20&62&88&62&20&2 \\
T^{10} &&&&1&5&10&10&5&1 \\
&&\\ \hline && \\
&(0)&(-1)&(3)&(-2)&(2)&(-2)&(1)&(0)&(0) \\
&& \\ \hline
\end{array}\right|
\nn \\ \sim \
\left|\begin{array}{c|ccccccccccccc}
\hline &&\\
&  q^3 & q^5 & q^7 & q^9 & q^{11} & q^{13} & q^{15}& q^{17} & q^{19} \\
&&\\ \hline && \\
1 &     \\
T & &\times    \\
T^2 &&&3\times   \\
T^3 &&\otimes&&2\times   \\
T^4 &&&3\otimes&&2\times   \\
T^5 &&\times&&2\otimes&&2\times   \\
T^6 &&&3\times&&2\boxed{\otimes} &&\times   \\
T^7 &&&&2\times&&2\otimes \\
T^8 &&&&&2\times&&{\otimes} \\
T^9 &&&&&&2\times \\
T^{10} &&&&&&&\times \\
&&\\ \hline && \\
&(0)&(-1)&(3)&(-2)&(2)&(-2)&(1)&(0)&(0) \\
&& \\ \hline
\end{array}\right|
\ee
Contribution of the middle diagonal to reduced polynomial is
\be
 \Big(q^5T^3(1+2q^2T+q^6T^3+q^8T^4)(1+q^2T)+q^{11}T^6\Big)\cdot q^{-2N}[N-1] \nn \\
\ \stackrel{(\ref{redur})}{\sim}\
q^{5-2N}T^3(1+2q^2T+q^6T^3+q^8T^4)\Big(q^{2-N}+q^NT\Big)\ + \ q^{11-2N}T^6[N-1]
\nn
\ee
\be
A_0^{6^2_3(v2)} = q^5T^2
\ee
Putting all the three contributions together and restoring the
common factor $q^{8(N-1)}/(q^NT)^2$, we obtain
\be
q^{4N+3}[N-1]T^4 +
q^{6N-8}T^{-2}\Big\{q^{8-N}T^5(1+q^2T)\Big(q^{2-N}+q^NT\Big)\Big(q^{3-N}+q^{N-1}T\Big)
+\nn \\ +q^{5-2N}T^3(1+2q^2T+q^6T^3+q^8T^4)\Big(q^{2-N}+q^NT\Big)
+ q^{5-3N}T^2\Big\} = \nn \\ \nn \\
= q^{4N+3}[N-1]T^4 + q^{3N-3}\Big\{1+q^2T + (q^{2N}+2q^4)T^2
+ (2q^{2N+2}+\overline{q^8})T^3 + \nn \\
+ (q^{2N+4}+\underline{q^{2N+6}+q^{10}}+\overline{q^{8}})T^4 + (q^{4N+2}+\underline{q^{2N+6}+q^{10}}
+q^{2N+6}+\underline{\underline{q^{2N+8}}} )T^5
+ (q^{4N+4}+\underline{\underline{q^{2N+8}}})T^6\Big\}
\label{623v2promr}
\ee
Thus
$$
{\cal P}^{6^2_3(v2)}_r =   q^{3N-3}\Big\{1+q^2T + (q^{2N}+2q^4)T^2
+ 2q^{2N+2} T^3 +q^{4+2N}T^4+ (q^{4N+2}+q^{2N+6})T^5 + q^{4N+4}T^6\Big\}+
$$
\vspace{-0.6cm}
\be
+ q^{4N+3}[N-1]T^4
\label{KR623v2r}
\ee
-- in agreement with \cite{CM}.

\bigskip

In unreduced case (\ref{623v2promr}) changes for
\be
q^{4N+3}[N][N-1]T^4 +
q^{6N-8}T^{-2}\Big\{q^{8-N}T^5(1+q^2T)\Big(q^{1-N}+q^{N+1}T\Big)\Big(q^{3-N}+q^{N-1}T\Big)[N-1]
+\nn \\ +q^{5-2N}T^3(1+2q^2T+q^6T^3+q^8T^4)\Big(q^{1-N}+q^{N+1}T\Big)[N-1]
+ q^{5-3N}T^2[N] \Big\}  = \nn \\ \nn \\
= q^{4N-4}+q^{4N+3}[N][N-1]T^4 + q^{3N-4}\Big\{1+q^2T + (q^{2N+2}+2q^4)T^2
+ (2q^{2N+4}+\overline{q^8})T^3 + \nn \\
+ (q^{2N+4}+\underline{q^{2N+8}+q^{10}}+\overline{q^{8}})T^4 + (q^{4N+4}+\underline{q^{2N+8}+q^{10}}
+q^{2N+6}+\underline{\underline{q^{2N+10}}} )T^5
+ (q^{4N+6}+\underline{\underline{q^{2N+10}}})T^6\Big\}[N-1]
\nn
\ee

$$
{\cal P}^{6^2_3(v2)} =   q^{3N-3}[N]+q^{3N-4}\Big\{q^2T + (q^{2N+2}+2q^4)T^2
+ 2q^{2N+4} T^3 +q^{2N+4}T^4+ (q^{4N+4}+q^{2N+6})T^5 + q^{4N+6}T^6\Big\}[N-1]+
$$
\vspace{-0.6cm}
\be
+ q^{4N+3}[N][N-1]T^4
\label{KR623v2ur}
\ee
This is again in accordance with the answers for $N=2,3$ in \cite{CM}.
As usual, there are some pairs in unreduced polynomial that can seem to be
cohomologically trivial -- but since they were not such in reduced case,
they survive in unreduced polynomial as well.

\subsection{A thick knot $9_{42}$
\label{sec942}}

This knot is especially interesting for superpolynomial calculus,
because it is surrounded by certain controversies since the very first
analysis in \cite{DGR}.
It is represented by the 4-strand braid
\be
(1,1,1,-2,-1,-1,3,-2,3):\ \ \
{\cal R}_1^3{\cal R}_2^{-1}{\cal R}_1^{-2}{\cal R}_3{\cal R}_2^{-1}{\cal R}_3
\ \ \ \ \ \ \begin{array}{cccccccccccccccc}
\bullet&& \bullet && \bullet &&&&\circ &&\circ &&\\
&&&&&&\circ &&&&&&\circ \\
&&&&&&&&&&\bullet &&&&\bullet
\end{array}
\nn
\ee
The function $A_3^{9_{42}} \sim (1+T)$ is cohomologically trivial,
thus we can start from $A_2$.
\be
A_2^{9_{42}} = \left|\begin{array}{c|ccccccccccccc}
\hline &&\\
& 1 & q^2 & q^4 & q^6 & q^8 & q^{10} & q^{12} & q^{14} & q^{16}& q^{18} \\
&&\\ \hline && \\
1 & 1& 4&6&4&1\\
T & 3 & 20 & 43 & 38 & 12 &     \\
T^2 & 3&34 & 134 & 181 & 92 & 14&  \\
T^3 & 1 & 24&185& 444 & 361 & 116 & 11   \\
T^4 & & 6 & 125 & 544 & 745 & 398 & 85 & 5 &  \\
T^5 & &&44& 356&805& 698 & 261& 37 & 1&  \\
T^6 & &&9& 131 &466& 634& 383 & 102& 9 &  \\
T^7 & &&1& 26 & 144 & 293 & 273 & 120& 22 & 1  \\
T^8 & &&& 2 &21 & 64 & 88 & 60 & 19 & 2 \\
T^9 & &&&&1 & 5 & 10 & 10 & 5 & 1 \\
&&\\ \hline && \\
& (0)&(0)&(1)&(-2) & (2) &(-2)& (1) &(0)&(0)&(0)\\
&& \\ \hline
\end{array}\right| \ \nn \\ \sim \
\left|\begin{array}{c|ccccccccccccc}
\hline &&\\
& 1 & q^2 & q^4 & q^6 & q^8 & q^{10} & q^{12} & q^{14} & q^{16}& q^{18} \\
&&\\ \hline && \\
1 & &&\times    \\
T & &&&2\times    \\
T^2 & &&\otimes&&2\times  \\
T^3 & &&&2\otimes&&2\times  \\
T^4 & &&\times && 2\otimes&&\times  \\
T^5 & &&&2\times &&2\otimes  \\
T^6 & && && 2\times &&\otimes  \\
T^7 & &&& && 2\times \\
T^8 & &&&&&& \times \\
T^9 & &&&&& && \\
&&\\ \hline && \\
& (0)&(0)&(1)&(-2) & (2) &(-2)& (1) &(0)&(0)&(0)\\
&& \\ \hline
\end{array}\right|
\nn
\ee
This time there are three possible choices of diagonal --
as usual, we take the middle one.
The resulting contribution to reduced KR polynomial is
\be
\!\!\!\!
q^4T^2 (1+q^4T^2) (1+q^2T)^2\cdot q^{-N}[N-1][N-2]
\ \stackrel{(\ref{redur2})}{\sim}\
q^{4-N}T^2 (1+q^4T^2)\Big(q^{2-N}+q^NT\Big)\Big(q^{3-N}+q^{N-1}T\Big)
\ee
The next function

\centerline{
$
\!\!\!\!\!\!\!
A_1^{9_{42}} = \left|\begin{array}{c|ccccccccccccc}
\hline &&\\
& q & q^3 & q^5 & q^7 & q^9 & q^{11}& q^{13} &q^{15} &q^{17} \\
&&\\ \hline && \\
1 &  1 & 3 & 3& 1&   \\
T &  &9& 18& 9&   \\
T^2 & &11&58 & 58 & 11&  \\
T^3 & &10&92&186&92&10&  \\
T^4 & &5&78&279&279&79&5&  \\
T^5 & &1&37&218&379&218&37&1  \\
T^6 & &&9&93&261&261&93&9  \\
T^7 & &&1&21&95&150&95&21&1\\
T^8 & &&&2&17&41&41& 17&2\\
T^9 & &&&&1&4&6&4&1\\
&&\\ \hline && \\
& (1) & (-1)& (0) & (-1)&(1)&(-1)&(1) & (0)&(0)\\
&& \\ \hline
\end{array}\right|
\ \sim \
\left|\begin{array}{c|ccccccccccccc}
\hline &&\\
& q & q^3 & q^5 & q^7 & q^9 & q^{11}& q^{13} &q^{15} &q^{17} \\
&&\\ \hline && \\
1 & \otimes    \\
T & & \otimes && \times    \\
T^2 && &\boxed{\otimes}&&\times \\
T^3 && &\boxed{\otimes}&\!\!2\boxed{\otimes}\ &&\times   \\
T^4 && &&\boxed{\otimes}&\otimes &&\times   \\
T^5 && && \times &&\otimes   \\
T^6 && &&&\times &&\otimes   \\
T^7 && &&&&\times \\
T^8 && &&&&&\times  \\
T^9 & \\
&&\\ \hline && \\
& (1) & (-1)& (0) & (-1)&(1)&(-1)&(1) & (0)&(0)\\
&& \\ \hline
\end{array}\right|
$
}

\bigskip

\noindent
The four crosses in boxes are non-minimal -- alternated sums allow to skip them.
Still, one of them completes the main diagonal and can look desirable.
In the next formulas we put their contributions in boxes to show what their role is.
The main diagonal contributes
\be
q\Big(1 +\boxed{q^4(T^2+T^3)}+ T^3q^6+T^5q^{10}\Big) \cdot q^{-2N}(1+q^2T)[N-1]\nn \\
\ \stackrel{(\ref{redur})}{\sim}\
q^{1-2N}\Big(1 +\boxed{q^4(T^2+T^3)}+ T^3q^6+T^5q^{10}\Big)\Big(q^{2-N}+q^NT\Big)
\nn
\ee
Finally,
\be
A^{9_{42}}_0 = q^7T^4
\ee
Collecting all the three contributions we obtain:
\be
\!\!\!\!\!\!
q^{N-5}T^{-4}\Big\{q^{4-N}T^2 (1+q^4T^2)\Big(q^{2-N}+q^NT\Big)\Big(q^{3-N}+q^{N-1}T\Big)
+\nn \\
+ q^{1-2N}\Big(1+\boxed{q^4(T^2+T^3)} + T^3q^6+T^5q^{10}\Big)\Big(q^{2-N}+q^NT\Big) + q^{7-3N}T^4\Big\} =
\nn \\ \nn \\
= \frac{1}{q^{2N+2}T^4} + \frac{1}{q^4T^3} + \frac{\overline{q^{4-2N}}}{T^2}+
(\overline{q^{4-2N}}+1+\underline{q^2})\frac{1}{T} +
(\widetilde{q^{2-2N}}+\underline{\underline{q^{8-2N}}}+\underline{q^2} + q^{2N-2}) + \nn \\
+ \boxed{ \frac{q^{2-2N}}{T^2} +  \widetilde{\frac{q^{2-2N}}{T}}+\frac{1}{T} + 1} \ \
+ (\underline{\underline{q^{8-2N}}}+q^4 + \overline{q^6})T + (\overline{q^6}+q^{2N+2})T^2
\label{P942pro}
\ee
If we neglected the terms in boxes, then
elimination of all the underlined and overlined cohomologically trivial pairs would give:
\be
\frac{1}{q^{2N+2}T^4} + \frac{1}{q^4T^3} +\frac{1}{T}
+(q^{2-2N}+q^{2N-2}) +q^4T + q^{2N+2}T^2
\ee
This is, however, inconsistent with \cite{DGR,GS}: in particular,
it does not reproduce properly the Jones-Khovanov polynomial at $N=2$.
To get the right answer we should take into account the boxed terms.
Then the tilded pair can also be eliminated and we obtain
a non-minimal answer
\be
\boxed{
{\cal P}^{9_{42}}_r = \frac{1}{q^{2N+2}T^4} + \frac{1}{q^4T^3} +
\frac{q^{2-2N}}{T^2}+\frac{2}{T}
+(1+q^{2N-2}) +q^4T + q^{2N+2}T^2
}
\label{KR942r}
\ee
which is in nice accordance with the old suggestion of \cite{DGR} and \cite{GS},
see eq.(B1) in  the last reference.

Thus we once again confirmed the subtlety of the $9_{42}$ case.
We showed once again, that KR polynomial -- if the four boxed items are included
into cohomologies -- just {\it coincides} with the superpolynomial of \cite{DGR,GS}
at $a=q^N$, no additional reductions are needed.
On one hand, we can not decide if addition of these items is indeed necessary --
before we explicitly construct and investigate the morphisms (which should be
the same for {\it all} link diagrams at once).
On another hand, if the boxed items are {\it not} added, then the well established
Jones-Khovanov polynomial at $N=2$ is {\it not} reproduced -- while it
should be a particular choice of KR polynomial in our construction.
Thus we believe that they should be added -- and then KR polynomial and
superpolynomial are just the same -- at least for $9_{42}$ and other examples
in this paper.
Still, at this level (without explicit morphisms) we refrain from a final judgement.


\bigskip

Anyhow, we can now immediately obtain the  unreduced counterpart of
(\ref{KR942r}) -- with additional terms in boxes.
Instead of (\ref{P942pro}) we now have:

\be
\!\!\!\!\!\!
q^{N-5}T^{-4}\Big\{q^{4-N}T^2 (1+q^4T^2)\Big(q^{1-N}+q^{N+1}T\Big)\Big(q^{3-N}+q^{N-1}T\Big)[N-1]
+\nn \\
+ q^{1-2N}\Big(1+\boxed{q^4(T^2+T^3)} + T^3q^6+T^5q^{10}\Big)\Big(q^{1-N}+q^{N+1}T\Big)[N-1]
+ q^{7-3N}T^4\underbrace{[N]}_{\!\!\!\!\!\!\!\!\!\!\!\!\!\!\!\!\frac{1}{q}[N-1]+q^{N-1}
\!\!\!\!\!\!\!\!\!\!\!\!\!\!\!\!}\Big\} =
\nn \\ \nn \\
= q^{1-N}+\Big\{\frac{1}{q^{2N+3}T^4} + \frac{1}{q^3T^3} + \frac{\overline{q^{3-2N}}}{T^2}+
(\overline{q^{3-2N}}+q^{-1}+\underline{q^3})\frac{1}{T} +
(\widetilde{q^{1-2N}}+\underline{\underline{q^{7-2N}}}+\underline{q^3} + q^{2N-1}) + \nn \\
+ \boxed{ \frac{q^{1-2N}}{T^2} +  \widetilde{\frac{q^{1-2N}}{T}}+\frac{q}{T} + q} \ \
+ (\underline{\underline{q^{7-2N}}}+q^3 + \overline{q^7})T + (\overline{q^7}+q^{2N+3})T^2\Big\}[N-1]
\nn
\ee
and
\be
{\cal P}^{9_{42}} = q^{1-N} + \left(\frac{1}{q^{2N+3}T^4} + \frac{1}{q^3T^3} +
\frac{q^{1-2N}}{T^2}+\frac{[2]}{T}
+(q+q^{2N-1}) +q^3T + q^{2N+3}T^2\right)[N-1]
= \nn \\
\boxed{ = \overbrace{[N]}^{\!\!\!\! \!\!\!\!q^{1-N}+q[N-1]
\!\!\!\!\!\!\!\!\!\!\!\!\!\!\!\!}
+ \left(\frac{1}{q^{2N+3}T^4} +\frac{q^{1-2N}}{T^2}+\frac{1}{qT} + q^3T \right)(1+q^{2N}T)[N-1]
}
\label{KR942ur}
\ee
Somewhat remarkably, in this expression there are no longer pairs,
which look cohomologically trivial: the item $q$, which could be combined with $q/T$,
is naturally absorbed into $[N]$.

\section{Beyond braids}

\setcounter{equation}{0}

Concise technique from the previous section
has a lot of advantages.
It overcomes explicit construction of spaces $v$ and works directly
with the primary polynomial $\P^{\cal L}$.
We remind that coefficients in front of powers of $T$ in $\P$
are spaces in the KR complex, which  are complicated sums over
vertices of the hypercube, i.e. complicated (big) direct sums
of the spaces $v$.
The potential drawback of this approach is that it requires
a construction of differentials, which does not directly refer
to elementary morphisms between the spaces $v$, which act along
the hypercube edges.
We demonstrated above that differentials possess some recognizable
properties, but it is unclear, whether these mnemonic rules
can be finally promoted to a self-sufficient theory,
not referring to morphisms at all.

Anyhow, whatever can be done directly with $\P$ has advantages,
because this primary polynomial is as simply calculable, as HOMFLY.
So far, including the previous section, we mostly relied
upon braid representation, what is justified, because from
\cite{MMMkn1,Anopaths} we know an absolutely explicit
and {\it practically working} construction
for arbitrary braid (moreover, in \cite{AnoMcabling} it is extended
to colored polynomials -- though we do not need them at present,
they can be important for the study of colored KR polynomials,
if they exist).
Still, it often happens that simple knots/links are represented
by rather complicated braids -- and then it can make sense to
use alternative constructions of $\P$.
In what follows we discuss just two examples.
The first will be the series of twist knots and antiparallel 2-strand braids --
a very simple one, but with a braid representation, where the number
of strands grows along the series, and this gets increasingly complicated.
The second  will be the skein relations -- since we are dealing with
the fundamental representation, they can be easily modified
for evaluation of $\P$ instead of HOMFLY -- and thus applied
to the same extent as the usual skein relations are.

\subsection{Twist knots}

For detailed description of knot polynomials for twist knots
see  sec.5.2 of \cite{evo} and especially sec.5.7 of \cite{DM3}.
We borrow an introductory piece from \cite{DM3} for a brief description.

Twist knots form one of the simplest 1-parametric families
(see, for example, sec.5.2 of \cite{evo}),
which includes unknot, trefoil and the figure-eight knot $4_1$.
They are made out of the $2$-strand braid, only
-- in variance with the torus knots -- anti-parallel:

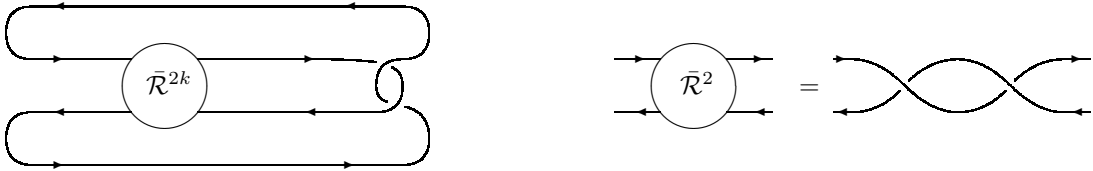
\begin{picture}(200,100)(-80,-50)
\put(0,0){\circle{30}}
\put(-50,-10){\line(1,0){37.5}}
\put(-50,10){\line(1,0){37.5}}
\put(80,-10){\line(-1,0){67.5}}
\put(60,10){\line(-1,0){47.5}}
\put(-50,-30){\line(1,0){140}}
\qbezier(-50,-10)(-60,-10)(-60,-20)
\qbezier(-50,-30)(-60,-30)(-60,-20)
\qbezier(91,-8)(100,-10)(100,-20)
\qbezier(90,-30)(100,-30)(100,-20)
\qbezier(-50,10)(-60,10)(-60,20)
\qbezier(-50,30)(-60,30)(-60,20)
\qbezier(90,10)(100,10)(100,20)
\qbezier(90,30)(100,30)(100,20)
\put(-50,30){\line(1,0){140}}
\qbezier(90,10)(80,10)(80,0)
\qbezier(80,0)(80,-5)(84,-6)
\qbezier(80,-10)(90,-10)(90,0)
\qbezier(90,0)(90,5)(86,7)
\qbezier(80,9)(70,10)(60,10)
\put(-40,30){\vector(-1,0){2}}
\put(-40,10){\vector(1,0){2}}
\put(-40,-10){\vector(-1,0){2}}
\put(-40,-30){\vector(1,0){2}}
\put(70,30){\vector(-1,0){2}}
\put(55,10){\vector(1,0){2}}
\put(55,-10){\vector(-1,0){2}}
\put(70,-30){\vector(1,0){2}}
\put(-7,-3){\mbox{$\bar {\cal R}^{2k}$}}
%
%
%
%
\put(200,0){\circle{30}}
\put(170,-10){\line(1,0){17.5}}
\put(170,10){\line(1,0){17.5}}
\put(230,-10){\line(-1,0){17.5}}
\put(230,10){\line(-1,0){17.5}}
\put(195,-3){\mbox{$\bar {\cal R}^2$}}
\put(240,-3){\mbox{$=$}}
\put(253,10){\line(1,0){7}}
\put(253,-10){\line(1,0){7}}
\put(340,10){\line(1,0){10}}
\put(340,-10){\line(1,0){10}}
\qbezier(260,10)(270,10)(280,0)
\qbezier(280,0)(290,-10)(300,-10)
\qbezier(300,10)(310,10)(320,0)
\qbezier(320,0)(330,-10)(340,-10)
\qbezier(260,-10)(270,-10)(278,-2)
\qbezier(281.5,2)(290,10)(300,10)
\qbezier(300,-10)(310,-10)(318,-2)
\qbezier(321.5,2)(330,10)(340,10)
\put(180,10){\vector(1,0){2}}
\put(180,-10){\vector(-1,0){2}}
\put(225,10){\vector(1,0){2}}
\put(225,-10){\vector(-1,0){2}}
\put(256,10){\vector(1,0){2}}
\put(256,-10){\vector(-1,0){2}}
\put(345,10){\vector(1,0){2}}
\put(345,-10){\vector(-1,0){2}}
\end{picture}

\noindent
Here  $k$ can be both positive and negative.
If the number of crossing in the antiparallel braid is
odd, this changes orientation at the two-vertex "locking block".
The corresponding knot diagrams
(after rotation by $90^\circ$) are:

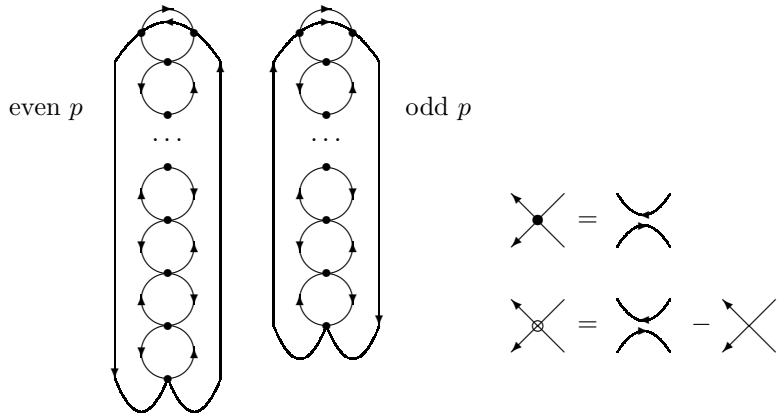
\begin{picture}(100,170)(-150,-90)
\put(-6,20){\mbox{$\ldots$}}
\put(0,0){\circle{20}}
\put(0,-20){\circle{20}}
\put(0,-40){\circle{20}}
\put(0,-60){\circle{20}}
\put(-10,0){\vector(0,1){2}}
\put(-10,-20){\vector(0,-1){2}}
\put(-10,-40){\vector(0,1){2}}
\put(-10,-60){\vector(0,-1){2}}
\put(10,0){\vector(0,-1){2}}
\put(10,-20){\vector(0,1){2}}
\put(10,-40){\vector(0,-1){2}}
\put(10,-60){\vector(0,1){2}}
\put(0,10){\circle*{3}}
\put(0,-10){\circle*{3}}
\put(0,-30){\circle*{3}}
\put(0,-50){\circle*{3}}
\put(0,-70){\circle*{3}}
\put(-60,30){\mbox{even $p$}}
\put(90,30){\mbox{odd $p$}}
\put(54,20){\mbox{$\ldots$}}
\put(60,0){\circle{20}}
\put(60,-20){\circle{20}}
\put(60,-40){\circle{20}}
\put(50,0){\vector(0,1){2}}
\put(50,-20){\vector(0,-1){2}}
\put(50,-40){\vector(0,1){2}}
\put(70,0){\vector(0,-1){2}}
\put(70,-20){\vector(0,1){2}}
\put(70,-40){\vector(0,-1){2}}
\put(60,10){\circle*{3}}
\put(60,-10){\circle*{3}}
\put(60,-30){\circle*{3}}
\put(60,-50){\circle*{3}}
%
%
\put(0,60){\circle{20}}
\put(-10,61){\circle*{3}}\put(10,61){\circle*{3}}
\put(0,70){\vector(1,0){2}}\put(0,65){\vector(-1,0){2}}
\qbezier(-20,50)(0,80)(20,50)
\put(0,40){\circle{20}}\put(0,50){\circle*{3}}\put(0,30){\circle*{3}}
\put(-10,40){\vector(0,-1){2}}\put(10,40){\vector(0,1){2}}
\qbezier(0,-70)(-10,-95)(-20,-70)
\put(-20,50){\vector(0,-1){120}}
\qbezier(0,-70)(10,-95)(20,-70)
\put(20,-70){\vector(0,1){120}}
\put(60,60){\circle{20}}
\put(50,61){\circle*{3}}\put(70,61){\circle*{3}}
\put(60,70){\vector(1,0){2}}\put(60,65){\vector(1,0){2}}
\qbezier(40,50)(60,80)(80,50)
\put(60,40){\circle{20}}\put(60,50){\circle*{3}}\put(60,30){\circle*{3}}
\put(50,40){\vector(0,-1){2}}\put(70,40){\vector(0,1){2}}
\qbezier(60,-50)(50,-75)(40,-50)
\put(80,50){\vector(0,-1){100}}
\qbezier(60,-50)(70,-75)(80,-50)
\put(40,-50){\vector(0,1){100}}
\put(150,-20){\vector(-1,1){20}}\put(150,0){\vector(-1,-1){20}}
\put(140,-10){\circle*{4}}
\put(155,-12){\mbox{$=$}}
\qbezier(170,0)(180,-16)(190,0) \put(180,-8){\vector(-1,0){2}}
\qbezier(170,-20)(180,-4)(190,-20) \put(180,-12){\vector(1,0){2}}
%
\put(150,-60){\vector(-1,1){20}}\put(150,-40){\vector(-1,-1){20}}
\put(140,-50){\circle{4}}
\put(155,-52){\mbox{$=$}}
\qbezier(170,-40)(180,-56)(190,-40) \put(180,-48){\vector(-1,0){2}}
\qbezier(170,-60)(180,-44)(190,-60) \put(180,-52){\vector(1,0){2}}
\put(198,-52){\mbox{$-$}}
\put(230,-60){\vector(-1,1){20}}\put(230,-40){\vector(-1,-1){20}}
%
\end{picture}

\noindent
The two pictures correspond respectively to the cases when the number $p$
of circles, and thus the number $p+2$ of vertices, is even and odd.
The two cases are essentially different,
already when all the vertices are black, i.e. at the
{\it main} Seifert vertex of the hypercube,
configurations are not the same:

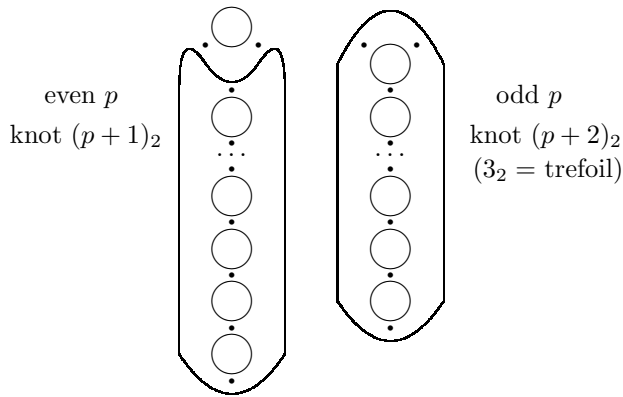
\begin{picture}(100,160)(-200,-80)

\put(60,50){\circle{15}}
\put(60,30){\circle{15}}
\put(54,15){\mbox{$\ldots$}}
\put(60,0){\circle{15}}
\put(60,-20){\circle{15}}
\put(60,-40){\circle{15}}
\put(40,-40){\line(0,1){90}}
\put(80,-40){\line(0,1){90}}
\qbezier(40,-40)(60,-70)(80,-40)
\qbezier(40,50)(60,90)(80,50)
\put(60,40){\circle*{2}}
\put(60,20){\circle*{2}}
\put(60,10){\circle*{2}}
\put(60,-10){\circle*{2}}
\put(60,-30){\circle*{2}}
\put(60,-50){\circle*{2}}
%
\put(50,57){\circle*{2}}
\put(70,57){\circle*{2}}
\put(-71,35){\mbox{even $p$}}
\put(100,35){\mbox{odd $p$}}
\put(-84,20){\mbox{knot $(p+1)_2$}}
\put(90,20){\mbox{knot $(p+2)_2$}}
\put(91,7){\mbox{($3_2=$ trefoil)}}
\put(0,64){\circle{15}}
\put(0,30){\circle{15}}
\put(-6,15){\mbox{$\ldots$}}
\put(0,0){\circle{15}}
\put(0,-20){\circle{15}}
\put(0,-40){\circle{15}}
\put(0,-60){\circle{15}}
\put(-20,-60){\line(0,1){100}}
\put(20,-60){\line(0,1){100}}
\qbezier(-20,-60)(0,-90)(20,-60)
\qbezier(-20,40)(-20,65)(-10,50)\qbezier(-10,50)(0,36)(10,50)\qbezier(20,40)(20,65)(10,50)
\put(0,40){\circle*{2}}
\put(0,20){\circle*{2}}
\put(0,10){\circle*{2}}
\put(0,-10){\circle*{2}}
\put(0,-30){\circle*{2}}
\put(0,-50){\circle*{2}}
\put(0,-70){\circle*{2}}
\put(-10,57){\circle*{2}}
\put(10,57){\circle*{2}}
\end{picture}

\noindent
Still, the number of cycles in the both  cases is $p+1$,
so that in the both cases the  hypercube vertex $b^{p+2}$ contributes
$N^{p+1}\longrightarrow [N]^{p+1}$.
When some vertex is changed from black to white, one subtracts a contribution
with a crossing at this vertex, what changes the number of cycles:
for example, when there is just one white vertex, subtraction contains $p$ cycles,
and the contribution of $b^{p+1}w$ vertex in the hypercube is $N^{p+1}-N^p
\longrightarrow [N]^p[N-1]$.

When all vertices are of the same color then the knot is
$(p+1)_2$ for even $p$ and $(p+2)_2$ for odd $p$.
If the two vertices at the top (two "horizontal" vertices)
have the opposite color to the $p$ vertical ones,
then the knot is $(p+2)_1$ for even $p$ and $(p+1)_1$ for odd $p$.
When the two horizontal vertices are of different colors,
we get an unknot. If some vertical vertices have different colors,
what matters is their algebraic sum, $p \equiv \sharp_b - \sharp_w$.
The answer for HOMFLY polynomials of the twist knots
is well known, see, for example, sec.5.2 of \cite{evo}:
\be
H_k = 1 +
F_k(A^2)\{Aq\}\{A/q\} =
1 + F_k\left(q^{2N} \right)\left(q^{2N}- q^2-q^{-2}+q^{-2N}\right)
\label{twisho}
\ee
with $F_k(A^2) = -A^{k+1}\{A^{k}\}/\{A\}$.
For $k=0$ and $F_0=0$ we get unknot,
for $k=1$ and $F_{1}=-A^2$ -- the trefoil $3_1$
and for $k=-1$ and $F_{-1}=1$ -- the figure eight knot $4_1$.
More generally, for positive $k$ we get the knots $(2k+1)_2$,
while for negative $k$ -- $(2-2k)_1$ in the Rolfsen notation,
see \cite{katlas}.
Note that trefoil $3_1$ gets its right place in the series
of twisted knots, if treated as $3_2$.

\bigskip

\be
\begin{array}{c|ccccccccccccccccccccccc}
p && 1 &\ \ & 2 &\ \ & 3 &\ \ & 4 &\ \ & 5 &\ \ & 6 &\ \ & 7 &\ \ & 8 &\ \ & 9 & \ \
& \ldots   \\  & \\
\begin{array}{ccc} \bullet\!\!\!\!\!\! &&\!\!\!\!\!\!\bullet\\ &\bullet &\\ &\bullet &\\ &
\ldots & \\ &\bullet & \end{array}
&& 3_2 && 3_2 && 5_2 && 5_2 && 7_2 && 7_2 && 9_2 && 9_2 && 11_2 &&  \ldots  \\
&k&&1&&&&2&&&&3&&&&4&&&&5&
\\
&m&&2&&&&3&&&&4&&&&5&&&&6&
\\&\\
& \\
\begin{array}{ccc} \bullet\!\!\!\!\!\! &&\!\!\!\!\!\!\bullet\\ &\circ &\\ &\circ &\\ &
\ldots & \\ &\circ & \end{array}
&&{\rm unknot} && 4_1 && 4_1 && 6_1 && 6_1 && 8_1 && 8_1 && 10_1 && 10_1 && \ldots \\
&k&&&&-1&&&&-2&&&&-3&&&&-4&
\\
&m&&&&3&&&&4&&&&5&&&&6&
\\
\end{array}
\nn
\label{twiknots}
\ee

\bigskip

\noindent
We  added to this table the lines with the values of $k$, which appears in (\ref{twisho}),
and of the minimal number $m$ of strands in the braid representation.
We see that $m$ grows linearly with $p$ -- this is what makes consideration
of the family of twist knots orthogonal to that of the families with given $m$,
which we concentrated on in the present paper.

However, for twist knots
we know the spaces $v$ (they were found in paper\footnote{
The second version of that paper is needed, the first version
 (and the journal one) was inadequate in the twisted-knot section 5.7.} \cite{DM3}
-- and this is enough to apply the technique that we developed here.
We denote by $i$ the number of white vertices on the vertical,
thus $0\leq i\leq p$.
There are $C^i_p$ vertices of each type, and
\be
v_{i}^{\bullet\bullet} = [N]^{p+1-i}[N-1]^i \nn \\
v_{i}^{\bullet\circ} =  v_i^{\circ\bullet} = [N]^{p-i}[N-1]^{i+1} \nn \\
v_{i}^{\circ\circ} = [2][N]^{p-i}[N-1]^{i+1}
\ee
for $0\leq i\leq p-1$ while for $i=p$ dimensions are more complicated.
Moreover, they are different,
depending on the parity of $p$:
for odd $p$
\be
v_{p}^{\bullet\bullet} = [N] [N-1]^p \nn \\
v_{p}^{\bullet\circ} =  v_i^{\circ\bullet} = [N] [N-1]\cdot Y_p \nn \\
v_{p}^{\circ\circ} = [2][N][N-1]\cdot Y_p
\ee
and for even $p$
\be
v_{p}^{\bullet\bullet} = [N]^2[N-1]\cdot Y_{p-1} \nn \\
v_{p}^{\bullet\circ} =  v_i^{\circ\bullet} = [N] [N-1]^2\cdot Y_{p-1} \nn \\
v_{p}^{\circ\circ} = [N] [N-1]^p+[N][N-1][N-2]\cdot Y_{p-1}
\ee
where
\be
Y_{k+2}=[N-1]^k[N-2]+Y_k = [2]+\sum_{i=0}^{\frac{k-1}{2}}[N-1]^{2i+1}[N-2]
\ee

These formulas provide primary polynomials
and it is a straightforward exercise to extract from them
the KR polynomials.
In the following section we use this knowledge for a closely related
family, which includes $6^2_1(v2)$ --
the last 2-component link  from the list of \cite{CM},
that we have not yet described.
In fact, a minor generalization is needed for $6^3_1(v1)$ -- and we complete
filling the list of \cite{CM} by making this example in s.\ref{3compl}.

\subsection{Two-component links from two antiparallel strands}

The knowledge of dimensions, relevant for twist knots allows one also to solve
an even simpler problem: of 2-component knots from the series
$(2p)^2_1$ with inverse orientation, i.e. $2^2_1$, $4^2_1(v1)$, $6^2_1(v2)$, $\ldots$
in notation of \cite{CM}.
As ordinary braids they are increasingly complicated -- the number of strands grows
typically as $2p$, and the {\it a posteriori} very simple example of $6^2_1$  would
require a 6-strand analysis.
If we instead use the technique, just applied to twist knots, the story gets nearly trivial.
Dimensions $v$ in this case can be just taken from the set $v^{\bullet\bullet}_i$,
with two factors of $[N]$ thrown away (and also we change $p$ for $2p$
to simplify some formulas).

The link diagram and associated pattern of $2p$ Seifert cycles,
obtained when all the vertices are black, look as follows:


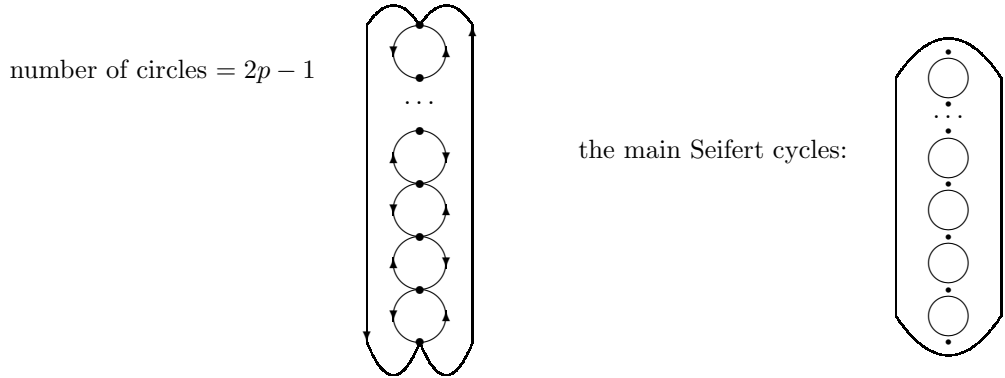
\begin{picture}(100,170)(-150,-90)
\put(-6,20){\mbox{$\ldots$}}
\put(0,0){\circle{20}}
\put(0,-20){\circle{20}}
\put(0,-40){\circle{20}}
\put(0,-60){\circle{20}}
\put(-10,0){\vector(0,1){2}}
\put(-10,-20){\vector(0,-1){2}}
\put(-10,-40){\vector(0,1){2}}
\put(-10,-60){\vector(0,-1){2}}
\put(10,0){\vector(0,-1){2}}
\put(10,-20){\vector(0,1){2}}
\put(10,-40){\vector(0,-1){2}}
\put(10,-60){\vector(0,1){2}}
\put(0,10){\circle*{3}}
\put(0,-10){\circle*{3}}
\put(0,-30){\circle*{3}}
\put(0,-50){\circle*{3}}
\put(0,-70){\circle*{3}}
\put(-155,30){\mbox{number of circles $=2p-1$ }}
\put(0,40){\circle{20}}\put(0,50){\circle*{3}}\put(0,30){\circle*{3}}
\put(-10,40){\vector(0,-1){2}}\put(10,40){\vector(0,1){2}}
\qbezier(0,50)(-10,65)(-20,50)
\qbezier(0,-70)(-10,-95)(-20,-70)
\put(-20,50){\vector(0,-1){120}}
\qbezier(0,50)(10,65)(20,50)
\qbezier(0,-70)(10,-95)(20,-70)
\put(20,-70){\vector(0,1){120}}
%
%
\put(200,30){\circle{15}}
\put(194,15){\mbox{$\ldots$}}
\put(200,0){\circle{15}}
\put(200,-20){\circle{15}}
\put(200,-40){\circle{15}}
\put(200,-60){\circle{15}}
\put(180,-60){\line(0,1){90}}
\put(220,-60){\line(0,1){90}}
\qbezier(180,-60)(200,-90)(220,-60)
\qbezier(180,30)(200,60)(220,30)
\put(200,40){\circle*{2}}
\put(200,20){\circle*{2}}
\put(200,10){\circle*{2}}
\put(200,-10){\circle*{2}}
\put(200,-30){\circle*{2}}
\put(200,-50){\circle*{2}}
\put(200,-70){\circle*{2}}
\put(60,0){\mbox{the main Seifert cycles:}}
\end{picture}

\bigskip

\noindent
From this picture it is clear that the cycle diagram for, say, $p=2$, is
\be
\begin{array}{ccccccccc}
&&&&2\\
&&3&&&&1 \\
&&&&2\\
&&3&&&&1 \\
&&&&2\\
4&&&&&&&\boxed{2}\\
&&&&2\\
&&3&&&&1\\
&&&&2\\
&&3&&&&1\\
&&&&2\\
\end{array}
\ \ \ \ \ \   \Longrightarrow \ \ \ \ \ \
\begin{array}{ccccccccc}
&&&&N^2(N-1)^2\\
&&N^3(N-1)&&&&N(N-1)^3 \\
&&&&N^2(N-1)^2\\
&&N^3(N-1)&&&&N(N-1)^3 \\
&&&&N^2(N-1)^2\\
N^4&&&&&&&\!\!\!\!\!\!\!\!\!\!\!\!\boxed{N(N-1)(N^2-3N+4)}\\
&&&&N^2(N-1)^2\\
&&N^3(N-1)&&&&N(N-1)^3\\
&&&&N^2(N-1)^2\\
&&N^3(N-1)&&&&N(N-1)^3\\
&&&&N^2(N-1)^2\\
\end{array}
\nn
\ee
The right table lists the classical dimensions of vector spaces
at the $2^{2p}$ vertices of the hypercube.
As usual for antiparallel braids the pattern is absolutely regular
with the single exception in the box.
Accordingly the table of quantum dimensions $v$ is
\be
\l[N] \times \ \ \ \left(
\begin{array}{ccccccccc}
&&&&[N][N-1]^2\\
&&[N]^2[N-1]&&&&[N-1]^3 \\
&&&&[N][N-1]^2\\
&&[N]^2[N-1]&&&&[N-1]^3 \\
&&&&[N][N-1]^2\\
\l[N]^3&&&&&&& \boxed{[N-1]\cdot Y_{3}}\\
&&&&[N][N-1]^2\\
&&[N]^2[N-1]&&&&[N-1]^3 \\
&&&&[N][N-1]^2\\
&&[N]^2[N-1]&&&&[N-1]^3 \\
&&&&[N][N-1]^2\\
\end{array}    \right)
\nn
\ee
where we took the common factor $[N]$ away -- so that these $v$
generate reduced primary polynomial, as everywhere in this paper.
For arbitrary $p$ the two entries of the central line are
$[N]^{2p-1}$ and $[N-1]\cdot Y_{2p-1}$.

Thus the primary polynomial for given $p$ is just
\be
\P_r^{(2p)^2_1(v1)} = q^{2p(N-1)}
\left(\sum_{j=0}^{2p-1} C^j_{2p}  [N]^{2p-j-1}[N-1]^j\cdot (qT)^j
+ [N-1]Y_{2p-1}\cdot (qT)^{2p}\right)
\ee
where
\be
Y_{2p-1} = [2]+\sum_{j=0}^{p-2}[N-1]^{2j+1}[N-2]
\ee

Minimization procedure, i.e. evaluation of cohomologies
gets clear from several examples.

\subsubsection{The Hopf link $2^2_1$ ($p=1$)}

\be
q^{2-2N}\P^{2^2_1(v1)} _r =  [N] +2qT[N-1]+[N-1]\overbrace{[2]}^{Y_1}(qT)^2 = q^{1-N} +
\underline{q[N-1](1+2T+T^2)}+q^3[N-1]T^2
\ee
The only tricks here are to reduce $[N]$ to $[N-1]$ via the usual
$[N]=q[N-1]+q^{1-N}$ and divide the contribution from $Y_1$ into two pieces.
This is the end of the story: we throw away the underlined cohomologically trivial
piece and obtain the answer (\ref{KRr22}).
Unreduced case requires this time just an additional multiplication by $[N]$,
exactly like in (\ref{KRr22}).
Thus, once again:
\be
{\cal P}^{2^2_1}_r = q^{N-1}\Big(1+q^{N+2}T^2[N-1]\Big),\nn \\
{\cal P}^{2^2_1} = q^{N-1}\Big([N]+q^{N+2}T^2[N][N-1]\Big)
\ee

\subsubsection{The Solomon link in opposite orientation $4^2_1(v1)$ ($p=2$)}

\be
\!\!\!\!\!q^{4-4N}\P^{4^2_1(v1)} _r =  [N]^3 +4qT[N]^2[N-1]+6(qT)^2[N][N-1]^2 + 4(qT)^3[N-1]^3
+ [N-1]\Big(\overbrace{[N-1][N-2]+[2]}^{Y_3}\Big)(qT)^4   =
\nn\\
\!\!\!\!\!\!\!\!\!= q^3[N-1]^3(1+4T+6T^2+4T^3)
+ q^{3-N}[N-1]^2(3+8T+6T^2)
+ q^{3-2N}[N-1](3+4T)
+ q^{3-3N} + q^4T^4[N-1]Y_3
\sim\nn\\
\!\!\!\!\!\!\!\sim q^3T^3[N-1]^3 + q^{3-N}T^2[N-1]^2+q^{3-2N}T[N-1]+\boxed{q^{3-3N}}
+ q^4T^4\Big([N-1]^2[N-2]+[2][N-1]\Big)
\nn
\ee
So far we reduced all $[N]$ to $[N-1]$ and eliminated
cohomologically trivial factors.
Now we pick up two terms and reduce $[N-1]$ to $[N-2]$ with the help
of $[N-1]=q[N-2]+q^{2-N}$:
\be
q^3T^3[N-1]^3 + q^4T^4[N-1]^2[N-2] = q^3T^3[N-1]^2\Big([N-1]+q[N-2]T\Big)
\sim q^{5-N}T^3[N-1]^2
\ee
Now we combine this new term with the next one:
\be
q^{5-N}T^3[N-1]^2 +q^{3-N}T^2[N-1]^2 =
q^{3-N}T^2(1+q^2T)[N-1]^2 \sim \nn \\
\sim q^{3-N}T^2[N-1]\Big(q^{2-N}+q^NT\Big)
= q^{5-2N}[N-1]T^2 + q^3T^3[N-1]
\ee
Now the first of these new terms is combined with the next one in the original expression:
\be
q^{5-2N}[N-1]T^2+q^{3-2N}T[N-1] \sim
q^{3-2N}T\Big(q^{2-N}+q^NT\Big) =
\boxed{q^{5-3N}T +q^{3-N}T^2}
\ee
while the second one -- with a part of the remaining piece from
$Y_3$, by splitting $[2]=\boxed{q}+q^{-1}$:
\be
q^3T^3[N-1]+q^4T^4[N-1]\cdot q^{-1} = q^3T^3[N-1](1+T)\sim 0
\ee
What remains are the three items in  boxes and one last fragment of $Y_3$,
associated with the piece of $[2]$, which was also boxed.
Putting all the four pieces together, we get:
\be
{\cal P}^{4^2_1(v1)}_r = q^{4N-4}\Big(q^{3-3N}+q^{5-3N}T +q^{3-N}T^2
+ q^5T^4[N-1]\Big) = \nn \\
=q^{N-1}\Big(1+q^2T+q^{2N}T^2+q^{3N+2}T^4[N-1]\Big)
\ee
what reproduces (\ref{KR421v1}) -- in fact, by a considerably simpler computation.
It is trivial to make the necessary corrections in unreduced case and get
\be
{\cal P}^{4^2_1(v1)} = q^{4N-4}\Big(q^{3-3N}[N]+(q^{4-3N}T +q^{4-N}T^2)[N-1]
+ q^5T^4[N][N-1]\Big) = \nn \\
= q^{N-1}\Big([N]+qT(1+q^{2N}T)[N-1]+q^{3N+2}T^4[N][N-1]\Big)
\ee
again in agreement with (\ref{KR421v1}) and \cite{CM}.

\subsubsection{The 2-component link $6^2_1(v2)$ or $L6a5$ ($p=3$)}

\be
\!\!\!\!\!q^{6-6N}\P^{6^2_1(v1)} _r =  [N]^5 +6qT[N]^4[N-1]+15(qT)^2[N]^3[N-1]^2
+ 20(qT)^3[N]^2[N-1]^3 +\nn \\
+15(qT)^4[N][N-1]^4 +6(qT)^4[N-1]^5
+ [N-1]\Big(\overbrace{[N-1]^2[N-2]+[N-1][N-2]+[2]}^{Y_5}\Big)(qT)^6   =
\nn\\ \nn \\
\!\!\!\!\!\!\!\!\!= q^5[N-1]^5(1+6T+15T^2+20T^3+15T^4+6T^5)
+ q^{5-N}[N-1]^4(5+24T+45T^2+40T^3+15T^4)+\nn \\
+ q^{5-2N}[N-1]^3(10+36T+45T^2+20T^3)+
+ q^{5-3N}[N-1]^2(10+24T+15T^2) + \nn \\
+q^{5-4N}[N-1](5+6T)
+ q^{5-3N} + q^4T^4[N-1]Y_5
\sim\nn\\ \nn \\
\!\!\!\!\!\!\!\sim q^5T^5[N-1]^5 + q^{5-N}T^4[N-1]^4+q^{5-2N}T^3[N-1]^3+\underline{q^{5-3N}T^2[N-1]^2}
+ \underline{\underline{q^{5-4N}T[N-1]}}+\boxed{q^{5-5N}}+\nn \\
+ q^6T^6\Big([N-1]^4[N-2]+[N-1]^2[N-2]+\underline{\underline{[2][N-1]}}\Big)
\nn
\ee
Now we pick up the first items from the last two lines:
\be
q^5T^5[N-1]^5 +q^6T^6[N-1]^4[N-2] =q^5T^5[N-1]^4\Big([N-1]+qT[N-2]\Big)
\sim q^{7-N}T^5[N-1]^4
\ee
Combining this with the second item of the first line,
\be
q^{7-N}T^5[N-1]^4+q^{5-N}T^4[N-1]^4 = q^{5-N}T^4(1+q^2T)[N-1]^4\sim \nn \\
\sim q^{5-N}T^4[N-1]^3\Big(q^{2-N}+q^NT\Big) =
q^{7-2N}T^4[N-1]^3 + q^5T^5[N-1]^3
\ee
we get {\it two} terms, and their fate is different.
The second is combined with the second term from $Y_5$:
\be
q^5T^5[N-1]^3+q^6T^6[N-1]^2[N-2] =
q^5T^5[N-1]^2([N-1]+qT[N-2])\sim \underline{q^{7-N}[N-1]^2T^5}
\ee
while the first one is combined with the third from the first line to provide
\be
q^{7-2N}T^4[N-1]^3+q^{5-2N}T^3[N-1]^3 =
q^{5-2N}T^3(1+q^2T)[N-1]^3 \sim \nn \\
\sim q^{5-2N}T^3[N-1]^2\Big(q^{2-N}+q^NT\Big) =
\underline{q^{7-3N}T^3[N-1]^2 + q^{5-N}[N-1]^2T^4}
\ee

Now we reached the next level, at which we have already four items --
they are now all in different places, and for convenience we underlined
all of them.
Of course, they nicely match:
\be
q^{5-3N}T^2[N-1]^2 + q^{7-N}[N-1]^2T^5 +
q^{7-3N}T^3[N-1]^2 + q^{5-N}[N-1]^2T^4 = \nn \\
= q^{5-3N}[N-1]^2(1+q^2T)\Big(q^2T^2+q^{2N}T^4\Big) \sim
\nn \\
\sim q^{5-3N}\Big(T^2+q^{2N}T^4\Big)[N-1]\Big(q^{2-N}+q^NT\Big)
= \Big(q^{7-4N}T^2+q^{5-2N}T^3+q^{7-2N}T^4+2q^{5}T^5\Big)[N-1]
\ee
To this one should a couple of terms linear in $[N-1]$,
underlined twice in original expression:
\be
\Big(q^{7-4N}T^2+q^{5-2N}T^3+q^{7-2N}T^4+2q^{5}T^5\Big)[N-1] +
q^{5-4N}T[N-1] + q^6T^6[2][N-1] = \nn \\
=q^{5-4N}T[N-1]\Big(1+q^2T+q^{2N}T^2+q^{2N+2}T^3+2\underline{q^{4N}T^4+q^{4N}T^5}\Big)
+\boxed{q^7T^6[N-1]}\sim \nn \\
\sim q^{5-4N}\Big(1+q^{2N}T^2\Big)(1+q^2T)[N-1] +\boxed{q^7T^6[N-1]}
\sim \boxed{q^{5-4N}\Big(1+q^{2N}T^2\Big)\Big(q^{2-N}+q^NT\Big) +q^7T^6[N-1]}
\nn
\ee
Collecting all the boxed terms we get:
\be
{\cal P}^{6^2_1(v2)}_r =
q^{6N-6}\left\{q^{5-5N} + q^{5-4N}\Big(1+q^{2N}T^2\Big)\Big(q^{2-N}+q^NT\Big) +q^7T^6[N-1]\right\}
= \nn \\
=q^{N-1}\left\{1 + \Big(q^NT+q^{3N}T^3\Big)\Big(q^{2-N}+q^NT\Big)
+q^{5N+2}T^6[N-1]\right\} = \nn \\
\boxed{
=q^{N-1}\Big(1+q^2T+q^{2N}T^2+q^{2N+2}T^3+q^{4N}T^4+q^{5N+2}T^6[N-1]\Big)
}
\label{KR621v2r}
\ee
and
\be
{\cal P}^{6^2_1(v2)} = q^{N-1}\left\{[N] + \Big(q^NT+q^{3N}T^3\Big)\Big(q^{1-N}+q^{N+1}T\Big)[N-1]
+q^{5N+2}T^6[N][N-1]\right\} = \nn \\
\boxed{
=q^{N-1}\left\{[N]+\Big(qT+q^{2N+1}T^2+q^{2N+1}T^3+q^{4N+1}T^4\Big)[N-1]+q^{5N+2}T^6[N][N-1]\right\}
}
\label{KR621v2ur}
\ee
These answers are in accordance with \cite{CM}.

\subsection{The 3-component link $L6a5$ ($6^3_1(v1)$ of \cite{CM}
\label{3compl}}

The link diagram $L6a5$ is

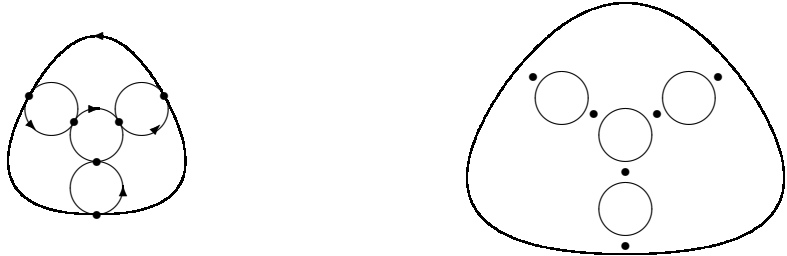
\begin{picture}(100,170)(-150,-90)
\put(0,0){\circle{20}}
\put(0,-20){\circle{20}}
\put(-17.1,10){\circle{20}}
\put(17.1,10){\circle{20}}
\qbezier(0,-30)(-50,-30)(-25.5,15)
\qbezier(0,-30)(50,-30)(25.5,15)
\qbezier(-25.5,15)(0,60)(25.5,15)
\put(0,-10){\circle*{3}}
\put(0,-30){\circle*{3}}
\put(-8.5,5){\circle*{3}}
\put(-25.5,15){\circle*{3}}
\put(8.5,5){\circle*{3}}
\put(25.5,15){\circle*{3}}
\put(0.5,37.5){\vector(-1,0){2}}
\put(-1,10){\vector(1,0){2}}
\put(10,-21){\vector(0,1){2}}
\put(22,2){\vector(1,1){2}}
\put(-25,4){\vector(1,-1){2}}
\put(200,0){\circle{20}}
\put(200,-28){\circle{20}}
\put(176,14){\circle{20}}
\put(224,14){\circle{20}}
\qbezier(200,-45)(110,-45)(155,20)
\qbezier(200,-45)(290,-45)(245,20)
\qbezier(155,20)(200,80)(245,20)
\put(200,-14){\circle*{3}}
\put(200,-42){\circle*{3}}
\put(188,8){\circle*{3}}
\put(165,22){\circle*{3}}
\put(212,8){\circle*{3}}
\put(235,22){\circle*{3}}
\put(0.5,37.5){\vector(-1,0){2}}
\put(-1,10){\vector(1,0){2}}
\put(10,-21){\vector(0,1){2}}
\put(22,2){\vector(1,1){2}}
\put(-25,4){\vector(1,-1){2}}
\end{picture}

\noindent
Orientation of the central circle is clockwise,
and counter-clockwise for the other four -- so that
the pattern of the main Seifert cycles (when all vertices are black)
is like in the picture on the right.

Therefore the cycle diagram is
\be
(5) \ \longrightarrow \ 6\times (4)  \ \longrightarrow \
15\times (3) \ \longrightarrow \ 20\times (2) \ \longrightarrow \
3\times (3) + 12\times (1) \ \longrightarrow \  6\times (2) \ \longrightarrow \ (3)
\nn
\ee
and classical dimensions at hypercube vertices  --
\be
N^5 \ \longrightarrow \ 6\times N^4(N-1)\ \longrightarrow \
15\times N^3(N-1)^2 \ \longrightarrow \ 20\times N^2(N-1)^3 \ \longrightarrow \
\nn \\
\ \longrightarrow \ 3\times N^2(N-1)(N^2-3N+4) + 12\times N(N-1)^4 \ \longrightarrow \
\nn \\
\ \longrightarrow \ 6\times N(N-1)^2(N^2-3N+4) \ \longrightarrow \
N(N-1)(N^3-5N^2+14N-12)
\nn
\ee
In the second line we easily recognize factors, which are classical values of
$Y_3 = [N-1][N-2]+[2]$, the factor in the third line is also
straightforward to quantize:
\be
Z =  [N-1][N-2]^2 + [2]^2[N-1]+[2][N-2]
\ee

Thus the primary reduced polynomial is
\be
q^{6-6N}\P_r =
[N]^4+6[N]^3[N-1]\cdot(qT)+15[N]^2[N-1]^2\cdot(qT)^2 +
20[N][N-1]^3\cdot(qT)^3+ \nn \\
+\Big(3[N][N-1]\big([N-1][N-2]+[2]\big) + 12[N-1]^4\Big)\cdot(qT)^4
+ 6[N-1]^2\Big([N-1][N-2]+[2]\Big)\cdot(qT)^5 + \nn \\
+[N-1]\Big([N-1][N-2]^2 + [2]^2[N-1]+[2][N-2]\Big)\cdot(qT)^6
\ \ \ \ \ \ \ \
\ee
It is easy to check that this quantization rule indeed
properly reproduces  HOMFLY polynomial at $T=-1$,
in particular, at  $N=2$ we obtain the right expression for Jones polynomial:
\be
q^{14}-q^{12}+3q^{10}-q^8+3q^6-2q^4+q^2
\ee

Evaluation of KR polynomial is now straightforward.

\bigskip

\underline{Level 4.} Substitute all $[N]$ by $q[N-1]+q^{1-N}$
and $T$-reduce whatever possible in all the coefficients:
\be
q^4[N-1]^4\Big(1+6T+15T^2+20T^3+12T^4\Big)
+ q^{4-N}[N-1]^3\Big(4+18T+30T^2+20T^3\Big) + \nn \\
+q^{4-2N}[N-1]^2\Big(6+18T+15T^2\Big)
+q^{4-3N}[N-1]\Big(4+6T\Big)
+\boxed{q^{4-4N}} + \nn \\
+ q^5T^4[N-1]^3[N-2]\Big(3+6T\Big) + 3q^{5-N}T^4[N-1]^2[N-2]
+ q^6T^6[N-1]^2[N-2]^2 + \nn\\
+q^5T^4[N-1]^2(3[2]+6[2]T+q[2]^2T^2)
+q^6T^6[2][N-1][N-2]
+ 3q^{5-N}T^4[2][N-1] \sim \nn \\ \nn \\
\sim 2q^4T^4[N-1]^4 + 3q^5T^5[N-1]^3[N-2]+q^6T^6[N-1]^2[N-2]^2 + \nn \\
+ 4q^{4-N}T^3[N-1]^3+ 3q^{5-N}T^4[N-1]^2[N-2]+ \nn \\
 +3q^{4-2N}T^2[N-1]^2
+q^5T^5[2]\Big(3+q[2]T\Big)[N-1]^2
+q^6T^6[2][N-1][N-2]+\nn \\
+2q^{4-3N}T[N-1] + 3q^{5-N}T^4[2][N-1] + \boxed{q^{4-4N}}
\ee
In the first line substitute
\be
2[N-1]^2+3qT[N-1][N-2]+q^2T^2[N-2]^2 \sim q^{2-N}(2[N-1]+qT[N-2])
\ee
and add the result to the content of the second line:
\be
q^4T^4[N-1]^2 \cdot q^{2-N}(2[N-1]+qT[N-2])
+ 4q^{4-N}T^3[N-1]^3+ 3q^{5-N}T^4[N-1]^2[N-2]=\nn \\
=  q^{4-N}T^3[N-1]^2\Big(4[N-1]+2q^2T[N-1]+ 3qT[N-2]+q^3T^2[N-2]\Big)
\label{lev3}
\ee
Thus we eliminated everything at the level 4.

\bigskip

\underline{Level 3:}

Now, in (\ref{lev3}) substitute $[N-1]$ by $q[N-2]+q^{2-N}$
and again eliminate cohomologically trivial combinations:
\be
q^{4-N}T^3[N-1]^2\Big(q[N-2](4+3T) + q^3T[N-2](2+T) + 4q^{2-N}+2q^{4-N}T\Big) \sim\nn \\
\sim q^{4-N}T^3[N-1]^2\Big(2q^{2-N}+(q[N-2]+2q^{2-N})(1+q^2T)\Big) \sim\nn \\
\sim 2q^{6-2N}T^3[N-1]^2 +q^{4-N}T^3[N-1](q[N-2]+2q^{2-N})(q^{2-N}+q^NT)
\ee
where at the last step we substituted $(1+q^2T)[N-1] \sim q^{2-N}+q^NT$.
The result belongs to levels 2 and 1: the primary polynomial is currently reduced to
\be
2q^{6-2N}T^3[N-1]^2 +q^{4-N}T^3[N-1](q[N-2]+2q^{2-N})(q^{2-N}+q^NT) + \nn \\
+3q^{4-2N}T^2[N-1]^2
+q^5T^5[2]\Big(3+q[2]T\Big)[N-1]^2
+q^6T^6[2][N-1][N-2]+\nn \\
+2q^{4-3N}T[N-1] + 3q^{5-N}T^4[2][N-1] + \boxed{q^{4-4N}} = \nn  \\ \nn \\
= [N-1]^2\Big(3q^{4-2N}T^2 +2q^{6-2N}T^3 +3q^5T^5[2] + q^6[2]^2T^6\Big)   +\nn\\
+ [N-1][N-2]\Big(q^{7-2N}T^3+q^5T^4+[2]q^6T^6\Big)+\nn \\
+[N-1]\Big(2q^{4-3N}T  +2q^{8-3N}T^3+2q^{6-N}T^4+ 3q^{5-N}T^4[2]\Big)
+ \boxed{q^{4-4N}}
\label{lev21}
\ee

\bigskip

\underline{Level 2:}

In the first line we substitute one of the factors $[N-1]$ by $q[N-2]+q^{2-N}$,
and obtain:
\be
\boxed{q^{4-4N}} \ + \
[N-1][N-2]\Big(3q^{5-2N}T^2(1+q^2T)+q^5T^4\underline{(1 +3q[2]T+q[2]T^2 + q^2[2]^2T^2)}\Big) + \nn \\
+[N-1]\Big(2q^{4-3N}T  +3q^{6-3N}T^2+4q^{8-3N}T^3+ 3q^{5-N}T^4[2]
+2q^{6-N}T^4+3q^{7-N}[2]T^5+q^{8-N}[2]^2T^6\Big)
\ee
In the underlined bracket we have:
\be
1 +3q[2]T+q[2]T^2 + q^2[2]^2T^2 =(1+T)(1+2T)+3q^2T(1+T)+q^4T^2 \sim \boxed{q^4T^2}
\ee
The remaining piece of the first line reduces to a contribution to the second line:
\be
[N-1][N-2]\cdot 3q^{5-2N}T^2(1+q^2T)\sim  3q^{5-2N}T^2\Big(q^{3-N}+q^{N-1}T)[N-1]
\ee
Thus at the first level we obtain
\be
q^{4-3N}T[N-1]\Big(2+3q^2T+3q^4T+4q^4T^2\Big)
+q^{4-N}T^3[N-1]\Big(3+(3q[2]+2q^2)T+3q^3[2]T^2+q^4[2]^2T^3\Big)
\ee

\bigskip

\underline{Level 1:}

In the first bracket we have:
\be
2+3q^2T+3q^4T+4q^4T^2\sim 2+q^2T+q^4T^2=(1+q^2T)(2+q^2T)
\ee
and in the second one --
\be
3+(3q[2]+2q^2)T+3q^3[2]T^2+q^4[2]^2T^3 = \ \ \ \ \ \ \ \ \ \ \ \ \ \ \nn \\
=3(1+T)+q^2T\Big(\boxed{3}+(2+T)(1+T)\Big) + 2q^4T^2(1+T)+q^4T^2(1+q^2T)
\sim \boxed{3q^2T} + q^4T^2(1+q^2T)
\ee
so that the full contribution from the first level to reduced KR polynomial is
\be
3q^{6-N}T^4[N-1] + \Big(q^{4-3N}T(2+q^2T)+q^{8-N}T^5\Big)\Big(q^{2-N}+q^NT\Big)
\ee
and, adding the two boxed items, obtained at the previous steps,
we finally obtain:
\be
{\cal P}^{6^3_1(v1)}_r = q^{6N-6}\Big\{q^{4-4N} +
\ \ \ \ \ \ \ \ \ \ \ \ \ \ \ \ \ \ \ \ \ \ \ \ \ \
\ \ \ \ \ \ \ \ \ \ \ \ \ \ \ \ \ \ \nn \\
\!\!\!\!\!\!\!\!\!\!\!\!\!\!\!\!\!
+\Big(q^{6-4N}T(2+q^2T)+q^{10-2N}T^5\Big)\underline{\Big(1+q^{2N-2}T\Big)} +
3q^{6-N}T^4[N-1] +q^9T^6[N-1][N-2]
\Big\} =
\label{KR631v1r}
\ee
\vspace{-0.4cm}
$$
\!\!\!\!\!\!\!\!\!\!\!\!\!\!
\boxed{
= q^{2N-2} + 2q^{2N}T+\Big(q^{2N+2}+2q^{4N-2}\Big)T^2+q^{4N}T^3
+ 3q^{5N}T^4[N-1]+q^{4N+4}T^5+q^{6N+2}T^6+q^{6N+3}T^6[N-1][N-2]
}
$$
for reduced KR polynomial.
As usual, to obtain unreduced KR polynomial, we substitute the underlined bracket
by $q^{-1}(1+q^{2N}T)[N-1]$, and multiply all other terms by $[N]$:
\be
{\cal P}^{6^3_1(v1)} = q^{6N-6}\Big\{q^{4-4N}[N] +
\ \ \ \ \ \ \ \ \ \ \ \ \ \ \ \ \ \ \ \ \ \ \ \ \ \
\ \ \ \ \ \ \ \ \ \ \ \ \ \ \ \ \ \ \nn \\
\!\!\!\!\!\!\!\!\!\!\!\!\!\!\!\!\!
+\Big(q^{5-4N}T(2+q^2T)+q^{9-2N}T^5\Big)\underline{\Big(1+q^{2N}T\Big)}[N-1] +
3q^{6-N}T^4[N][N-1] +q^9T^6[N][N-1][N-2]
\Big\} = \nn \\ \nn \\
= q^{2N-2}[N] +
\Big(2q^{2N-1}T+\big(q^{2N+1}+2q^{4N-1}\big)T^2+q^{4N+1}T^3
+q^{4N+3}T^5+q^{6N+3}T^6\Big)[N-1]+\nn \\
+ 3q^{5N}T^4[N][N-1]+q^{6N+3}T^6[N][N-1][N-2]
\label{KR631v1ur}
\ee
As usual, this is in agreement with the results of \cite{CM}.

\subsection{Skein relations for primary polynomials $\P$}

Our main idea in this paper is to construct the primary polynomial
just in the same way as the fundamental HOMFLY, only
substituting ${\cal R}$-matrix and its inverse at black and white
vertices of the link diagram by
\be
\R =q^{N-1}\Big(I+q[2]T\pi\Big) \ \ \ \ {\rm for}
\ \ \ \ \nearrow\!\!\!\!\!\!\bullet\!\!\!\!\!\!\nwarrow
\ee
and
\be
\R^- =q^{-N}T^{-1}\Big([2]\pi+qT\cdot I\Big)\ \ \ \ {\rm for}
\ \ \ \ \nearrow\!\!\!\!\!\!\circ\!\!\!\!\!\!\nwarrow
\ee
In order to simplify formulas we write just $I$ instead of
more adequate $\ I\otimes I$.
Operator $\pi$ here can be just considered as expressed
through the ordinary ${\cal R}$-matrix:
\be
\pi= \frac{\frac{1}{q}\cdot I  \ -\ q^{-N}{\cal R}}{[2]}
\ee
and alternatively write
\be
\R=  - T\cdot{\cal R}  \ +\ q^{N-1}(1+T)\cdot I, \ \ \ \ \
\R^-= -\frac{1}{T}\cdot{\cal R}^{-1} \ + \ q^{1-N}(1+T^{-1})\cdot I
\ee

The fundamental ${\cal R}$-matrix satisfies
\be
q^{-N}{\cal R}-q^N{\cal R}^{-1} = q^{-1}-q
\ee
and this implies skein relations for HOMFLY polynomials.
Just in the same way one can get skein relations for the primary
polynomials $\P$.
Namely, since
\be
\frac{1}{Tq^N}\R - Tq^N\,\R = \left(\frac{1}{qT}-qT\right)\cdot I
\ee
when calculating the primary polynomial $\P$
we can apply the following identity at any vertex of arbitrary link diagram.
\be
\frac{1}{Tq^N} \ \  \nearrow\!\!\!\!\!\!\bullet\!\!\!\!\!\!\nwarrow  \ \ - \
Tq^N \ \ \nearrow\!\!\!\!\!\!\circ\!\!\!\!\!\!\nwarrow   \ \ \ =
\ \left(\frac{1}{qT}-qT\right) \ \ \uparrow \, \uparrow
\label{modskein}
\ee
After that one can proceed, as one usually does with skein relations --
iteratively reducing arbitrary link diagram to a collection of unknots,
each of them finally substituted by $[N]$.

The only thing one should be careful with is that $\R$ and $\R^-$ are no longer inverse of each other, so that
instead of the second one we have
\be
\R\cdot\R^- = I + \frac{(1+T)(1+q^2T)}{qT}\cdot \frac{q^{1-N}\R- q^{N-1}\R^-}{qT-\frac{1}{qT}}
\label{2ndRed}
\ee

\begin{picture}(100,85)(-50,-50)
\put(0,20){\circle*{5}}
\put(0,-20){\circle{5}}
\qbezier(0,-20)(-20,0)(0,20)
\qbezier(0,-20)(20,0)(0,20)
\put(0,20){\vector(1,1){15}}
\put(0,20){\vector(-1,1){15}}
\put(0,-20){\line(-1,-1){15}}
\put(0,-20){\line(1,-1){15}}

\put(30,-2){\mbox{$=$}}

\put(55,-20){\vector(0,1){40}}
\put(65,-20){\vector(0,1){40}}

\put(90,-2){\mbox{$+$}}

\put(120,-2){\mbox{$\frac{(1+T)(1+q^2T)}{qT\left(qT-\frac{1}{qT}\right)}\ \times$}}

\put(200,-2){\mbox{$q^{1-N}$}}
\put(260,-2){\mbox{$-\ \ \  \ \ \ q^{N-1}$}}

\put(240,0){\circle*{5}}
\put(330,0){\circle{5}}
\put(230,-10){\vector(1,1){20}}
\put(250,-10){\vector(-1,1){20}}
\put(320,-10){\vector(1,1){20}}
\put(340,-10){\vector(-1,1){20}}

\qbezier(195,-15)(185,0)(195,15)
\qbezier(350,-15)(360,0)(350,15)
\end{picture}

\noindent
The first Reidemeister relation is now different for $\R$ and $\R^-$:

\begin{picture}(100,170)(-50,-100)
\put(0,20){\circle*{5}}
\put(-30,-10){\vector(1,1){40}}
\put(10,10){\vector(-1,1){40}}
\qbezier(10,30)(40,50)(40,20)
\qbezier(10,10)(40,-10)(40,20)

\put(75,18){\mbox{$=$}}
\put(110,-10){\vector(0,1){60}}
\put(130,18){\mbox{$\times \ \ \ \Big(1\ + \ q^N[N-1](1+T)\Big)$}}

\put(0,-60){\circle{5}}
\put(-30,-90){\vector(1,1){40}}
\put(10,-70){\vector(-1,1){40}}
\qbezier(10,-50)(40,-30)(40,-60)
\qbezier(10,-70)(40,-90)(40,-60)

\put(75,-62){\mbox{$=$}}
\put(110,-90){\vector(0,1){60}}
\put(130,-62){\mbox{$\times \ \ \ \Big(1\ + \ \frac{1}{q^NT}[N-1](1+T)\Big)$}}

\end{picture}

as well as the third one (the Yang-Baxter equation). Say, for positive crossings is deforms as:
\be
\R_1\R_2\R_1-\R_2\R_1\R_2=(Tq^2+1)(\R_1-\R_2)
\nn\ee
We will study these deformed skein relations in the subsequent publications.
\section{Table of KR polynomials}

In this section we collect the answers for KR polynomials.
This is essentially the same table as in \cite{CM},
only we always obtain answers for arbitrary $N$ and
they naturally come in peculiar form,
not always seen in \cite{CM},
thus we can not simply refer to that paper. For links, we present the names according
to \cite{katlas, indiana} (the first column) in addition to Rolfsen names used in \cite{CM}
(the second column). The numbers in figure brackets in the first column label links orientations:
there are $2^{n-1}$ possible ones for an $n$-component link, i.e., 2 ones for a 2-component link
and 4 for a 3-component link. However, for some of these orientations may turn topologically
equivalent; all links that we consider possess at most two inequivalent orientations
labeled as $v1$ and $v2$ in the second column\footnote{Note, that all mutual orientations of Borromean
rings are topologically equivalent -- thus there
is only one orientation, and only one "version"
in terms of \cite{CM}; it was named $v2$ in that
paper.
}. For torus knots and links, we also present the alternative notation $[n,m]$.

The answers are given for KR polynomials for generic $N$,
for small values of $N$ (actually, for $N<m$, where $m$ is the number of
strands in the braid), cohomologies are sometimes further diminished,
in particular {\bf comparison with the results for Khovanov polynomials (at $N=2$)
is not always straightforward}.
The typical examples are torus knots, starting from 3 strands,
see sec.\ref{torus34} for detailed explanation.

\be
\begin{array}{|c|c|c|c|c|}
\hline
\multicolumn{2}{|c|}{}&\multicolumn{3}{c|}{}\\
\multicolumn{2}{|c|}{\mbox{Name}} 
&\multicolumn{3}{|c|}{\mbox{Our\ answer}}\\
\multicolumn{2}{|c|}{}&\multicolumn{3}{c|}{}\\
\hline\hline&&&&\\[-2mm]
L2a1& 2_1^2
&q^{N-1} + q^{2N+1}T^2[N-1]&red&(\ref{KRr22})\\[1mm]
\cline{3-5}&&&&\\[-2mm]
\{0\}\&\{1\}&&q^{N-1}[N] + q^{2N+1}T^2[N][N-1]&unr&(\ref{KRur22})\\[1mm]
\hline\hline&&&&\\[-2mm]
&3_1
& q^{2N-2}\Big(1+q^4T^2+q^{2N+2}T^3\Big)&red&(\ref{KR23})\&(\ref{KR32r})\\[1mm]
\cline{3-5}&&&&\\[-2mm]
&[2,3]&q^{2N-2}[N] +q^{2N+1} T^2(1+q^{2N}T)[N-1]&unr&(\ref{KR[23]ur})\&(\ref{KR32})\\[1mm]
\hline\hline&&&&\\[-2mm]
&4_1
& q^{-2N}T^{-2} + q^{-2}T^{-1} + 1 + q^2T + q^{2N}T^2&red&(\ref{KR41r}) \\[1mm]
\cline{3-5}&&&&\\[-2mm]
&&[N]
+[N-1]\left(q^{-2N-1}T^{-2} + (qT)^{-1} + qT + q^{2N+1}T^2\right)&unr&(\ref{KR41ur})\\[1mm]
\hline\hline&&&&\\[-2mm]
L4a1&4_1^2
&q^{N-1}\Big( 1 + q^2T+ q^{2N}T^2+q^{3N+2}[N-1]T^4\Big)&red&(\ref{KR421v1})\\
\cline{3-5}&&&&\\[-2mm]
\{0\}&(v1)&q^{N-1}\Big\{[N] + \Big(qT+ q^{2N+1}T^2\Big)[N-1]+q^{3N+2}[N][N-1]T^4\Big\}
&unr&(\ref{KR421v1})\\[1mm]
\hline&&&&\\[-2mm]
\{1\}&(v2)& q^{3N-3}\Big(1 + q^4T^2 + q^{2N+2}T^3 + q^{N+6}T^4 [N-1]\Big)&red&(\ref{KRo24})
\&(\ref{KR421})\\[1mm]
\cline{3-5}&&&&\\[-2mm]
&[2,4]&q^{3N-3}\Big\{[N] + \Big(q^3T^2 + q^{2N+3}T^3\Big)[N-1] + q^{N+6}T^4 [N][N-1]\Big\}
&unr&(\ref{KRo24})\&(\ref{KR421v2ur})\\[1mm]
\hline\hline&&&&\\[-2mm]
&5_1
&q^{4(N-1)}\Big( 1 + q^4T^2 +q^{2N+2}T^3 +
q^8T^4+q^{2N+6}T^5\Big)&red&(\ref{KR51})\\[1mm]
\cline{3-5}&&&&\\[-2mm]&[2,5]&q^{4(N-1)}\left\{[N] + \Big(q^3T^2 +
q^7T^4 +q^{2N+3}T^3+q^{2N+7}T^5\Big)[N-1]\right\}&unr&(\ref{KR51})\\[1mm]
\hline\hline&&&&\\[-2mm]
&5_2&
q^{2N-2}\Big( 1 + q^2T + (q^4+q^{2N})T^2 + q^{2N+2}T^3 + q^{2N+4}T^4+q^{4N+2}T^5\Big)
&red&(\ref{KR52})\\[1mm]
\cline{3-5}&&&&\\[-2mm]&&q^{2N-2}[N] +[N-1]\cdot&unr&(\ref{KR52})\\[1mm]&&
\cdot\Big(q^{2N-1}T + (q^{2N+1}+q^{4N-1})T^2
+q^{4N+1}(T^3+T^4) +q^{6N+1}T^5\Big)&&\\[1mm]
\hline\hline&&&&\\[-2mm]
L5a1&5_1^2
&q^{-N-1}T^{-2}+ q^{N-3}T^{-1} +&red&(\ref{KR521r})\\[1mm]&& + \Big(q[N-1]+q^{N-1}\Big) + q^{N+1}T +
\Big(q^{N+3}+q^{3N-1}\Big)T^2 + q^{3N+1}T^3
&& \\[1mm]
\cline{3-5}&&&&\\[-2mm]\{0\}\&\{1\}
&&\Big(q[N-1] + q^{N-1}\Big)[N]+[N-1]\cdot&unr& (\ref{KR521ur})\\[1mm]&&
\cdot\left\{q^{-N-2}T^{-2}+ q^{N-2}T^{-1} + q^NT +
\Big(q^{N+2}+q^{3N}\Big)T^2 + q^{3N+2}T^3\right\}&&\\[1mm]
\hline
\end{array}
\nn\ee
\be
\begin{array}{|c|c|c|c|c|}
\hline
\multicolumn{2}{|c|}{}&\multicolumn{3}{c|}{}\\
\multicolumn{2}{|c|}{\mbox{Name}} 
&\multicolumn{3}{c|}{\mbox{Our\ answer}}\\
\multicolumn{2}{|c|}{}&\multicolumn{3}{c|}{}\\
\hline\hline&&&&\\[-2mm]
&6_1&  q^{-2N}T^{-2} + q^{-2}T^{-1} + 2
+(q^2+q^{2N-2})T+q^{2N}T^2+q^{2N+2}T^3+q^{4N}T^4&red&(\ref{KR61r})\\[1mm]
\cline{3-5}&&&&\\[-2mm]
&&[N] + [N-1]\Big(q^{-2N-1}T^{-2}+(qT)^{-1}+q^{-1}+q^{2N-1}T+&unr&(\ref{KR61ur})\\[1mm]
&&+qT+q^{2N+1}T^2+q^{2N+1}T^3+q^{4N+1}T^4
\Big)&&\\[1mm]
\hline\hline&&&&\\[-2mm]
&6_2& q^{-2}T^{-2}+q^{2N-4}T^{-1} + (q^2+q^{2N-2}) + 2q^{2N}T +
(q^{4N-2}+q^{2N+2})T^2
+&red&(\ref{KR62r})\\[1mm]&& + (q^{4N} + q^{2N+4})T^3 +q^{4N+2}T^4&&\\[1mm]
\cline{3-5}&&&&\\[-2mm]
&&q^{N-1}+[N-1]\Big(q^{-3}T^{-2}+q^{2N-3}T^{-1} + (q^{2N-1}+q)
+ q^{2N}[2] T +&unr&(\ref{KR62ur})\\[1mm]&& +  (q^{4N-1}+ {q^{2N+1}})T^2 +(q^{4N+1}
+ q^{2N+3})T^3+ q^{4N+3}T^4\Big)&&\\[1mm]
\hline\hline&&&&\\[-2mm]
&6_3& q^{-2N-2}T^{-3} + 3 + q^{2N+2}T^3\ +&red&(\ref{KR63r})\\[1mm]&& + \
(q^{N-2}+q^{2-N})\left(q^{-N-2}T^{-2} +q^{-N}T^{-1}+q^NT  + q^{N+2}T^2\right)&&\\[1mm]
\cline{3-5}&&&&\\[-2mm]
&&[N]\ + \ [N-1]\Big\{q^{-2N-3}T^{-3} + [2] + q^{2N+3}T^3 +&unr&(\ref{KR63ur})\\[1mm]&&+
(q^{N-1}+q^{1-N})\Big(q^{-N-2}T^{-2} +q^{-N}T^{-1}+q^NT  + q^{N+2}T^2\Big)\Big\}&&
\\[1mm]
\hline\hline&&&&\\[-2mm]
L6a3& 6_1^2&q^{5(N-1)}\Big( 1 + q^4T^2 +q^{2N+2}T^3 +
q^8T^4+q^{2N+6}T^5 + q^{N+10}T^{6}[N-1]\Big)&red&(\ref{KR[26]})\\[1mm]
\cline{3-5}&&&&\\[-2mm]
\{0\}&(v1)&q^{5(N-1)}\Big\{ [N] + \Big(q^3T^2 +q^{2N+3}T^3 +
q^7T^4+q^{2N+7}T^5   \Big)[N-1]+&unr&(\ref{KR[26]})\\[1mm]&[2,6]&+q^{N+10}T^{6}[N][N-1]\Big\}&&\\[1mm]
\hline&&&&\\[-2mm]
\{1\}&(v2)
& q^{N-1}\Big(1+q^2T+q^{2N}T^2+q^{2N+2}T^3+q^{4N}T^4+q^{5N+2}T^6[N-1]\Big)&red&(\ref{KR621v2r})\\[1mm]
\cline{3-5}&&&&\\[-2mm]
&&q^{N-1}\Big\{[N]+\Big(qT+q^{2N+1}T^2+q^{2N+1}T^3+q^{4N+1}T^4\Big)[N-1]+&unr&(\ref{KR621v2ur})
\\[1mm]&&+q^{5N+2}T^6[N][N-1]\Big\}&&\\[1mm]
\hline\hline&&&&\\[-2mm]
L6a5&6_1^3&
q^{6N-6}\Big\{q^{4-4N}
+\Big(q^{6-4N}T(2+q^2T)+q^{10-2N}T^5\Big)\Big(1+q^{2N-2}T\Big) +&red&(\ref{KR631v1r})\\[1mm]&&+
3q^{6-N}T^4[N-1] +q^9T^6[N-1][N-2]
\Big\}&&\\[1mm]
\cline{3-5}&&&&\\[-2mm]
\{0,0\}&(v1)&q^{2N-2}[N]+[N-1]\cdot&unr&(\ref{KR631v1ur})\\[1mm]&&
\cdot\Big(2q^{2N-1}T+\big(q^{2N+1}+2q^{4N-1}\big)T^2+q^{4N+1}T^3
+q^{4N+3}T^5+q^{6N+3}T^6\Big)+&&\\[1mm]&&
+ 3q^{5N}T^4[N][N-1]+q^{6N+3}T^6[N][N-1][N-2]&&\\[1mm]
\hline&&&&\\[-2mm]
\{0,1\}\&\{1,0\} &(v2)& (qT)^{-2} + q^{2N-4}T^{-1} + \Big(q^{2N-2} + 2q^N[N-1]\Big)
+&red&(\ref{KR631r})\\[1mm]\&\{1,1\}&&
+q^{2N}T +\Big(q^{4N-2} + 2q^{2N+2} + q^{2N+1}[N-1][N-2]\Big)T^2+&&\\[1mm]&&
+ 2q^{4N}T^3 + q^{3N+4}[N-1]T^4&&\\[1mm]
\cline{3-5}&&&&\\[-2mm]
&&q^{2N-2}[N] +\Big(2q^N+q^{3N+4}T^4\Big)[N][N-1] +q^{2N+1}[N][N-1][N-2]T^2+&unr&(\ref{KR631ur})\\[1mm]&&
+[N-1]\cdot\Big\{q^{-3}T^{-2} + q^{2N-3}T^{-1}
+q^{2N-1}T+&&\\[1mm]&&
+\Big(q^{4N-1} + 2q^{2N+1}\Big)T^2
+ 2q^{4N+1}T^3 \Big\}&&\\[1mm]
\hline
\end{array}
\nn\ee
\be
\begin{array}{|c|c|c|c|c|}
\hline
\multicolumn{2}{|c|}{}&\multicolumn{3}{c|}{}\\
\multicolumn{2}{|c|}{\mbox{Name}} 
&\multicolumn{3}{c|}{\mbox{Our\ answer}}\\
\multicolumn{2}{|c|}{}&\multicolumn{3}{c|}{}\\
\hline\hline&&&&\\[-2mm]
L6a2&6_2^2&
q^{3N-3}\Big\{1+ q^2T + (q^{2N}+q^4)T^2 + (q^{2N+2}+q^6)T^3 +2q^{2N+4}T^4
+&red&(\ref{KR622r})\\[1mm]&&+ q^{4N+2}T^5 + q^{3N+6}[N-1]T^6\Big\}&&\\[1mm]
\cline{3-5}&&&&\\[-2mm]
\{0\}\&\{1\}&&q^{3N-3}\Big\{[N]+ [N-1]\Big(qT + (q^{2N+1}+q^3)T^2 + (q^{2N+3}+q^5)T^3
+&unr&(\ref{KR622ur})\\[1mm]&&
+q^{2N+4}[2]T^4 + q^{4N+3}T^5\Big) + q^{3N+6}[N][N-1]T^6\Big\}&&\\[1mm]
\hline\hline&&&&\\[-2mm]
L6a4&6_2^3&
q^{-2N-2}T^{-3} + (q^{-4}+2q^{-2N})T^{-2} +
2q^{-2}T^{-1}+\Big([N-1]+q^{N-2}\Big)\cdot &red&(\ref{KRborr_r})\\[1mm]&&
\cdot\Big([N-1]+q^{2-N}\Big)
+ 2q^2T+ (2q^{2N}+q^4)T^2  + q^{2N+2}T^3&&\\[1mm]
\cline{3-5}&&&&\\[-2mm]
\{0,0\}\&\{0,1\}\&&&\left(2q^{-1}T^{-1}+q^{-3}T^{-2}\right)\left(1+q^{-2N}T^{-1}\right)[N-1]
+ \Big([N-1]+q^{N-2}\Big)\cdot&unr&(\ref{KRborr_ur})\\[1mm]\{1,0\}\&\{1,1\}\phantom{\&}
&&\cdot\Big([N-1]+q^{2-N}\Big)[N]
+ \Big(2qT+q^3T^2\Big)\Big(1+q^{2N}T\Big)[N-1]&&\\[1mm]
\hline\hline&&&&\\[-2mm]
L6a1&6_3^2
&[N-1]q^{-4N-1}T^{-4}+ q^{-3N-1}T^{-3} +
\left(2q^{-3N+1}+q^{-N-3}\right)T^{-2}+&red&(\ref{KR623v1r1})\\[1mm]&&+
2q^{-N-1}T^{-1} +2q^{1-N}+\Big(q^{3-N}+q^{N-1}\Big)T +q^{N+1}T^2&&\\[1mm]
\cline{3-5}&&&&\\[-2mm]
\{0\}&(v1)&[N][N-1]q^{-4N-1}T^{-4} \ +\ q^{1-N}[N] \ +[N-1]\cdot&unr&(\ref{KR623v1ur1})\\[1mm]&&
\cdot\left(q^{-3N-2}T^{-3} + 2q^{-3N}T^{-2}+q^{-N} +q^{2-N}T\right)(1+q^{2N}T)&&\\[1mm]
\hline\hline&&&&\\[-2mm]
\{1\}&(v2) &  q^{3N-3}\Big\{1+q^2T + (q^{2N}+2q^4)T^2
+ 2q^{2N+2} T^3 +q^{4+2N}T^4+&red&(\ref{KR623v2r})\\[1mm]&&+ (q^{4N+2}+q^{2N+6})T^5
+ q^{4N+4}T^6\Big\}+q^{4N+3}[N-1]T^4&&\\[1mm]
\cline{3-5}&&&&\\[-2mm]&&q^{3N-3}[N]+q^{3N-4}\Big\{q^2T + (q^{2N+2}+2q^4)T^2
+ 2q^{2N+4} T^3 +q^{2N+4}T^4+&unr&(\ref{KR623v2ur})\\[1mm]&&+ (q^{4N+4}+q^{2N+6})T^5
+ q^{4N+6}T^6\Big\}[N-1]+
q^{4N+3}[N][N-1]T^4&&\\[1mm]
\hline\hline&&&&\\[-2mm]
L6n1&6_3^3& 1 + q^{N-1}[2][N-1] + q^2T + \Big(q^{2N}+q^{2N+1}[N-1][N-2]\Big)T^2 +&red&
(\ref{KR633v2})\\[1mm]&& \phantom{[N-1][N-2]}+ q^{3N+2}[N-1]T^4&&\\[1mm]
\cline{3-5}&&&&\\[-2mm]
\{0,0\}&(v2)& [N] + [N-1]\Big\{q^{N-1}[2][N] + qT +&unr&(\ref{KR633v2})\\[1mm]&&+
q^{2N+1}\Big(1+[N][N-2]\Big)T^2 + q^{3N+2}[N]T^4\Big\}&&\\[1mm]
\hline\hline&&&&\\[-2mm]
\{1,0\}\&\{0,1\}&(v1)
& q^{4N-4}\Big\{1 +
q^4T^2+q^{2N+2}T^3 + \Big(q^{N+4}+2q^{N+6}\Big)[N-1]\,T^4+&red&(\ref{KR33true})\&\\[1mm]
\&\{1,1\}&&+q^{2N+7}[N-1][N-2]\,T^6\Big\}&&(\ref{KR33true1})\phantom{\&}\\[1mm]
\cline{3-5}&&&&\\[-2mm]&[3,3]&q^{4N-4}\Big\{[N] + \Big(q^3\,T^2 + q^{2N+3}\,T^3\Big)[N-1]
+&unr&(\ref{KR33true})\&\\[1mm]&& \Big(q^{N+4}+2q^{N+6}\Big)[N][N-1]\,T^4
+ q^{2N+7}[N][N-1][N-2]\,T^6\Big\}&&(\ref{KR33true1})\phantom{\&}\\[1mm]
\hline
\end{array}
\nn\ee

\be
\begin{array}{|c|c|c|c|c|}
\hline
\multicolumn{2}{|c|}{}&\multicolumn{3}{c|}{}\\
\multicolumn{2}{|c|}{\mbox{Name}} 
&\multicolumn{3}{c|}{\mbox{Our\ answer}}\\
\multicolumn{2}{|c}{}&\multicolumn{3}{|c|}{}\\
\hline\hline&&&&\\[-2mm]
9_{42}&&q^{-2N-2}T^{-4} + q^{-4}T^{-3} +
q^{2-2N}T^{-2}+2T^{-1}
+(1+q^{2N-2}) +q^4T + q^{2N+2}T^2&red&(\ref{KR942r})\\[1mm]
\cline{3-5}&&&&\\[-2mm]
&&q^{1-N} +[N-1]\cdot\Big(q^{-2N-3}T^{-4} + q^{-3}T^{-3} +&unr^*&(\ref{KR942ur}) \\[1mm]&&+
q^{1-2N}T^{-2}+[2]T^{-1}
+(q+q^{2N-1}) +q^3T + q^{2N+3}T^2\Big)&&\\[1mm]
\hline\hline&&&&\\[-2mm]
8_{19}&[3,4]&q^{6N-6}\Big\{ 1 + q^4T^2 +q^{2N+2}T^3 + [2]q^7T^4 + [2]q^{2N+5}T^5+&red&
(\ref{KRT35})\& \\[1mm]&&
+ q^{12}T^6 + [2]q^{2N+9}T^7 + q^{4N+6}T^8\Big\}&&(\ref{KRT34})\phantom{\&}\\[1mm]
\cline{3-5}&&&&\\[-2mm]
&&q^{6N-6}\Big\{ [N] + \Big(q^3T^2 +q^6[2]T^4 +q^{11}T^6+q^{2N+7}T^7 \Big)
(1+q^{2N}T)[N-1]\Big\}&unr^*&(\ref{KRT34ur}) \\[1mm]&&
&&\\[1mm]
\hline\hline&&&&\\[-2mm]
10_{124}&[3,5]&q^{8N-8}\Big\{ 1 + q^4T^2\Big(1+q^3[2]T^2+q^6T^4\Big)(1+q^{2N-2}T)
+ &red&(\ref{KRT35ra})\&\\[1mm]&&+ q^{12}T^6\Big(1+q^4T^2\Big) (1+q^{2N-2}T)(1+q^{2N-4}T)\Big\}
&&(\ref{KRT35r})\phantom{\&}\\[1mm]
\cline{3-5}&&&&\\[-2mm]
&&q^{8N-8}\Big\{[N]
+\Big(q^3T^2(1 +q^3[2]T^2
+q^{6}T^4)+&unr^*&(\ref{KRT35ura})\&\\[1mm]&&
+ q^{11}T^6(1+q^4T^2)(1+q^{2N-4}T)\Big)(1+q^{2N}T)[N-1]\Big\}&&(\ref{KRT35ur})\phantom{\&}\\[1mm]
\hline\hline&&&&\\[-2mm]
&[3,6]&q^{10N-10}\Big\{ 1 + q^4T^2 + q^{2N}T^3 + [2]q^7T^4
+ [2]q^{2N+5}T^5 +&red&(\ref{KR36r})\&\\[1mm]&&+ [2]q^{11}T^6
+ [3]q^{2N+10}T^7 + (q^{4N+6}+q^{16})T^8 +[2]q^{2N+13}T^9 +&&(\ref{KR36ra})\phantom{\&}\\[1mm]
&&+ q^{4N+10}T^{10}
+ (2q^2+1)\Big(q^{N+10}T^8 + q^{N+16}T^{10}+q^{3N+12}T^{11}\Big)[N-1] +&&\\[1mm]&&+
q^{2N+19}T^{12}[N-1][N-2]\Big\}&&\\[1mm]
\cline{3-5}&&&&\\[-2mm]
&&q^{10N-10}\Big\{ [N] +[N-1]\cdot &unr^*&(\ref{KR36})\&\\[1mm]&&\cdot\Big(q^3T^2 + q^{2N+3}T^3
 + [2]q^{6}T^4
+ [2]q^{2N+6}T^5 + [2]q^{10}T^6
+ [3]q^{2N+9} T^7+ &&(\ref{KR36ura})\phantom{\&}\\[1mm]&&
+ (q^{4N+7}+q^{15})T^8
+(q^{11}+q^{15})q^{2N}T^9+ q^{4N+11}T^{10}
\Big) +&&\\[1mm]&&
+ (2q^2+1)\Big(q^{N+10}T^8 + q^{N+16}T^{10}+q^{3N+12}T^{11}\Big)[N][N-1] +&&\\[1mm]&&+
q^{2N+19}T^{12}[N][N-1][N-2]\Big\}&&\\[1mm]
\hline
\end{array}
\nn\ee

\be
\begin{array}{|c|c|c|c|}
\hline
&\multicolumn{3}{c|}{}\\
\mbox{Torus}&\multicolumn{3}{|c|}{\mbox{Our answer}}\\
&\multicolumn{3}{|c|}{}\\[1mm]
\hline\hline&&&\\[-2mm]
[2,2k+1]&q^{2k(N-1)}\left(1 + \Big(1+q^{2N-2}T\Big)\sum_{j'=1}^{k} (q^2T)^{2j'}\right)&
red&(\ref{KR2strknotr})\\
\cline{2-4}&&&\\[-2mm]
&q^{2k(N-1)}\left([N] + \frac{1}{q}\sum_{j'=1}^{k} (q^2T)^{2j'}\Big(1+q^{2N}T\Big)[N-1]\right)&
unr^*&(\ref{KR2strknotur})\\[1mm]
\hline\hline&&&\\[-2mm]
[2,2k]&q^{(2k-1)(N-1)}\left(1
+\Big(1+q^{2N-2}T\Big)\sum_{j' = 1}^{k-1}
 (q^{2}T)^{2j'} + (q^2T)^{2k}q^{N-2}[N-1]\right)&red&(\ref{KR2strlinkr})\\[1mm]
\cline{2-4}&&&\\[-2mm]
&q^{(2k-1)(N-1)}\left([N]
+\frac{1}{q}\sum_{j' = 1}^{k-1} (q^{2}T)^{2j'} \Big(1+q^{2N}T\Big)[N-1]
+ (q^2T)^{2k}q^{N-2}[N][N-1]\right)&unr^*&(\ref{KR2strknotur})\\[1mm]
\hline
\end{array}
\nn\ee

Relation to superpolynomials  is via the {\it rule} (\ref{KRtosup0}).
Remarkably, application of this rule
converts somewhat different reduced and unreduced KR polynomials
into superpolynomials, which differ just by a factor of $[N]$ --
exactly as it happens for HOMFLY.
Since the rule (\ref{KRtosup0}) is easy to apply
in both directions, this fact actually removes the need
to evaluate unreduced KR polynomials ones
reduced are known (and vice versa).

\section{On properties of morphisms}

\setcounter{equation}{0}

In this section we briefly summarize what we implicitly learnt about the structure
of cut-and-join morphisms, acting along the edges of hypercube.

The main new step in the present paper -- as compared to \cite{DM3} --
is explicit construction of the graded vector spaces, standing  at the
hypercube vertices -- we call them $v$.
The spaces of KR complexes are deducible combinations  of these spaces
(depending on the coloring of the link diagram, i.e. on the initial
vertex of the hypercube).
As graded spaces they can be represented as polynomials in $q$,
which we further represented as columns of our tables.
Differential of KR complex is after that a linear mapping between
adjacent columns, which decreases the power of $q$ by one.
Differentials are actually made from morphisms between the particular
constituent spaces $v$ -- and if the morphisms are known, the differentials
are known as well.
The spaces $v$ in unreduced and reduced cases differ by a common factor
of $[N]$, but the difference between morphisms is a little more involved.
Differentials are almost one-to-one maps, with some exceptions:
elements in the columns which are either mapped to zero or are not the images
of the one-to-one maps, belong to cohomologies of the differential.
Cohomologies form the KR polynomial.

Coming one step back. in this paper we provided an explicit construction for
columns, but differentials -- and thus morphisms -- are described only at
mnemonic level.
However, we confirmed a number of properties of the morphisms,
advocated in \cite{DM3}.

\bigskip

1) Differentials and thus morphisms have a block form: they preserve decomposition
of the spaces $v$ into a $m$-linear ($m$ is the number of strands) combination
of factors $[N-p]$  with $p=1,\ldots,m-1$ and $q^{-N}$ -- and entire $N$-dependence
is concentrated in these factors.
The fact that $p\neq 1$ reflects the slight deviation of morphisms from just
elementary shift operations $\alpha_i$ and $\beta_i$: the small tails of the
spaces $[N-1]$ are rearranged in a somewhat more complicated way --
as was already depicted in \cite{DM3}.

This property allows one to make our tables $N$-independent: polynomials in columns
are further decomposed into poly-linear combinations of above factors
and coefficients are smaller polynomials in $q$, which do not depend on $N$.
At least technically, this property provides a distinguished role to the
alternated sums in lines: to constituents of peculiar decomposition of
Euler characteristic of the KR complex, i.e. of the HOMFLY polynomial.
This decomposition looks like an analogue of decomposition in characters
\cite{DMMSS,MMMkn1}, which captures the $A$-dependence of HOMFLY and superpolynomials.
It is interesting that the decomposition basis for KR polynomials is
related, but different -- and this effectively eliminates the mysterious
$\gamma$-factors \cite{DMMSS}, appearing in MacDonald decomposition of superpolynomials.
Extremely interesting is also an obvious analogy with differential
decomposition of \cite{diff} -- if there is any, this could open a route
for generalizations to colored polynomials.

\bigskip

2) Cohomologies possess a clear nested structure: in building them one can
hierarchically eliminate the factors $[N-p]$ one after another
(there are $m$ such factors for unreduced and $m-1$ for reduced polynomials,
also for $l$-component links some "minor" contributions begin to survive $l$ steps
before the end of reduction).

Technically this property was reflected in our chain of reductions of the tables --
from initial one, dictated by decomposition into $v$-spaces and describing the
primary polynomial $\P^{\cal L}$, to a final one,
describing KR polynomial ${\cal P}^{\cal L}$.
This property confirms the suggestion of \cite{DM3} that morphisms are naturally
decomposed into shifts in various $m$ ($m-1$) directions.
Another outcome is that the difference between unreduced and reduced polynomials
occurs only at the very last reduction -- and this explains both the difference
between the two and their otherwise-unexpected similarity.

\bigskip

To these two basic observations we add a number of "smaller" ones,
which can still be equally important.

\bigskip

3)
Due to projector property $P^2=P$
the variety of spaces $v$ is in fact not as big as the number of different
link diagrams.
It is in fact nicely ordered by the choice of the strand number $m$.
For example, {\it all} the 3-strand diagrams are described by just the
2-parameter set  $v_{n,k}$.

\bigskip

4) At each step of the reduction hierarchy cohomologies are
concentrated "near" one-dimensional diagonals of two-dimensional tables
so that the cohomologies are made predominantly from some "standard blocks",
like "increasing sequences" and twin pairs with the same powers of $q$
in adjacent lines $[N-k]$ and $[N-k-1]$.

This fact seems closely related to the well-known fact, that
superpolynomials are "almost" obtained from HOMFLY by the change
of variables, like $A^2\longrightarrow A^2T$.

\bigskip

5) In most cases, the cohomologies at each reduction step {\it nearly}
saturate the alternated sums in lines,but some time extra
"cohomologically trivial" pairs should be added to guarantee the
formation of proper twin pairs in adjacent lines.

\bigskip

6) There is usually a non-trivial cohomology in the third column
(numeration starts from zero) -- for all-black initial vertex
this would be order $T^3$.

This rule should be easy to justify from analysis of morphisms:
in the third column they still do not deviate too much from
the naive shifts $\alpha_i$ and $\beta_i$ and thus can probably be
analyzed in enough generality.
At the same time, this property is crucial for explaining the
breakdown of naive minimality of KR polynomials:
in many cases this term has a counterpart with another power of $T$
and could naively be eliminated -- but this is forbidden by the
actual structure of the morphisms and differentials.

\bigskip

These mnemonic rules, though not fully formalized, provide a very
strong support to existence of the KR calculus on the simple lines
of \cite{DM3}.
The next three steps: full classification of the spaces $v$,
explicit construction of morphisms between their combinations
at adjacent numbers of white vertices (there should be morphisms between
$v_n$ and a direct sum of $n$ spaces $v_{n-1}$),
and the proof of Reidemeister invariance,
should now be  within reach.


\section{Conclusion}

\setcounter{equation}{0}

By definition, Khovanov polynomials is the Poincare polynomial of a complex,
associated with Abelian quiver, which is made from a hypercube of
resolutions of the given link/knot diagram.
It is a natural $T$-deformation of HOMFLY polynomial, because the latter
turns to be the Euler characteristic of the same complex, which
arises from Poincare polynomial at $T=-1$.
The hypercube appears in the story whenever the ${\cal R}$ matrix is
split in two parts -- thus the choice of the splitting is the starting point
of entire construction.

Original Khovanov construction at $N=2$ uses the splitting associated with
geometric resolution, and vector spaces over the vertices of the hypercube
are naturally associated with the cycles of resolved link diagram,
while morphisms are provided by cut and join operators (see \cite{Kho}
for original presentation and \cite{BN,DM1} for detailed reviews).
Khovanov-Rozansky (KR) construction \cite{KhR} for arbitrary $N$ substitutes
vector spaces by certain cyclic complexes, inspired by matrix-factorization
theory, which is rather sophisticated technically
(beyond the capacity of modern computers already in rather simple examples) --
and the whole construction, while extremely elegant,
becomes technically unfeasible.
This makes the theory of superpolynomials (a further universalization
of KR polynomials) more an artistic guesswork than a solidly based science.

In \cite{DM3} it was suggested to deduce KR polynomials from a literal
generalization of Khovanov construction, returning to the use
of vector spaces, associated with cycles of resolved diagrams,
without any reference to matrix factorization.
Already in \cite{DM3} it was demonstrated that this is practical approach,
and in this paper we provided a lot more evidence to support this claim.
Moreover, here we made explicit the underlying splitting of the ${\cal R}$-matrix
and reduce the story to the study of associated "primary" $T$-deformation $\P$
of HOMFLY. This stage is calculatingly no more complicated than evaluation
of ordinary HOMFLY.
Afterwards $\P$ should be substituted by its minimal positive residue
${\cal P}$ w.r.t. division over $T+1$ -- and this ${\cal P}$ is the KR polynomial.
The only problem with this approach is certain ambiguity in minimization
procedure, and we formulated some mnemonical rules to fix it, at least partly.
We are still far from providing a rigorous formulation of this construction,
but already at this stage it appears quite powerful technically --
providing a list of previously unfeasible examples,
including series like 2-, 3- and 4-strand torus knots.
All  examples from the maximal existing regular list \cite{CM} are
also reproduced.

At the same time our current presentation sheds certain light on
the problem of colored KR polynomials.
Fundamental representation is distinguished because the ${\cal R}$-matrix
and its inverse are expressed via a single non-trivial projector
(onto representation $[11]\subset [1]^{\otimes 2}$).
In colored case things are at least not so simple.
It is even possible that a separate deformation parameter can be associated
with each item $Q$ in decomposition of representation product
$R\otimes R= \oplus\ Q$.

\bigskip

To conclude and to prevent misunderstanding, we repeat once again that
{\bf the KR polynomials in this paper are  {\it not fully} derived}
on the lines of \cite{DM3}: most of them are actually chosen from
a finite set of choices.
Full derivation provides accurate description of morphisms,
suggested in \cite{DM3}, and evaluation of their cohomologies --
what requires additional work and is left for the future publications.
In this paper we just showed {\it what} these morphisms should be
in each particular case to provide the right answers -- and it is
made quite clear that they obey strict rules, what provides very
strong evidence that they indeed exist.
Thus this paper is an important step in sharpening and justification
of the claims in \cite{DM3} and, most important, in providing a
working tool for evaluation of arbitrary KR polynomials in the
fundamental representation.

\section*{Acknowledgements}

\section*{Acknowledgements}

We are indebted to I.Danilenko and V.Dolotin for numerous discussions
and to N.Carqueville for  help with \cite{CM}.
We thank the referee of this paper for very valuable requests and comments
and S.Arthamonov for pointing a number of misprints in the original
version of the text.

Our work is partly supported by grants
RFBR grants 12-01-00482, 14-02-31446-mol-a  (A.A.),
13-02-00478 (A.M.),
by NSh -1500.2014.02
and by joint grants
13-02-91371-ST-a, 14-01-92691-Ind-a.

\end{document}